\documentclass[a4paper,fleqn,usenatbib]{mnras}

\usepackage{mathptmx}

\usepackage[T1]{fontenc}
\usepackage{ae,aecompl}

\usepackage{graphicx}	
\usepackage{amsmath}	
\usepackage{amssymb}	
\usepackage{color}

\newcommand{\tE}{\mbox{$t_\mathrm{E}$}}
\renewcommand{\vec}[1]{\pmb{#1}}

\definecolor{lscb}{rgb}{0, 0.75, 1}
\definecolor{lscc}{rgb}{0, 0.4, 0.7}
\definecolor{cpta}{rgb}{1, 0.9, 0}
\definecolor{cptb}{rgb}{1, 0.7, 0}
\definecolor{cptc}{rgb}{0.7, 0.5, 0}
\definecolor{coja}{rgb}{0.8, 0, 0.8}
\definecolor{cojb}{rgb}{0.6, 0, 0.6}
\definecolor{fts}{rgb}{0.2, 0.7, 0.2}
\definecolor{lsccv}{rgb}{1, 0.6, 0.6}
\definecolor{dk154}{rgb}{0, 0, 1.0}

\title[Microlensing planet or close binary source?]{OGLE-2014-BLG-1186: gravitational microlensing providing evidence for a planet orbiting the foreground star or for a close binary source?}

\author[M. Dominik et al.]{M. Dominik,$^{1}$\thanks{E-mail: md35@st-andrews.ac.uk}\thanks{Royal Society University Research Fellow}
E. Bachelet,$^{2,3}$
V. Bozza,$^{4,5}$
R. A. Street,$^{2}$
C. Han,$^{6}$ \newauthor
M. Hundertmark,$^{1,7,8}$  
A. Udalski,$^{9}$
D. M. Bramich,$^{10}$
K. A. Alsubai,$^{3}$ \newauthor
S. Calchi Novati,$^{11}$ 
S. Ciceri,$^{12}$ G. D'Ago,$^{4,13,14}$ R. Figuera Jaimes,$^{1,2,15}$ \newauthor
T. Haugb{\o}lle,$^{7}$ 
T. C. Hinse,$^{16}$  
K. Horne,$^{1}$ U. G. J{\o}rgensen,$^{7}$ D. Juncher,$^{7}$ \newauthor
N. Kains,$^{17}$  H. Korhonen,$^{18}$ 
L. Mancini,$^{19,12,20}$  J. Menzies$^{21}$, A. Popovas,$^{7}$ \newauthor
M. Rabus,$^{12,22}$ 
S. Rahvar,$^{23}$ 
G. Scarpetta,$^{4,13}$ R. Schmidt,$^{8}$  J. Skottfelt,$^{7,24}$  \newauthor
 C. Snodgrass,$^{25,26}$
J. Southworth,$^{27}$ D. Starkey,$^{1}$  I. A. Steele$^{28}$, J. Surdej,$^{29}$ 
\newauthor
Y. Tsapras,$^{2,8,30}$ 
J. Wambsganss,$^{8}$  O. Wertz,$^{29}$
P. Pietrukowicz,$^9$ M. K. Szyma{\'n}ski,$^9$ \newauthor 
P. Mr{\'o}z,$^9$  J. Skowron,$^9$ I. Soszy{\'n}ski,$^9$ K. Ulaczyk,$^{9,31}$
R. Poleski,$^{9,32}$ 
\newauthor
 {\L}. Wyrzykowski,$^9$ and S. Koz{\l}owski$^9$ \\
$^{1}$Centre for Exoplanet Science, SUPA School of Physics \& Astronomy, University of St Andrews, North Haugh, St Andrews, KY16 9SS\\
$^{2}$Las Cumbres Observatory Global Telescope Network, 6740 Cortona Drive, suite 102, Goleta, CA 93117, USA\\
$^{3}$Qatar Environment and Energy Research Institute (QEERI), HBKU, Qatar Foundation, Doha, Qatar\\
$^{4}$Dipartimento di Fisica ``E. R. Caianiello'', Universit\`a di Salerno, Via Giovanni Paolo II 132, 84084 Fisciano (SA), Italy\\
$^{5}$Istituto Nazionale di Fisica Nucleare, Sezione di Napoli, Italy\\
$^{6}$Department of Physics, Chungbuk National University, Cheongju 28644, Republic of Korea\\
$^{7}$Niels Bohr Institute \& Centre for Star and Planet Formation, University of Copenhagen, {\O}ster Voldgade 5, 1350 Copenhagen K, Denmark\\
$^{8}$Astronomisches Rechen-Institut, Zentrum f{\"u}r Astronomie der Universit{\"a}t Heidelberg (ZAH), 69120 Heidelberg, Germany\\
$^{9}$Warsaw University Observatory, Al. Ujazdowskie 4, 00-478 Warszawa, Poland\\
$^{10}$New York University Abu Dhabi, Saadiyat Island, Abu Dhabi, PO Box 129188, United Arab Emirates\\
$^{11}$IPAC, Mail Code 100-22, Caltech, 1200 E. California Blvd., Pasadena, CA 91125, USA\\
$^{12}$Max Planck Institute for Astronomy, K\"onigstuhl 17, 69117 Heidelberg, Germany\\
$^{13}$International Institute for Advanced Scientific Studies (IIASS), Via G. Pellegrino 19, 84019 Vietri sul Mare (SA), Italy\\
$^{14}$INAF - Osservatorio Astronomico di Capodimonte, Salita Moiarello 16, 80131 Napoli, Italy\\
$^{15}$European Southern Observatory, Karl-Schwarzschild-Str. 2, 85748 Garching bei M\"unchen, Germany\\
$^{16}$Korea Astronomy \& Space Science Institute, 776 Daedukdae-ro, Yuseong-gu, 305-348 Daejeon, Republic of Korea\\
$^{17}$Space Telescope Science Institute, 3700 San Martin Drive, Baltimore, MD 21218, USA\\
$^{18}$Dark Cosmology Centre, Niels Bohr Institute, University of Copenhagen, Juliane Maries Vej 30, 2100 Copenhagen, Denmark\\ 
$^{19}$Department of Physics, University of Rome Tor Vergata, Via della
Ricerca Scientifica 1, 00133 Roma, Italy\\
$^{20}$INAF -- Astrophysical Observatory of Turin, Via Osservatorio 20,
10025 Pino Torinese, Italy \\
$^{21}$South African Astronomical Observatory, PO Box 9, Observatory 7935, South Africa\\
$^{22}$Instituto de Astrof\'isica, Pontficia Universidad Cat\'olica de Chile, Av. Vicu{\~n}a Mackenna 4860, 7820436 Macul, Santiago, Chile\\
$^{23}$Department of Physics, Sharif University of Technology, P.O. Box 11155-9161 Tehran, Iran\\
$^{24}$Centre for Electronic Imaging, Department of Physical Sciences, The Open University, Milton Keynes, MK7 6AA \\
$^{25}$Planetary and Space Science, Department of Physical Sciences, The Open University, Milton Keynes, MK7 6AA \\
$^{26}$Max-Planck-Institut f\"ur Sonnensystemforschung, Justus-von-Liebig-Weg 3, 37077 G\"ottingen\\
$^{27}$Astrophysics Group, Keele University, Keele ST5 5BG \\
$^{28}$Astrophysics Research Institute, Liverpool John Moores University, Liverpool CH41 1LD\\
$^{29}$Institut d'Astrophysique et de G\'eophysique, All\'ee du 6 Ao\^ut 19c, Sart Tilman, 4000 Li\`ege, Belgium\\
$^{30}$School of Physics and Astronomy, Queen Mary University of London, Mile End Road, London, E1 4NS\\
$^{31}$Department of Physics, University of Warwick, Gibbet Hill Road, Coventry, CV4 7AL\\
$^{32}$Department of Astronomy, Ohio State University, 140 W. 18th Ave., Columbus, OH 43210, USA}

\date{Accepted XXX. Received YYY; in original form ZZZ}

\pubyear{2016}

\begin{document}
\label{firstpage}
\pagerange{\pageref{firstpage}--\pageref{lastpage}}
\maketitle

\begin{abstract}
Using the gravitational microlensing event OGLE-2014-BLG-1186 as an instructive example, we present a systematic methodology for identifying the nature of localised 
deviations from single-lens point-source light curves, which ensures that
1) the claimed signal is substantially above the noise floor,
2) the inferred properties are robustly determined and their estimation not subject to confusion with systematic noise in the photometry,
3) there are no alternative viable solutions within the model framework that might have been missed. Assessing the photometric noise by means of an effective model significantly increases the sensitivity arising from an analysis of the total microlensing data set to more subtle perturbations, and thereby in particular to low-mass planets. With a time-scale $t_\mathrm{E} \sim 300~\mbox{d}$ and the brightness being significantly above baseline for four years, OGLE-2014-BLG-1186 is particularly long.
Consequently, annual parallax and binarity could be separated and robustly measured  from the wing and the peak data, respectively. While we were able to establish the presence of binarity, we find model light curves matching the features indicated by the acquired data (within the estimated noise) that involve either a binary lens or a binary source. Our binary-lens models indicate a planet of mass $M_2 =  (45 \pm 9)~M_\oplus$, orbiting a star of mass $M_1 =  (0.35 \pm 0.06)~M_{\odot}$, located at a distance $D_\mathrm{L}  =  (1.7 \pm 0.3)~\mbox{kpc}$ from Earth, whereas our binary-source models suggest a brown-dwarf lens of $M  =  (0.046 \pm 0.007)~M_{\odot}$, located at a distance $D_\mathrm{L}  =  (5.7 \pm 0.9)~\mbox{kpc}$, with the source potentially being a (partially) eclipsing binary involving stars predicted to be of similar colour given the ratios between the luminosities and radii. The ambiguity in the interpretation would be resolved in favour of a lens binary by observing the luminous lens star separating from the source at the predicted proper motion of $\mu = (1.6 \pm 0.3)~\mbox{mas}\;\mbox{yr}^{-1}$, whereas it would be resolved in favour of a source binary if the source could be shown to be a (partially) eclipsing binary matching the obtained model parameters. We experienced that close binary source stars pose a challenge for claiming the detection of planets by microlensing in events where the source trajectory passes close to the central caustic near the lens star hosting the planet.
\end{abstract}





\section{Introduction}

The vast majority of claimed microlensing planet detections are based on a pretty obvious signal in the acquired photometric data \citep[e.g.][]{OB03235,OB05071,OB05390,OB07368,Gaudi:doubleplanet,KB09266}.
 This makes one wonder why detections from less obvious signals \citep[e.g.][]{KB07400,KB08310} are scarce, given that more subtle features should be quite common. 
Clearly, if more subtle features are discarded altogether, we lose out on the significance of the planet population statistics arising from the acquired data, and we lose sensitivity particularly to low-mass companions. Moreover, sampling events more densely than necessary can be quite a waste of telescope resources, and strongly diminish the overall detection efficiency of follow-up campaigns \citep[e.g.][]{Horne:metric, Dominik:PLANET, SIGNALMEN, MiNDSTEp,RoboNet-II}. The detection efficiency \citep[e.g\ ][]{GS:deteff,BenRhie:deteff} is a crucial characteristic, with planets probabilistically escaping their detection through microlensing even with perfectly sampled and precise photometric light curves \citep{MP91}, depending on where they happen to be located along their orbit during the course of a microlensing event.

If we assume a photometric time series composed of $N$ data points $(t_i, F_i, \sigma_i)$ with
measured fluxes $F_i$ and estimated uncertainties  $\sigma_i$, as well as a theoretical light curve $F(t_i)$, one finds the sum of the squared standardised residuals as
\begin{equation}
\chi^2 \equiv \sum_{i = 1}^{N} \left(\frac{F_i-F(t_i)}{\sigma_i}\right)^2\,.
\end{equation}

As compared to gravitational microlensing by a single isolated lens star \citep{Ein36,Pac1}, a quasi-static binary lens system (e.g. a star with a single planet) is characterised by an additional three parameters \citep{MP91}. Moreover, 
a planetary signature also usually reveals the angular size of the source star, described by a further parameter. For such a signature, one therefore finds only a small probability ${\cal P}_4(\Delta \chi^2 \geq 20) = 4\times 10^{-4}$ for a difference in $\chi^2$ in excess of 20 for 4 additional degrees of freedom. This means that a likelihood ratio test suggests a clear signal for e.g. as few as 5 data points at the 2-$\sigma$ level, under the provision that the measurement uncertainties are accurately estimated, uncorrelated, and follow a Gaussian profile.

However, in reality it cannot be tacitly assumed that these conditions hold, and we rather need to be careful about false positives lurking in the actual noise of the photometric measurements. Even a high detection threshold does not provide an insurance policy on this because correlated noise (or ``red noise'') can lead to ``pseudo-detections'' at arbitrarily large $\Delta \chi^2$ if just the cadence of the photometric time series is high enough. In fact, in at least one case, the careful analysis of an observed gravitational microlensing event arrived at the conclusion that a putative planetary signal is likely due to red noise \citep{Bachelet:noise}.

A consistent interpretation of data requires to demonstrate that putative signals are not likely to arise from noise, and adequate criteria are required to distinguish signals from the noise floor. It would be obviously inconsistent to claim a detection of a signal from data that show deviations that are similar to what is being considered ``noise'' for other data. It is therefore indicated to establish a suitable ``noise'' model and estimate some ``noise'' statistics.

Blind searches in high-dimensional non-linear parameter spaces bear a substantial risk of confusing true signals in the data with noise. It is rather straightforward to find a good match between noise patterns and models describing small localised deviations, as previous analyses of microlensing events explicitly demonstrated \citep[e.g.][]{OB08510}.

Signals of low-mass planets and satellites may be subtle, but fortunately these are well localised. In other words, the vast majority of photometric data provide no relevant constraint to the model parameters that describe the anomaly. Moreover, all the other parameters can usually be well determined from the data not containing the anomaly. This permits splitting up parameter space into two subspaces with disjoint associated data sets. Looking at the effect of the anomaly region on the anomaly-independent parameters provides a valuable consistency check, while the data not covering the putative anomaly can be used to infer parameters describing noise statistics
that do not depend on any assumptions about the anomaly. It should however be noted that while such an approach works well for weak anomaly features, strong features (e.g.\ due to caustic passages) can be highly sensitive to the track of the source relative to the lens system, thereby substantially affecting a large number of model parameters.

In this article, we discuss the microlensing event OGLE-2014-BLG-1186, which not only is of exceptionally long duration, but also shows a putative anomaly in the form of a close double peak. 
We explicitly demonstrate how this anomaly can be systematically and robustly identified and present viable interpretations of its physical nature.
Gravitational microlensing events that show a photometric light curve involving two peaks can result from either (or both) a lens binary \citep{MP91,GL92,GS98} or a source binary \citep{GriHu92}. 
\citet{Gaudi:source} discussed an ambiguity between planetary binary-lens and binary-source models for putative planetary signatures that arise from the source passing close to one of the `planetary caustics' (see Sect.~\ref{sec:BLspace}), so that the light ray passes close to the planet \citep{Erdl}. In the case of OGLE-2014-BLG-1186, we are however facing a different situation, where the source passes close to the central caustic of the putative binary-lens system, located near the position of the planet's host star.


In Sect.~\ref{sec:data}, we describe our data acquisition and original identification of a putative anomaly over the peak of the light curve,  while Sect.~\ref{sec:model} is devoted to a detailed account of our modelling efforts. We discuss the physical nature of the lens and source objects and the wider significance of our findings in Sect.~\ref{sec:interpretation}. We draw final conclusions in Sect.~\ref{sec:conclusions}.


\section{Data acquisition}
\label{sec:data}

\subsection{Survey and follow-up}
Soon after \citet{MP91} demonstrated that the gravitational microlensing effect could be used to detect extra-solar planets, \citet{GL92} argued that
a combination of survey and follow-up would be an efficient way to do so. With the implementation of the ``Early Warning System" (EWS; \citealt{OGLE:EWS}) by the Optical Gravitational Lensing Experiment (OGLE) team, the real-time detection of microlensing events became public information, enabling a wider scientific community to engage in harvesting the scientific returns of these transient phenomena.

In 2014, the fourth phase of OGLE (OGLE-IV; \citealt{OGLE4}) was in operation, using the 1.3m Warsaw University Telescope at Las Campanas Observatory in Chile and a mosaic camera of 32~E2V44-82 $2048 \times 4102$ CCD chips with $I$- and $V$-band filters, delivering a total field of view of 1.4 square degrees at 0.26\arcsec/pixel.\footnote{{\tt http://ogle.astrouw.edu.pl/main/OGLEIV/mosaic.html}}
The current implementation of the OGLE-IV Early Warning System, using a photometric data pipeline based on Difference Image Analysis (DIA) photometry  \citep{AlardLupton:DIA,Alard:DIA,Wozniak:DIA} assesses about 380 million stars in 85 Galactic bulge fields, leading to 2049 microlensing events announced in 2014.

\subsection{The RoboNet campaign}

The RoboNet microlensing campaign makes use of the Las Cumbres Observatory (LCO) network\footnote{{\tt https://lco.global}} of globally distributed 1m and 2m telescopes, operated by LCOGT Inc.\ (Goleta, California). Three of the southern 1m telescopes are owned by the University of St Andrews, which in turn holds a respective fraction of observing time on the network. LCO's 1m telescopes are organised in clusters at 4 sites in the network. Due to the location of the Galactic bulge, we are using only the 3 telescopes at the Cerro-Tololo Interamerican Observatory (CTIO, Chile), the 3 at the South African Astronomical Observatory (SAAO, South Africa) and 2 installed alongside LCO's 2m telescope (Faulkes Telescope South, FTS) at the Siding Spring Observatory (SSO, Australia).   

All of the telescopes are robotically operated. At the time of these observations, most 1m telescopes
hosted  SBIG STX-16803 cameras with Kodak KAF-16803 front illuminated $4096 \times 4096$~pix CCDs.  These
instruments have a field of view of $15.8\arcmin$ square and a pixel scale of 0.464$\arcsec$/pix when used in
the standard bin $2 \times 2$ mode.  Two 1m telescopes in Chile supported Sinistro cameras, which
consist of $4096 \times 4096$~pix Fairchild CCD486 back-illuminated CCDs operated in bin $1 \times 1$ mode to
produce a 26.5\arcmin{} square field with a pixel scale of 0.387"/pix.  The 1m telescopes are designed
to be as identical as possible to facilitate networked observations and all feature the same complement
of filters.  The majority of these observations were made in SDSS-$i^\prime$, with some images taken in
Bessell-$V$ and -$R$.  

Observations on the 2m network telescopes made use of the Spectral imagers, which are also $4096 \times 4096$~pix
Fairchild CCD486 CCDs but have a field of view of $10.5\arcmin$ square, and a pixel scale of 0.304$\arcsec$/pix in bin $2 \times 2$
mode. 

LCOGT operates a network-wide scheduler, which dynamically allocates resources to meet observation
requests in real time.  The advantage of this system lies in its robust and graceful accommodation of
outages due to weather or technical problems at any given telescope.  Observations are immediately and
automatically re-assigned to an alternative telescope wherever possible.  

The RoboNet microlensing programme exploits this flexibility in real-time with a system of software designed to respond automatically to digital alerts of transient phenomena \citep{RoboNet-II}. Based on all available data (from both surveys and follow-up campaigns), the SIGNALMEN anomaly detector \citep{SIGNALMEN}, part of the Automated Robotic Terrestrial Exoplanet Microlensing Search (ARTEMiS) system \citep{ARTEMiS2,ARTEMiS1}, quasi-continuously produces up-to-date point-source-single-lens models of all microlensing events, updates being triggered by any new incoming data, while departures of data from such models are flagged as microlensing `anomalies'. Using a metric to determine the expected return of observing any specific event \citep{Horne:metric,MiNDSTEp}, a TArget Prioritisation algorithm
(TAP; \citealt{TAP}) then selects those events that are most valuable, giving special attention to anomalies flagged by SIGNALMEN, while considering the time available and the capabilities of the resources. The Observation Control (ObsControl) software interprets TAP's target recommendations into network observing requests and also handles the returned stream of imaging data, preparing them for reduction.  This stage is also fully robotic, depending on LCOGT's ORAC-based pipeline to remove the instrumental signatures from the images prior to Difference Image Analysis performed by a pipeline based around DanDIA \citep{Bramich2008}.  The resulting photometric light curves were immediately made available to the community to facilitate event analysis.  

\subsection{The MiNDSTEp campaign}
The MiNDSTEp observations were performed from the Danish 1.54m telescope at ESO's La Silla observatory in Chile. The telescope is equipped with a two-colour $512 \times 512$ pixel EMCCD camera \citep{DanishCam1,DanishCam2} with  $0.09\arcsec$/pixel, corresponding to a $45\arcsec \times 45\arcsec$ field of view on the sky. A dichroic beam splitter sends light shortward and longward of $655~\mbox{nm}$ to a ``visual'' and a ``red'' camera, respectively, allowing simultaneous two-colour photometry. A second beam splitter sends the light shortward of $466~\mbox{nm}$ into a continuous focusing camera. In order to obtain maximum intensity, and since microlensing is achromatic, there are no filters. In this way the visual and the red colours are determined by the sensitivity function of the CCD plus the combined throughput of the atmosphere and the telescope. \citet{Evans2016} provide the final sensitivity function, a comparison with the Sloan and Johnson systems, as well as the calibration toward stellar parameters. During the 2014 microlensing observations, the camera was operated at 10~\mbox{Hz} with a gain setting of $300~\mbox{e}^-$/photon, which typically results in photometric accuracy of the order 1 per cent per 2~min spools. The individual frames in each spool are re-centred during the on-line reduction (corresponding to a ``tip-and-tilt'' hardware compensation for the atmospheric turbulence in adaptive optics), and then sorted into 10~quality classes according to point spread function (PSF). Under good weather conditions, the best PSF groups approach the diffraction limit of the telescope. These are used as templates for the reduction of the full set of exposures, which is performed by use of the DanDIA pipeline \citep{Bramich2008}. While real-time photometric data immediately become publicly available, final data sets are prepared after more careful manual inspection of the process and the tuning of parameters in order to optimise the data quality.

Despite the fact that an observer is present for the operation of the Danish 1.54m telescope, the monitoring of the sequence of microlensing events during the night is fully automated, with the observer just pressing a `start microlensing' button on the telescope control system. The telescope then directly follows the target recommendations provided by the ARTEMiS system \citep{ARTEMiS2,ARTEMiS1}, according to the adopted MiNDSTEp strategy \citep{MiNDSTEp} and incorporating any suspected or detected anomalies identified by the SIGNALMEN detector \citep{SIGNALMEN}.

\subsection{Monitoring the OGLE-2014-BLG-1186 microlensing event}

On 2014 June 20 UTC, the OGLE survey announced the discovery of event OGLE-2014-BLG-1186, at $\mbox{RA} = 17\fh41\fm59\fs63$, $\mbox{Dec}=-34\fdg17\farcm18\farcs1$ (J2000), in tile BLG509 of its low-cadence zone  (about 1 observation every 1--2 nights).  The event brightened relatively slowly given a rather long event time-scale of $\tE \sim 100~\mbox{d}$ (predicted at that time) as compared to a median of $\tE \sim 20~\mbox{d}$ across all Galactic bulge microlensing events. OGLE-2014-BLG-1186 achieved a sufficient priority to make it into the list of events to be monitored by RoboNet and MiNDSTEp consistently both on 2014 September 20 UTC. At that time of the year, the Galactic bulge remains low above the horizon from the observing sites, limiting the target visibility to at most  $\sim 4~\mbox{hrs}$ per night.

The SIGNALMEN anomaly detector first spotted behaviour not matching the predictions based on real-time RoboNet data on 2014 September 22 UTC, and consequently an e-mail alerting all teams carrying out regular Galactic bulge microlensing observations was circulated. On 2014 September 27 UTC, SIGNALMEN then concluded that a microlensing anomaly was in progress, automatically triggering more intense follow-up from the RoboNet and MiNDSTEp campaigns, as well as fully-automated real-time binary-lens model analysis of the light curve data by the RTmodel system\footnote{{\tt \scriptsize http://www.fisica.unisa.it/GravitationAstrophysics/RTModel.htm}}, run at the University of Salerno and based on the VBBinaryLensing contour integration code \citep{Bozza:contour}. Rather than just providing a single best-fitting model, RTmodel produces a range of alternatives, which narrows down as the anomaly progresses. While initially following the SIGNALMEN trigger, a large variety of models appeared to match the data reasonably well, by 2014 October 6 UTC, it was only models with a mass ratio corresponding to a planet orbiting the lens star that remained feasible (V.~Bozza, private communication). An independent assessment (C.~Han, private communication) arrived at the same conclusion by 2014 October 20 UTC.
However, these preliminary analyses left us with substantial apparent discrepancies between the models and some of the acquired data, and most notably, OGLE and RoboNet data appeared to favour different scenarios. We therefore had to consider the possibility that the putative planetary ``signal'' was due to systematic noise in the data. Consequently, this prompted a more careful analysis of the photometric noise in order to be able to consistently claim a signal and to ensure a meaningful interpretation (or to rather reject such a claim).

As it turned out, SIGNALMEN concluded anomalous behaviour being in progress based on the prominent annual parallax signature  (due to the Earth's revolution), causing an asymmetry between the rising and falling wing of the light curve, rather than on binarity. Unfortunately, 2014 September 28 UTC was the last night of the annual observing season with the Danish 1.54m telescope, so that the MiNDSTEp observations missed the binary signature and provided data only on the rising part of the light curve. By the end of the 2014 observing season, the light curve of event OGLE-2014-BLG-1186 was still within the falling wing, about 2~mag above the ($I$-band) baseline magnitude. While a substantial part of the falling wing was missed due to lack of observability of the target from our sites during the southern summer, a further fading was measured over the full course of the 2015 observing season, and it was only in 2016 that the event reached its baseline magnitude, from which it started to depart already in 2013.

Table~\ref{tab:data} provides an overview of the photometric data acquired for microlensing event OGLE-2014-BLG-1186.

\begin{table*}
\begin{center}
\begin{tabular}{llcccrrr}
\hline
 & & & & & \multicolumn{3}{c}{Number of data points} \\
Site & Telescope & Filter & Team & Label & off-peak & peak & total \\
\hline
Las Campanas Observatory & Warsaw 1.3m & $I$ & OGLE & OGLE (I) & 642 & 3 & 645\\
Cerro Tololo Inter-American Observatory (CTIO) & LCO 1m, Dome B & $I$ & RoboNet & \color{lscb}LSC B (I)\color{black} & 40 &4 & 44\\
Cerro Tololo Inter-American Observatory (CTIO) & LCO 1m, Dome C & $I$ & RoboNet & \color{lscc}LSC C (I)\color{black} & 32 &11 & 43\\
South African Astronomical Observatory (SAAO) & LCO 1m, Dome A & $I$ & RoboNet & \color{cpta}CPT A (I)\color{black} & 41 &4 & 45\\
South African Astronomical Observatory (SAAO) & LCO 1m, Dome B & $I$ & RoboNet & \color{cptb}CPT B (I)\color{black} & 46 &6 & 52\\
South African Astronomical Observatory (SAAO) & LCO 1m, Dome C & $I$ & RoboNet & \color{cptc}CPT C (I)\color{black} & 67 &18 & 85\\
Siding Spring Observatory (SSO) & LCO 1m, Dome A & $I$ & RoboNet & \color{coja}COJ A (I)\color{black} & 78 &45 & 123\\
Siding Spring Observatory (SSO) & LCO 1m, Dome B & $I$ & RoboNet & \color{cojb}COJ B (I)\color{black} & 54 &39 & 93\\
Haleakala Observatory & Faulkes North 2m & $I$ & RoboNet & \color{fts}FTS (I)\color{black} & 35 &79 & 114\\
Cerro Tololo Inter-American Observatory (CTIO) & LCO 1m, Dome C & $V$ & RoboNet & \color{lsccv}LSC C (V)\color{black} & 24 &0 & 24\\
ESO La Silla Observatory & Danish 1.54m & $Z$ & MiNDSTEp & \color{dk154}Dk1.54m (Z)\color{black} & 41 &0 & 41\\
\hline
\multicolumn{5}{l}{total} & 1100 & 209 & 1309 \\
\hline
\end{tabular}
\caption{Number of data points acquired with the various telescopes on gravitational microlensing event OGLE-2014-BLG-1186. The ``peak region'' is defined as
the epoch range $6928.8 \leq \mbox{HJD}-2\,450\,000 \leq 6934.0$.}
\label{tab:data}
\end{center}
\end{table*}

\section{Modelling the photometric light curve}
\label{sec:model}
\subsection{Methodology}

Our preliminary assessment obviously showed that OGLE-2014-BLG-1186 is strongly affected by annual parallax, and there is a putative further deviation near the peak, potentially caused by a planet orbiting the lens star. However, we also found that the data show some substantial systematic noise. Clearly, we must not take noise for a planetary signal, nor must we let noise corrupt the parallax measurement, which provides valuable information on the properties of the lens star and its planet (should there be one). 

Given that previous studies have shown that low-level deviations could be due to red noise instead of real signal \citep{Bachelet:noise}, we decided to conduct a similar study on the RoboNet data acquired for OGLE-2014-BLG-1186, which correlates and corrects common brightness patterns of stars in the field of view with various quantities (airmass, CCD position etc...). Using a Python implementation of \citet{BramFreud},\footnote{{\tt https://github.com/ebachelet/RoboNoise}} we found that any systematics are at least one magnitude smaller than the deviations around the peak.

We also should not confuse features in the putative anomaly over the peak with features due to parallax.
Given the long event time-scale, the parallax signal is clearly evident in the wings of the light curve, and measuring it from the wings alone should give pretty much the same result as measuring it from the full data set. The wing region however is not affected by binarity, considered to cause a visible anomaly over the peak. If we were to find a model for the full light curve that successfully describes the peak region, but suggests a significantly different parallax measurement than the wing region does, we would find a clear indication for our interpretation being inconsistent. 

We therefore divide the data set into `peak' and an `off-peak' subsets, with visual inspection suggesting to define the `peak' region as the epoch range $6928.8 \leq \mbox{HJD}-2\,450\,000 \leq 6934.0$. Moreover, we adopt an effective noise model, involving a global systematic error and an error bar scaling factor, while a robust fitting procedure prevents parameter estimates being driven by data outliers. We find it fair to assume that the off-peak region is well described by a point-source single-lens model with annual parallax, so that we can construct an effective model for the data residuals with respect to such a model and subsequently apply it to the peak region. With an established model for the noise, we can then assess the significance of a putative anomaly over the peak. Successively determining dominant model parameters, we therefore find full viable models describing event OGLE-2014-BLG-1186 as follows:
\begin{enumerate}
\item rough estimation of point-source single-lens parameters from off-peak OGLE data,
\item measurement of parallax parameters from off-peak data by means of robust fitting and simultaneous estimation of global systematic error and error bar scaling factor for each data set,
\item application of the estimated global systematic error and error bar scaling factor to the peak data,
\item assessment whether putative peak anomaly is significantly above noise floor and check for consistency between data sets.
\end{enumerate}
If there is evidence for the putative peak anomaly, we consider binary-lens or binary-source interpretations by
\begin{enumerate}
\addtocounter{enumi}{4}
\item grid search for model parameters characterising a binary lens and establishment of a complete set of all potential viable solutions,
\item robust fitting of point-source binary-lens models to all data,
\item fitting of finite-source binary-lens models to all data,
\item fitting of binary-point-source single-lens models to all data,
\item fitting of binary-finite-source single-lens models to all data.
\end{enumerate}

\subsection{Parallax measurement and noise model}

\subsubsection{Ordinary microlensing light curves}

A light ray passing a body of mass $M$ at the impact distance $\xi$ experiences a gravitational bending by the angle \citep{Einstein:bending}
\begin{equation}
\alpha(\xi) = \frac{4GM}{c^2\,\xi}\,,
\end{equation}
where $G$ is the universal gravitational constant, and $c$ is the vacuum speed of light.
If we observe a background object (`source') at distance $D_\mathrm{S}$ in close angular proximity to the deflecting body (`lens') at distance $D_\mathrm{L}$, it appears at angular image positions $x_i\,\theta_\mathrm{E}$, measured relative to the lens position,
rather than its true angular position $u\,\theta_\mathrm{E}$, related by
\begin{equation}
u(x) = x - \frac{1}{x}\,,
\label{eq:lenseq}
\end{equation}
with $\theta_E$ being the 
angular Einstein radius 
\begin{equation}
\theta_\mathrm{E}=\sqrt{\frac{4GM}{c^2}\,\frac{\pi_\mathrm{LS}}{1~\mbox{AU}}}\,,
\label{eq:thetaE}
\end{equation}
where
\begin{equation}
\pi_\mathrm{LS} = 1~\mbox{AU}\,\left(D_\mathrm{L}^{-1}-D_\mathrm{S}^{-1}\right)
\end{equation}
is the relative parallax of lens and source with respect to the observer.

Gravitational microlensing events show a transient brightening of an observed source star that results from the gravitational bending of its light by an intervening object, which
follows from Eq.~(\ref{eq:lenseq}) as
\begin{equation}
A(u) = \sum_i \left|\frac{u(x_i)}{x_i}\;\frac{\mathrm{d}u}{\mathrm{d}x}\left(x_i\right)\right|^{-1}\,.
\label{eq:magnifgen}
\end{equation}
For single point-like source and lens stars, one finds two images
\begin{equation}
x_{1/2} = \frac{1}{2} \left(u \pm \sqrt{u^2+4}\right)\,,
\end{equation}
so that the observed magnification, Eq.~(\ref{eq:magnifgen}),
 evaluates to the analytic expression \citep{Ein36}
\begin{equation}
A(u) = \frac{u^2+2}{u\,\sqrt{u^2+4}}\,.
\label{eq:ptsrcmag}
\end{equation}
If we assume a uniform relative proper motion $\mu$ between lens and source star, the separation parameter $u$ becomes \citep{Pac1}
\begin{equation}
u(t;t_0,u_0,t_\mathrm{E}) = \sqrt{u_0^2 + \left(\frac{t-t_0}{t_\mathrm{E}}\right)^2}\,,
\label{eq:trajectory}
\end{equation}
where $t_\mathrm{E} = \theta_\mathrm{E}/\mu$ is the event time-scale, and the closest angular approach $u_0\,\theta_\mathrm{E}$ is realised at time $t_0$.

With $F_\mathrm{S}^{[j]}$ being the unmagnified flux of the observed target star, and $F_\mathrm{B}^{[j]}$ the flux contributed by other light sources, corresponding to a specific detector and labelled by the index $m$, the total observed flux becomes
\begin{eqnarray}
F^{[j]}(t) & = & F_\mathrm{S}^{[j]}\, A[u(t;\vec{p})] + F_\mathrm{B}^{[j]} \nonumber \\
   & = & F_\mathrm{S}^{[j]} \left\{A[u(t;\vec{p})] - 1\right\} + F_\mathrm{base}^{[j]}\,,
\end{eqnarray}
where $F_\mathrm{base}^{[j]}= F_\mathrm{S}^{[j]} + F_\mathrm{B}^{[j]}$ is the baseline flux and $\vec{p}$ denotes the set of parameters characterising the magnification function $A[u(t;\vec{p})]$.
The total flux can also be written as
\begin{equation}
F^{[j]}(t) = F_\mathrm{base}^{[j]} \;A^{[j]}_\mathrm{obs}(t;\vec{p})\,,
\end{equation}
where
\begin{equation}
A^{[j]}_\mathrm{obs}[u(t;\vec{p})] =  \frac{A[u(t;\vec{p})]+g^{[j]}}{1+g^{[j]}}
\end{equation}
is the observed magnification, with
\begin{equation}
g^{[j]} =  F_\mathrm{B}^{[j]}/F_\mathrm{S}^{[j]} = F_\mathrm{base}^{[j]}/F_\mathrm{S}^{[j]} - 1
\end{equation}
being the blend ratio for the given detector.

Because of $A(u)$ monotonically increasing as $u \to 0$, the light curves of ordinary microlensing events, assuming a single isolated lens star and a point-like source star as well as uniform relative proper motion, reach a peak at
$t_0$, where the closest angular approach between lens and source $u(t_0) = u_0$ is realised, and are symmetric in time with respect to this peak. They are fully characterised by $\vec{p} = (t_0,u_0,t_\mathrm{E})$ and the set  of $(F_\mathrm{base}^{[j]},g^{[j]})$ for each detector. While $(F^{[j]}_\mathrm{base},F^{[j]}_\mathrm{S})$ follow analytically from linear regression, the magnification function $A[u(t;\vec{p})]$ is generally non-linear
in the parameters $\vec{p}$.

\subsubsection{Annual parallax}


An annual parallax effect is caused by the revolution of the Earth, leading to a change of the line of sight, which alters the observed microlensing magnification.
Let $\vec{\gamma}(t)\,(1~\mbox{AU})$ denote the projection of the Earth's orbit onto a plane perpendicular to the direction towards the source star.
With $\vec{\mu}_\mathrm{S}$ and $\vec{\mu}_\mathrm{L}$ denoting the proper motions of the source and lens stars, respectively, while $\pi_\mathrm{S}$ and $\pi_\mathrm{L}$ denote their parallaxes, the apparent geocentric positions of source and lens star may be written  \citep[c.f.\ ][]{An2002,Gould:parallax}
\begin{eqnarray}
\vec{\theta}_\mathrm{S}(t) & = & \vec{\theta}_{\mathrm{S},0} + (t-t_0)\,\vec{\mu}_\mathrm{S} - \pi_\mathrm{S}\,\vec{\gamma}(t) \nonumber\,, \\
\vec{\theta}_\mathrm{L}(t) & = & \vec{\theta}_{\mathrm{L},0} + (t-t_0)\,\vec{\mu}_\mathrm{L} - \pi_\mathrm{L}\,\vec{\gamma}(t) \,,
\end{eqnarray}
so that
\begin{equation}
\vec{\theta}(t) \equiv \vec{\theta}_\mathrm{S}(t) - \vec{\theta}_\mathrm{L}(t) = (\vec{\theta}_\mathrm{S}-\vec{\theta}_\mathrm{L})_0 - (t-t_0)\,\vec{\mu}_\mathrm{LS} + \pi_\mathrm{LS}\,\vec{\gamma}(t)\,,
\end{equation}
with $\vec{\mu}_\mathrm{LS} \equiv \vec{\mu}_\mathrm{L} -\vec{\mu}_\mathrm{S}$ and $\pi_\mathrm{LS} \equiv \pi_\mathrm{L}-\pi_\mathrm{S}$ denoting the relative proper motion and relative parallax between lens and source, while $(\vec{\theta}_\mathrm{S}-\vec{\theta}_\mathrm{L})_0 \equiv \vec{\theta}_{\mathrm{S},0} - \vec{\theta}_{\mathrm{L},0}$.

Hence, for $\vec{u}(t) = \vec{\theta}(t)/\theta_\mathrm{E}$ we find with the microlensing parallax parameter $\pi_\mathrm{E} \equiv \pi_\mathrm{LS}/\theta_\mathrm{E}$,
\begin{equation}
\vec{u}(t) = \vec{u}_0 + (t-t_0)\,\dot{\vec{u}}_0 + \pi_\mathrm{E}\,\delta\vec{\gamma}(t) \,,
\label{eq:parallax}
\end{equation}
where
\begin{eqnarray}
 \vec{u}_0 \equiv  \vec{u}(t_0) & = &  \frac{(\vec{\theta}_\mathrm{S}-\vec{\theta}_\mathrm{L})_0}{\theta_\mathrm{E}} + \pi_\mathrm{E}\,\vec{\gamma}(t_0)\,, 
 \\
  \dot{\vec{u}}_0 \equiv  \dot{\vec{u}}(t_0)  & = & -\frac{\vec{\mu}_\mathrm{LS}}{\theta_\mathrm{E}} + \pi_\mathrm{E}\,\dot{\vec{\gamma}}(t_0)\,,\\
  \delta\vec{\gamma}(t) & =& \vec{\gamma}(t)-\vec{\gamma}(t_0)-(t-t_0)\,\dot{\vec{\gamma}}(t_0)\,.
  \label{eq:pargam}
\end{eqnarray}
Given that by construction $\delta\vec{\gamma}(t_0) = 0$ and $\delta\dot{\vec{\gamma}}(t_0) = 0$,
one explicitly sees that for epochs near $t_0$, the lowest-order local effect of the annual parallax distorting the symmetric light curve of a single lens arises from the Earth's acceleration along its orbit, corresponding to the curvature of the effective source trajectory $\vec{u}(t)\,\theta_\mathrm{E}$.

With $(\hat{\vec{e}}_\mathrm{n},\hat{\vec{e}}_\mathrm{e})$ denoting unit vectors in the direction of ecliptic north and east, respectively, 
\begin{equation}
\delta\vec{\gamma}(t) = \delta\gamma_\mathrm{n}(t)\,\hat{\vec{e}}_\mathrm{n} + \delta\gamma_\mathrm{e}(t)\,\hat{\vec{e}}_\mathrm{e}\,,
\end{equation}
while $\vec{u}(t)$ can be written in terms of components parallel and perpendicular to the effective source trajectory as
\begin{eqnarray}
u_\parallel(t;t_0,t_\mathrm{E},\vec {\pi}_\mathrm{E}) & = & \frac{t-t_0}{t_\mathrm{E}} +  \pi_{\mathrm{E},\mathrm{N}}\,\delta\gamma_\mathrm{n}(t)\, +  \pi_{\mathrm{E},\mathrm{E}} \,\delta\gamma_\mathrm{e}(t)\,,\nonumber \\
u_\perp(t;u_0,\vec {\pi}_\mathrm{E}) & = & u_0 -   \pi_{\mathrm{E},\mathrm{E}} \,\delta\gamma_\mathrm{n}(t) + \pi_{\mathrm{E},\mathrm{N}}\,\delta\gamma_\mathrm{e}(t) \,,
\end{eqnarray} 
where $t_\mathrm{E} = \theta_\mathrm{E}/|\vec \mu|$ with
\begin{equation}
\vec{\mu} = \dot{\vec{u}}_0\,\theta_\mathrm{E} =  - \vec{\mu}_\mathrm{LS} + \pi_\mathrm{E}\,\theta_\mathrm{E}\,\dot{\vec{\gamma}}(t_0)
\end{equation}
and
\begin{equation}
\pi_\mathrm{E}  = \sqrt{\pi_{\mathrm{E},\mathrm{N}}^2+\pi_{\mathrm{E},\mathrm{E}}^2}\,,
\end{equation}
so that $(\pi_{\mathrm{E},\mathrm{N}},\pi_{\mathrm{E},\mathrm{E}})$ form components of a vector $\vec{\pi}_\mathrm{E}$.

Hence, accounting for annual parallax, the microlensing light curve due to a single lens star can be characterised by the parameters
$\vec{p} = (t_0,u_0,t_\mathrm{E},\pi_{\mathrm{E},\mathrm{N}},\pi_{\mathrm{E},\mathrm{E}})$,
with the magnification given by Eq.~(\ref{eq:ptsrcmag}) and
\begin{equation}
u(t;t_0,u_0,t_\mathrm{E},\vec {\pi}_\mathrm{E})
=\sqrt{\big[u_\parallel(t;t_0,t_\mathrm{E},\vec {\pi}_\mathrm{E})\big]^2 + \big[u_\perp(t;u_0,\vec {\pi}_\mathrm{E})\big]^2}\,.
\label{eq:trajectoryparallax}
\end{equation}

\subsubsection{Noise model for photometric measurements and robust fitting}

Let us consider $M$ data sets, one for each detector, labelled by the index $j \in \{1,\ldots,M\}$, containing $N^{[j]}$ data points, respectively,  labelled by the index $i \in \{1,\ldots,N^{[j]}\}$,
so that the data tuple $(t_i^{[j]}, F_i^{[j]}, \sigma_i^{[j]})$ denotes the the time the measurement was taken, the measured flux, and the uncertainty of the measured flux.

In order to describe the measurement uncertainties of our photometric data, we adopt a model that combines error bar rescaling
with a robust-fitting procedure that applies weights to effectively correct for outliers and wide tails.

Similar to \citet{Tsapras2003}, we adopt a scaling factor $\kappa^{[j]}$ for the reported uncertainty $\sigma_i^{[j]}$, 
as well as a constant fractional systematic uncertainty $s_0^{[j]}$ in the reported flux $F_i^{[j]}$ (equivalent to a constant systematic uncertainty in the reported magnitude),
so that 
\begin{equation}
\tilde{\sigma}_i^{[j]} \left( \sigma_i^{[j]}, \kappa^{[j]}, s_0^{[j]}   \right) = \sqrt{\left(\kappa^{[j]} \sigma_i^{[j]}\right)^2 + \left(s_0^{[j]} F_i^{[j]}\right)^2}\,,
\label{eq:scaling}
\end{equation}
is assumed to represent the 
standard deviation of a Gaussian distribution.
This leads to the standardised residuals
\begin{equation}
r_i^{[j]} \left(F^{[j]}\big(t_i^{[j]}\big), F_i^{[j]},\tilde{\sigma}_i^{[j]}\right) = \frac{F_i^{[j]} - F^{[j]}\big(t_i^{[j]}\big)}{\tilde{\sigma}_i^{[j]}}\,.
\end{equation}

With the modified uncertainties $\tilde{\sigma}_i^{[j]}$ depending on the parameters $\kappa^{[j]}$ and $s_0^{[j]}$, a maximum-likelihood estimate is
then obtained by minimising 
\begin{equation}
\tilde{\chi}^2 = \sum\limits_{i = 1}^{M}  \sum\limits_{j=1}^{N^{[j]}}  \left[\left(r_i^{[j]} \right)^2  +  2 \,\ln \tilde{\sigma}_i^{[j]}  \right]\,,
\end{equation}
which is a modification of the ordinary $\chi^2$, which differs by an additional term due to the non-constant $\tilde{\sigma}_i^{[j]}$ 
and does not follow $\chi^2$ statistics.

Accounting for scaling factors $\kappa^{[j]}$ and systematic uncertainties  $s_0^{[j]}$  according to Eq.~(\ref{eq:scaling}) does not account for the distribution of 
the standardised residuals being more tail-heavy than a Gaussian distribution. While this could be achieved by using Student's t-distribution (with an additional parameter), 
we adopt a procedure that uses a pseudo-Gaussian distribution involving a weight factor, similar to that used by the SIGNALMEN anomaly detector \citep{SIGNALMEN}.
Robust fitting procedures \citep*[e.g.][]{robustbook,Huber} enforce the model function $F^{[j]}(t)$ to follow the bulk of the data rather than being substantially effected by 
outliers in the data set. Like \citet{SIGNALMEN}, we determine 
the median of the absolute standardised residuals ${\tilde r}^{[j]}$, and apply a 
bi-square weight
\begin{equation}
w_i^{[j]} = \left\{\begin{array}{ccl}
\left[1-\left(\frac{r_i^{[j]}}{K\,{\tilde r}^{[j]}}\right)^2\right]^2
& \mbox{for} & |r_i^{[j]}| < K\,{\tilde r}^{[j]} \\
0 & \mbox{for} & |r_i^{[j]}| \geq K\,{\tilde r}^{[j]}
\end{array}\right.
\label{eq:weight}
\end{equation}
to each data point, where we adopt $K = 6$ for the tuning constant. In principle, we could have chosen $\beta \equiv K^{-1}$ as a further free parameter,
with $\beta = 0$ corresponding to a Gaussian without any data downweighting, i.e. $w_i^{[j]} = 1$ for all $n$. However, $\beta$ is not strictly constrained
by our data, and thus the exact choice does not make a significant difference, and we can accept that our procedure would enforce downweighting even to data that perfectly match a Gaussian distribution.
We explicitly choose a continuous weight function in order to ensure that our numerical minimisation procedures behave well rather than getting
confused by discontinuities.
The weight $w_i^{[j]}$ becomes zero for data points 
whose absolute standardised residuals exceeds $K$ times their
median.

With the weights $w_i^{[j]}$, we estimate model parameters by minimising
\begin{equation}
\tilde{\chi}^2 = \sum\limits_{i = 1}^{M} \sum\limits_{j=1}^{N^{[j]}} w_i^{[j]}  \left[ \left(r_i^{[j]} \right)^2  +  2 \,\ln \tilde{\sigma}_i^{[j]}  \right]\,,
\label{eq:robmini}
\end{equation}
which is repeated for subsequent sets of standardised residuals until $\tilde{\chi}^2$ converges.

\subsubsection{Off-peak parallax model for OGLE-2014-BLG-1186}

\begin{figure*}
\centering{\begin{minipage}{6.64cm}
\centering{$u_0 < 0$}
\end{minipage}
\hspace*{5mm}
\begin{minipage}{6.64cm}
\centering{$u_0 > 0$}
\end{minipage}
}\\[0.5mm]
\centering{\begin{minipage}{3cm}
\centering{CDF of residuals}
\end{minipage}
\hspace*{5mm}
\begin{minipage}{3.14cm}
\centering{CDF of weights}
\end{minipage}
\hspace*{5mm}
\begin{minipage}{3cm}
\centering{CDF of residuals}
\end{minipage}
\hspace*{5mm}
\begin{minipage}{3.14cm}
\centering{CDF of weights}
\end{minipage}
}\\[1mm]
\centering{ 
\begin{minipage}{3cm}
\resizebox{2.7cm}{!}{\includegraphics{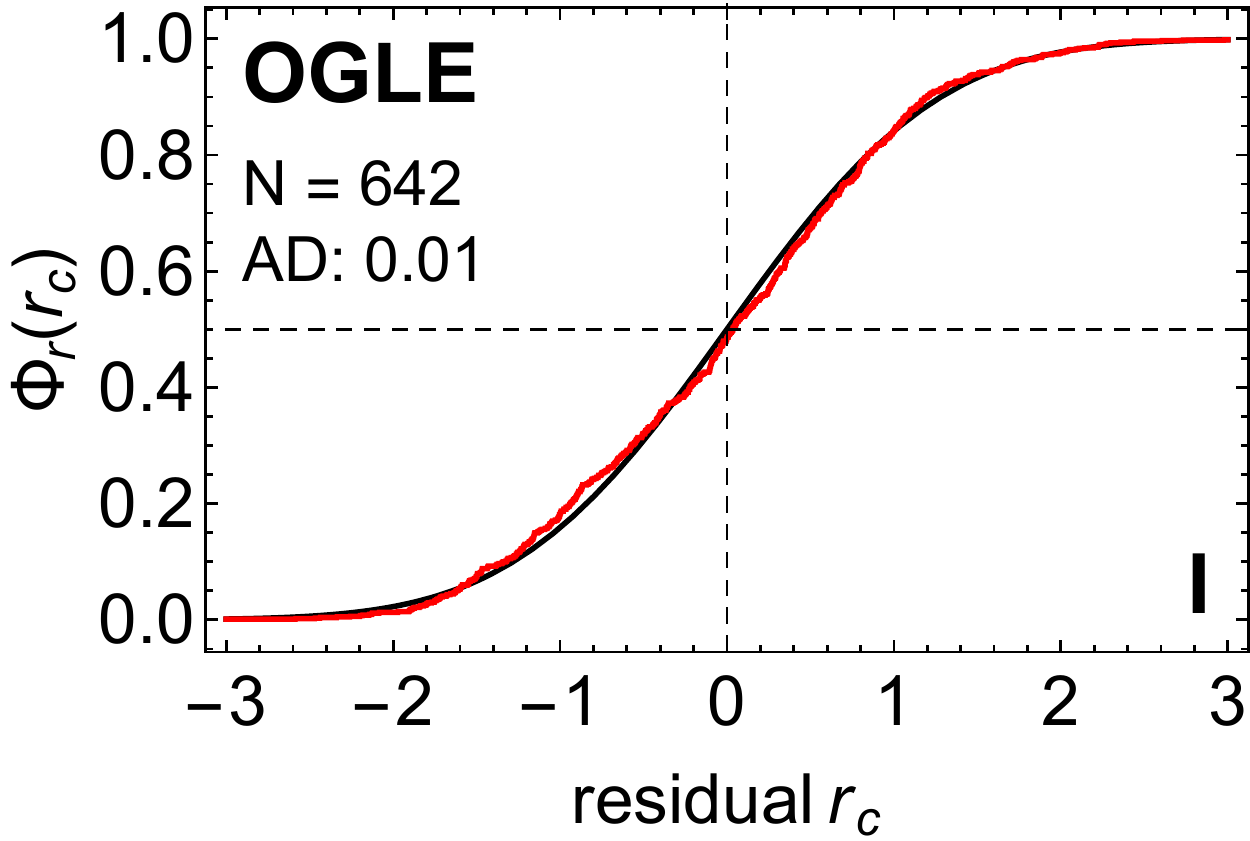}}
\end{minipage}
\hspace*{5mm}
\begin{minipage}{3.14cm}
\resizebox{2.826cm}{!}{\includegraphics{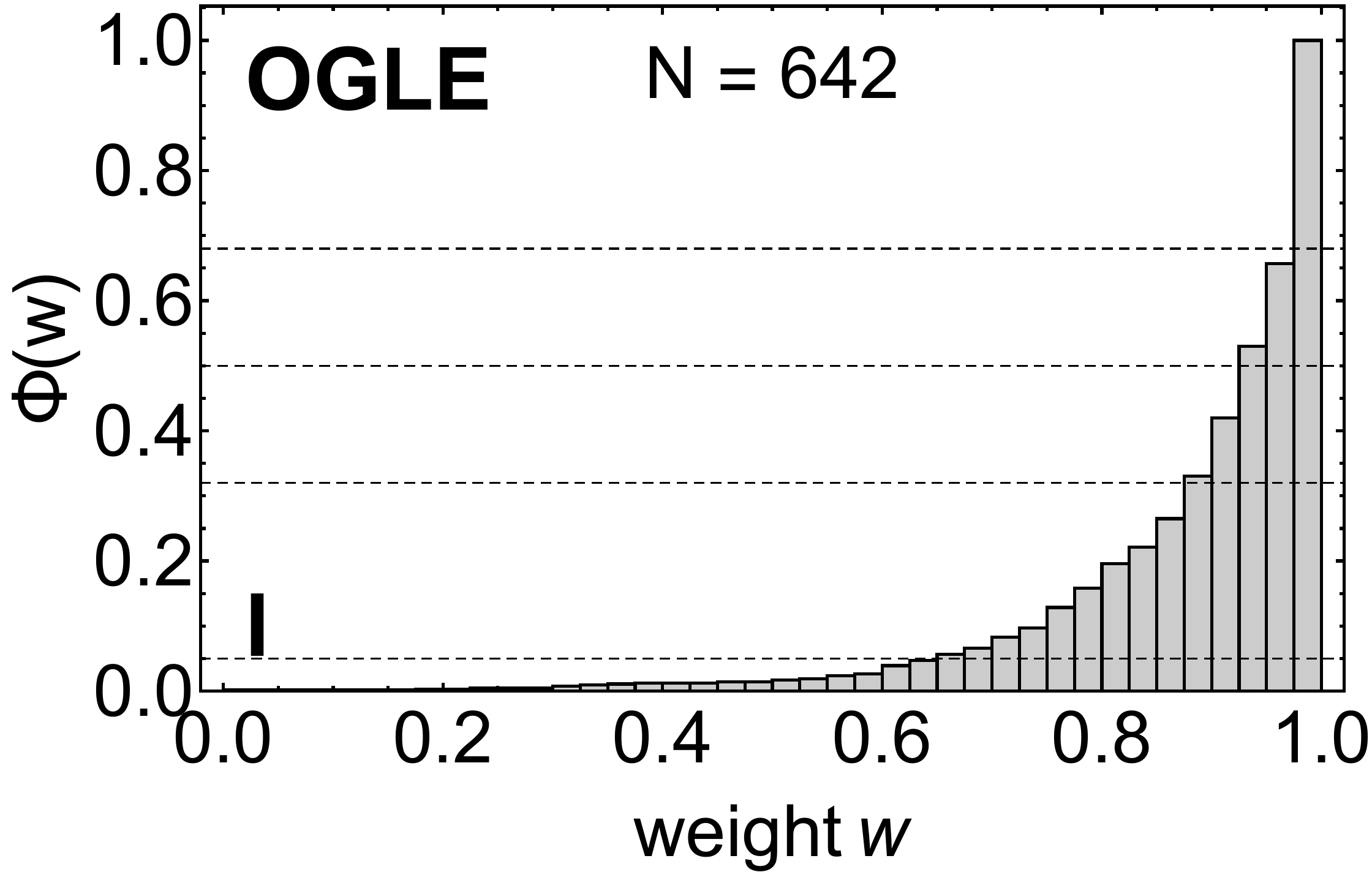}}
\end{minipage}
\hspace*{5mm}
\begin{minipage}{3cm}
\resizebox{2.7cm}{!}{\includegraphics{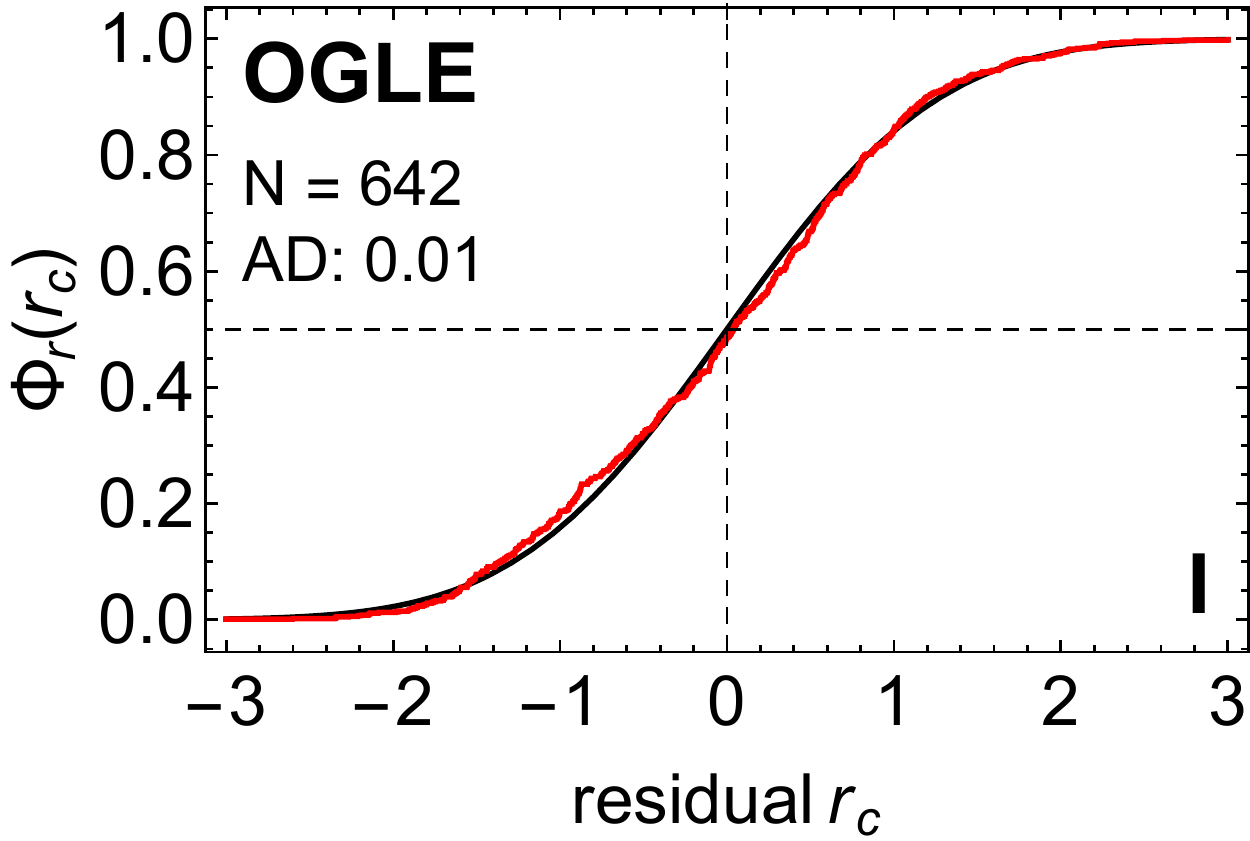}}
\end{minipage}
\hspace*{5mm}
\begin{minipage}{3.14cm}
\resizebox{2,826cm}{!}{\includegraphics{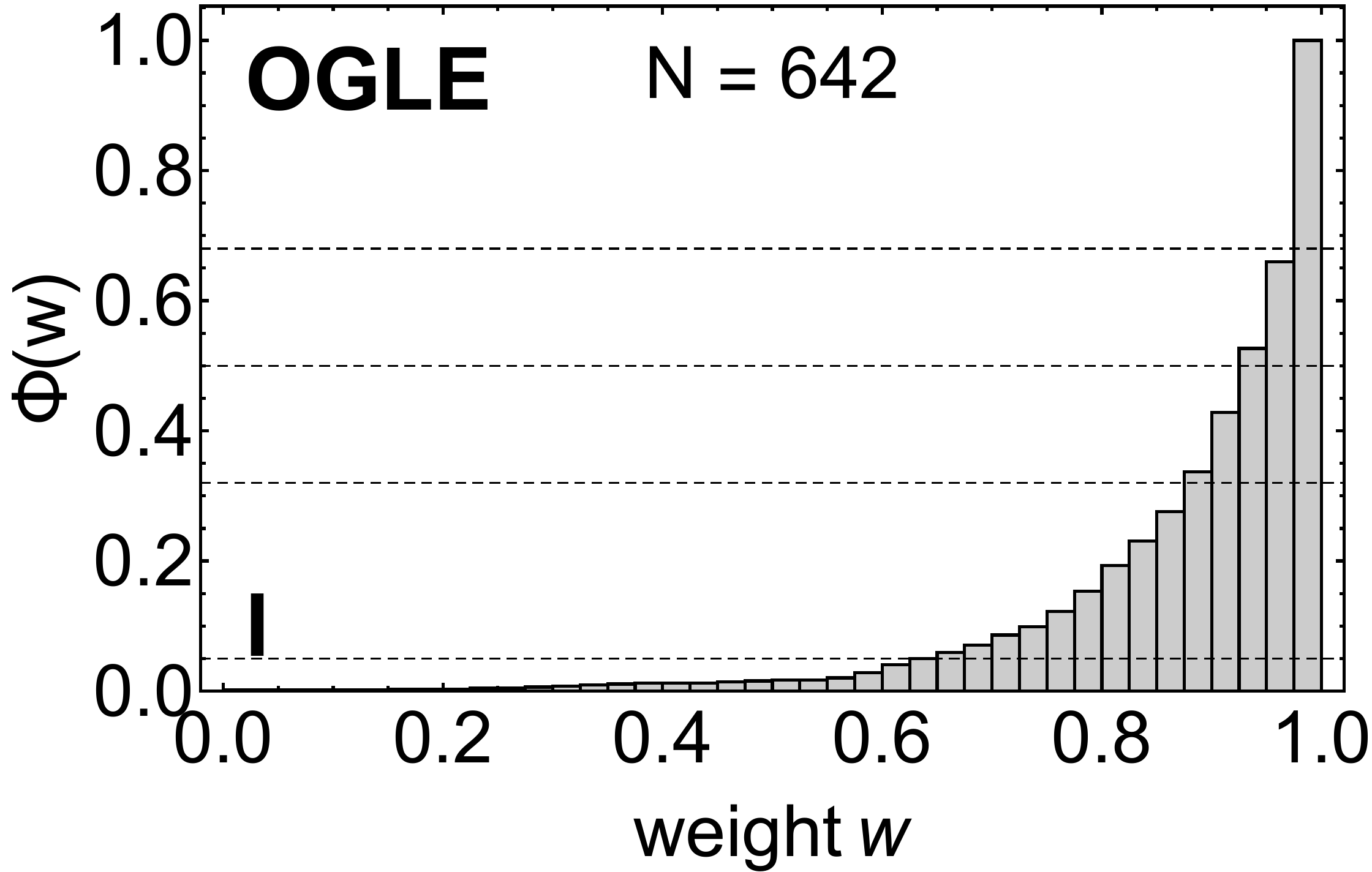}}
\end{minipage}
}\\[1mm]
\centering{\begin{minipage}{3cm}
\resizebox{2.7cm}{!}{\includegraphics{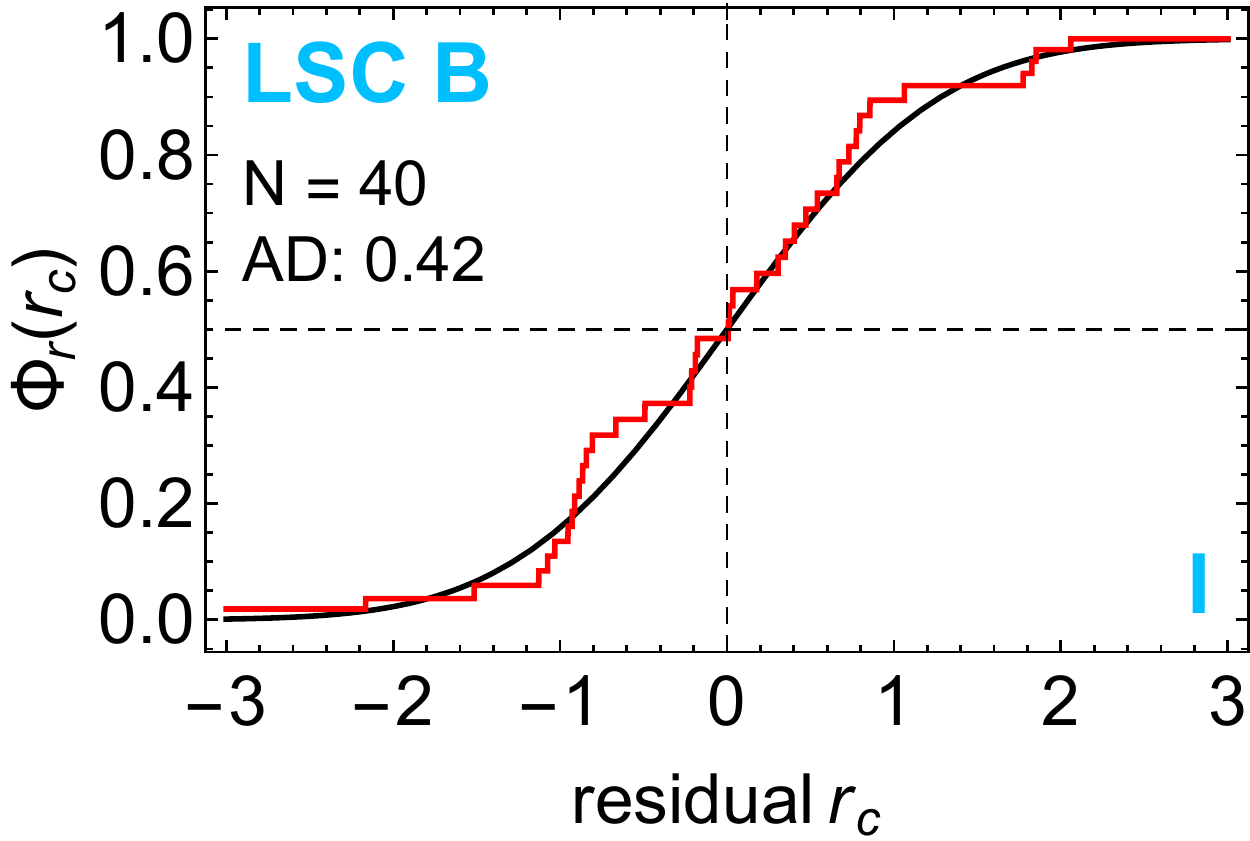}}
\end{minipage}
\hspace*{5mm}
\begin{minipage}{3.14cm}
\resizebox{2.826cm}{!}{\includegraphics{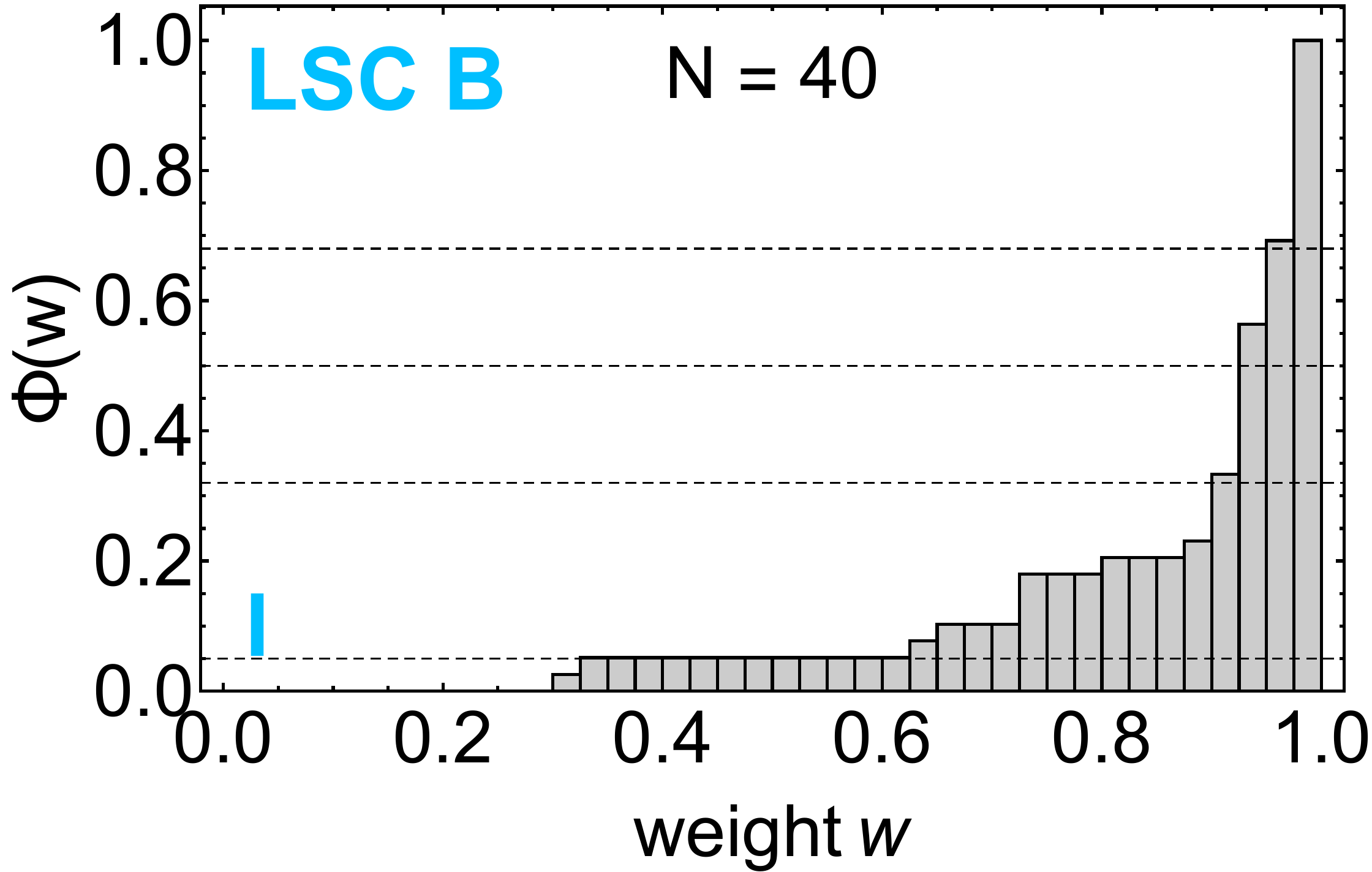}}
\end{minipage}
\hspace*{5mm}
\begin{minipage}{3cm}
\resizebox{2.7cm}{!}{\includegraphics{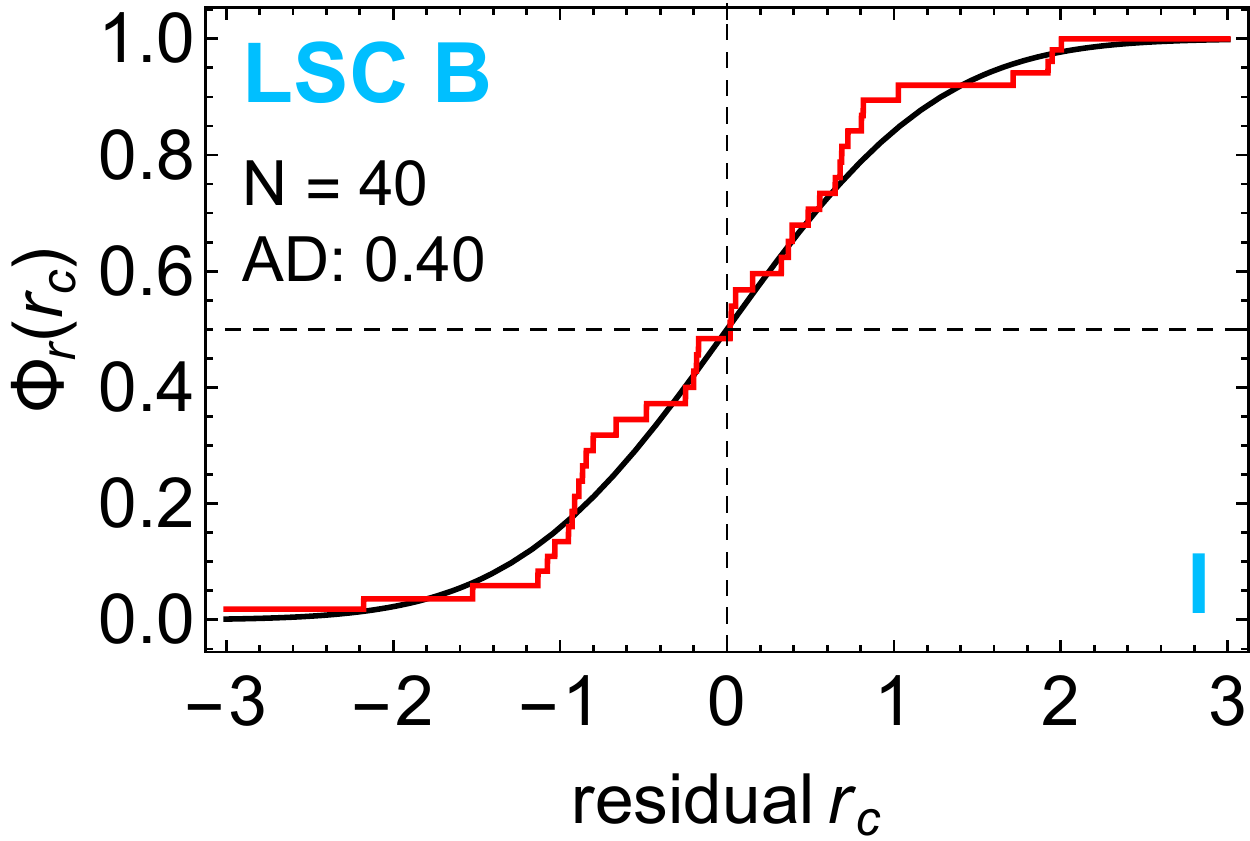}}
\end{minipage}
\hspace*{5mm}
\begin{minipage}{3.14cm}
\resizebox{2.826cm}{!}{\includegraphics{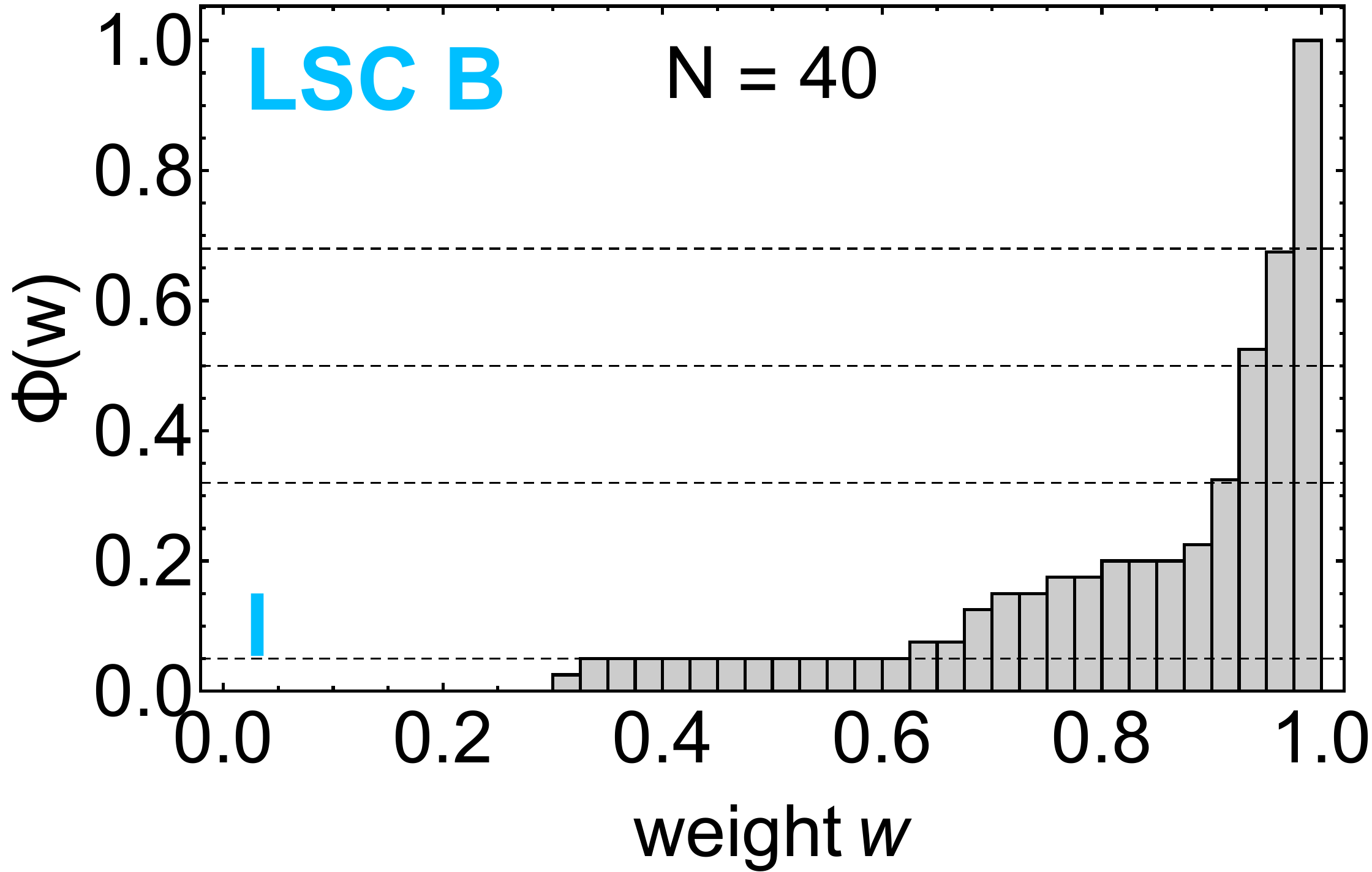}}
\end{minipage}
}\\[1mm]
\centering{\begin{minipage}{3cm}
\resizebox{2.7cm}{!}{\includegraphics{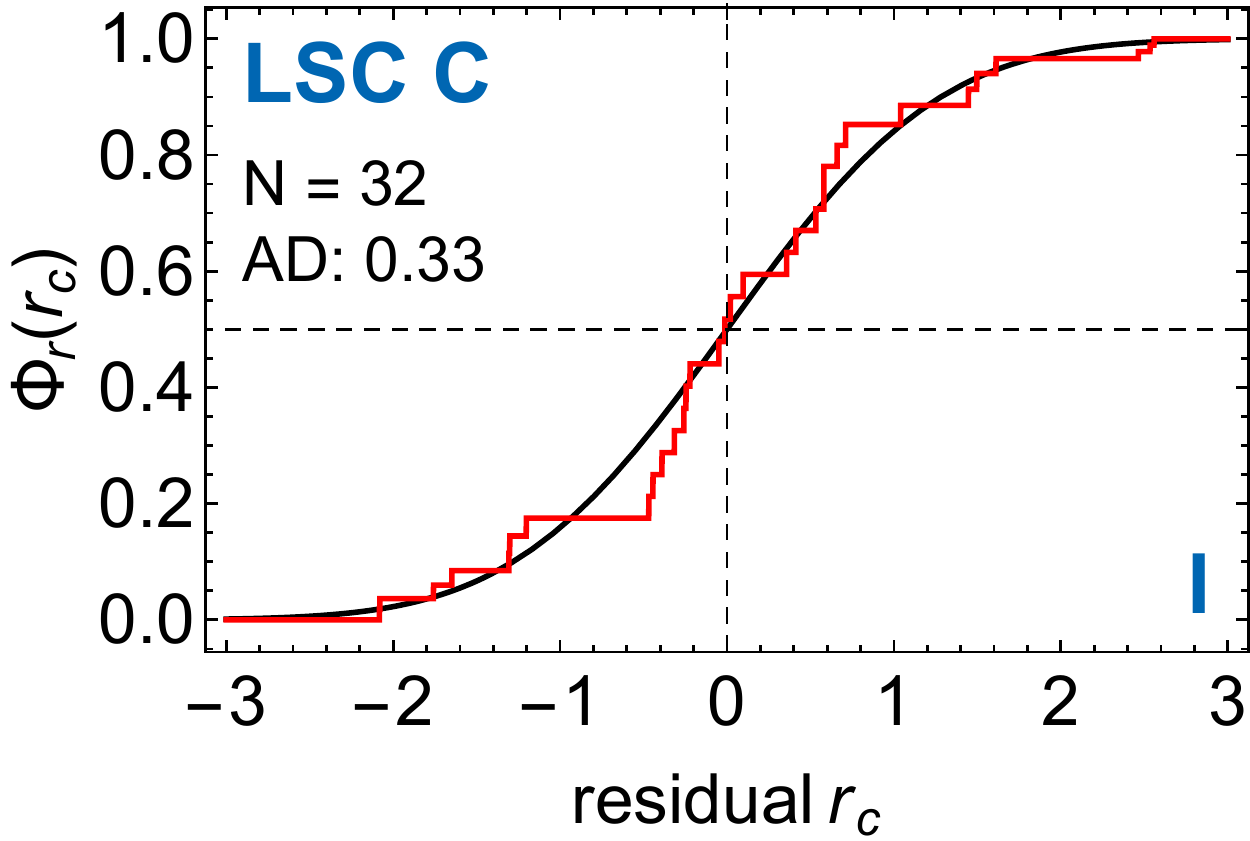}}
\end{minipage}
\hspace*{5mm}
\begin{minipage}{3.14cm}
\resizebox{2.826cm}{!}{\includegraphics{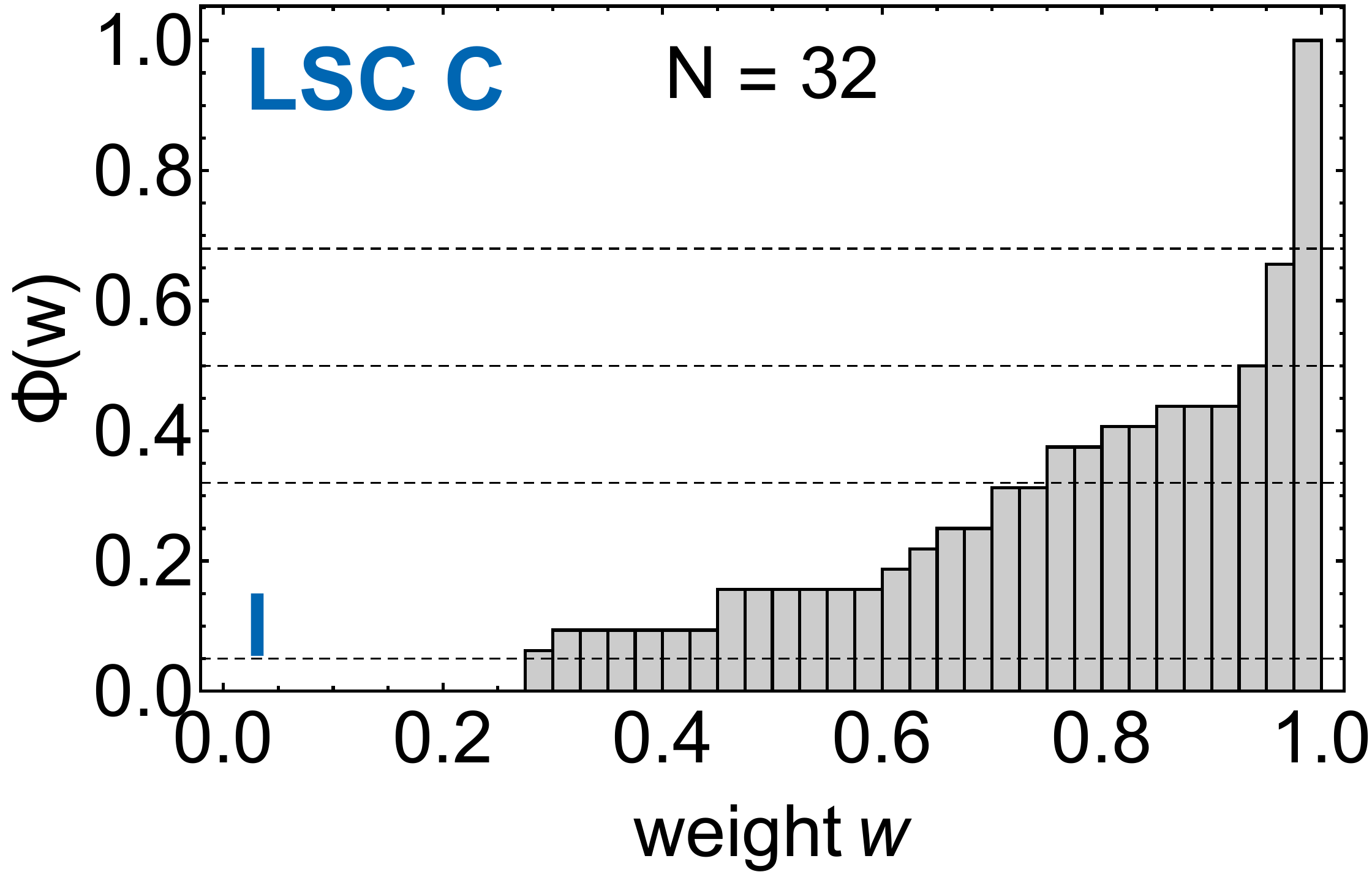}}
\end{minipage}
\hspace*{5mm}
\begin{minipage}{3cm}
\resizebox{2.7cm}{!}{\includegraphics{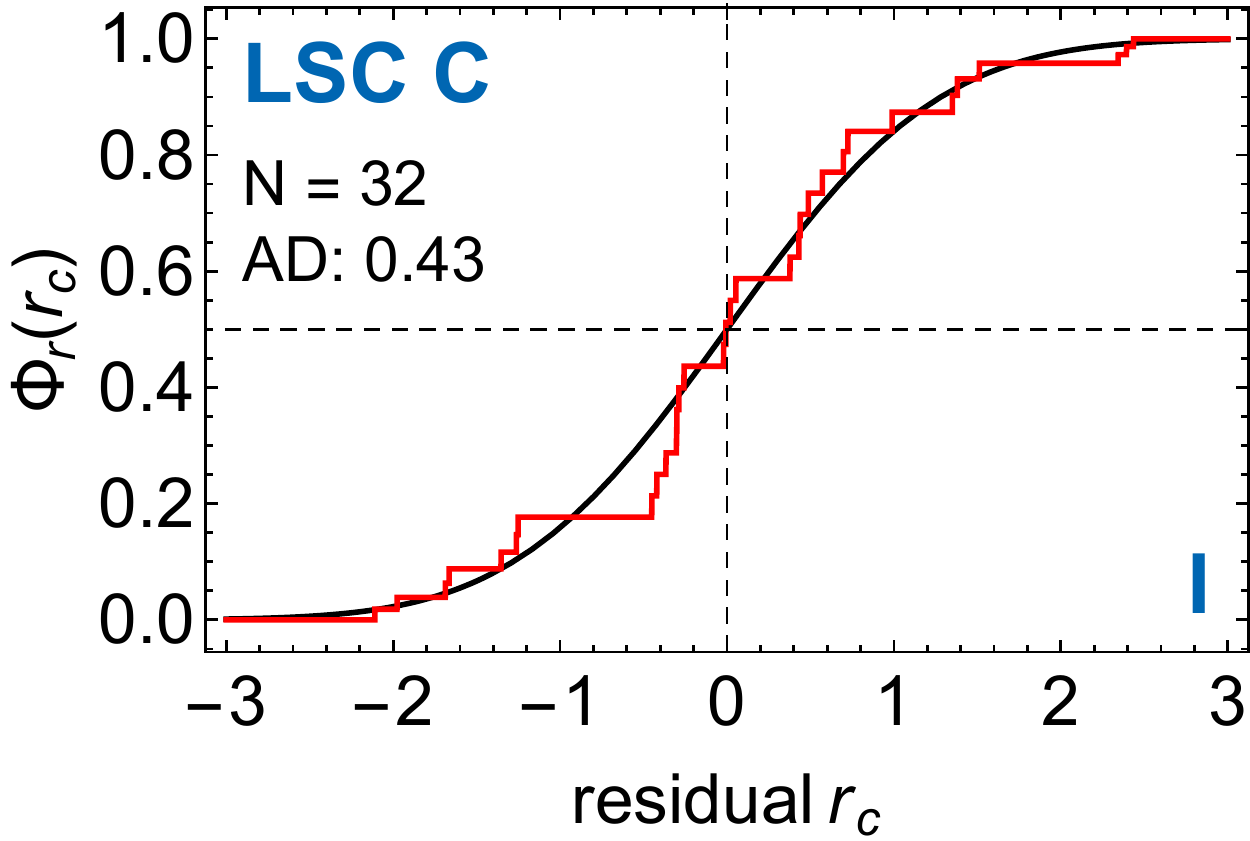}}
\end{minipage}
\hspace*{5mm}
\begin{minipage}{3.14cm}
\resizebox{2.826cm}{!}{\includegraphics{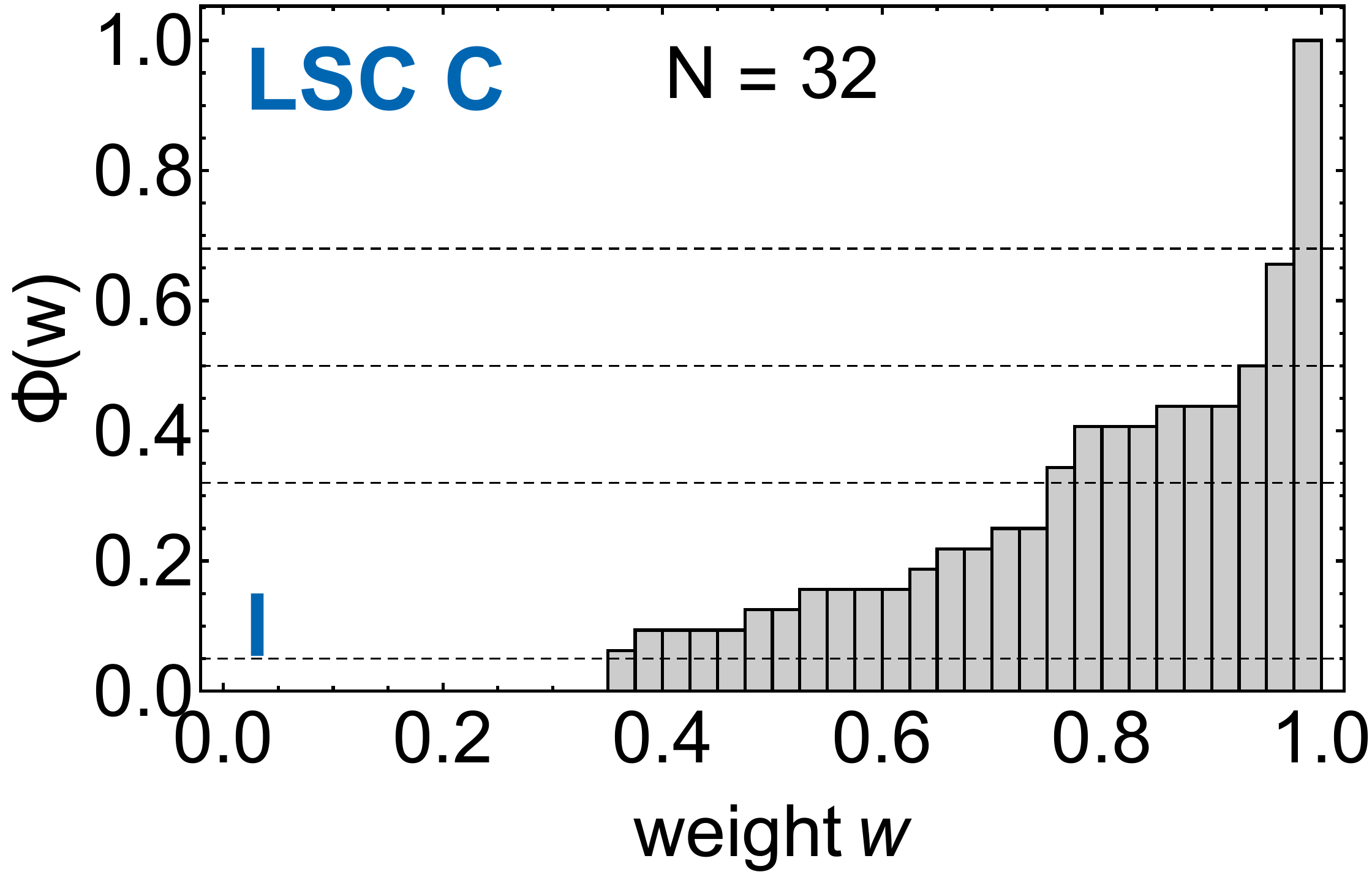}}
\end{minipage}
}\\[1mm]
\centering{\begin{minipage}{3cm}
\resizebox{2.7cm}{!}{\includegraphics{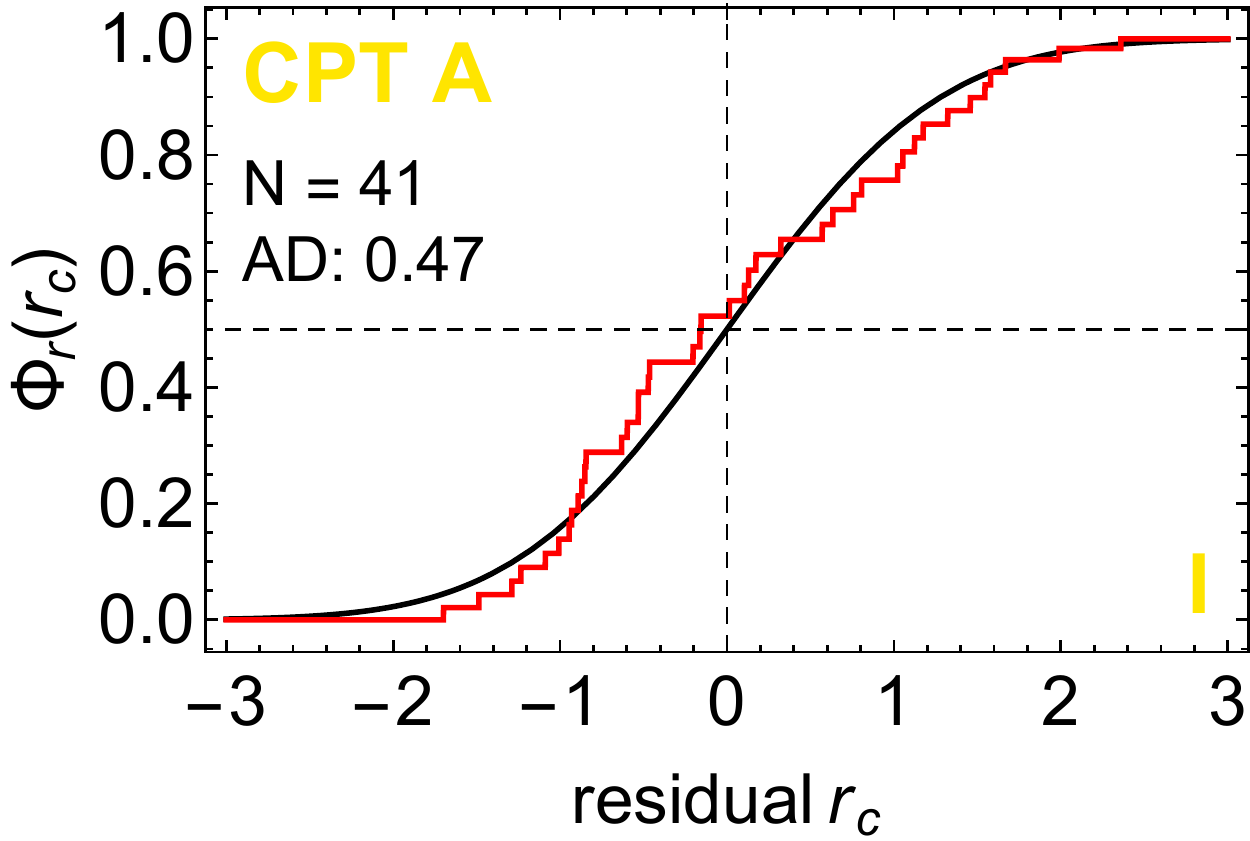}}
\end{minipage}
\hspace*{5mm}
\begin{minipage}{3.14cm}
\resizebox{2.826cm}{!}{\includegraphics{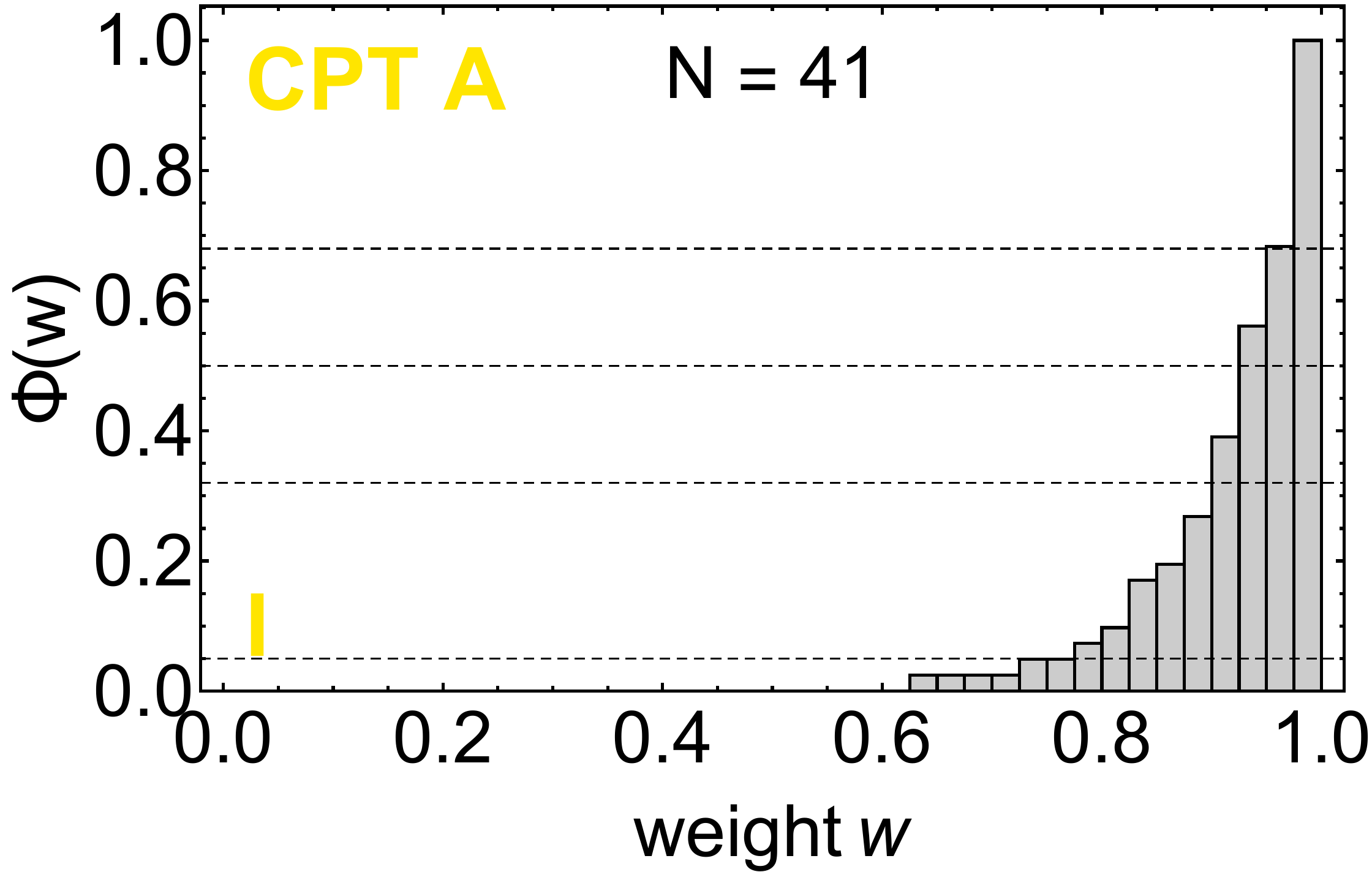}}
\end{minipage}
\hspace*{5mm}
\begin{minipage}{3cm}
\resizebox{2.7cm}{!}{\includegraphics{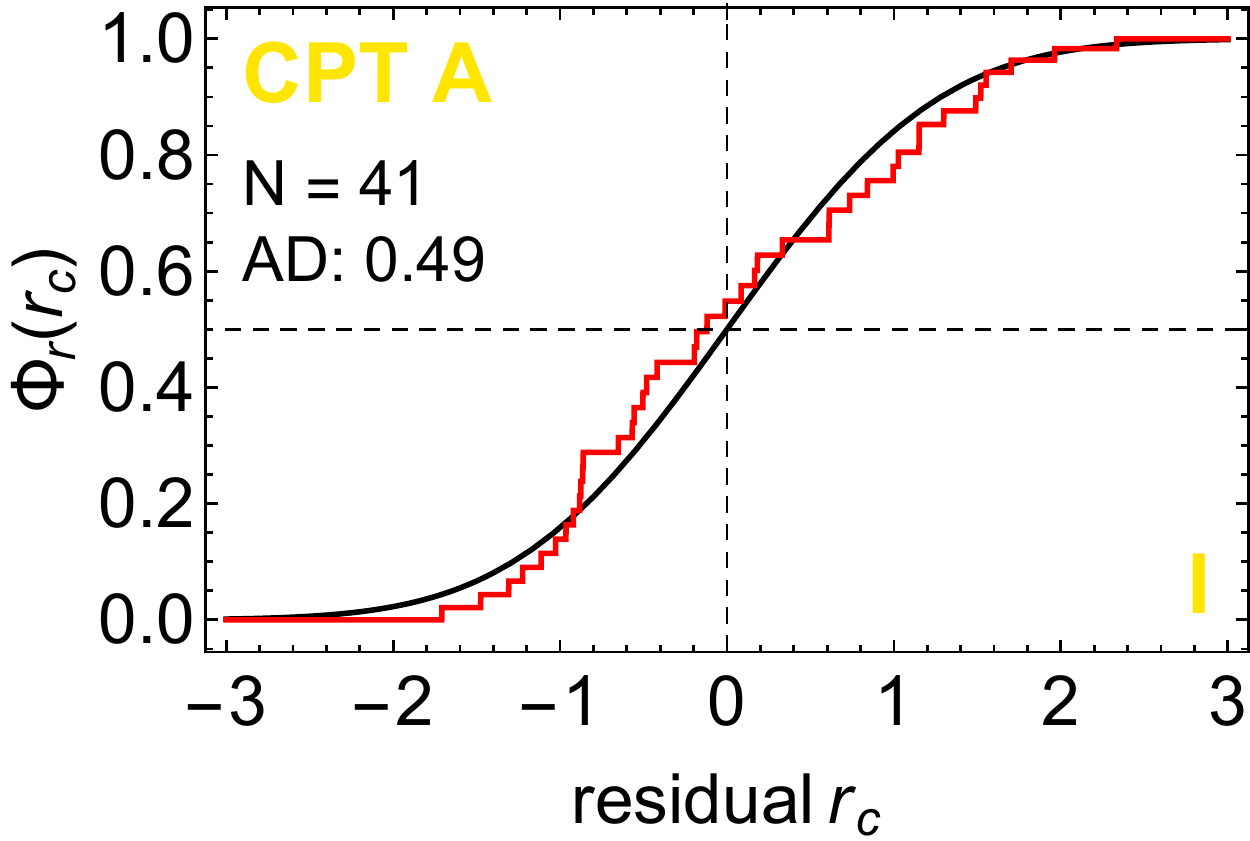}}
\end{minipage}
\hspace*{5mm}
\begin{minipage}{3.14cm}
\resizebox{2.826cm}{!}{\includegraphics{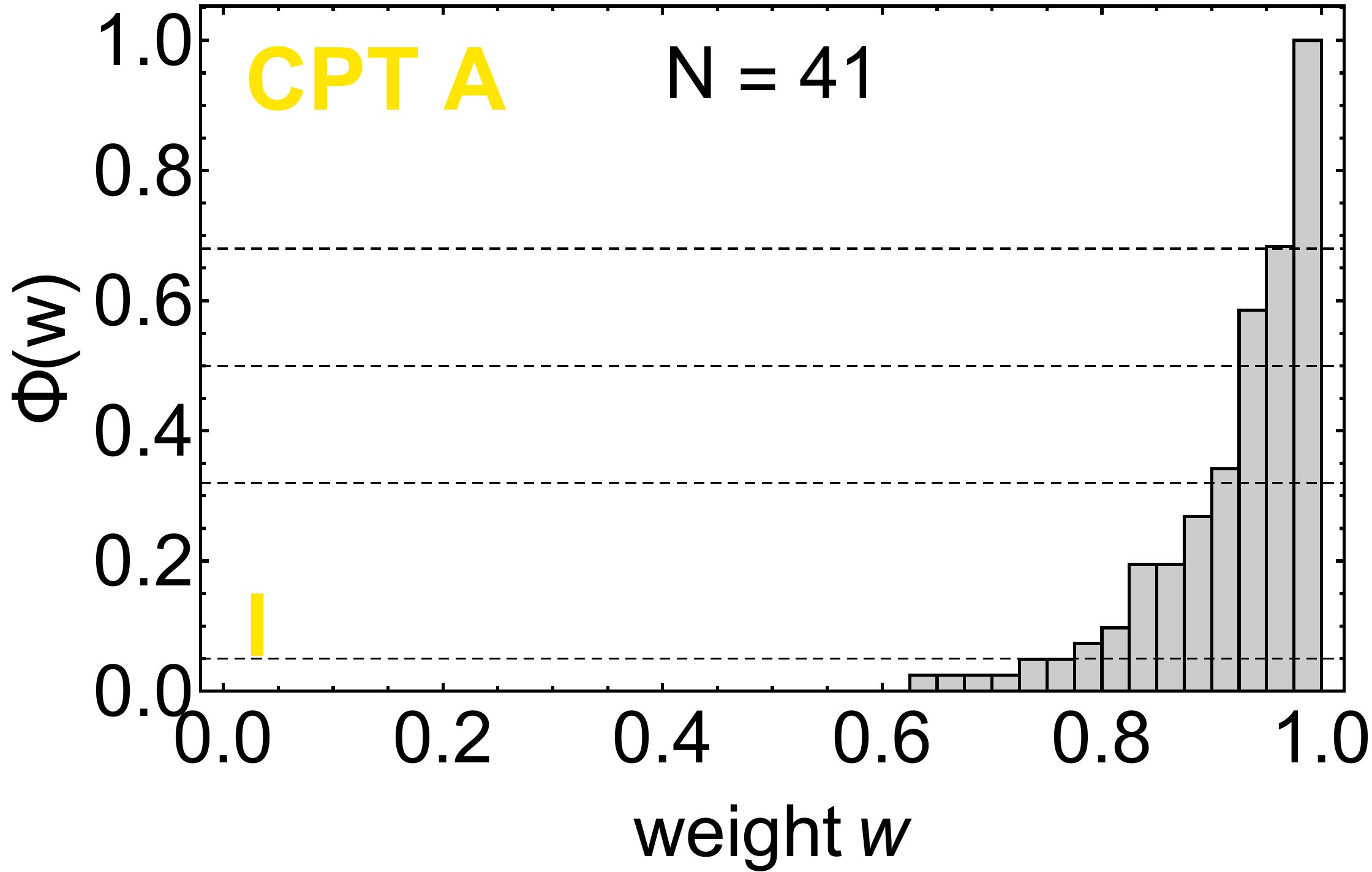}}
\end{minipage}
}\\[1mm]
\centering{\begin{minipage}{3cm}
\resizebox{2.7cm}{!}{\includegraphics{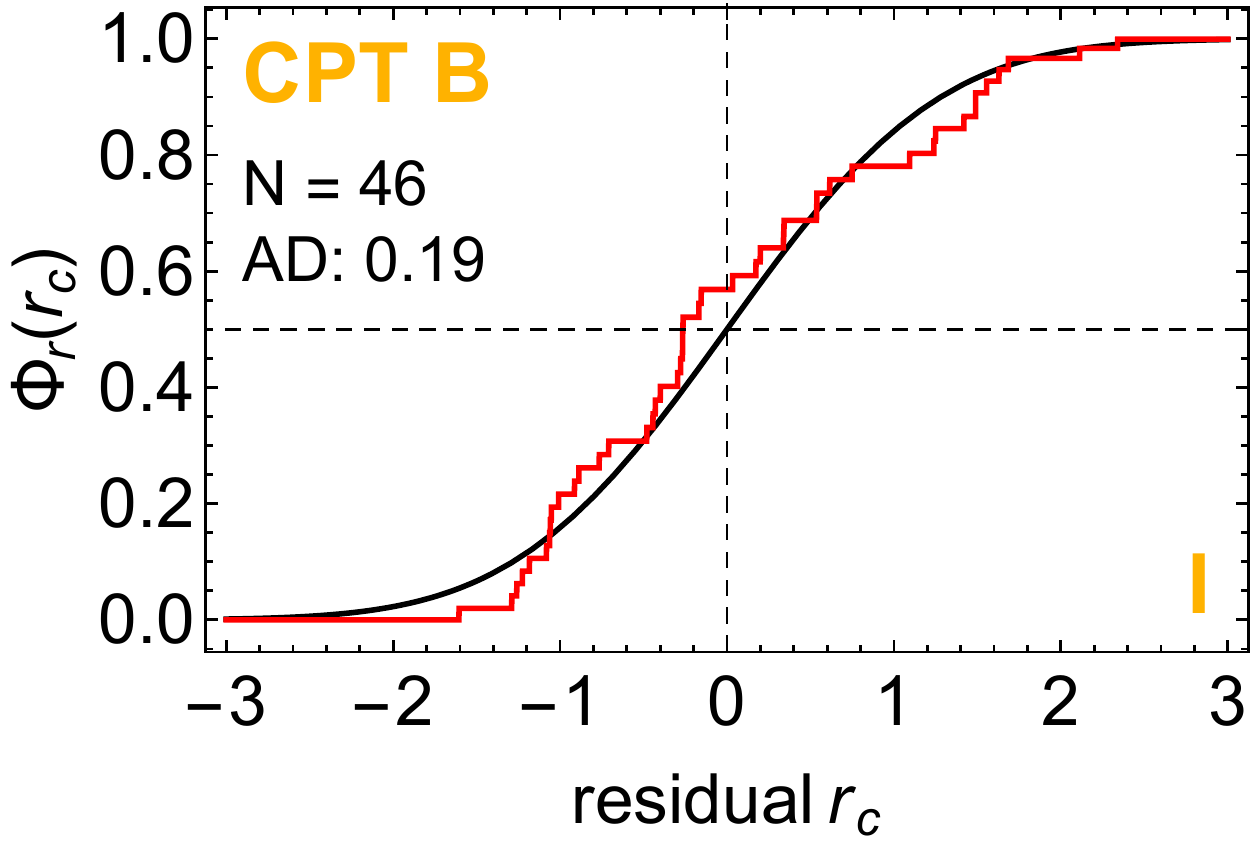}}
\end{minipage}
\hspace*{5mm}
\begin{minipage}{3.14cm}
\resizebox{2.826cm}{!}{\includegraphics{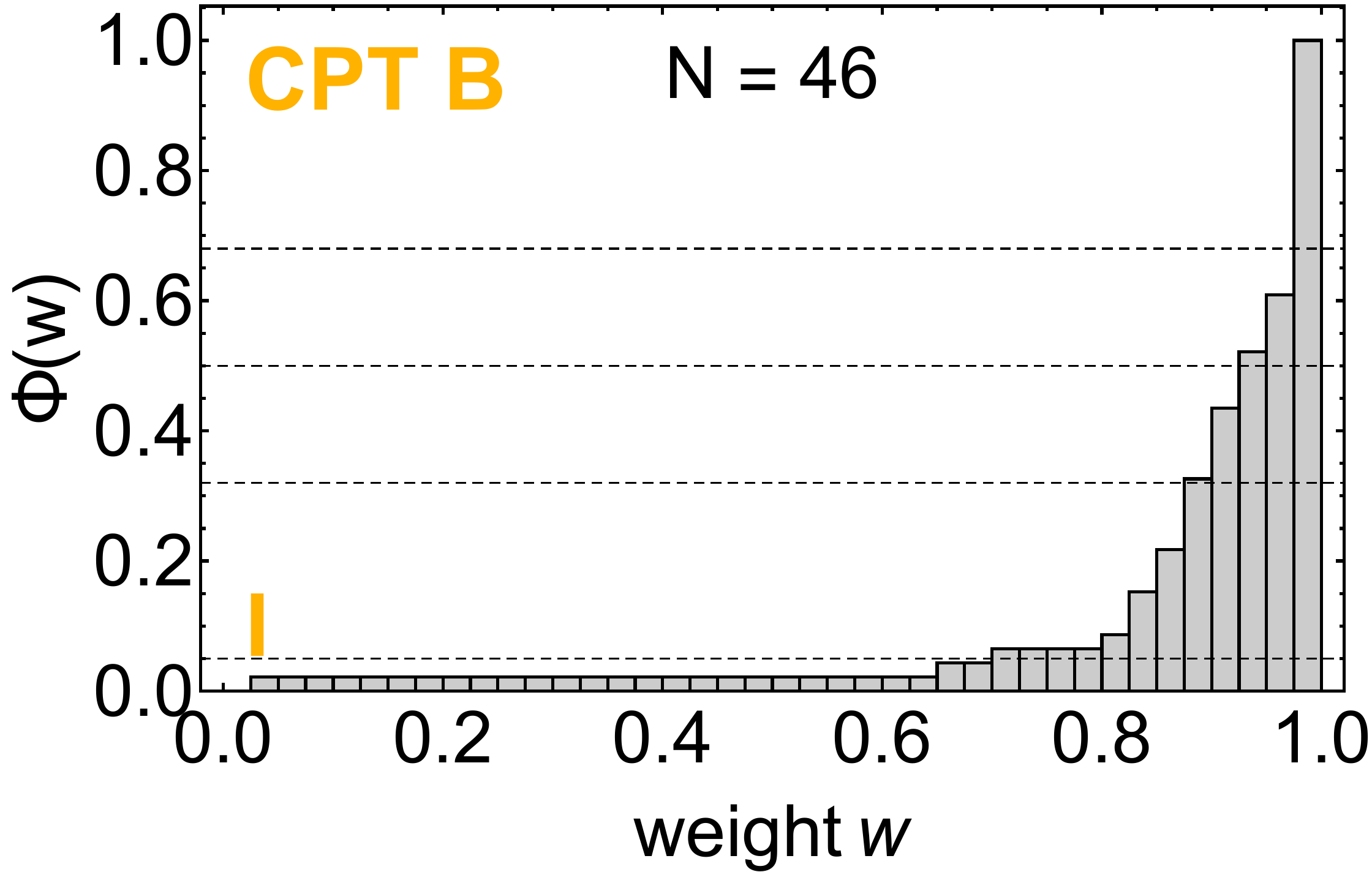}}
\end{minipage}
\hspace*{5mm}
\begin{minipage}{3cm}
\resizebox{2.7cm}{!}{\includegraphics{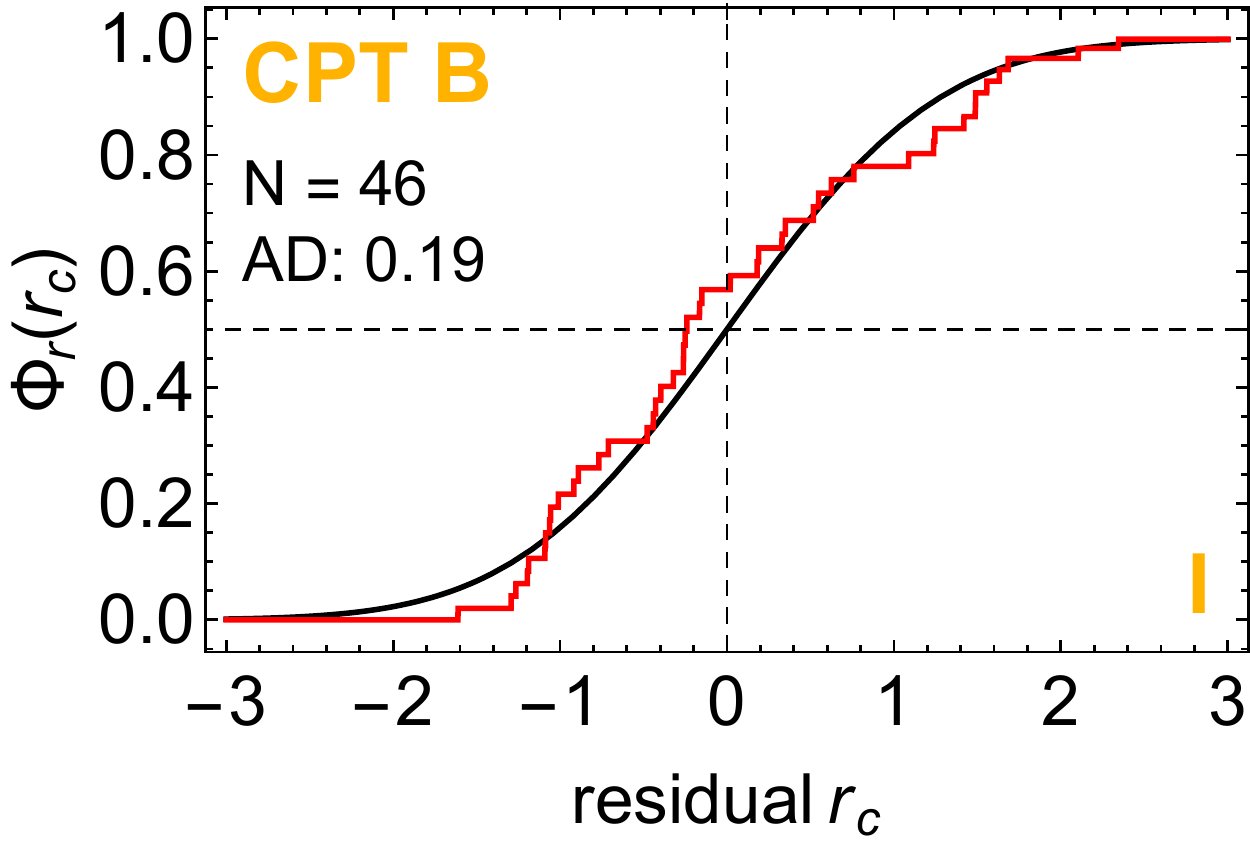}}
\end{minipage}
\hspace*{5mm}
\begin{minipage}{3.14cm}
\resizebox{2.826cm}{!}{\includegraphics{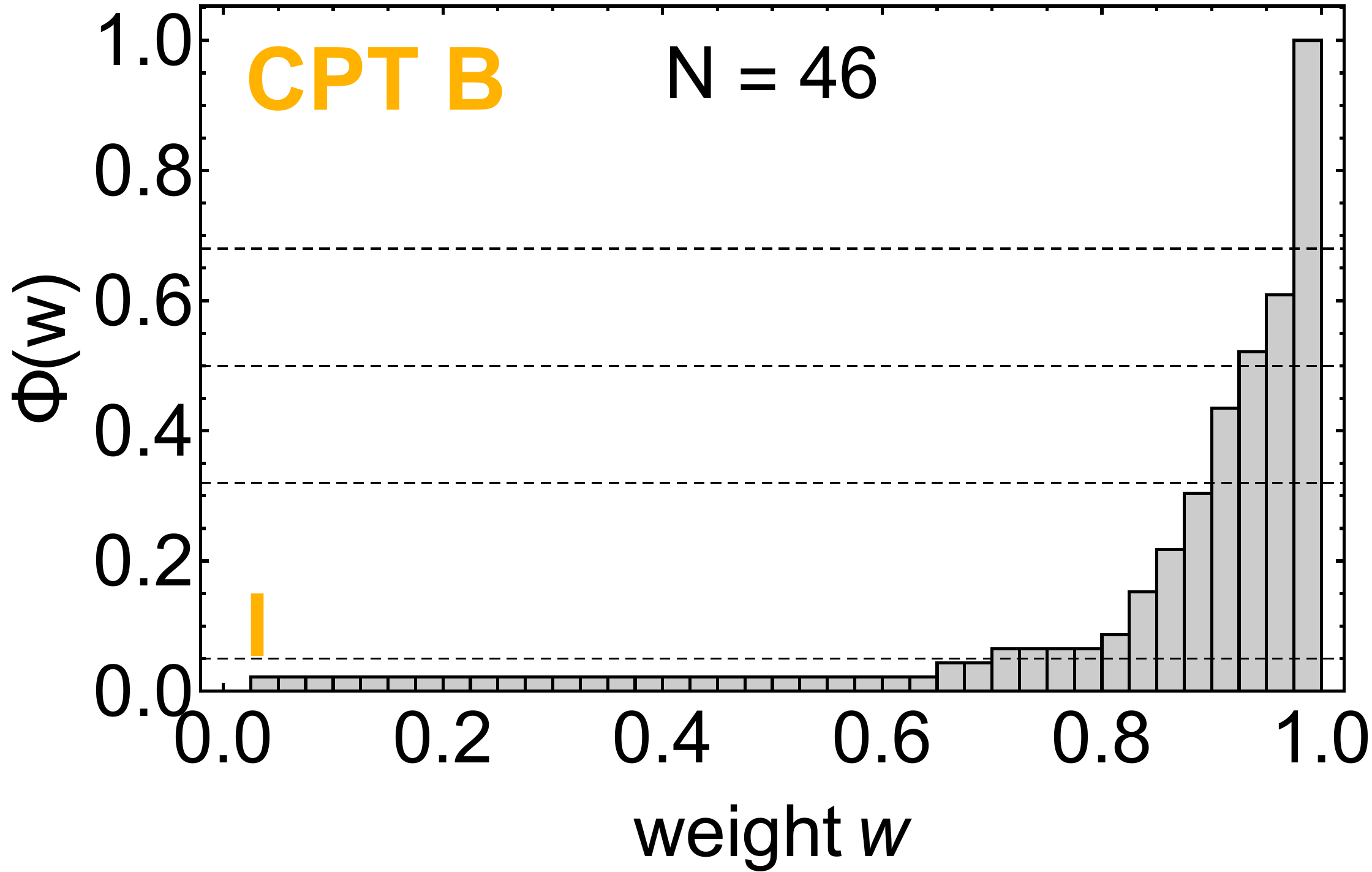}}
\end{minipage}
}\\[1mm]
\centering{\begin{minipage}{3cm}
\resizebox{2.7cm}{!}{\includegraphics{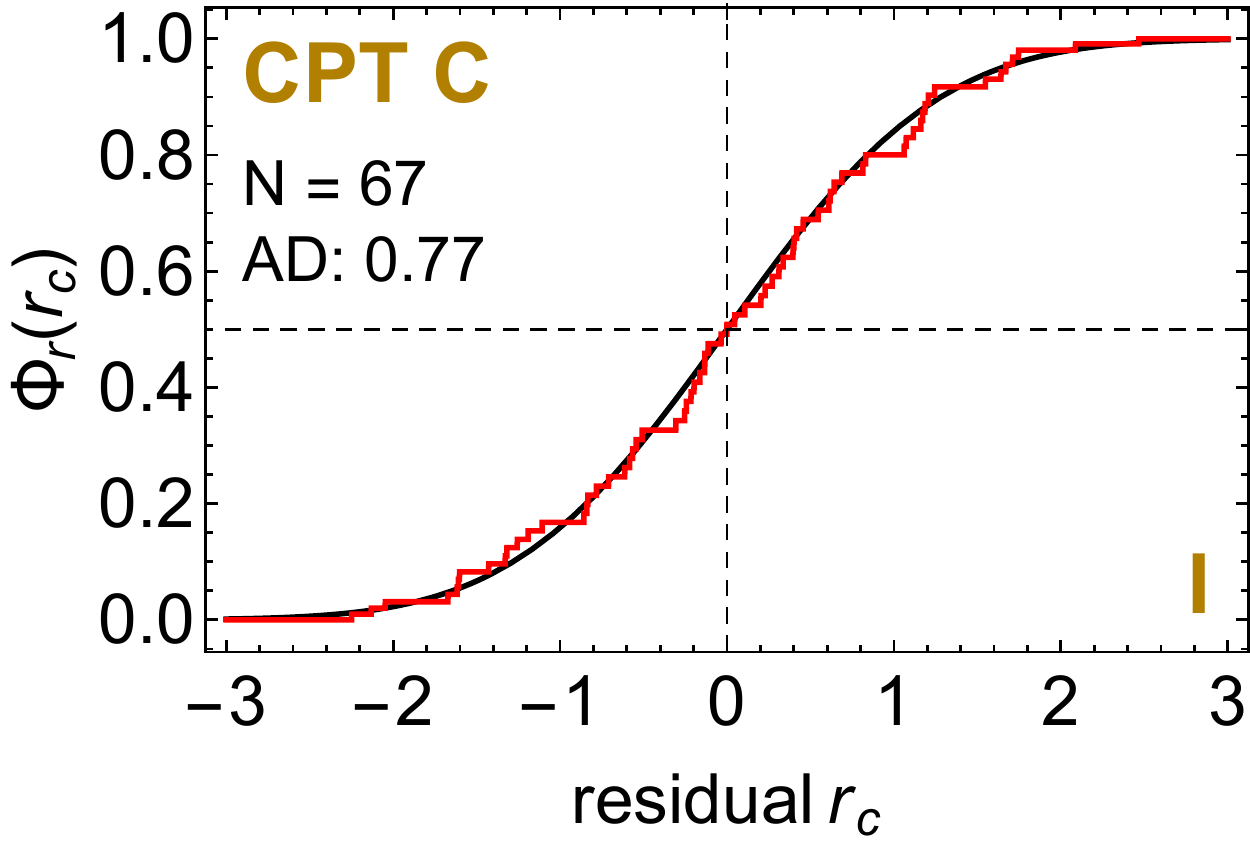}}
\end{minipage}
\hspace*{5mm}
\begin{minipage}{3.14cm}
\resizebox{2.826cm}{!}{\includegraphics{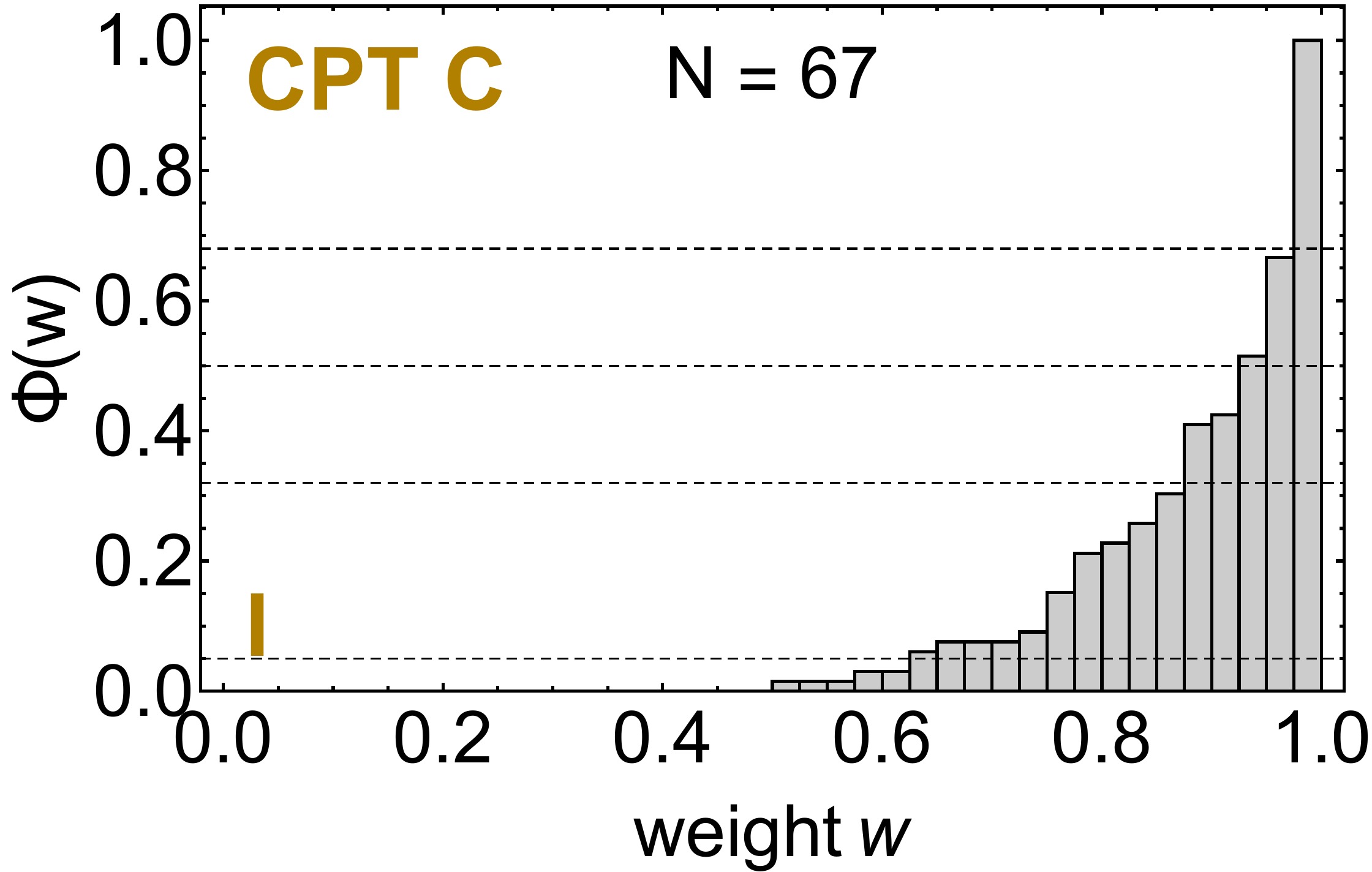}}
\end{minipage}
\hspace*{5mm}
\begin{minipage}{3cm}
\resizebox{2.7cm}{!}{\includegraphics{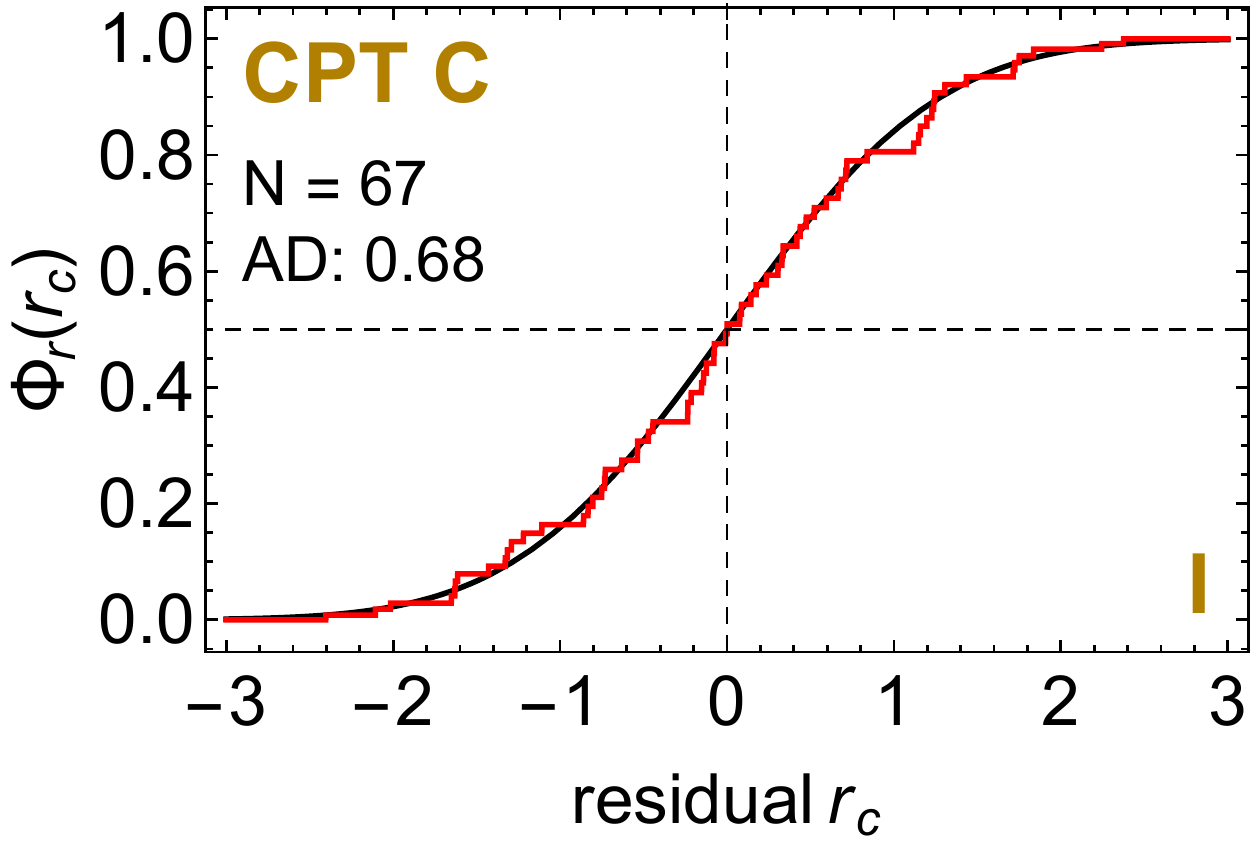}}
\end{minipage}
\hspace*{5mm}
\begin{minipage}{3.14cm}
\resizebox{2.826cm}{!}{\includegraphics{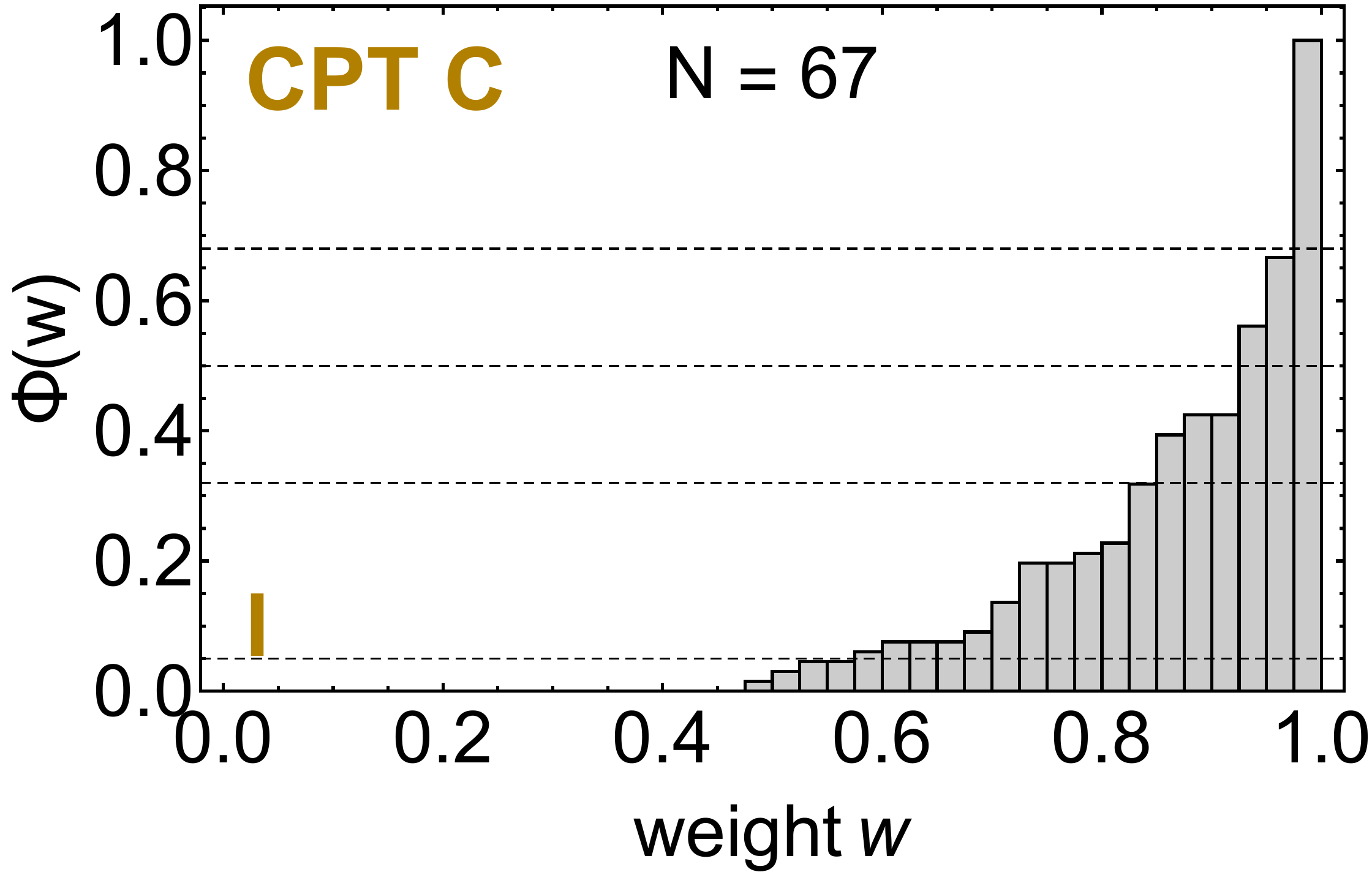}}
\end{minipage}
}\\[1mm]
\centering{\begin{minipage}{3cm}
\resizebox{2.7cm}{!}{\includegraphics{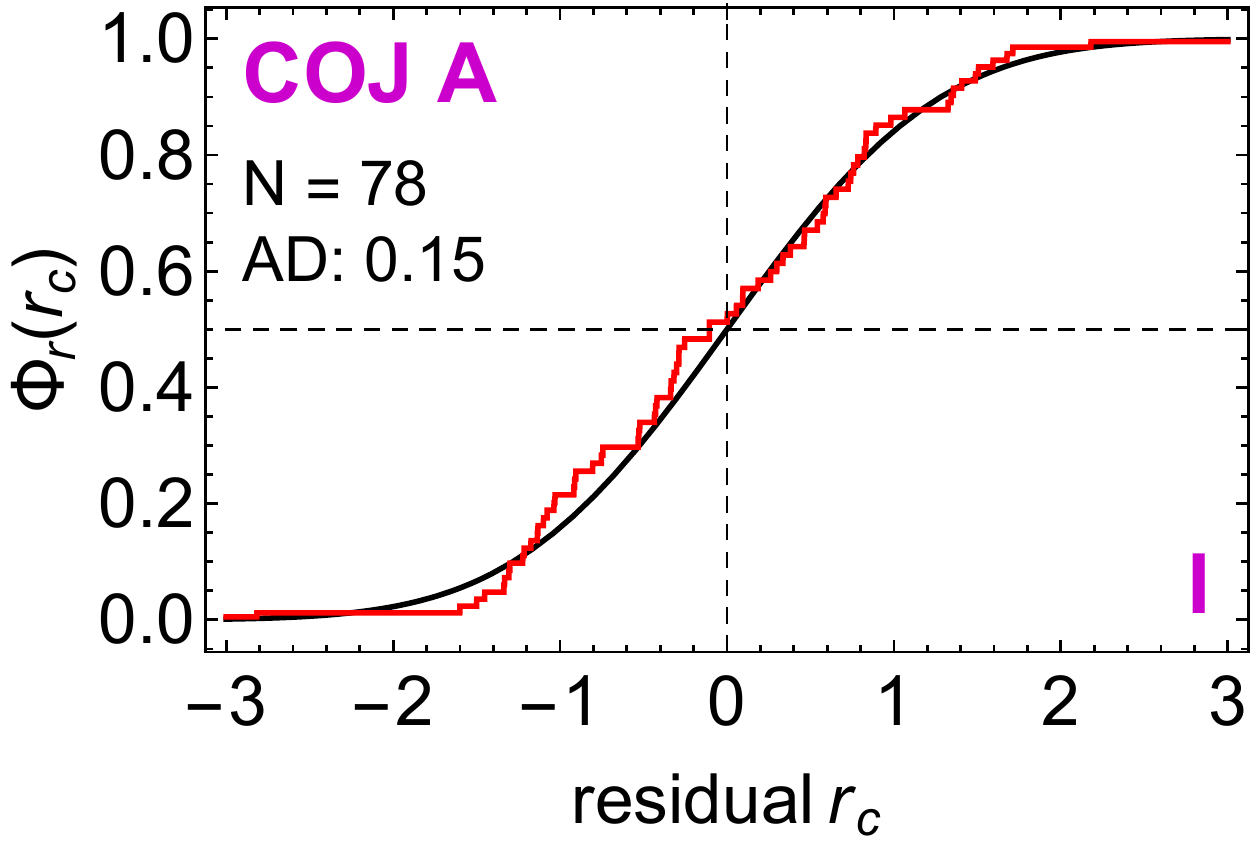}}
\end{minipage}
\hspace*{5mm}
\begin{minipage}{3.14cm}
\resizebox{2.826cm}{!}{\includegraphics{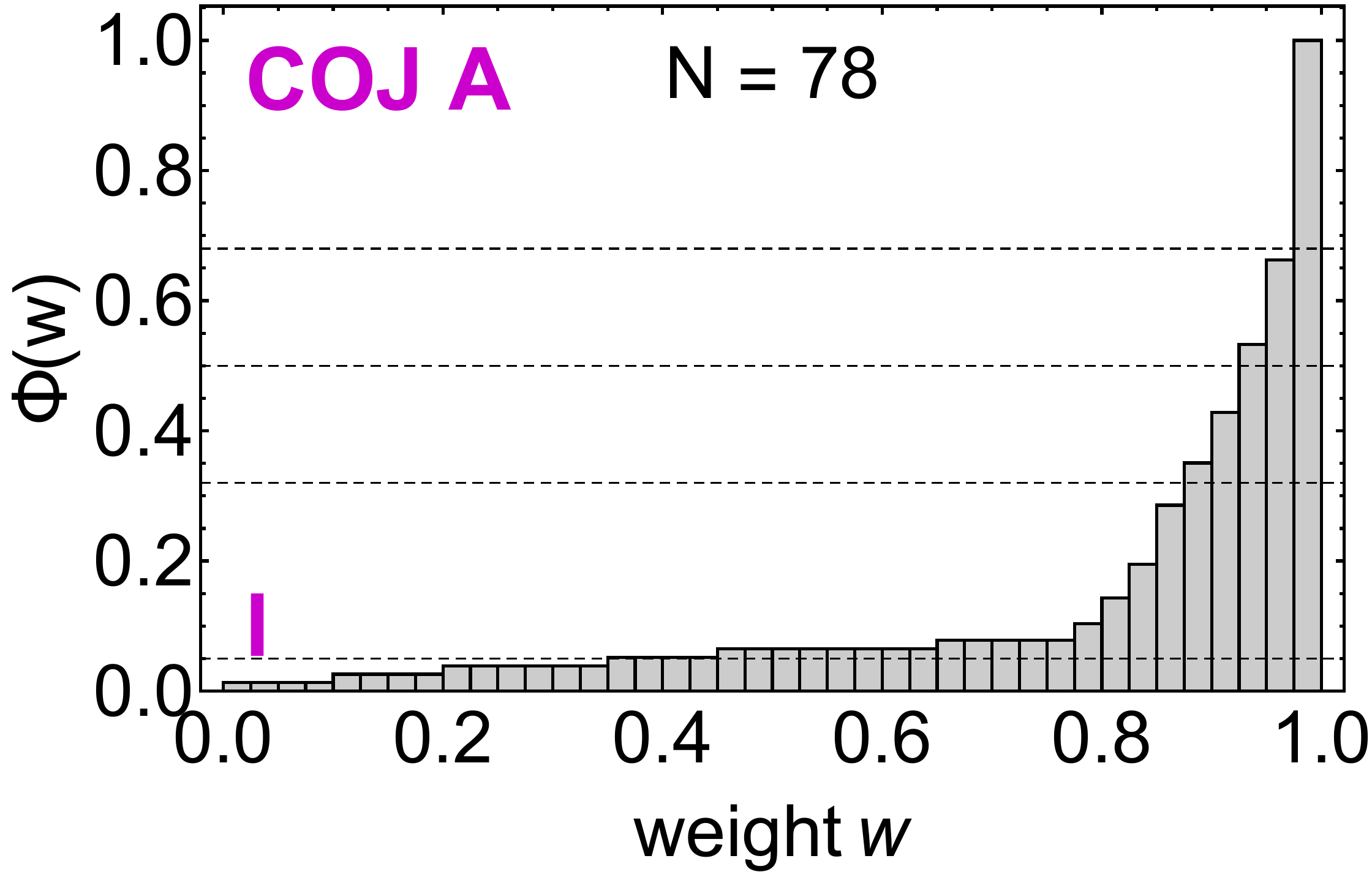}}
\end{minipage}
\hspace*{5mm}
\begin{minipage}{3cm}
\resizebox{2.7cm}{!}{\includegraphics{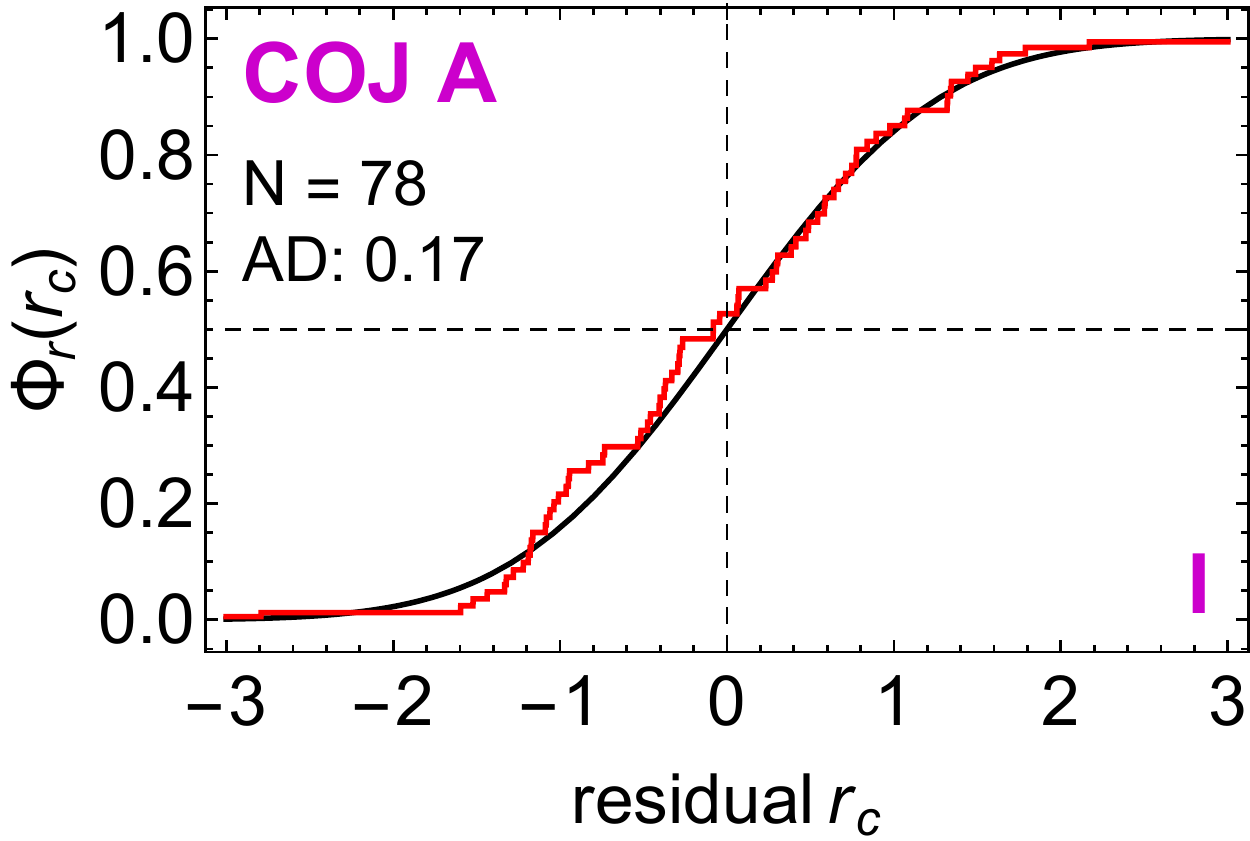}}
\end{minipage}
\hspace*{5mm}
\begin{minipage}{3.14cm}
\resizebox{2.826cm}{!}{\includegraphics{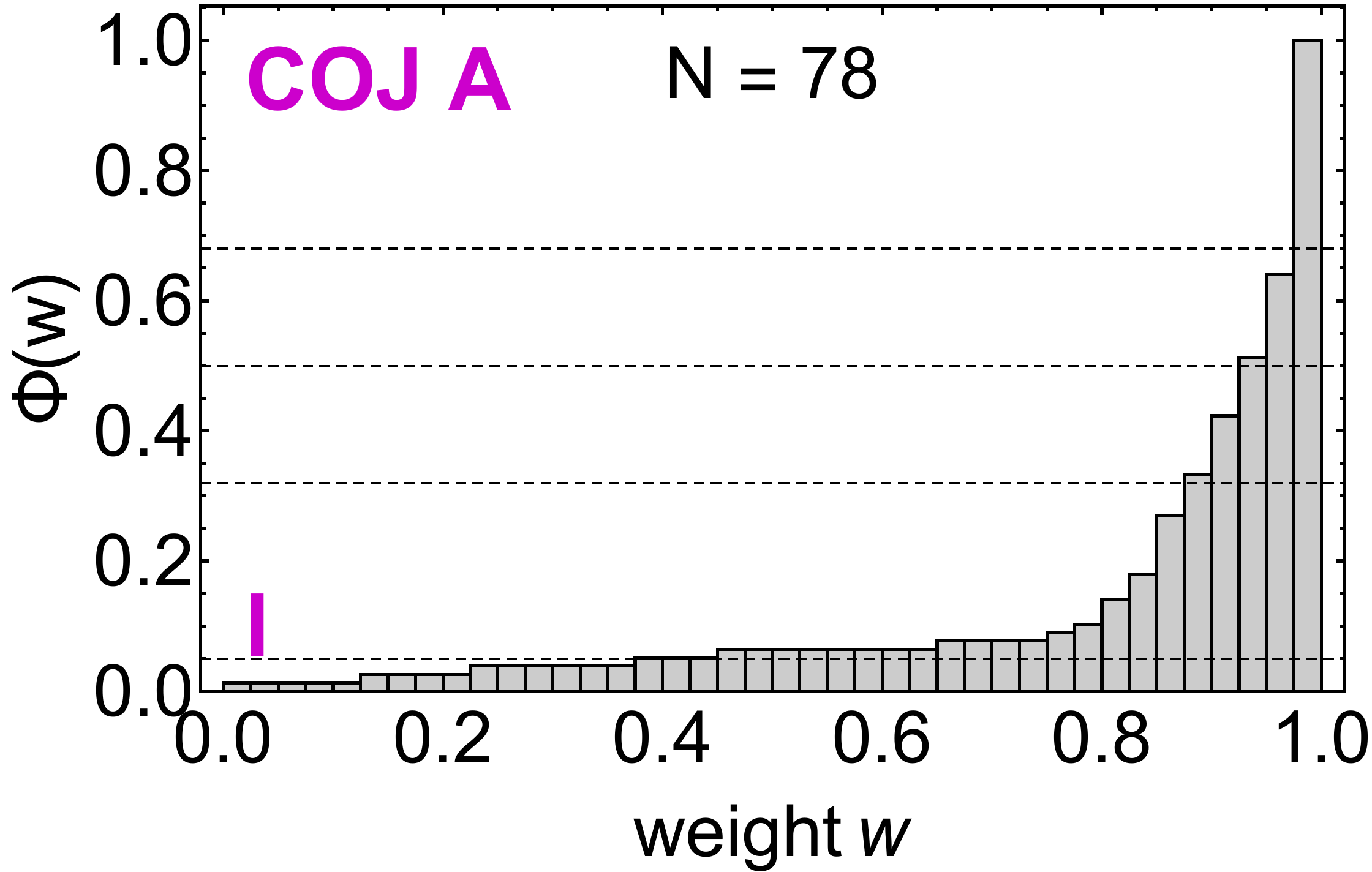}}
\end{minipage}
}\\[1mm]
\centering{\begin{minipage}{3cm}
\resizebox{2.7cm}{!}{\includegraphics{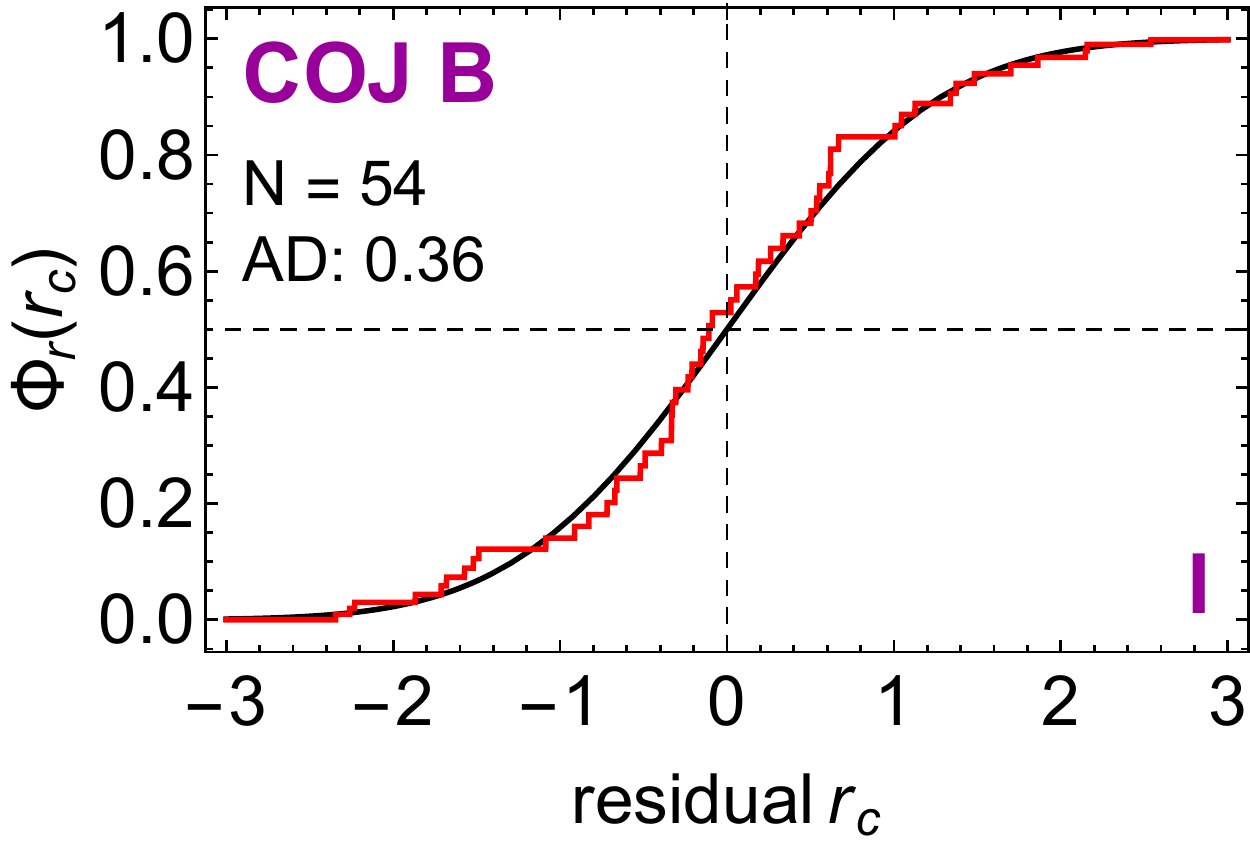}}
\end{minipage}
\hspace*{5mm}
\begin{minipage}{3.14cm}
\resizebox{2.826cm}{!}{\includegraphics{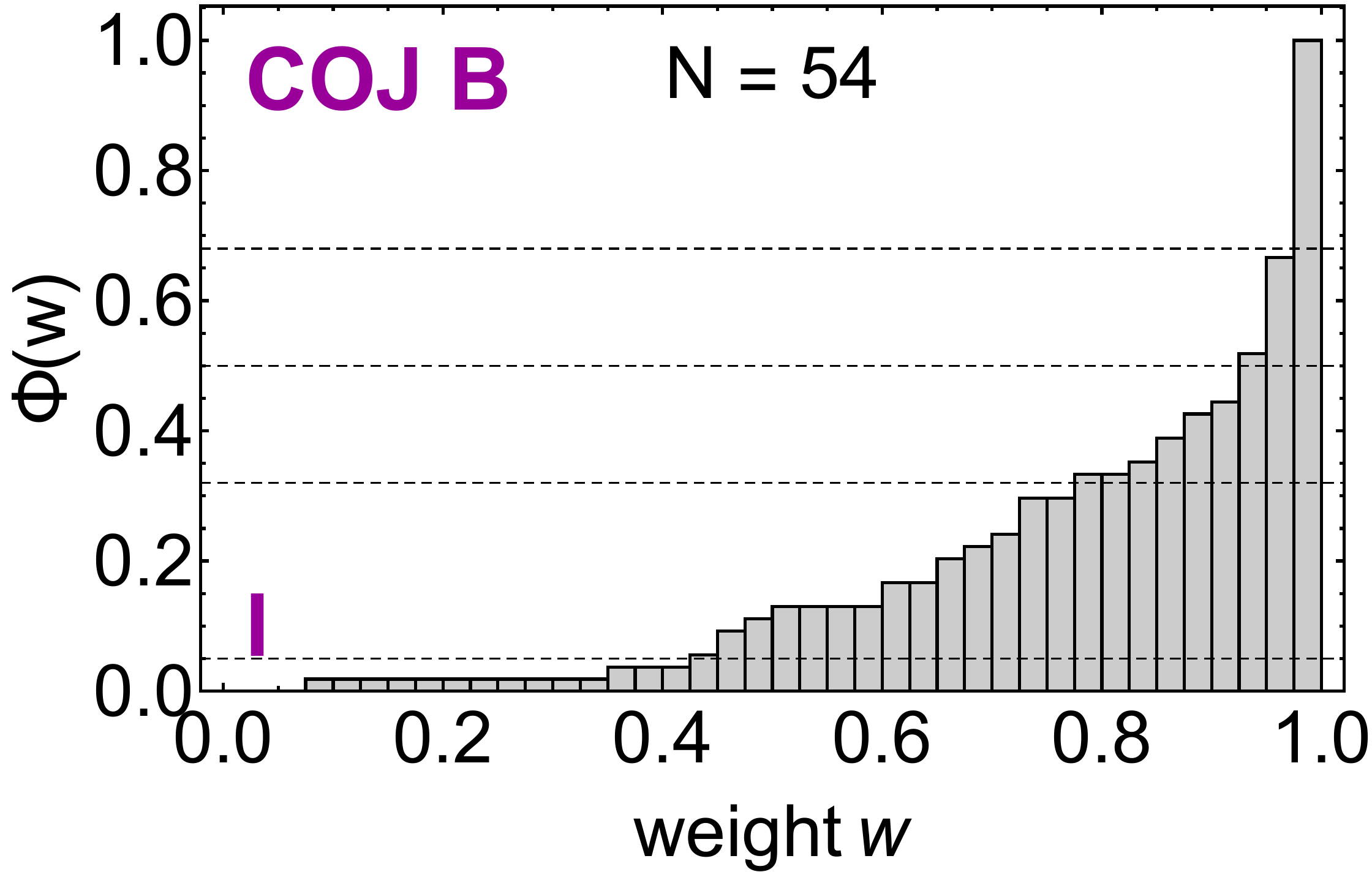}}
\end{minipage}
\hspace*{5mm}
\begin{minipage}{3cm}
\resizebox{2.7cm}{!}{\includegraphics{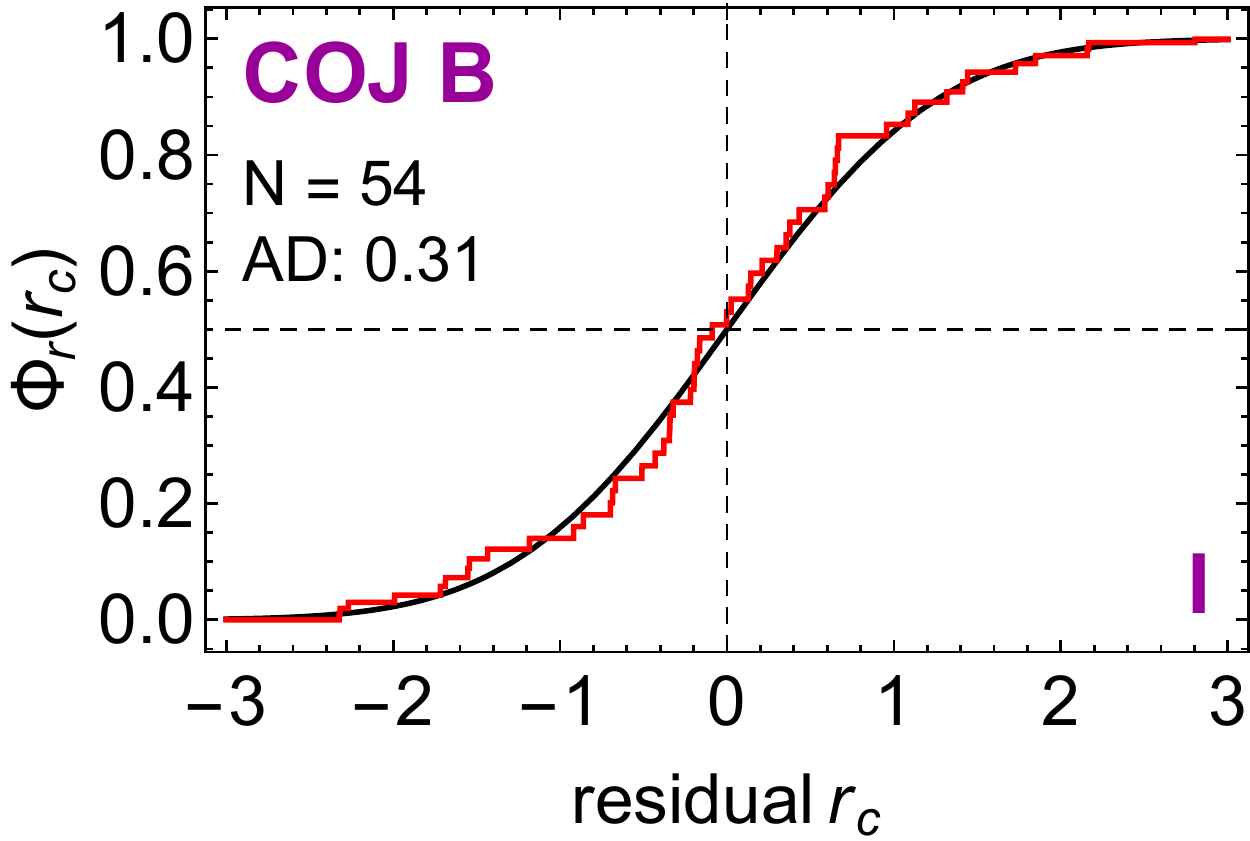}}
\end{minipage}
\hspace*{5mm}
\begin{minipage}{3.14cm}
\resizebox{2.826cm}{!}{\includegraphics{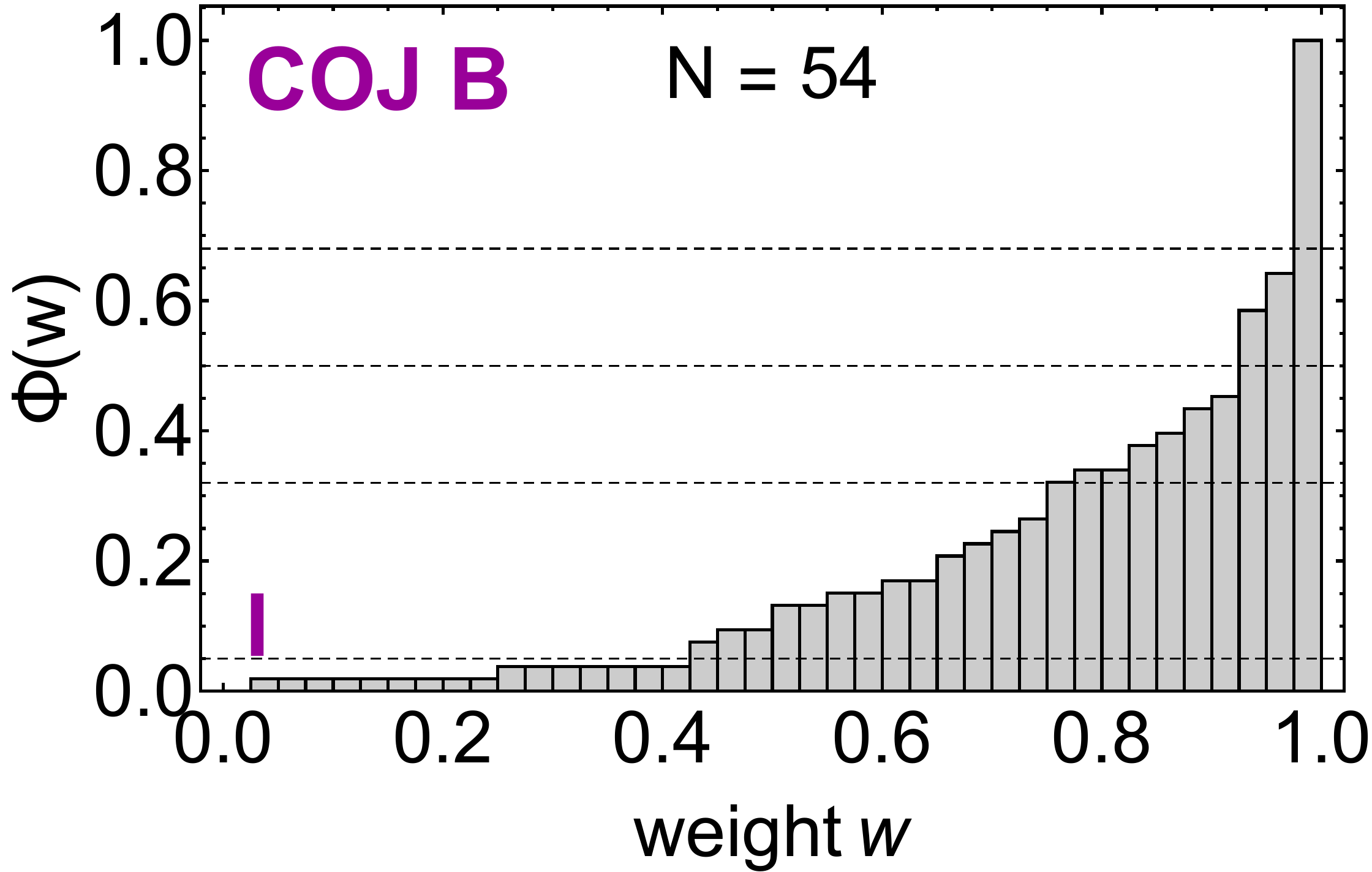}}
\end{minipage}
}\\[1mm]
\centering{\begin{minipage}{3cm}
\resizebox{2.7cm}{!}{\includegraphics{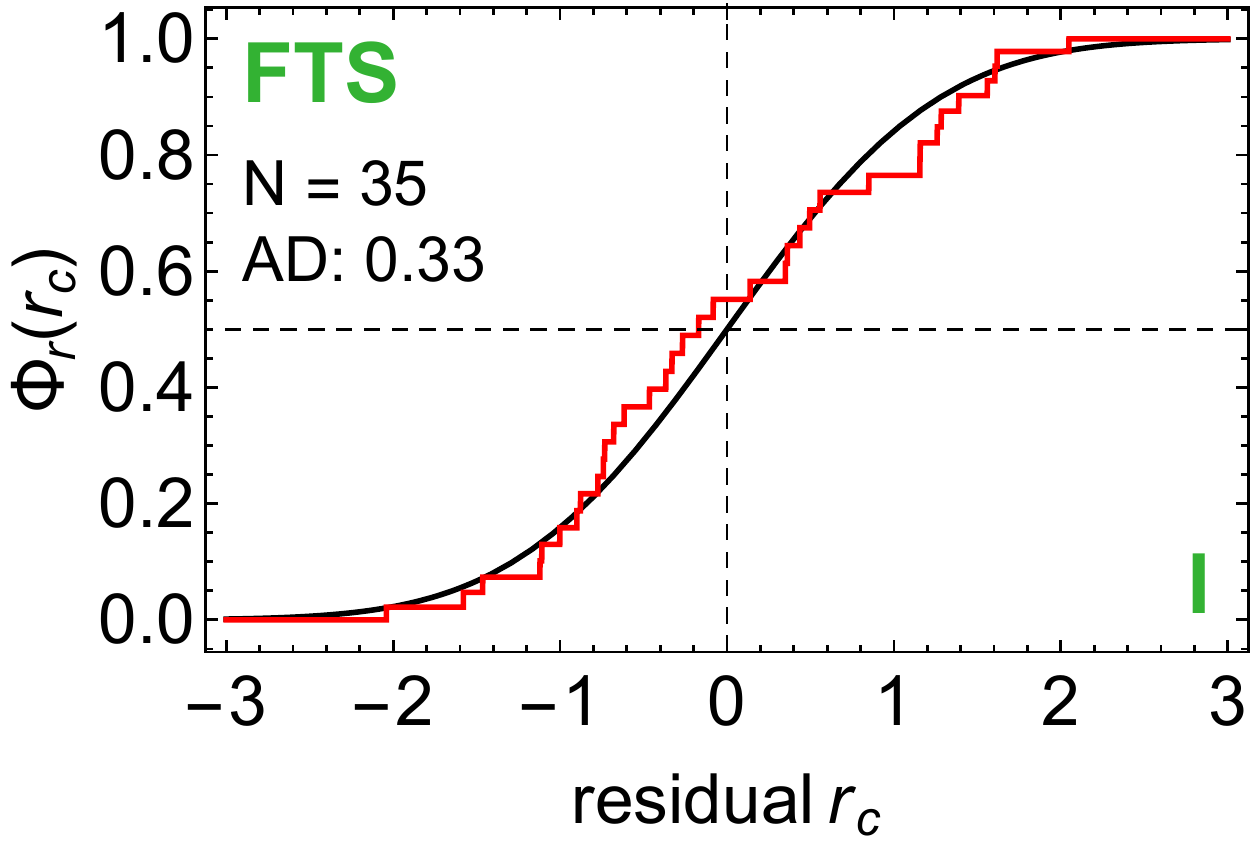}}
\end{minipage}
\hspace*{5mm}
\begin{minipage}{3.14cm}
\resizebox{2.826cm}{!}{\includegraphics{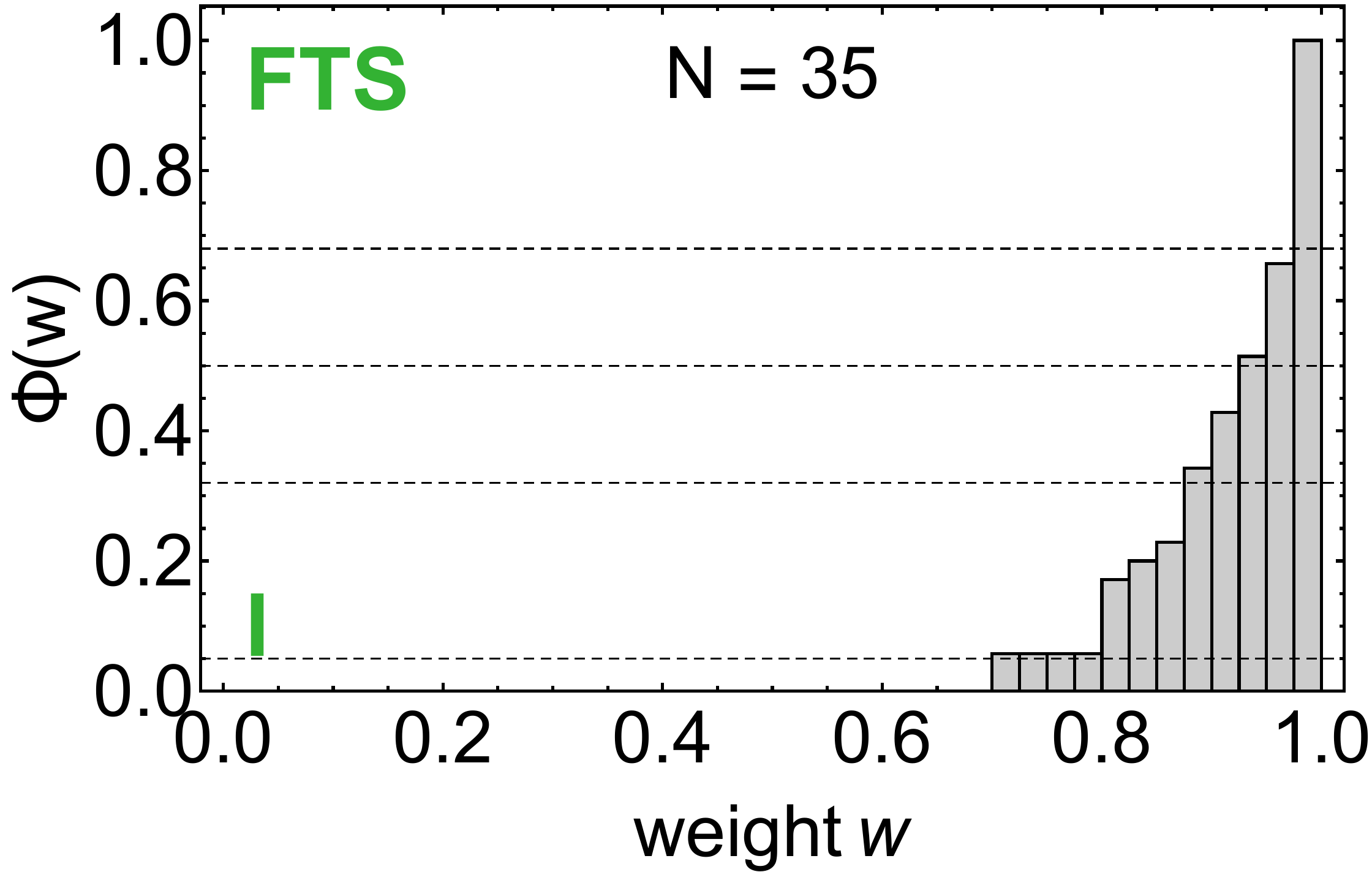}}
\end{minipage}
\hspace*{5mm}
\begin{minipage}{3cm}
\resizebox{2.7cm}{!}{\includegraphics{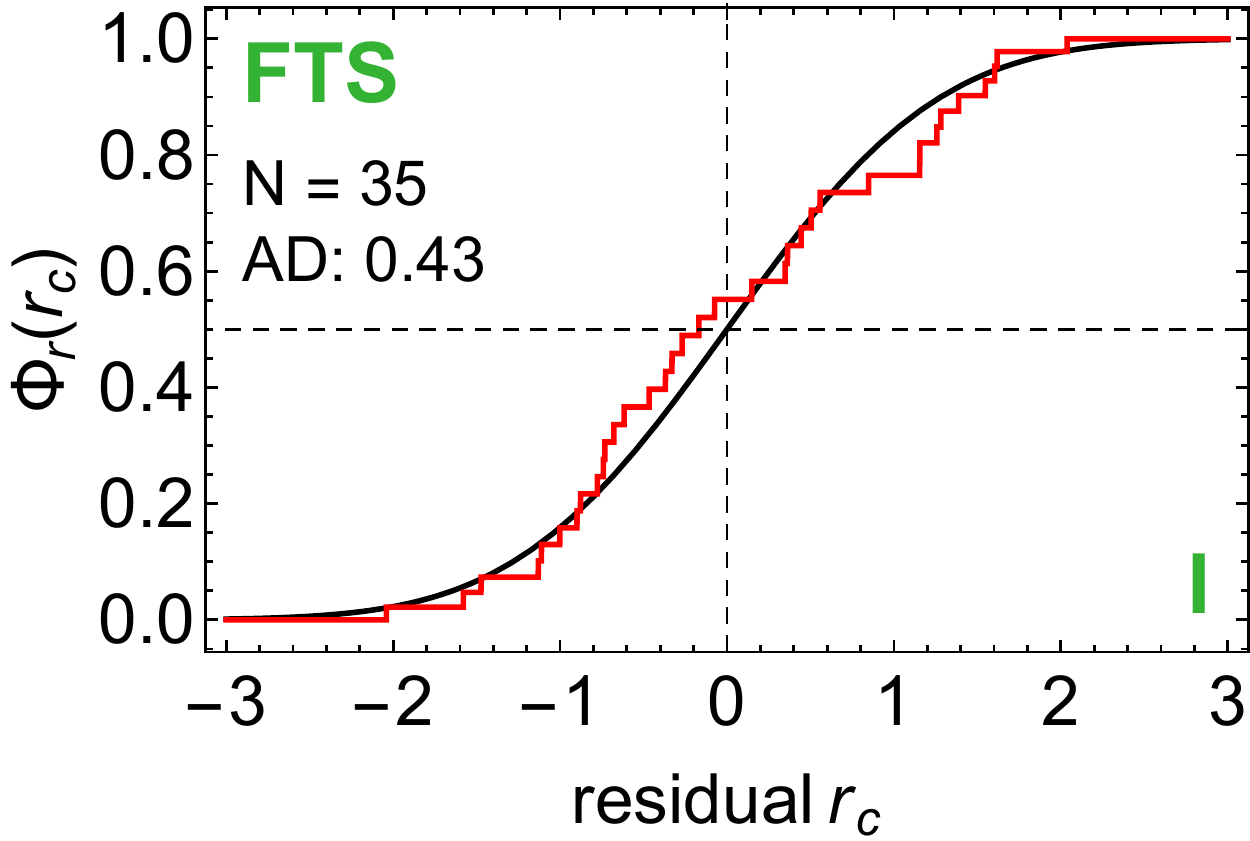}}
\end{minipage}
\hspace*{5mm}
\begin{minipage}{3.14cm}
\resizebox{2.826cm}{!}{\includegraphics{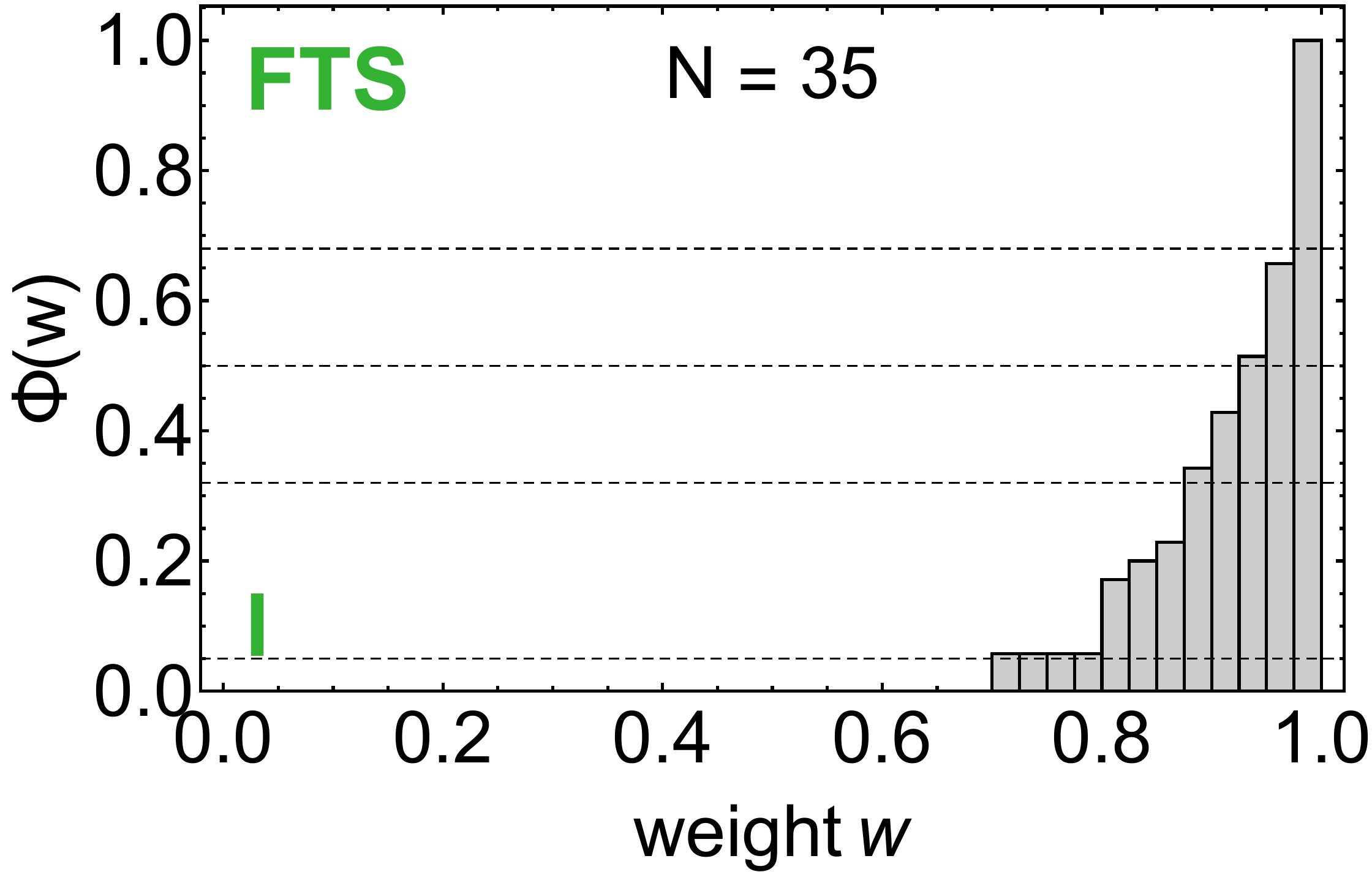}}
\end{minipage}
}\\[1mm]
\centering{\begin{minipage}{3cm}
\resizebox{2.7cm}{!}{\includegraphics{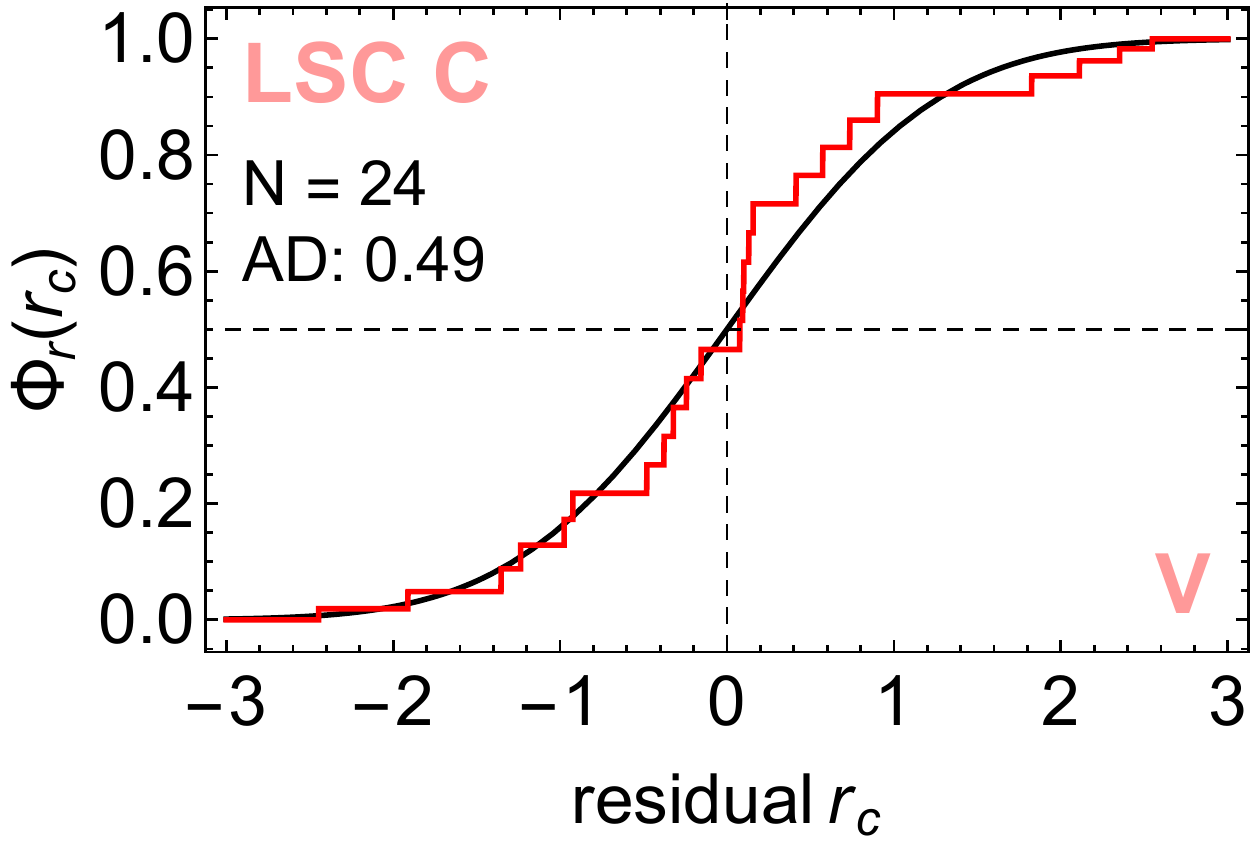}}
\end{minipage}
\hspace*{5mm}
\begin{minipage}{3.14cm}
\resizebox{2.826cm}{!}{\includegraphics{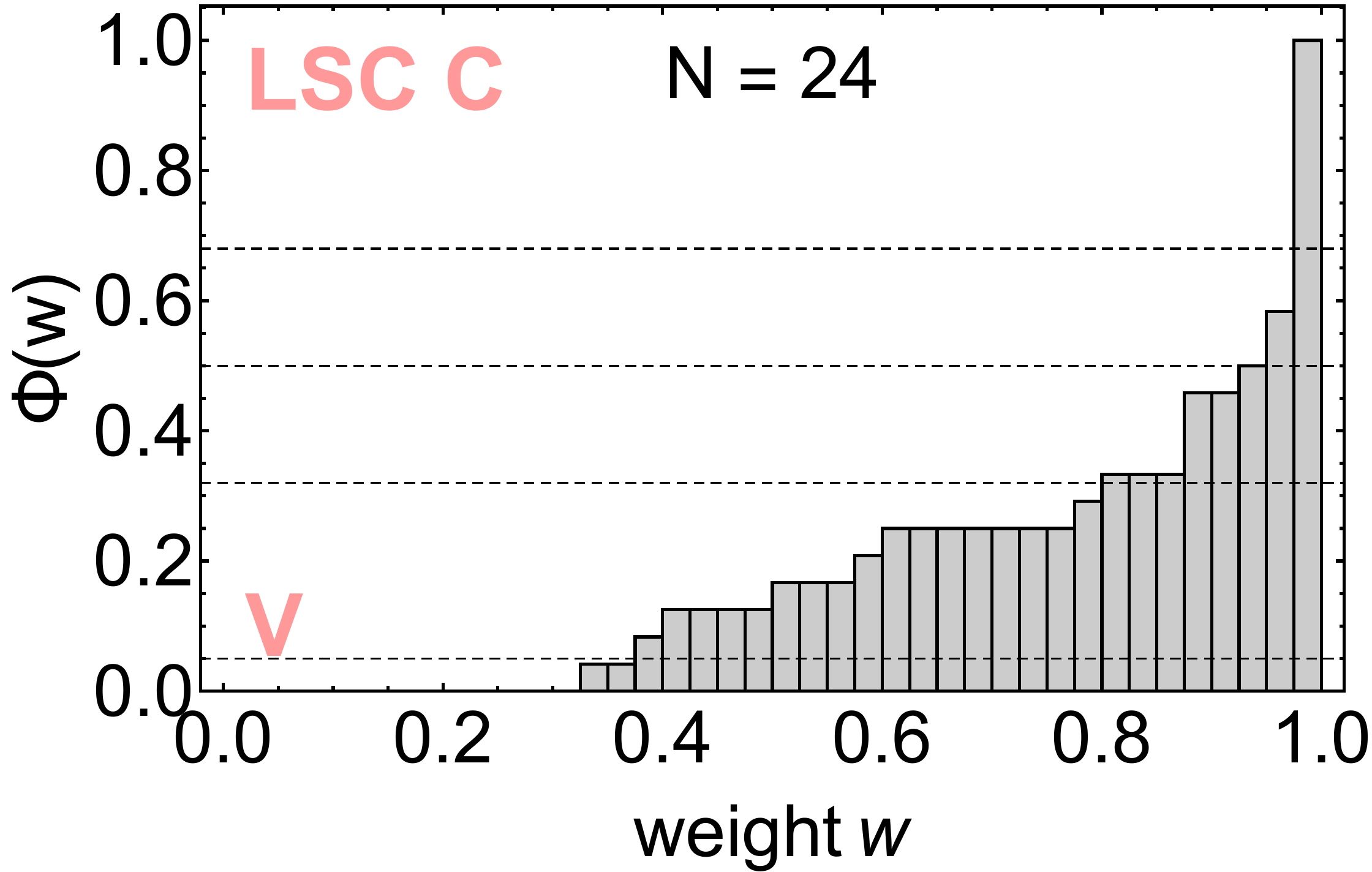}}
\end{minipage}
\hspace*{5mm}
\begin{minipage}{3cm}
\resizebox{2.7cm}{!}{\includegraphics{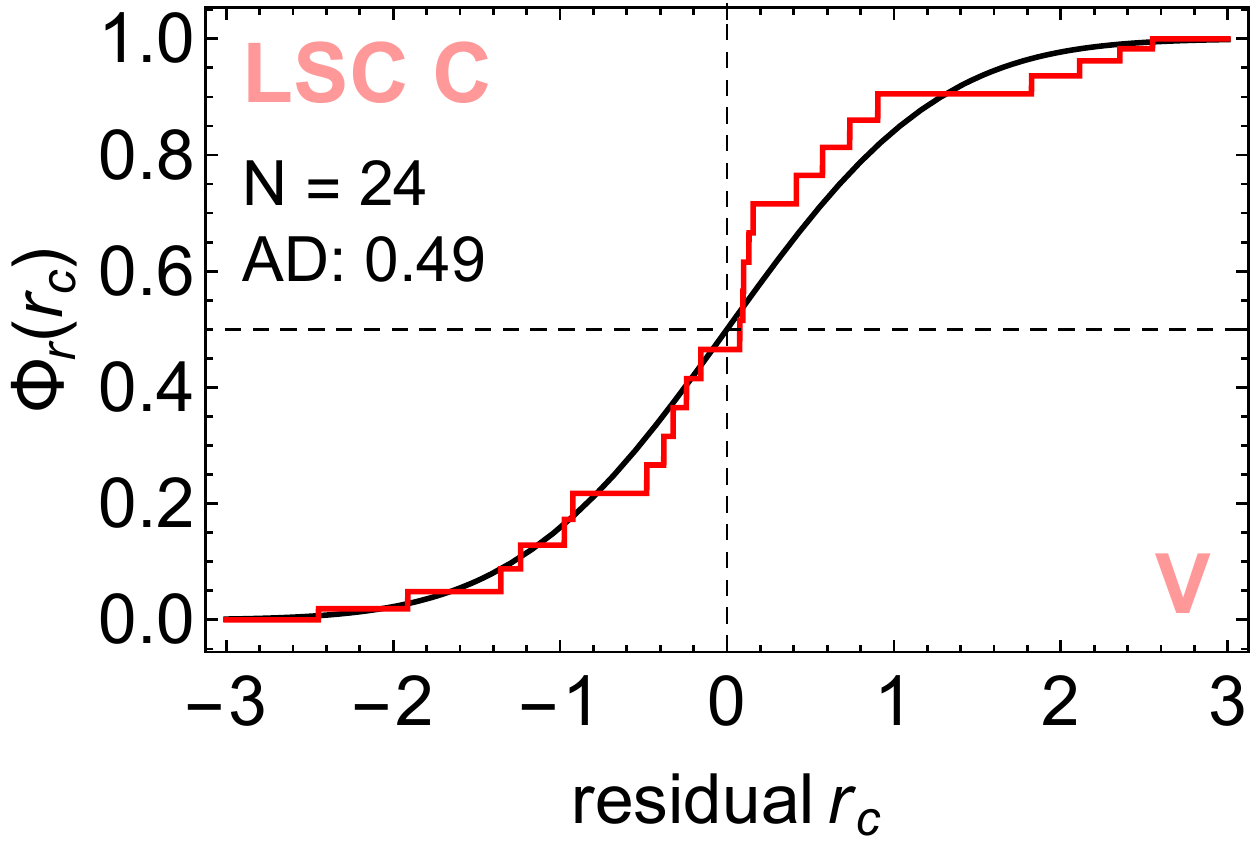}}
\end{minipage}
\hspace*{5mm}
\begin{minipage}{3.14cm}
\resizebox{2.826cm}{!}{\includegraphics{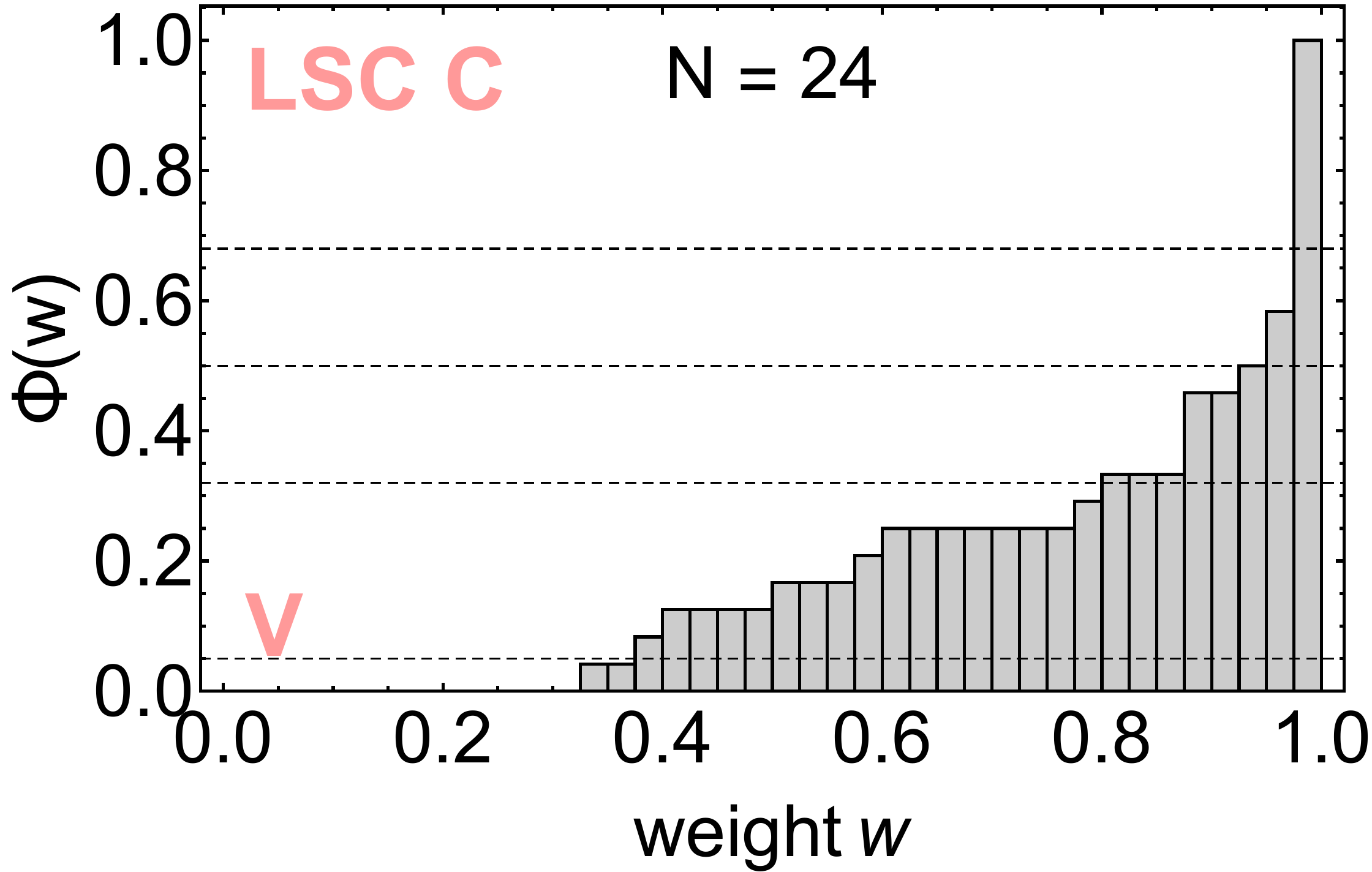}}
\end{minipage}
}\\[1mm]
\centering{\begin{minipage}{3cm}
\resizebox{2.7cm}{!}{\includegraphics{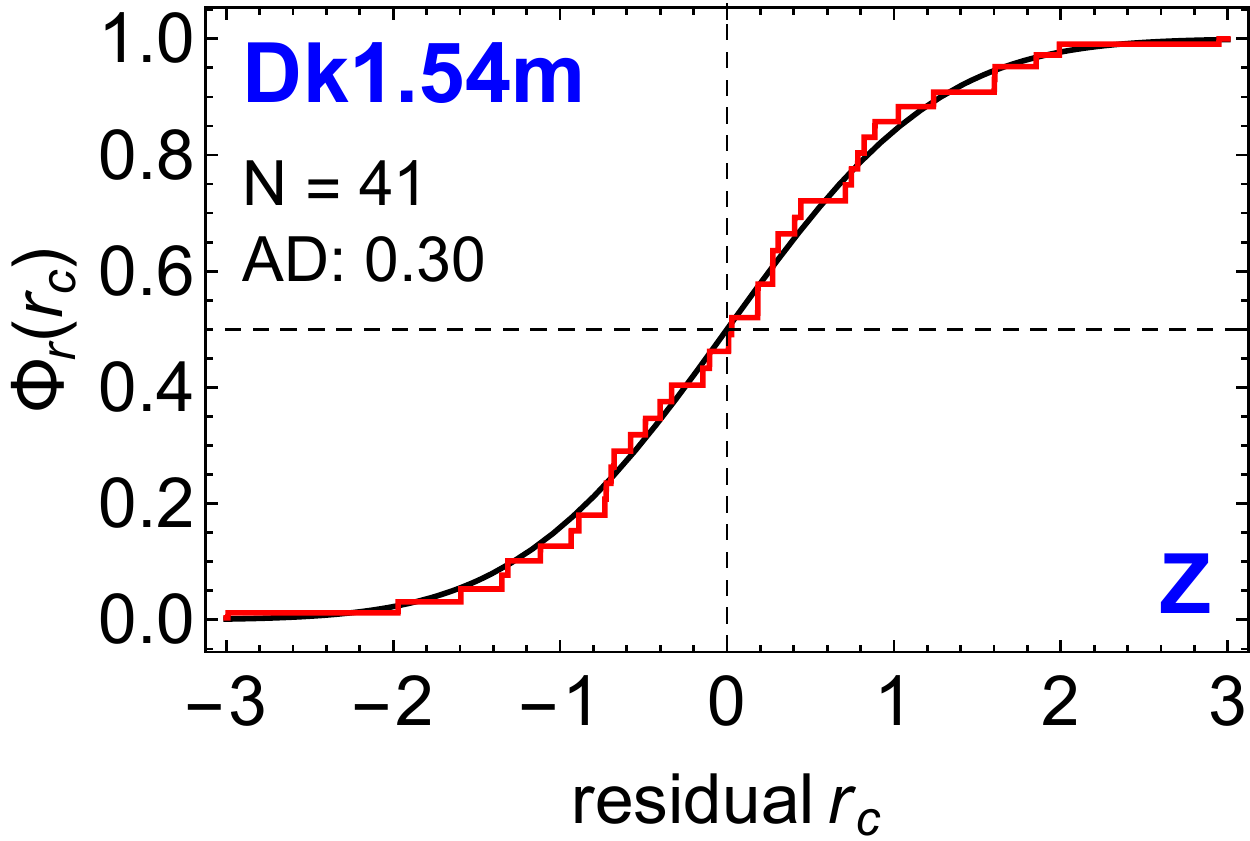}}
\end{minipage}
\hspace*{5mm}
\begin{minipage}{3.14cm}
\resizebox{2.826cm}{!}{\includegraphics{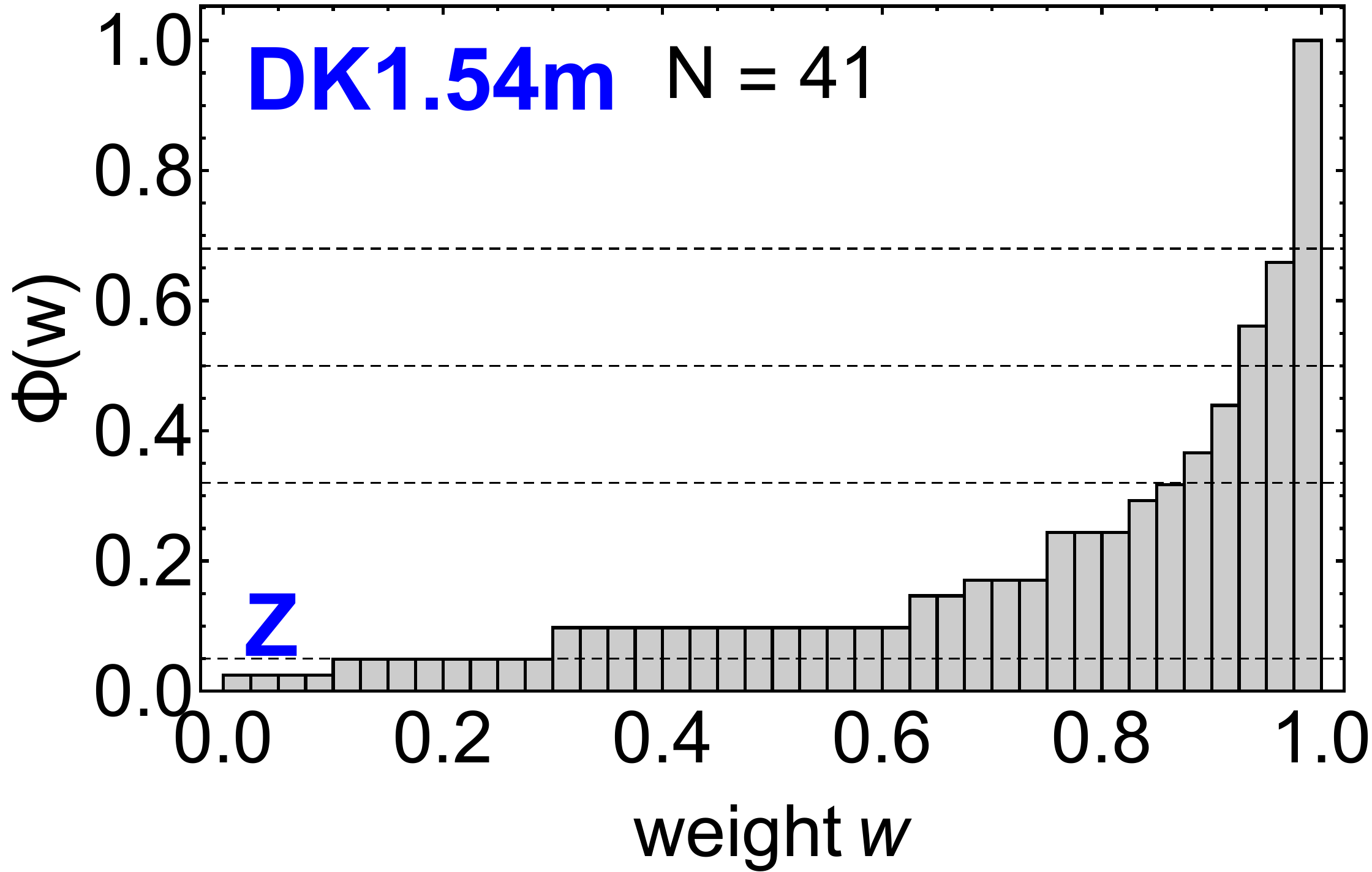}}
\end{minipage}
\hspace*{5mm}
\begin{minipage}{3cm}
\resizebox{2.7cm}{!}{\includegraphics{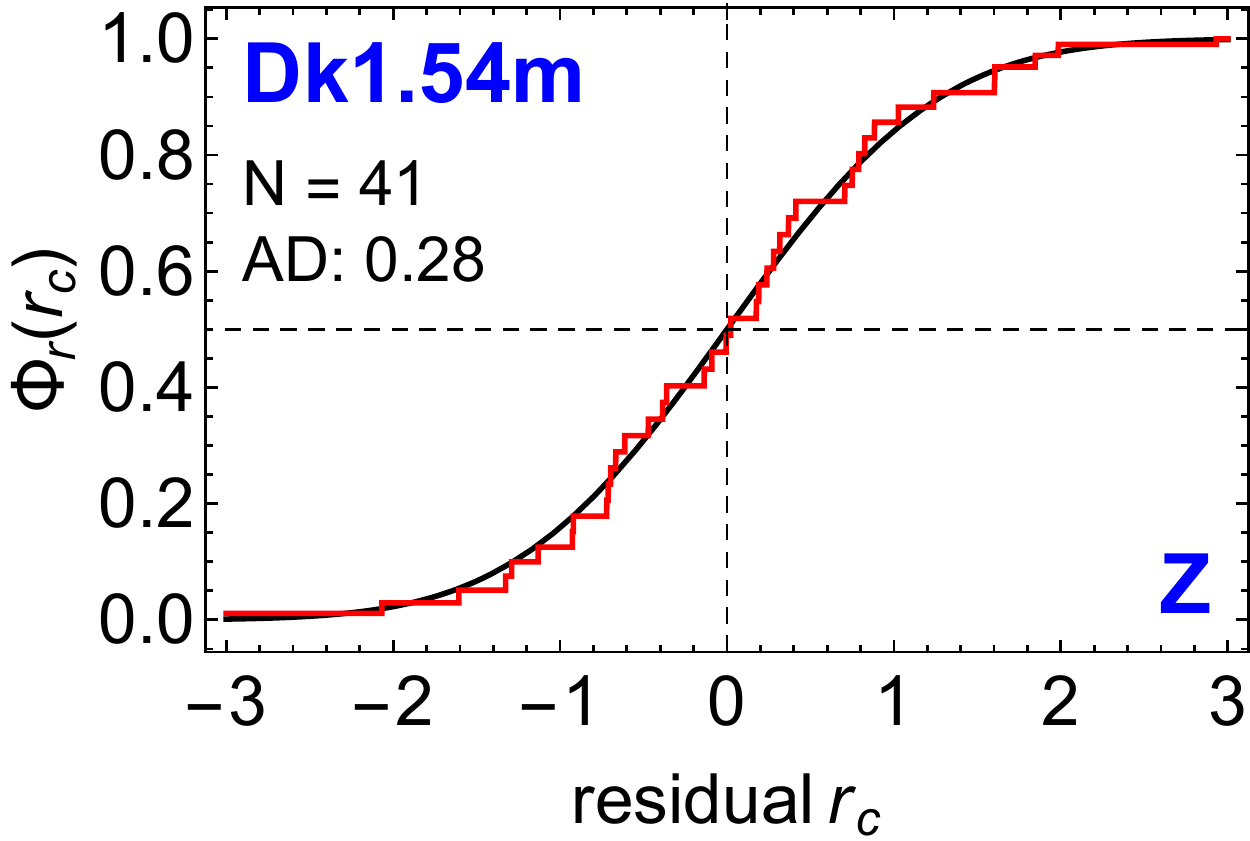}}
\end{minipage}
\hspace*{5mm}
\begin{minipage}{3.14cm}
\resizebox{2.826cm}{!}{\includegraphics{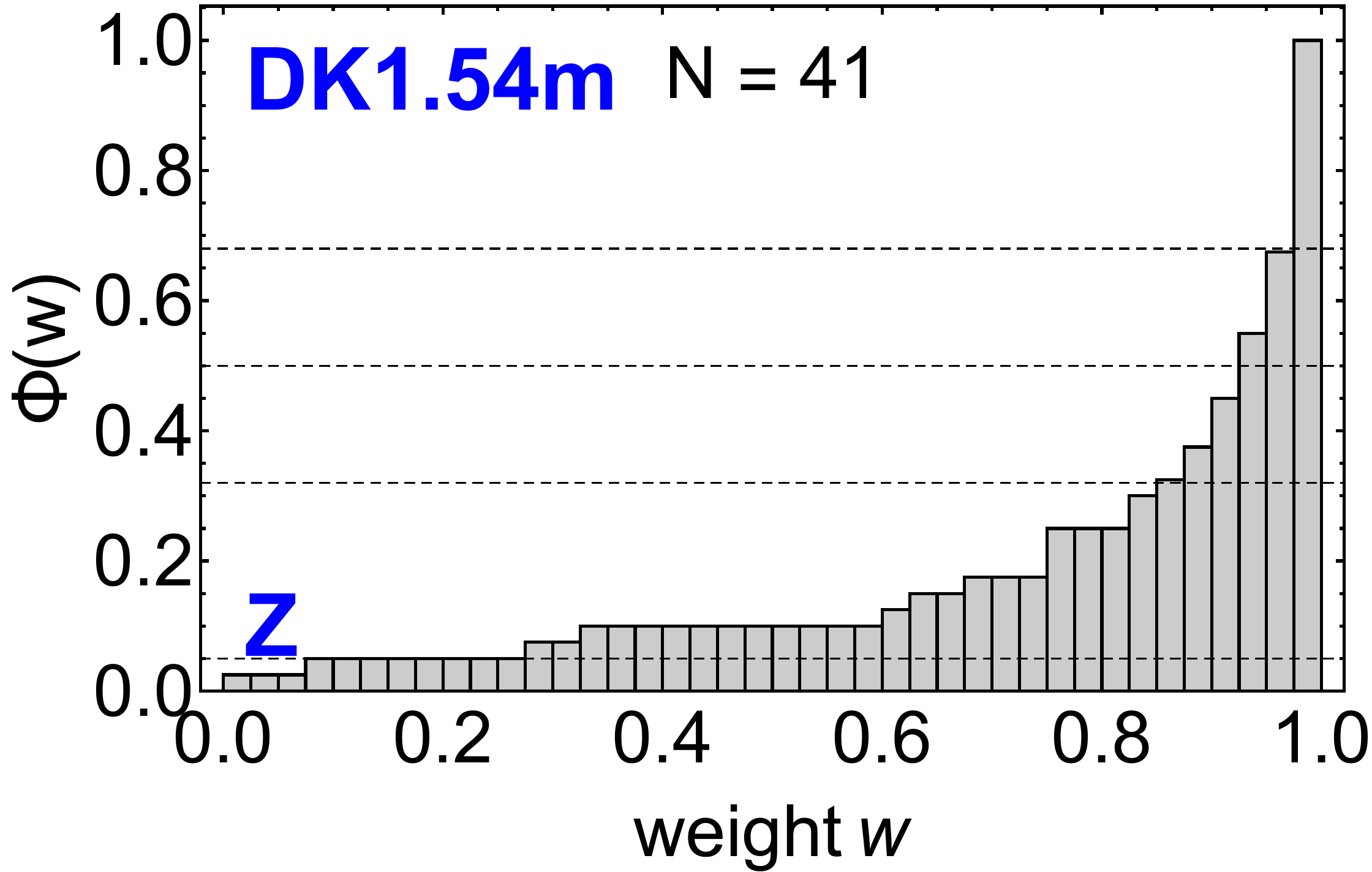}}
\end{minipage}
}\\[0.2mm]

 \caption{Weighted cumulative distribution functions (CDF) of the standardised residuals and CDF of data weights for the various off-peak data sets using the described robust-fitting procedure with single-lens point-source parallax models (see Tables~\ref{Tab:Models1} and~\ref{Tab:Models2}) for $u_0 < 0$ and $u_0 > 0$, respectively. The distribution of the standardised residuals is compared to a standard Gaussian distribution, quoting the $p$-value of an Anderson-Darling (AD) test \citep{AndersonDarling}. For the distribution of weights, cumulative probabilities of 5\%, 32\%, 50\%, and 68\% are indicated.}
 \label{Fig:statdata}
\end{figure*}

\begin{table*}
\begin{center}
\begin{tabular}{cccccc}
\hline
 & & \multicolumn{2}{c}{$u_0 < 0$} & \multicolumn{2}{c}{$u_0 > 0$}\\
 &  & $\kappa$ & $s_0$  & $\kappa$ & $s_0$ \\
\hline
OGLE & I & $0.99 \pm 0.07$ & $0.021 \pm  0.006$ & $0.99 \pm 0.07$ & $0.021 \pm  0.005$\\
LSC B & I & $3.8 \pm  0.5$ & $10^{-5}~(^\star)$ & $3.8 \pm  0.5$ & $10^{-5}~(^\star)$\\
LSC C & I & $0.1~(^\star)$& $0.022 \pm  0.003$ & $0.1~(^\star)$& $0.023 \pm  0.003$\\
CPT A & I &$0.1~(^\star)$ & $0.032 \pm  0.004$ &$0.1~(^\star)$ & $0.032 \pm  0.004$  \\
CPT B & I & $0.1~(^\star)$ & $0.0124 \pm  0.0014$ & $0.1~(^\star)$ & $0.0123 \pm  0.0014$\\
CPT C & I & $1.10 \pm 0.16$ & $0.004 \pm  0.003$ & $1.08  \pm 0.16$ & $0.004 \pm  0.003$\\
COJ A & I & $1.5  \pm 0.2$  & $0.002  \pm 0.003$ & $1.5  \pm 0.2$  & $0.002  \pm 0.004$\\
COJ B & I &  $1.51 \pm  0.16$ & $10^{-5}~(^\star)$ &  $1.45 \pm  0.15$ & $10^{-5}~(^\star)$ \\
FTS & I &  $0.1~(^\star)$ & $0.0094 \pm   0.0012$ &  $0.1~(^\star)$ & $0.0094 \pm   0.0012$\\
LSC C & V & $0.30 \pm  0.05$ &  $10^{-5}~(^\star)$ & $0.30 \pm  0.05$ &  $10^{-5}~(^\star)$\\
Dk1.54m & Z & $0.8 \pm 0.3$ &  $0.003 \pm 0.002$  & $0.8\pm  0.3$ &  $0.004 \pm 0.002$\\
\hline
\end{tabular}
\caption{Adopted error bar scaling factor  $\kappa$ and systematic error $s_0$ for the various data sets, as defined by Eq.~(\ref{eq:scaling}), determined from the standardised residuals arising for the point-source single-lens parallax models to all off-peak data (except for Danish 1.54m) for $u_0<0$ or $u_0 >0$, respectively, whose parameters are listed in Table~\ref{Tab:Models1} and~\ref{Tab:Models2}. Range constraints $\kappa \geq 0.1$ and $s_0 \geq 10^{-5}$ have been adopted, and the asterisk ($^\star$) marks bouncing against the range boundary. Several data sets do not hold sufficient information to constrain both $\kappa$ and $s_0$, leaving us with parameter ambiguities for our effective noise model.} 
\label{Tab:Scales}
\end{center}
\end{table*}

We used the modelling capabilities of the SIGNALMEN anomaly detector \citep{SIGNALMEN}, which itself calls the CERN library routine MINUIT \citep{MINUIT} for non-linear minimisation, in order to fit a point-source single-lens parallax model to the off-peak data while establishing an effective noise model of our data.

A rough estimate of the fundamental parameters $(t_0,u_0,t_E)$ can be obtained from simple maximum-likelihood fitting of a point-source single-lens model to the OGLE data, starting at any seed that roughly locates the peak, e.g. $(t_0,u_0,t_\mathrm{E}) = (6932.0,0.3,20~\mbox{d})$. This gave us the parameters listed in the first column of Tables~\ref{Tab:Models1} and~\ref{Tab:Models2}, which were then used to construct
seeds for models including the annual parallax, where, in order to account for potential ambiguities, we used all permutations of signs for the parameters $(u_0,\pi_{\mathrm{E},\mathrm{N}},\pi_{\mathrm{E},\mathrm{E}})$, specifically $(u_0,\pi_{\mathrm{E},\mathrm{N}},\pi_{\mathrm{E},\mathrm{E}}) = (\pm 0.009275,\pm 0.1,\pm 0.1)$. Using the robust fitting procedure with the noise model outlined above, i.e. by minimising $\tilde{\chi}^2$ as defined by Eq.~(\ref{eq:robmini}), we found two classes of local minima, corresponding to a `good' fit with $\chi^2 \sim 1050$ for 645 data points with $t_\mathrm{E} \sim 300~\mbox{d}$ and a `bad' fit with $\chi^2 \sim 3050$ for 645 data points with $t_\mathrm{E} \sim 180~\mbox{d}$. We accepted the former, and rejected the latter due to not reasonably matching the data. This left us with the two viable options $(u_0,\pi_{\mathrm{E},\mathrm{N}},\pi_{\mathrm{E},\mathrm{E}}) = (-0.0052\pm 0.0018,-0.367\pm 0.012, -0.143\pm 0.015)$ and $(u_0,\pi_{\mathrm{E},\mathrm{N}},\pi_{\mathrm{E},\mathrm{E}})= (0.0054\pm 0.0017,-0.354\pm 0.010, -0.138\pm 0.014)$, distinguished by the sign of $u_0$.

While the OGLE data provides a coverage of all event phases (except for the epochs that correspond to the gaps in between the annual seasons) and therefore should provide a good estimate of the parallax parameters, other data sets cover the event more densely over substantial parts of the wings, but all data might suffer from some systematics. With all data sets, except for the Danish 1.54m (which cover only the rising part and therefore lack of relevant information), we find
$(u_0,\pi_{\mathrm{E},\mathrm{N}},\pi_{\mathrm{E},\mathrm{E}}) = (-0.0065\pm 0.0004,-0.354\pm 0.009, -0.178\pm 0.008)$ and $(u_0,\pi_{\mathrm{E},\mathrm{N}},\pi_{\mathrm{E},\mathrm{E}})= (0.0061\pm 0.0004,-0.343\pm 0.009, -0.165\pm 0.009)$, so that the parallax appears to be robustly measured, with the further data giving a tighter constraint. We determined the error bar rescaling for the Danish 1.54m data based on these models.

\begin{figure*}
\hspace*{1,4mm}\resizebox{14.75cm}{!}{\includegraphics{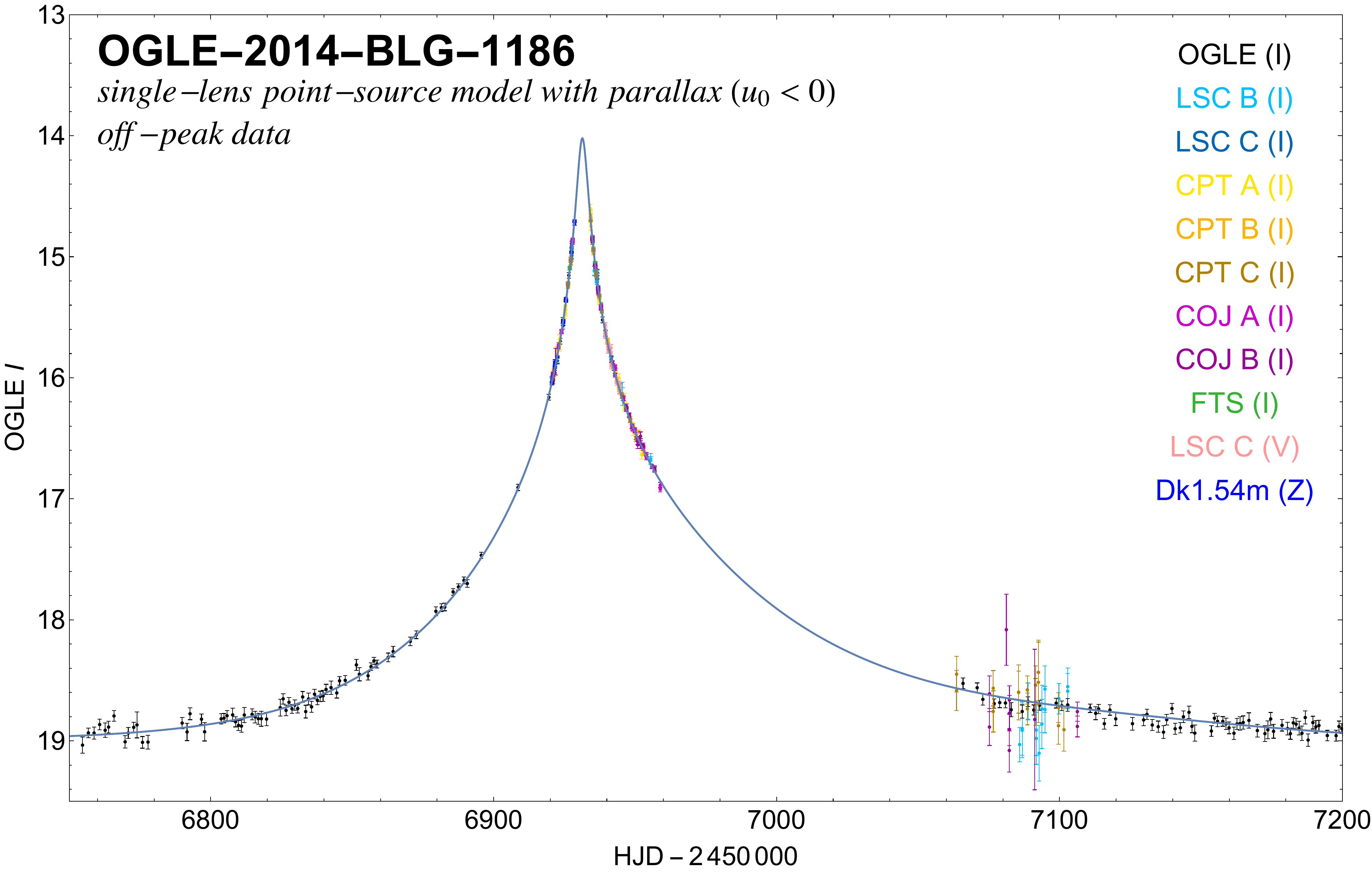}}\\[4mm]
\resizebox{15cm}{!}{\includegraphics{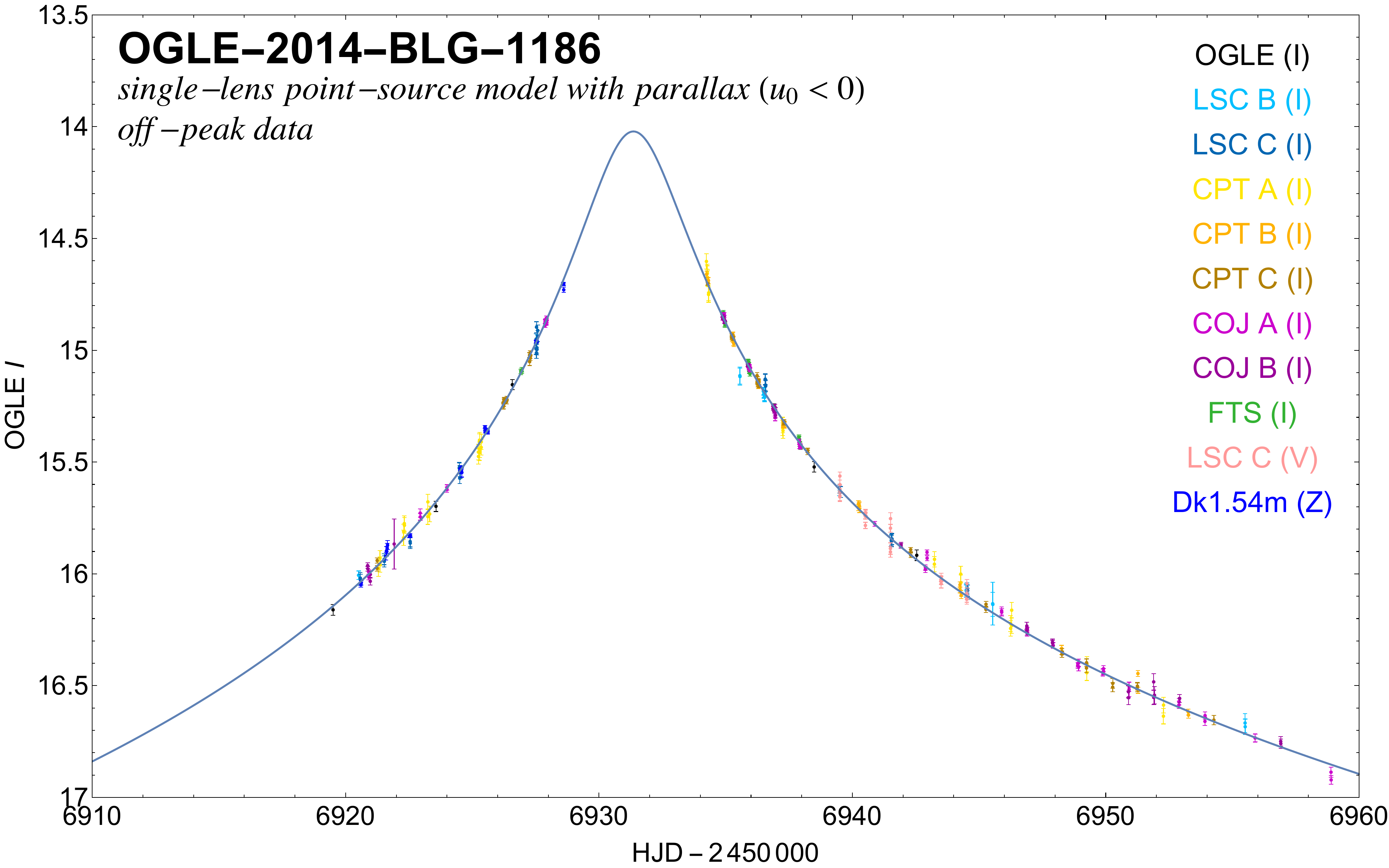}}
 \caption{Acquired off-peak data on event OGLE-2014-BLG-1186 with the various telescopes together with a model light curve that assumes an isolated single lens as well as a point-like source and accounts for the annual parallax, where $u_0 < 0$ (see Table~\ref{Tab:Models1}). The error bars displayed include a systematic error $s_0$ and scaling factor $\kappa$, as listed in Table~\ref{Tab:Scales} and determined with respect to the adopted model.}
 \label{Fig:BLC1}
\end{figure*}

\begin{figure*}
\hspace*{1,4mm}\resizebox{14.75cm}{!}{\includegraphics{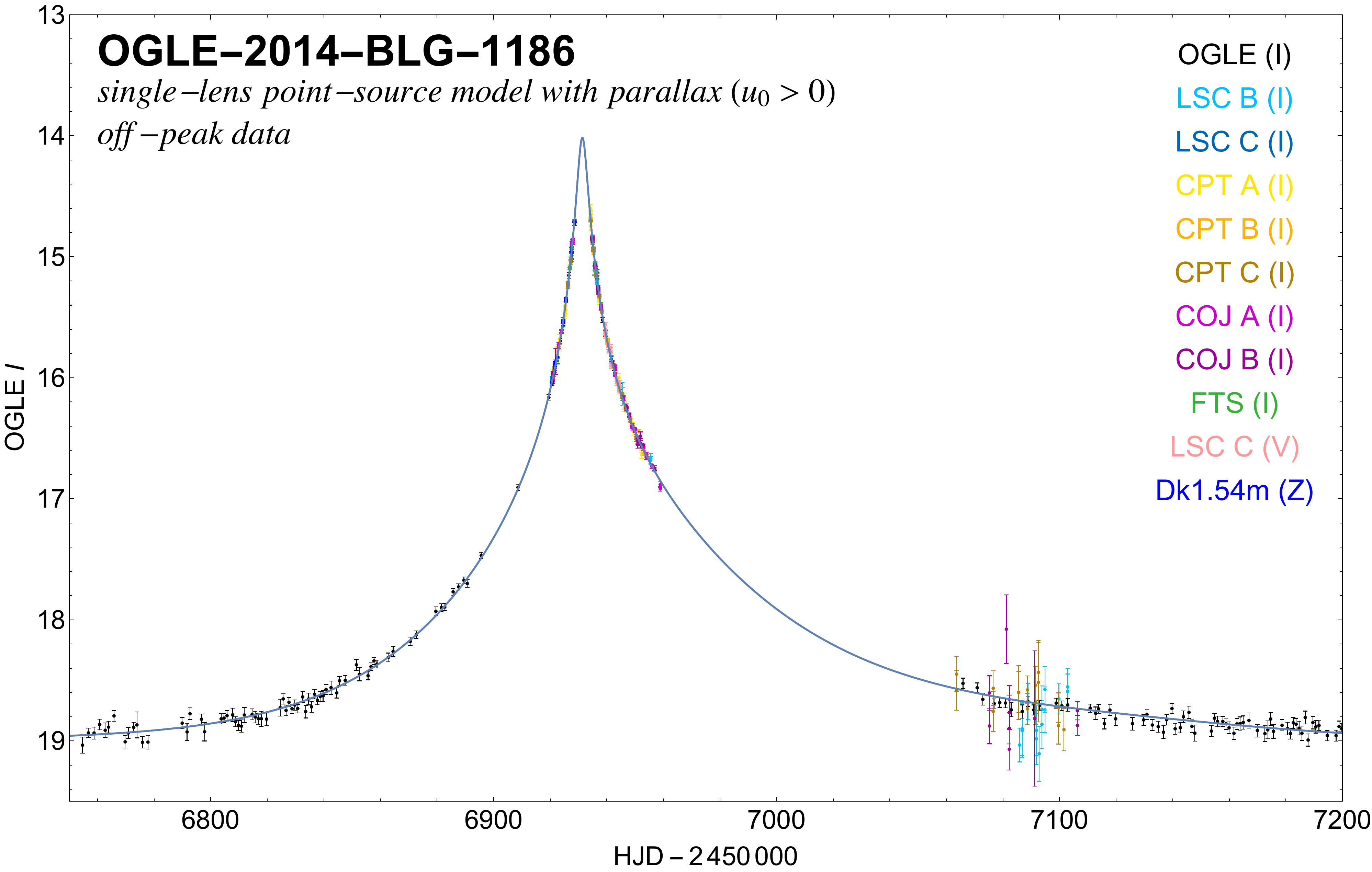}}\\[4mm]
\resizebox{15cm}{!}{\includegraphics{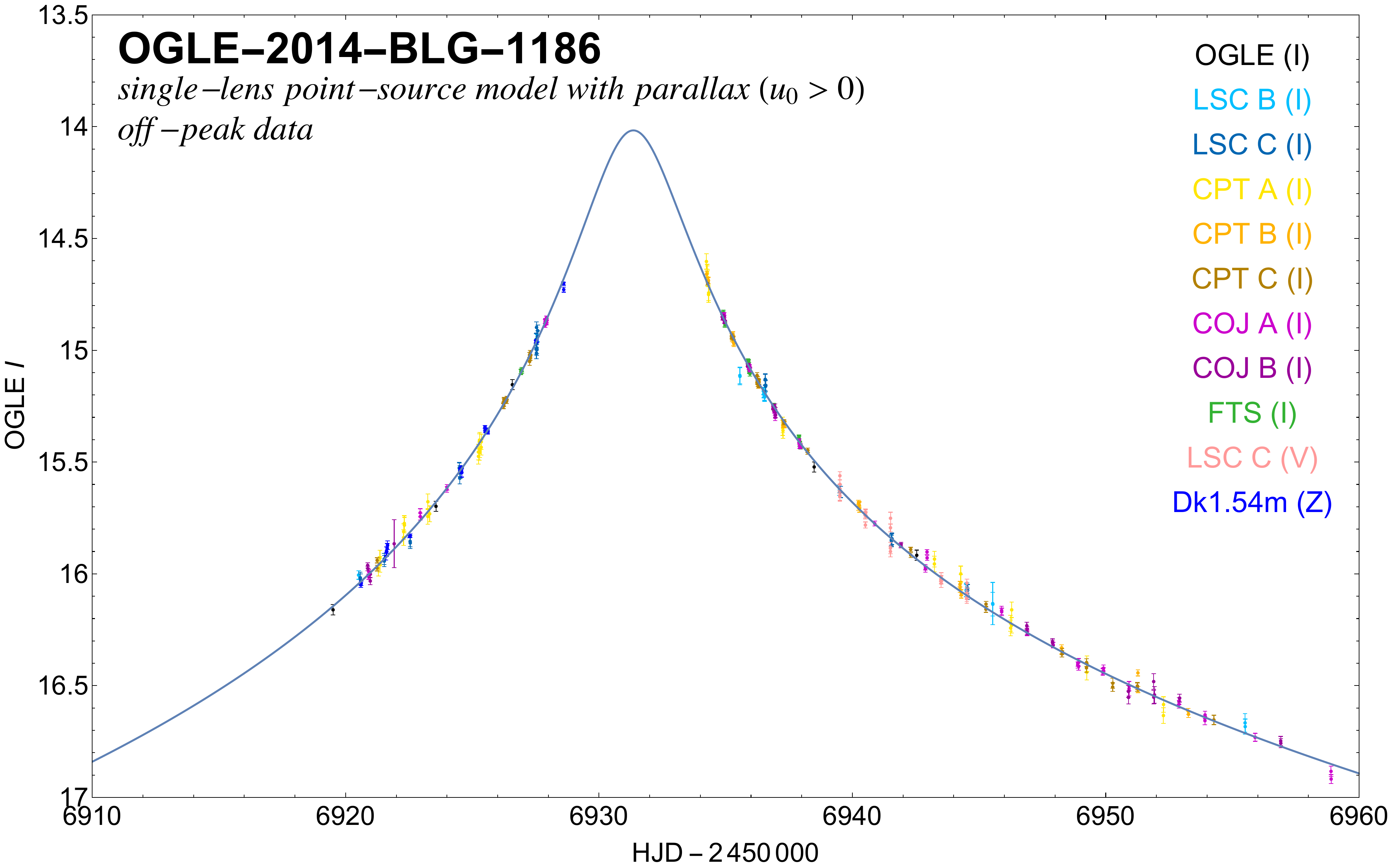}}
 \caption{Acquired off-peak data on event OGLE-2014-BLG-1186 with the various telescopes together with a model light curve that assumes an an isolated single lens as well as a point-like source and accounts for the annual parallax, similar to Fig.~\ref{Fig:BLC1}, but now for $u_0 > 0$ (see Table~\ref{Tab:Models2}). The error bars displayed include a systematic error $s_0$ and scaling factor $\kappa$, as listed in Table~\ref{Tab:Scales} and determined with respect to the adopted model.}
 \label{Fig:BLC2}
\end{figure*}

In Table~\ref{Tab:Scales}, we report the inferred systematic errors $s_0^{[j]}$ and scaling factors $\kappa^{[j]}$ for the various data sets, based on the standardised residuals of the two robust single-lens point-source models with parallax to all data (except for the Danish 1.54m), while Fig.~\ref{Fig:statdata} shows the weighted cumulative distribution functions (CDF) of the standardised residuals and CDF of the data weights, quoting $p$-values of an Anderson-Darling test \citep{AndersonDarling} comparing the weighted distribution of standardised residuals with a standard Gaussian. Some of the reported uncertainties on $s_0^{[j]}$ and $\kappa^{[j]}$ are large, and for some of the data sets, we find an ambiguity between the systematic error and the scaling factor. In fact, if the reported error bars on the magnitude do not vary much, there is no difference between adding a systematic error in quadrature and scaling the error bars by a common factor. For some data sets, the photometric uncertainty can pretty much be described just by a constant systematic error, regardless of the reported error bar, while for some other data sets, a systematic error is rejected, but a substantial scaling factor is suggested. For most data sets, the small number of data points prevents the establishment of a noise model that is more detailed than a simple effective model, particularly given the small number of large absolute standardised residuals (which are relevant in order to provide such statistics). Comparing the CDF of the weighted standardised residuals with a Gaussian distribution (see Fig.~\ref{Tab:Scales}) shows that our effective model provides a reasonable description. The distribution of the weights reveals that the distribution of the standardised residuals is generally more tail-heavy than a Gaussian distribution, where the weight of the tail differs amongst the data sets. Hence, a Gaussian profile with just an increased error bar would not be a good description. However, a Student-t distribution would provide an alternative to our adopted weight function.

The respective model light curves for the two single-lens point-source models with parallax to all data along with the data with modified error bars are shown in Fig.~\ref{Fig:BLC1} for $u_0 <0$ and Fig.~\ref{Fig:BLC2} for $u_0>0$, respectively, whereas Table~\ref{Tab:Models1} and Table~\ref{Tab:Models2} list the corresponding model parameters.

\begin{table*}
\begin{center}
\begin{tabular}{ccccccc}
\hline
Model & single & single, parallax & single, parallax & binary, parallax &  binary, parallax \\
Data selection & off-peak & off-peak & off-peak & all  & all \\
Data sets & OGLE (I) & OGLE (I) & all except Dk1.54m & all & all\\
Data scaling & none & none & none & $u_0 < 0$ off-peak &  $u_0 < 0$ off-peak   \\
Minimisation & ML & ML robust rescale & ML robust rescale & ML robust  & ML robust\\
Option & --- & $u_0 < 0$ & $u_0 < 0$ & $u_0 < 0$, close & $u_0 < 0$, wide\\
\hline
$t_0$ & $6931.685 \pm 0.005$ & $6931.39 \pm 0.09$ & $6931.359 \pm 0.006$ & $6931.429 \pm 0.003$ & $6931.477 \pm 0.003$\\
$t_E$ [d]& $179.13 \pm 0.39$  & $300 \pm 20$ & $287 \pm 16$ & $286 \pm 18$ & $279 \pm 7$\\
$u_0$ & $0.009275 \pm 0.000011$ & $-0.0052 \pm 0.0018$& $-0.0065 \pm 0.0004$ & $-0.0067 \pm 0.0004$ & $-0.0067 \pm 0.0002$\\
$\pi_{\mathrm{E},\mathrm{N}}$ & --- & $-0.367  \pm 0.012$& $-0.354 \pm 0.009$ &  $-0.364 \pm 0.009$&  $-0.353 \pm 0.007$\\
$\pi_{\mathrm{E},\mathrm{E}}$ & ---&  $-0.143\pm 0.015$ & $-0.178 \pm 0.008$ & $-0.171\pm 0.009$& $-0.171 \pm 0.006$\\
$d$ & ---& ---& ---& $0.713\pm 0.006$&  $1.428 \pm 0.009$\\
$q$& ---& ---&  --- & $(3.6 \pm 0.3) \times 10^{-4}$& $(3.8 \pm 0.2) \times 10^{-4}$\\
$\alpha$&---& ---& ---& $4.023 \pm 0.002$& $4.022 \pm 0.002$\\
\hline
\end{tabular}
\caption{Successive construction of models for $u_0 < 0$ in 5 steps: 1) rough maximum-likelihood estimation of $t_0$, $t_\mathrm{E}$, and $u_0$ from the off-peak OGLE data on the basis of the reported error bars and a single-lens point source model, 2) Measurement of parallax parameters from the off-peak OGLE data (assuming $u_0 < 0$) by means of robust fitting and simultaneous estimation of global systematic error and error bar scaling factor, with refinement of $t_0$, $t_\mathrm{E}$, and $u_0$ estimates, 3) Confirmation of robustness of parallax measurement and refinement of parameters by including all off-peak data (except for Danish 1.54m), followed by determination of the systematic error and error bar scaling factor for the Danish 1.54m data based on the arising model parameters, 4) Inclusion of the peak data using the established modification of error bars, and robust fitting of a binary-lens point-source model to all data (including Danish 1.54m), with seed values for the binary parameters $(d,q,\alpha)$ arising from a grid search with the other parameters fixed, 5) Finding a corresponding solution with a wide binary lens ($d > 1$ rather than $d < 1$) by using the previously determined parameter values as seed, and just flipping the separation parameter $d \leftrightarrow d^{-1}$.}
 \label{Tab:Models1}
\end{center}
\end{table*}

\begin{table*}
\begin{center}
\begin{tabular}{ccccccc}
\hline
Model & single & single, parallax & single, parallax & binary, parallax & binary, parallax \\
Data selection & off-peak & off-peak & off-peak & all  & all \\
Data sets & OGLE (I) & OGLE (I) & all except Dk1.54m & all & all\\
Data scaling & none & none & none & $u_0 > 0$ off-peak &  $u_0 > 0$ off-peak   \\
Minimisation & ML & ML robust rescale & ML robust rescale & ML robust  & ML robust\\
Option & --- & $u_0 > 0$ & $u_0 > 0$ & $u_0 > 0$, close & $u_0 > 0$, wide\\
\hline
$t_0$ & $6931.685 \pm 0.005$ & $6931.37 \pm 0.09$ & $6931.356 \pm 0.006$ & $6931.444 \pm 0.004$ & $6931.516 \pm 0.005$\\
$t_E$ [d]& $179.13 \pm 0.39$ & $310 \pm 20$& $289 \pm 19$ & $288 \pm 18$ & $292 \pm 18$\\
$u_0$ & $0.009275 \pm 0.000011$ & $0.0054 \pm 0.0017$& $0.0061 \pm 0.0004$ & $0.0063 \pm 0.0004$ & $0.0059 \pm 0.0004$ \\
$\pi_{\mathrm{E},\mathrm{N}}$ & --- & $-0.354 \pm 0.010$& $-0.343 \pm 0.009$ & $-0.354 \pm 0.009$ & $-0.352 \pm 0.009$\\
$\pi_{\mathrm{E},\mathrm{E}}$ & ---& $-0.138\pm 0.014$& $-0.165 \pm 0.009$ & $-0.160 \pm 0.009$ & $-0.157 \pm 0.008$\\
$d$ & ---& ---& ---& $0.681 \pm 0.006$ & $1.483 \pm 0.013$\\
$q$& ---& ---& ---& $(4.3 \pm 0.3) \times 10^{-4}$ & $(4.3 \pm 0.3) \times 10^{-4}$ \\
$\alpha$&---& ---& ---& $2.308 \pm 0.003$ & $2.305 \pm 0.002$\\
\hline
\end{tabular}
\caption{Successive construction of models for $u_0 > 0$, analogous to the $u_0 < 0$ case presented in Table~\ref{Tab:Models1}. Step 1 is identical to the procedure for $u_0 < 0$ (given that it the single-lens point-source light curve without parallax depends on $|u_0|$ only), whereas for the other steps the opposite sign for $u_0$ has been enforced, leading to a flip in sign of the trajectory angle $\alpha$ (or respectively $\alpha \leftrightarrow \alpha \pm \pi$), while all other parameters differ slightly.}
 \label{Tab:Models2}
\end{center}
\end{table*}

\subsubsection{Significance of putative anomaly}

\begin{figure*}
\resizebox{\columnwidth}{!}{\includegraphics{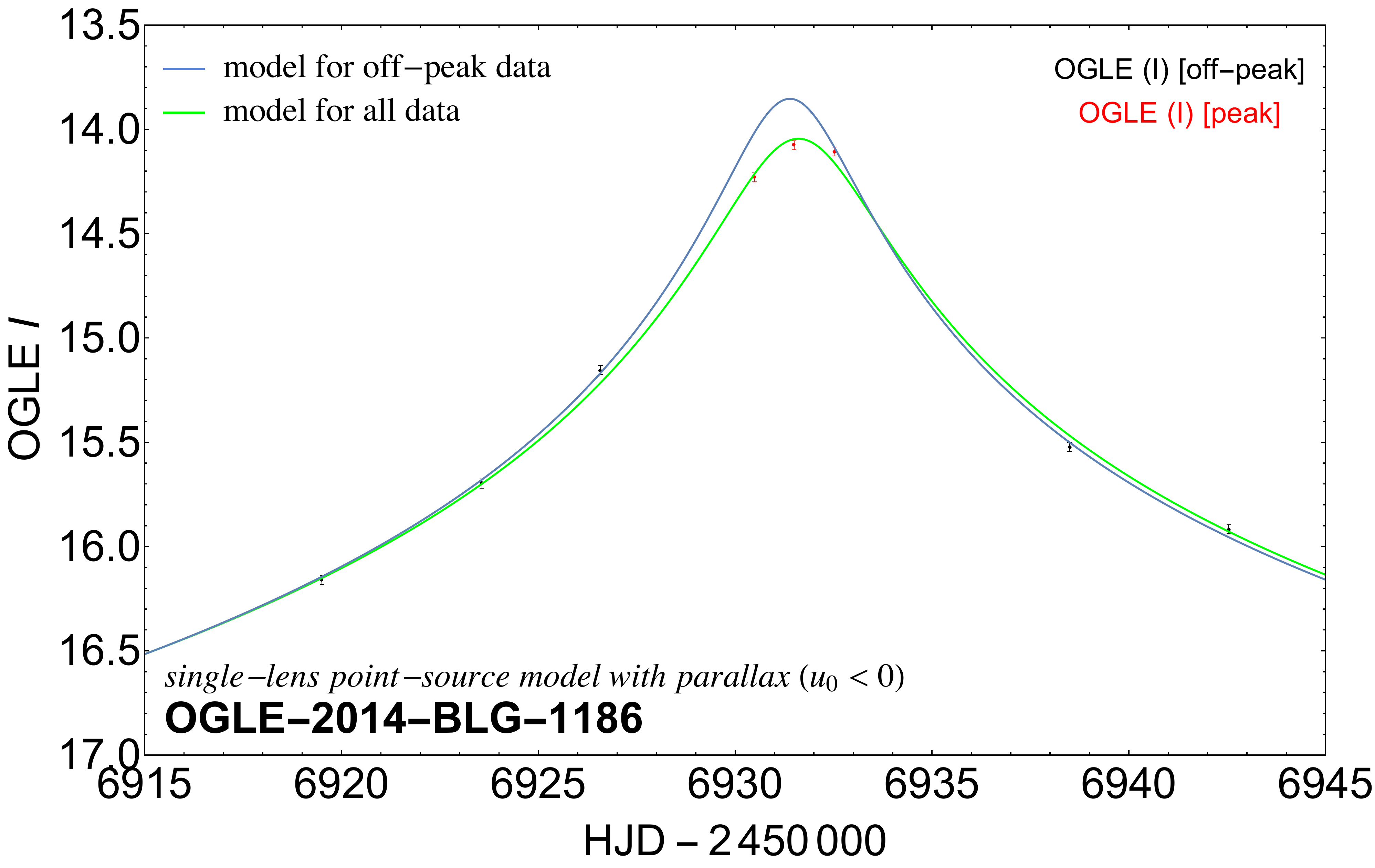}}\hfill
\resizebox{\columnwidth}{!}{\includegraphics{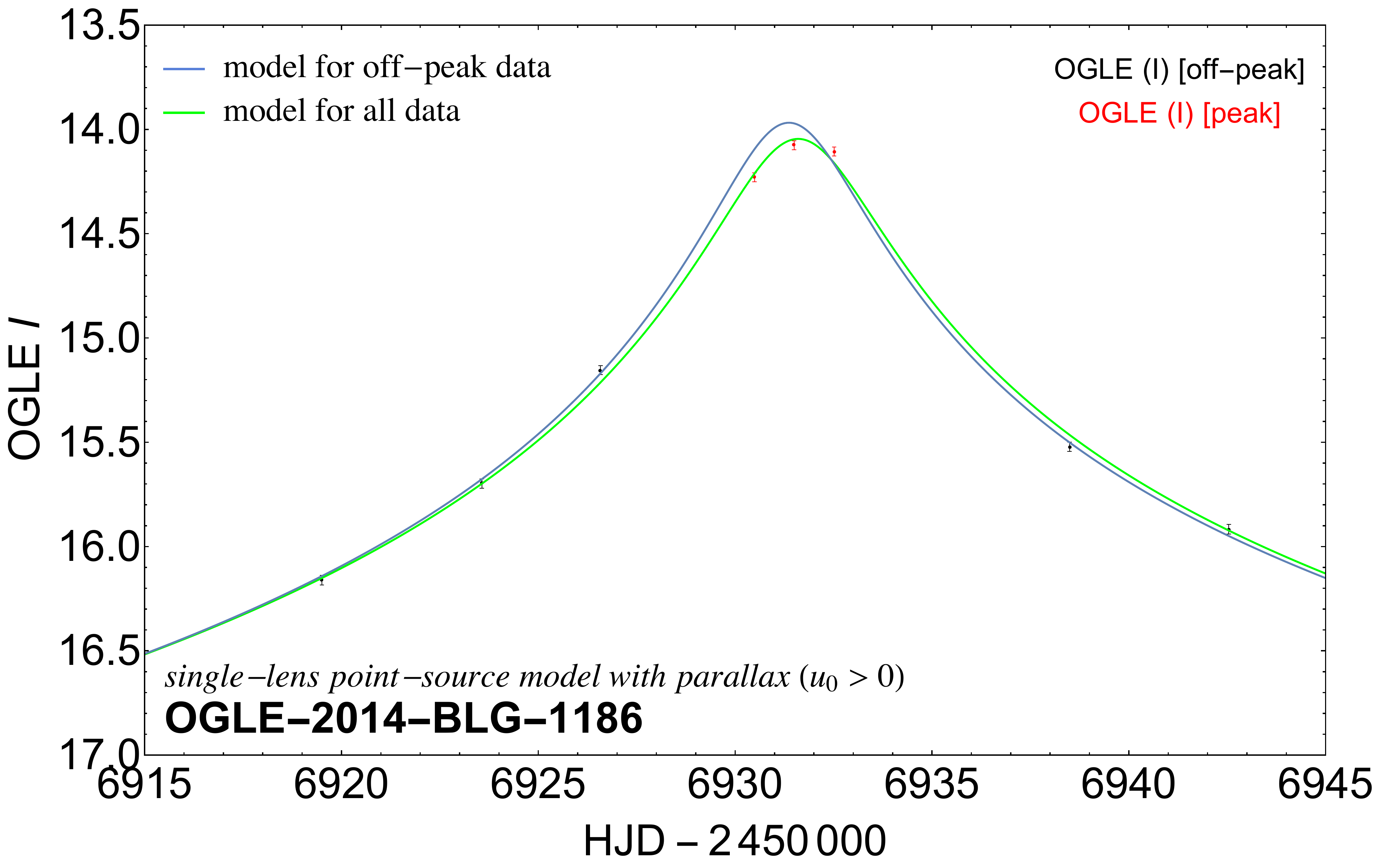}}
\caption{Single-lens point-source models with parallax fitted to OGLE data only, using either the off-peak data only (blue) or the full data set (green). The OGLE data do not obviously hint at an anomaly in event OGLE-2014-BLG-1186.}
 \label{Fig:OGLEmodels}
\end{figure*}

Given our robust measurement of parallax and our noise model from the off-peak data, we can assess the putative anomaly in the peak region, assuming that the inferred systematic errors and scale factors reasonably apply to the peak data as well. If we consider only OGLE data, there is no obvious hint of an anomaly, as illustrated in Fig.~\ref{Fig:OGLEmodels}, which shows single-lens point-source models with parallax for all OGLE data for the two cases $u_0 <0$ and $u_0 > 0$, respectively.

\begin{figure*}
\resizebox{\columnwidth}{!}{\includegraphics{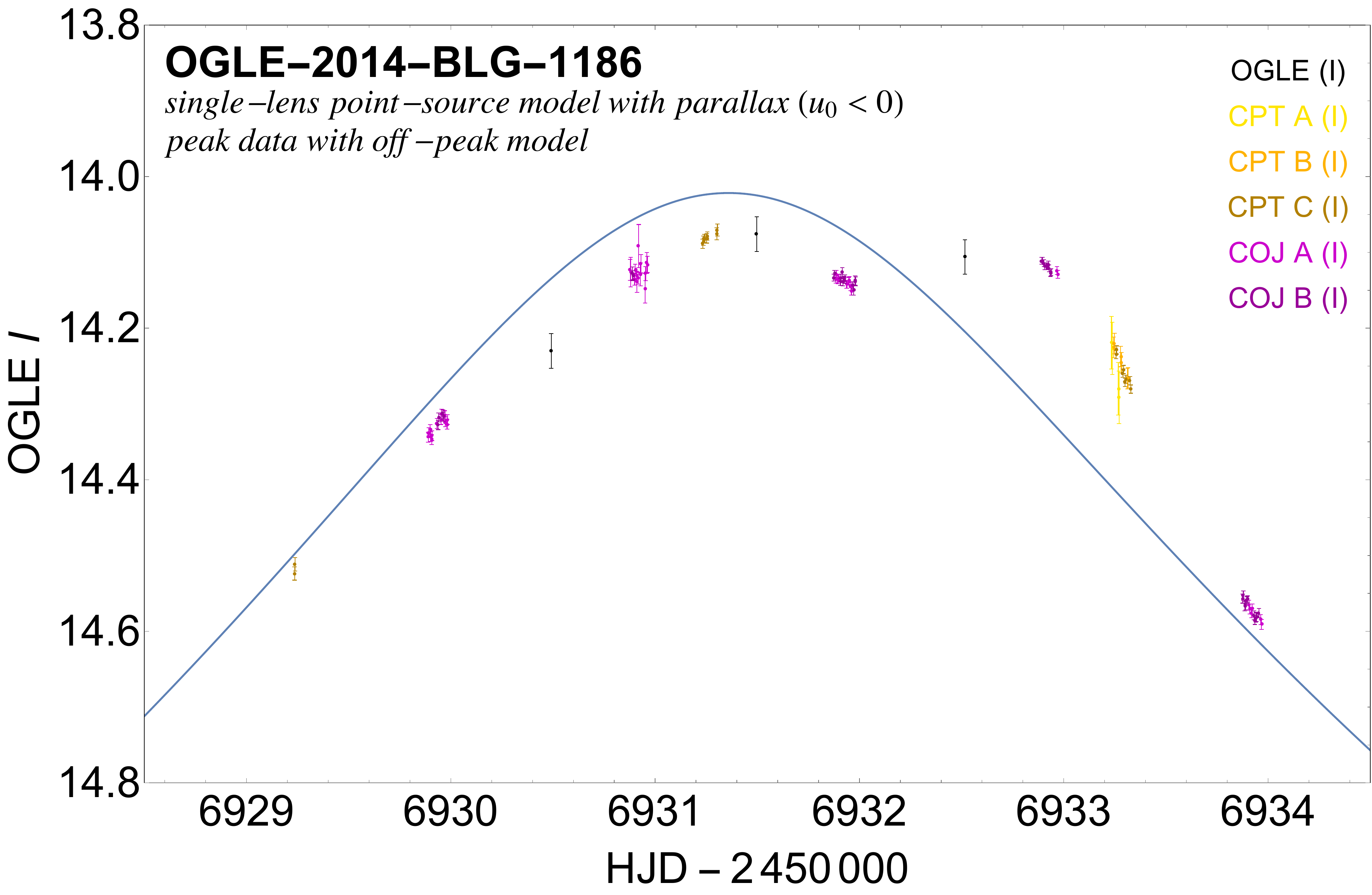}}\hfill
\resizebox{\columnwidth}{!}{\includegraphics{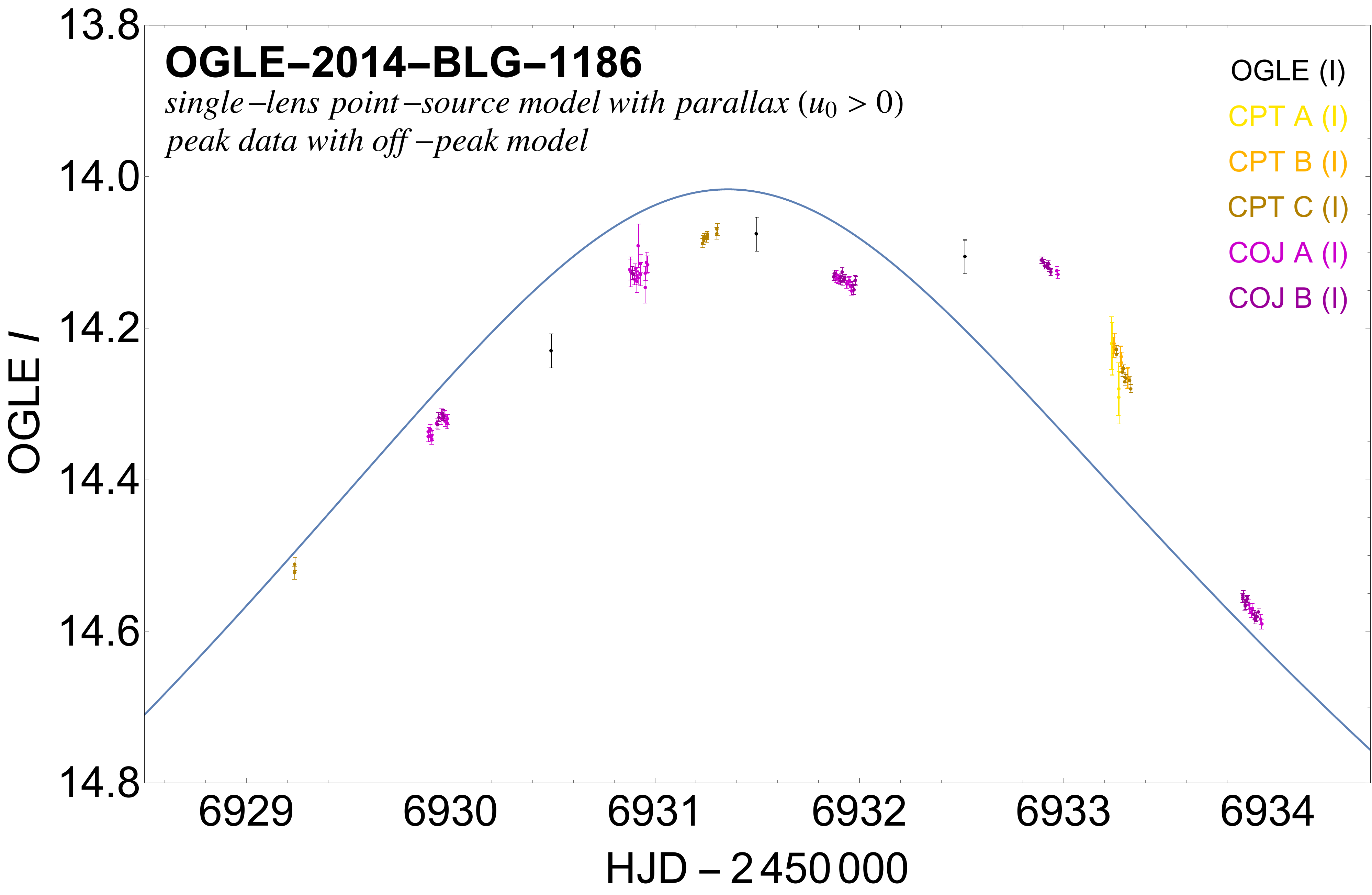}}
\\[2mm]
\resizebox{\columnwidth}{!}{\includegraphics{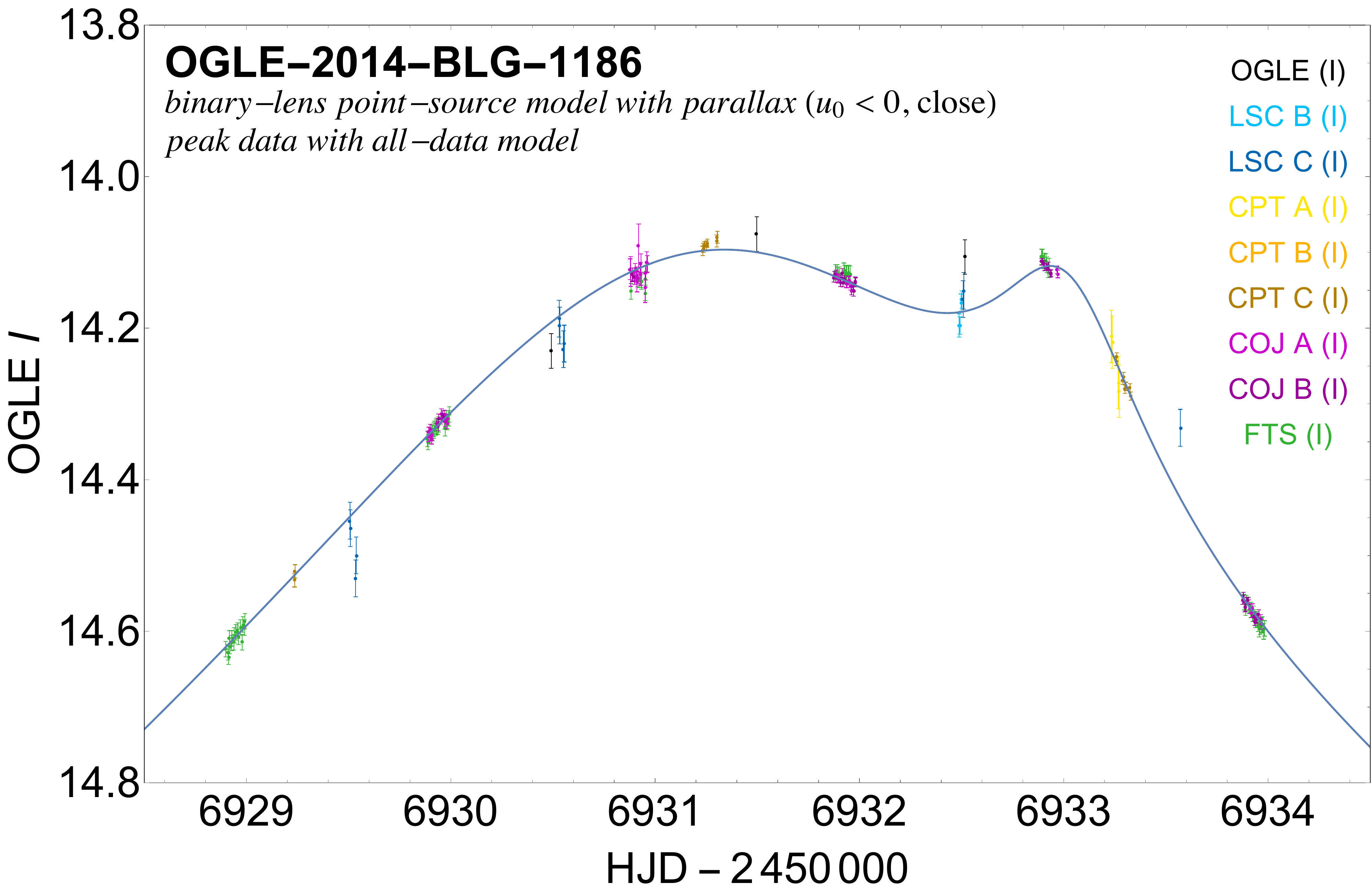}}\hfill
\resizebox{\columnwidth}{!}{\includegraphics{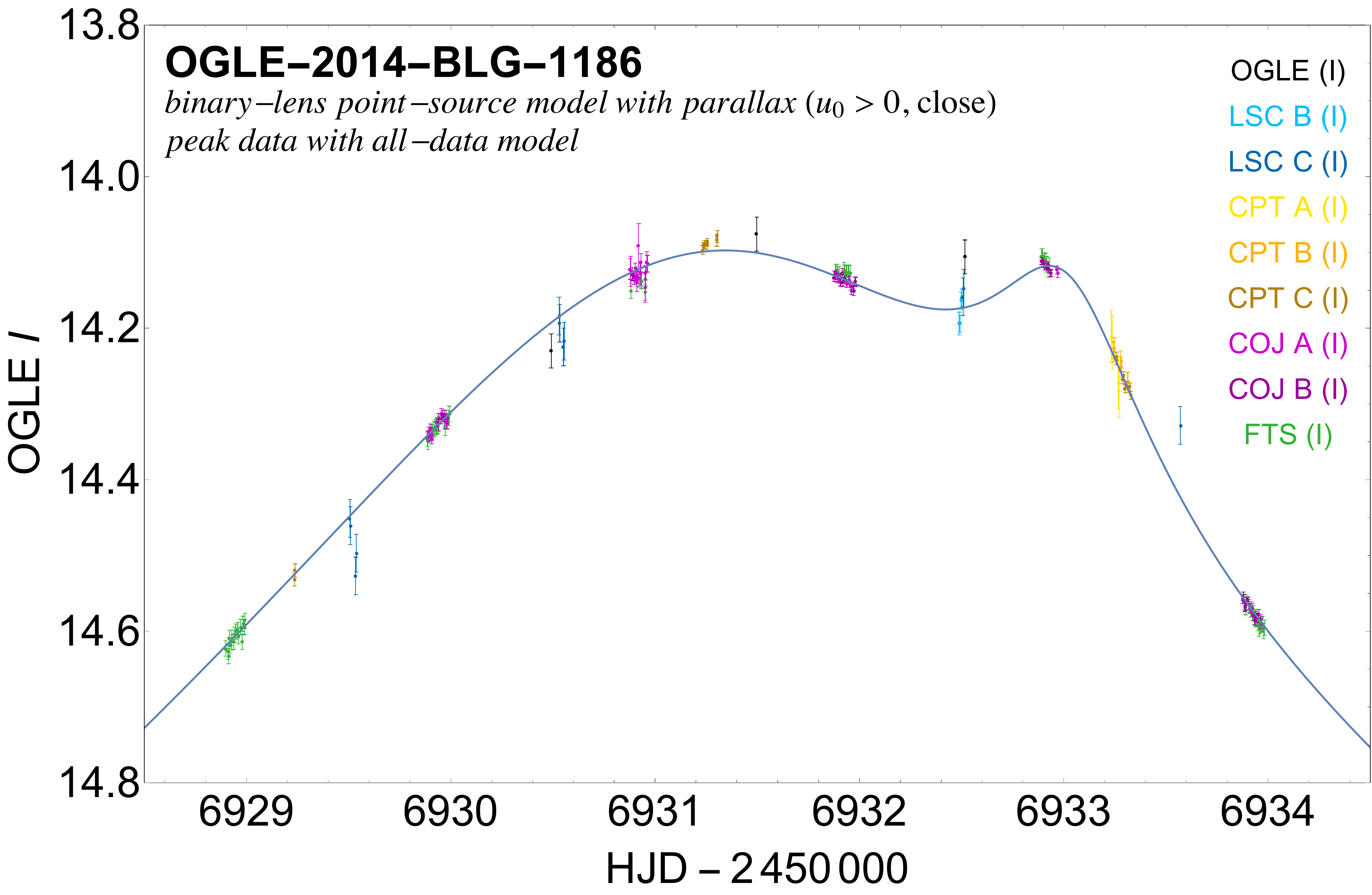}}
\\[2mm]
\resizebox{\columnwidth}{!}{\includegraphics{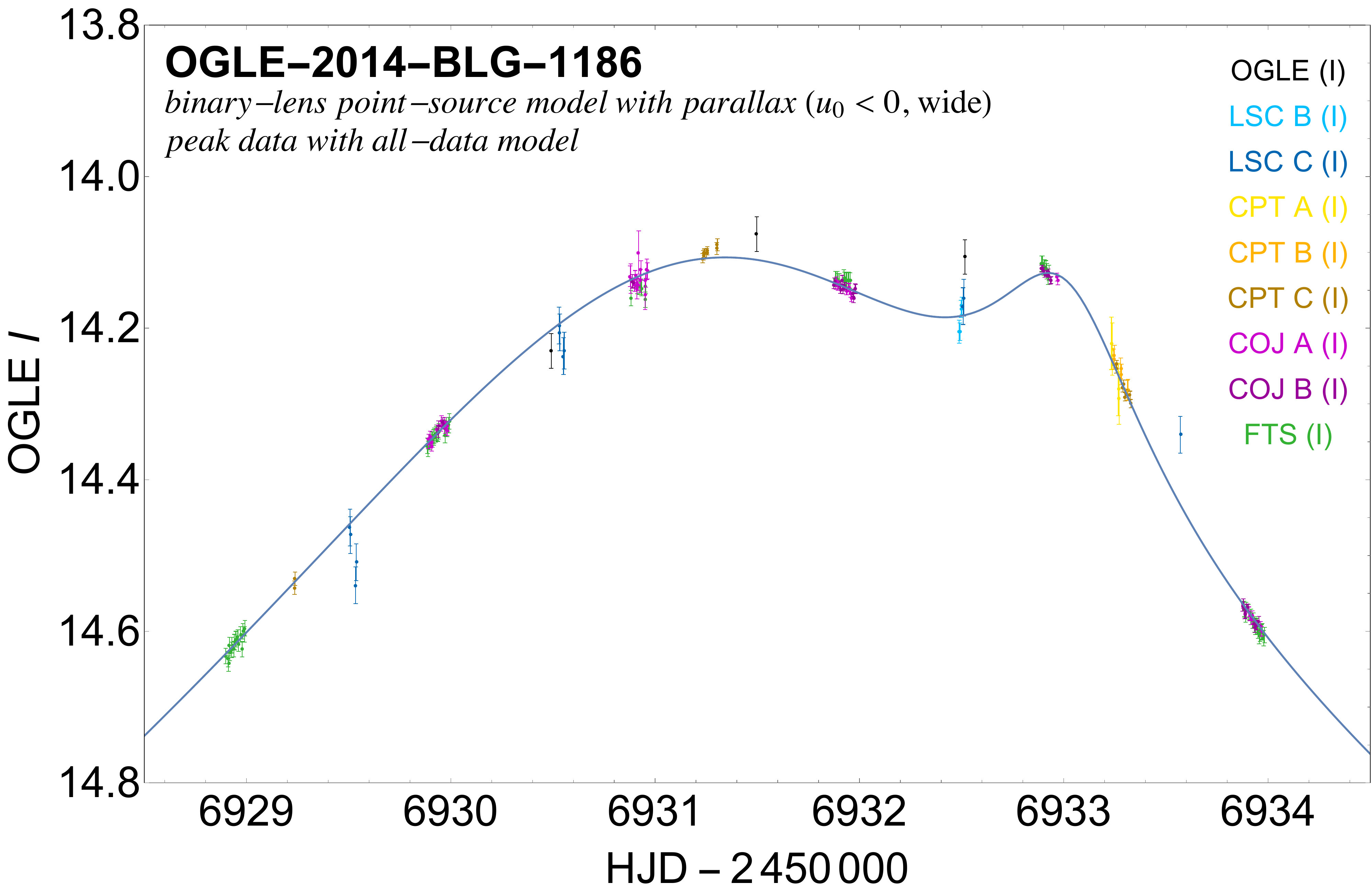}}\hfill
\resizebox{\columnwidth}{!}{\includegraphics{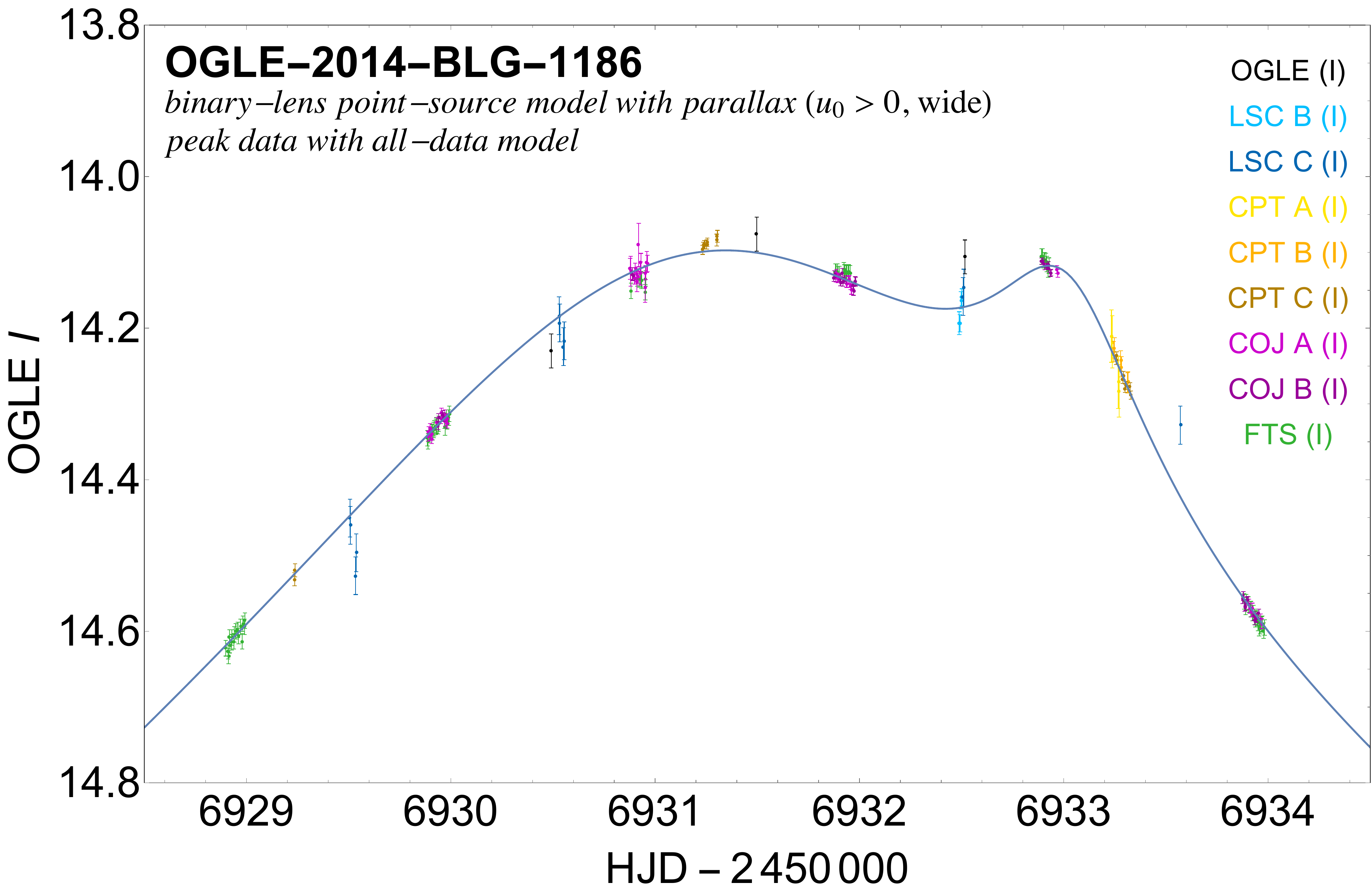}}
 \caption{Peak anomaly of microlensing event OGLE-2014-BLG-1186. The upper panels show the single-lens point-source parallax models to the off-peak data for $u_0 <0$ or $u_0 > 0$, respectively (c.f.\ Tables~\ref{Tab:Models1} and~\ref{Tab:Models2}) with the peak data, aligned according to the baseline fluxes $\vec{F}_\mathrm{base}$ and blend ratios $\vec g$ suggested by the models for the off-peak data. Only those data sets for which these parameters could be well determined are shown. These align very well, giving a clear and consistent picture of the anomaly over the peak. The middle and lower panels compare binary-lens point-source parallax models for the four cases $u_0 <0$ or $u_0 > 0$ as well as $d<1$ or $d>1$ (c.f.\ Tables~\ref{Tab:Models1} and~\ref{Tab:Models2})  with the acquired peak data, showing that such models can account for the major features of the double-peak light curve, but most notably do not match the slope indicated by the COJ~A and FTS data over the second peak.}
 \label{Fig:BPLC}
\end{figure*}

The situation however becomes dramatically different once one considers the RoboNet data. The top panels of Fig.~\ref{Fig:BPLC} show the respective single-lens point-source model with parallax for the off-peak data only, along with the peak data, for which the baseline magnitude $F_\mathrm{base}^{[j]}$ and blend ratio $g^{[j]}$ also follow the fit to the off-peak data only. Apparently, the RoboNet data over the peak from three telescopes in South Africa and two telescopes in Australia, for which the baseline magnitude and blend ratio are well determined (in contrast to the FTS and Chilean data), consistently line up to very high precision without the modelling process ever having involved these data. Moreover, a microlensing anomaly is clearly visible, much above the noise level. 

\subsection{Binary-lens models}

\subsubsection{Constraining binary-lens parameter space}
\label{sec:BLspace}

With the presence of a real anomaly over the peak firmly established, let us systematically find all potentially viable binary-lens models, which include the case of a star orbited by a planet (with the effect of other planets neglected).


Given that the peak anomaly lasts only about 5~days, we can at first neglect the binary orbital motion, assuming that the orbital period is much longer. With regard to its effect on the gravitational bending of light, a binary lens composed of constituents with masses $M_1$ and $M_2$ is then fully characterised by its total mass $M = M_1+M_2$, the mass ratio $q = M_2/M_1$, and the separation
parameter $d$,  where $d\,\theta_\mathrm{E}$ is the angle on the sky between the primary and the secondary as seen from the observer with the angular Einstein radius $\theta_\mathrm{E}$, as given by Eq.~(\ref{eq:thetaE}), referring to the total mass $M$.

Let us choose a coordinate frame with the origin at the centre of mass of the lens system and the coordinate axes $(\vec{e}_1,\vec{e}_2)$ spanning a plane orthogonal to the line of sight so that $\vec{e}_1 \perp \vec{e}_2$ and $\vec{e}_1 \times \vec{e}_2$ points towards the observer. With $\vec{e}_1$ being along the orthogonally projected separation vector from $M_2$ to $M_1$, the primary of mass $M_1$ is at the angular coordinate $[d\,q/(1+q),0]\,\theta_\mathrm{E}$ and the secondary of mass $M_2$ is at  the angular coordinate $[-d/(1+q),0]\,\theta_\mathrm{E}$. 

In contrast to a single lens, the microlensing light curve depends on the orientation of the source trajectory, where we measure the trajectory angle $\alpha$ from the axis $\vec{e}_1$. We can then 
describe the source trajectory by \begin{equation}
\vec{u}(t) = u_0 \left( \begin{array}{c} -\sin \alpha \\ \cos \alpha \end{array} \right) + \frac{t-t_0}{t_\mathrm{E}}\,\left( \begin{array}{c} \cos \alpha \\ \sin \alpha \end{array}\right)\,,
\end{equation}
where the source most closely approaches the centre of mass of the lens system at epoch $t_0$ and angular separation $u_0\,\theta_\mathrm{E}$.

For weak gravitational fields, one finds a linear superposition of the deflection terms that arise for each point-like deflector with mass $M_k$ at angular position $\vec{x}^{(k)}\,\theta_\mathrm{E}$, so that the relation between the source and image positions (c.f.\ Eq.~(\ref{eq:lenseq})) becomes
\begin{equation}
\vec{u}(\vec{x}) = \vec{x} - \sum_k \frac{M_k}{M}\,\frac{\vec{x}-\vec{x}^{(k)}}{\left|\vec{x}-\vec{x}^{(k)}\right|^2}\,,
\end{equation}
while the magnification is given by
\begin{equation}
A(\vec u) = \sum_i \left| \det\left(\frac{\partial \vec{u}}{\partial \vec{x}}\right)\left(\vec{x}_i\right)\right|^{-1}\,,
\end{equation}
where the sum is taken over all images at angular positions $\vec{x_i}\,\theta_\mathrm{E}$. Binary (and multiple) lenses create line caustics $\cal{C}$, defined by
\begin{equation}
{\cal{C}} = \left\{\vec{u}(\vec{x}^\prime)\;\Big | \;\det\left(\frac{\partial \vec{u}}{\partial \vec{x}}\right)\left(\vec{x}^\prime\right) = 0\ \right\}
\end{equation}
on which the point-source magnification diverges, $A(\vec{u}) \to \infty$. The features of the diverse morphologies of microlensing light curves arising for binary (and multiple) lenses are characterised by the track of the source relative to the caustics, providing a type classification \citep{Liebig}.

The possible topologies of caustics are the same for all binary lenses \citep{Erdl}, discriminated by the separation parameter $d$ for any given mass ratio $q$. For small mass ratios $q$, the intermediate topology with a single caustic curve with 6 cusps, occupies only a small range near $d \sim 1$, essentially leaving a close-binary ($d < 1$) and a wide-binary ($d > 1$) case \citep{GS98,Do:extreme}. In both of these cases, one finds a `central caustic' around the centre of mass of the binary (i.e.\ factually near the host star for a star-planet system), which has two cusps along the binary axis, and a further two cusps symmetrically above and below. As $q \to 0$, the central caustics for pairs of close- and wide-binary models with $d \leftrightarrow d^{-1}$ become identical, which causes a model ambiguity. Moreover, near a location that has an image under gravitational lensing by the star at the position of the planet, one finds `planetary caustics'. In the case of a wide binary, there is a single diamond-shaped caustic with four cusps (two on the star-planet axis, and two above and below), whereas a close binary has two off-axis triangular-shaped caustics with 3 cusps each, where the longest side is close to parallel to the star-planet axis.

The magnification function $A[\vec{u}(t;\vec p)]$ for a binary lens, where $\vec p = (t_0,u_0,t_E,d,q,\alpha)$, neglecting the finite extent of the source star, is no longer an analytic function, but can be numerically evaluated by solving a 5th order complex polynomial for the image positions \citep{WM95,SG12}.

\begin{figure*}
\includegraphics[width=4cm]{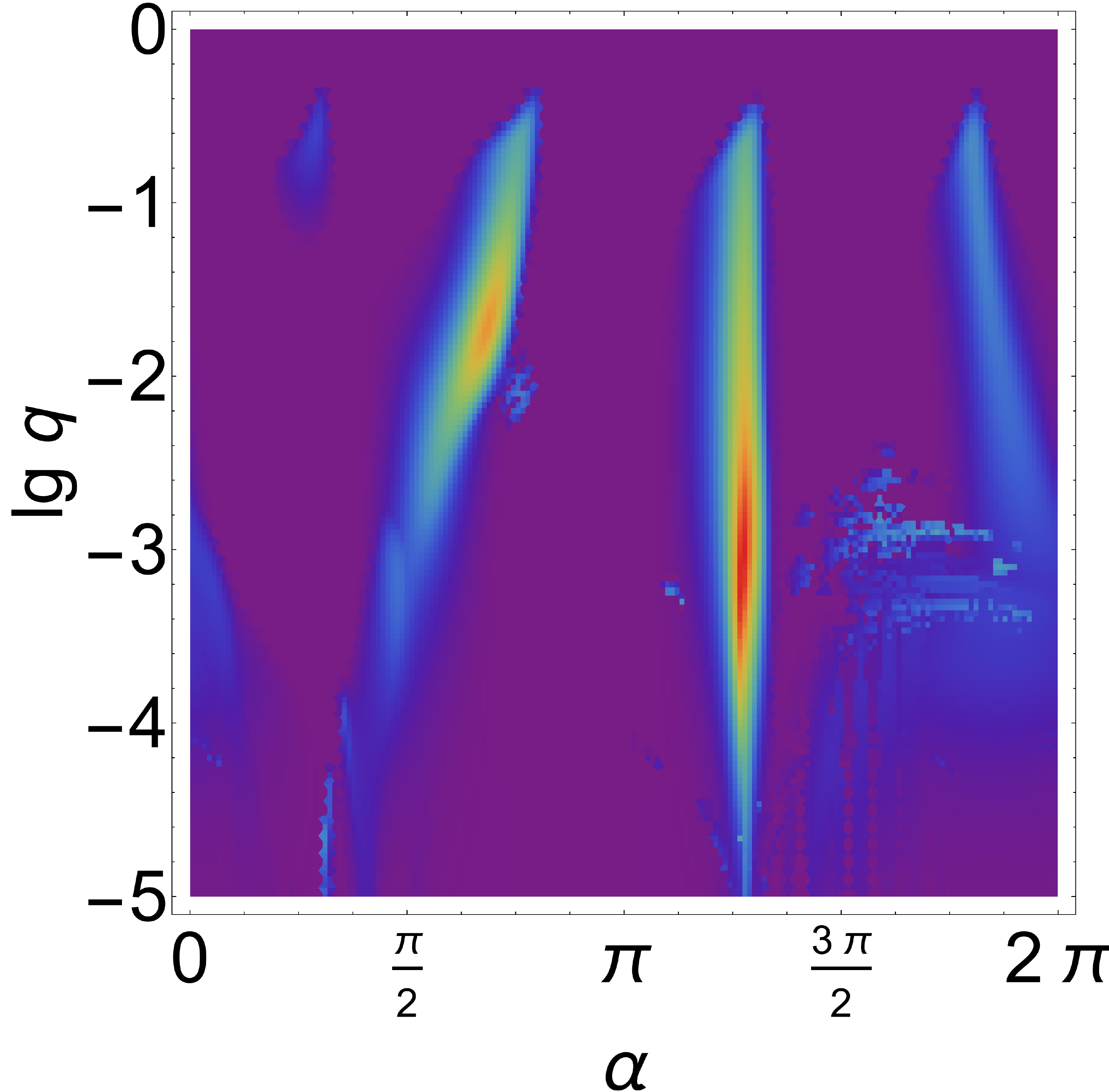}
\includegraphics[width=4cm]{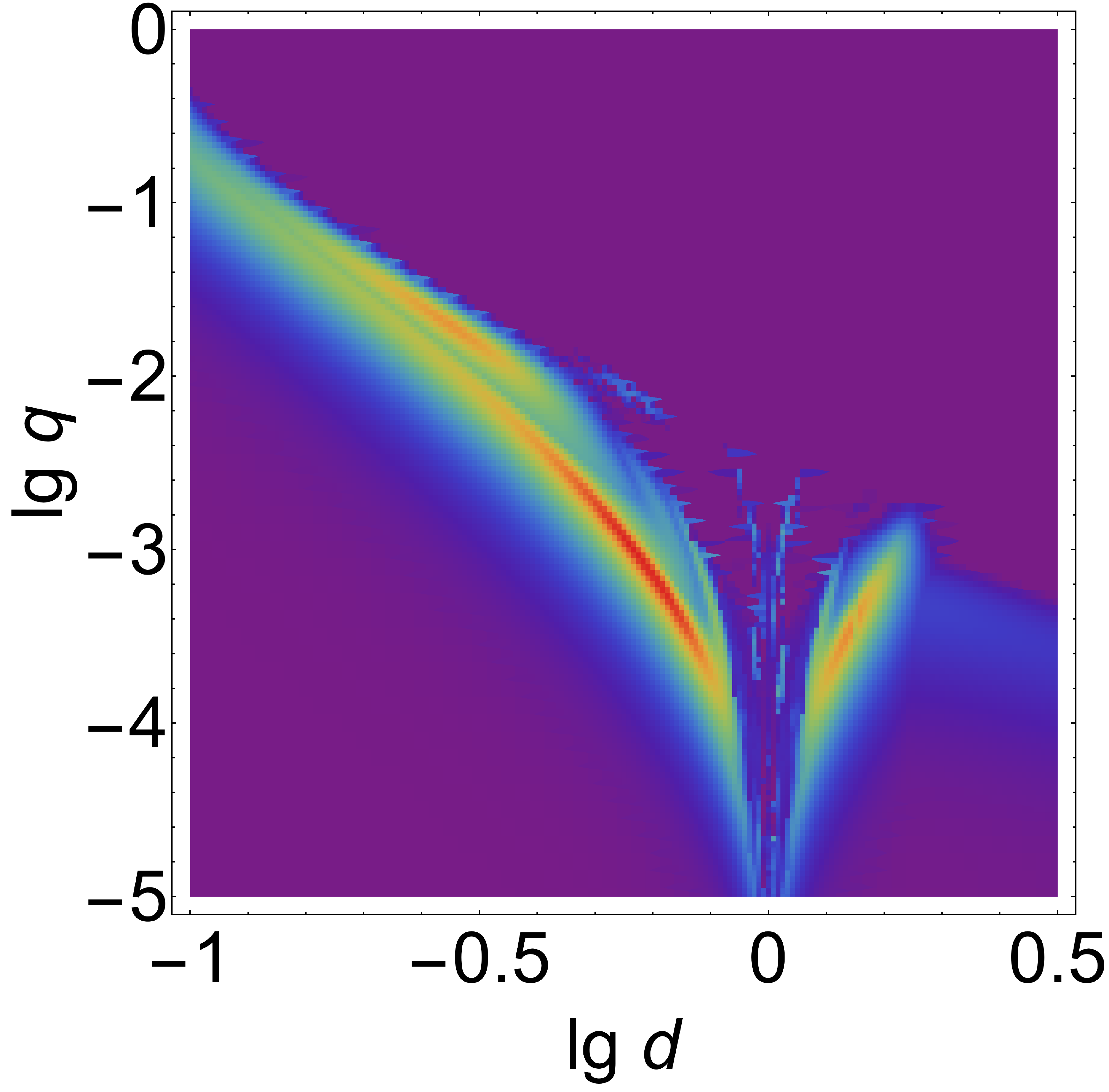}
\hfill
\includegraphics[width=4cm]{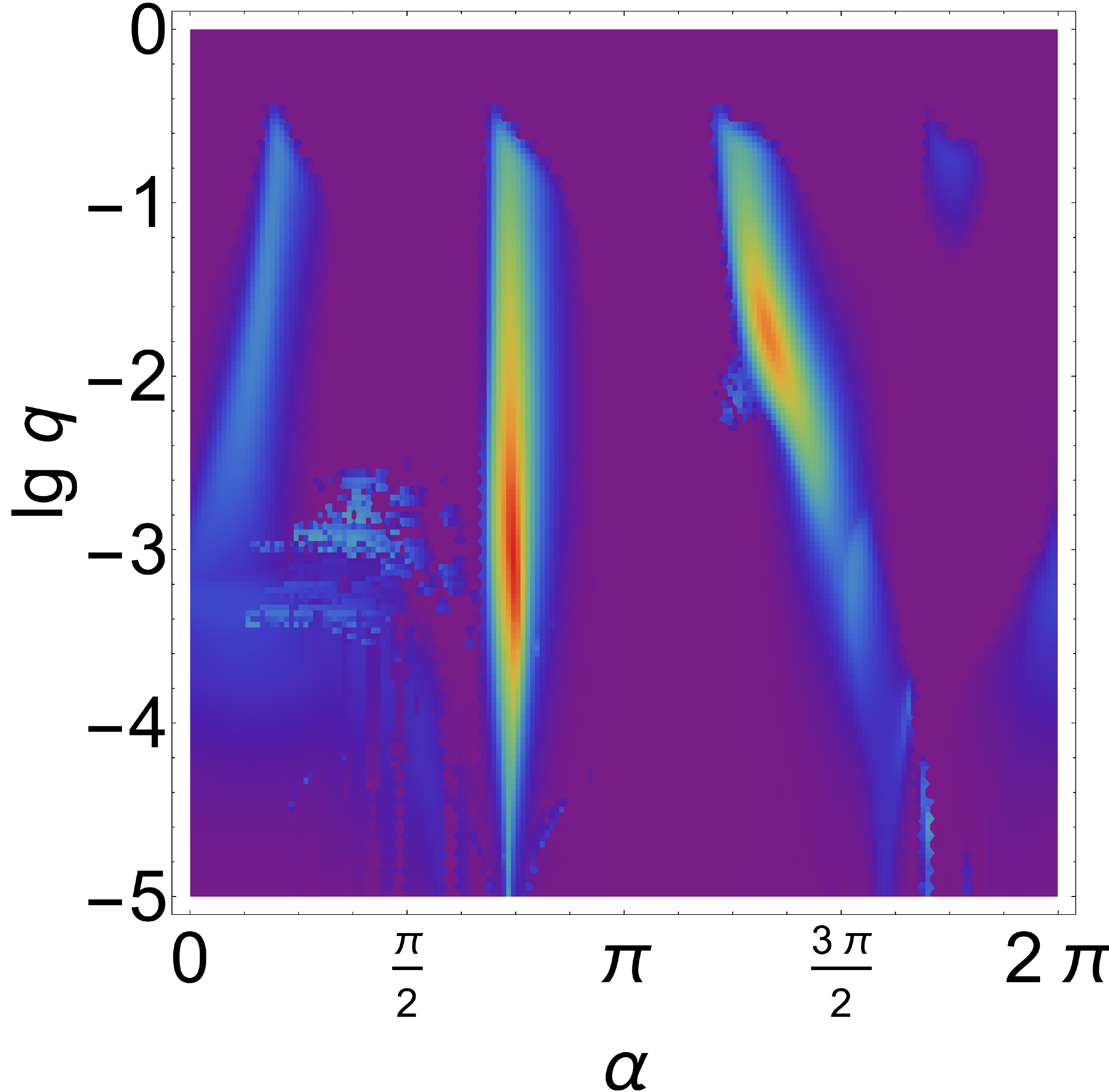}
\includegraphics[width=4cm]{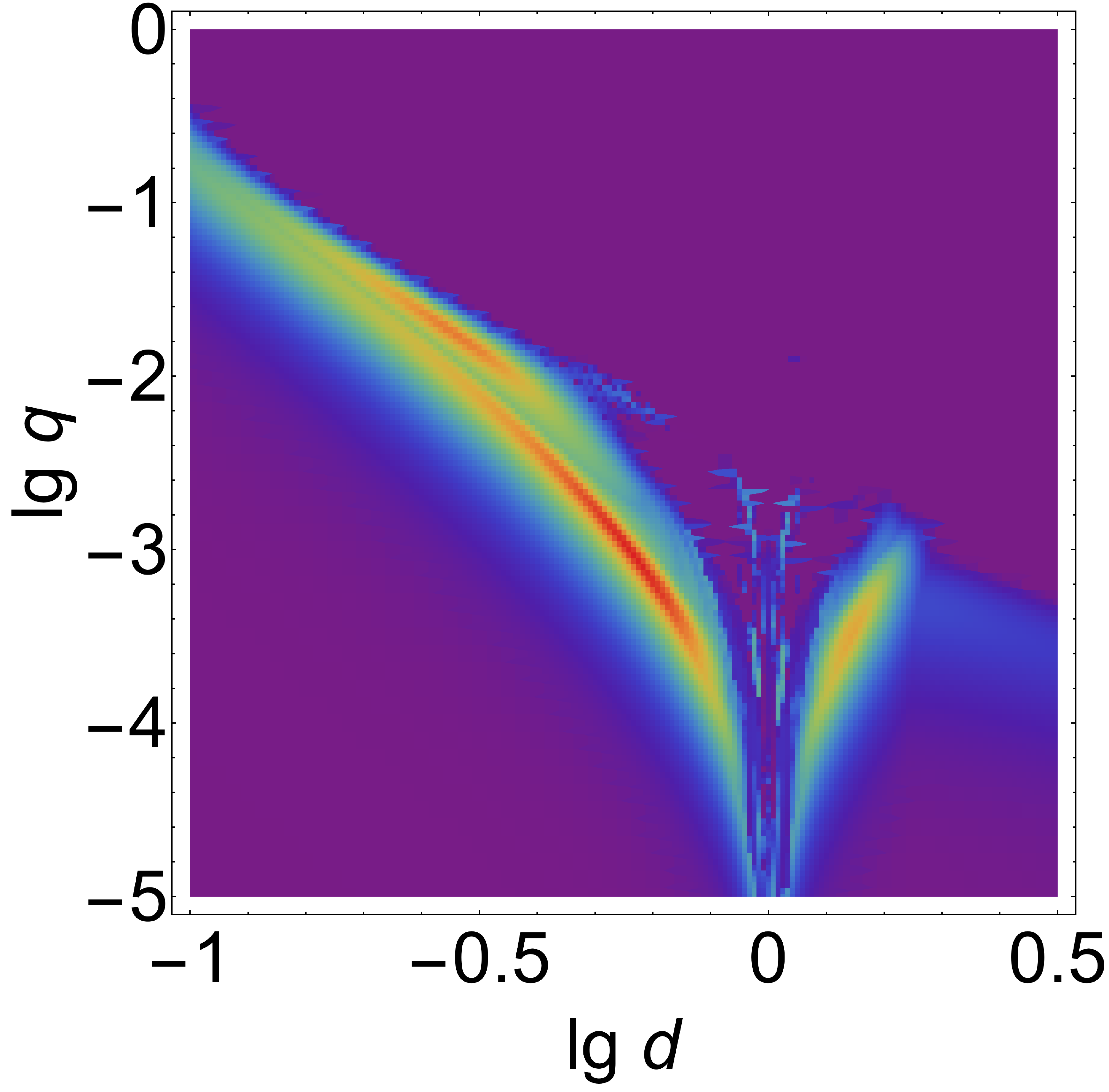}\\[1mm]
\begin{minipage}{4cm}
\hspace*{0.6cm}
\includegraphics[width=3.3cm]{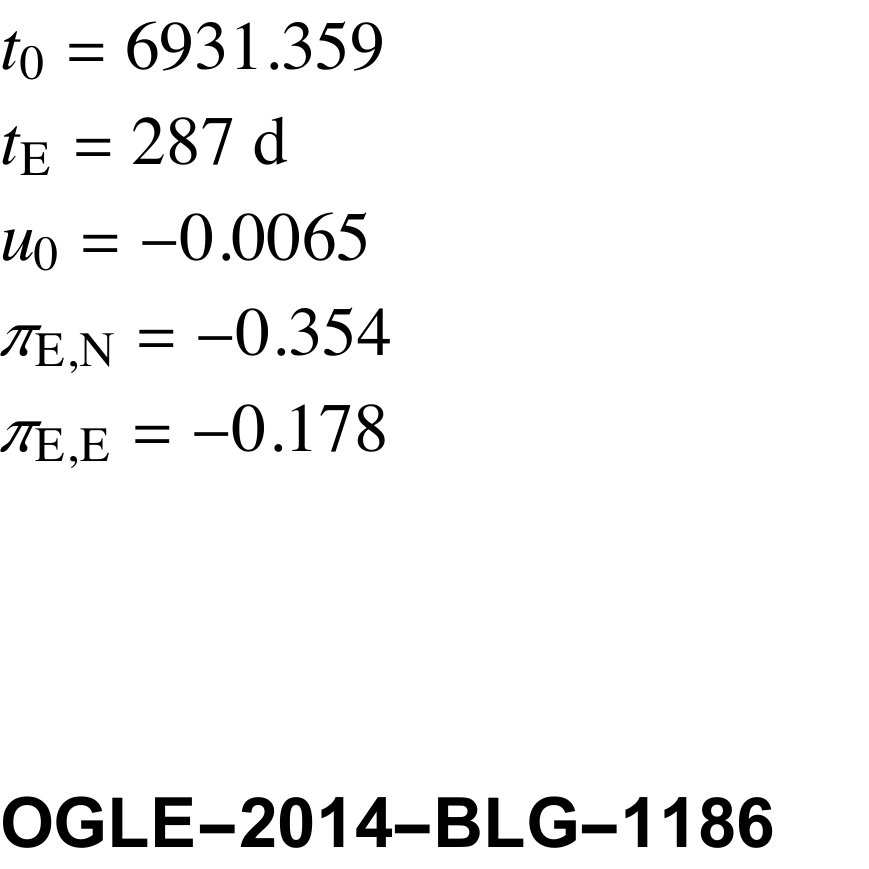}
\hspace*{-1.6cm}
\raisebox{5.3mm}{\includegraphics[height=3.34cm]{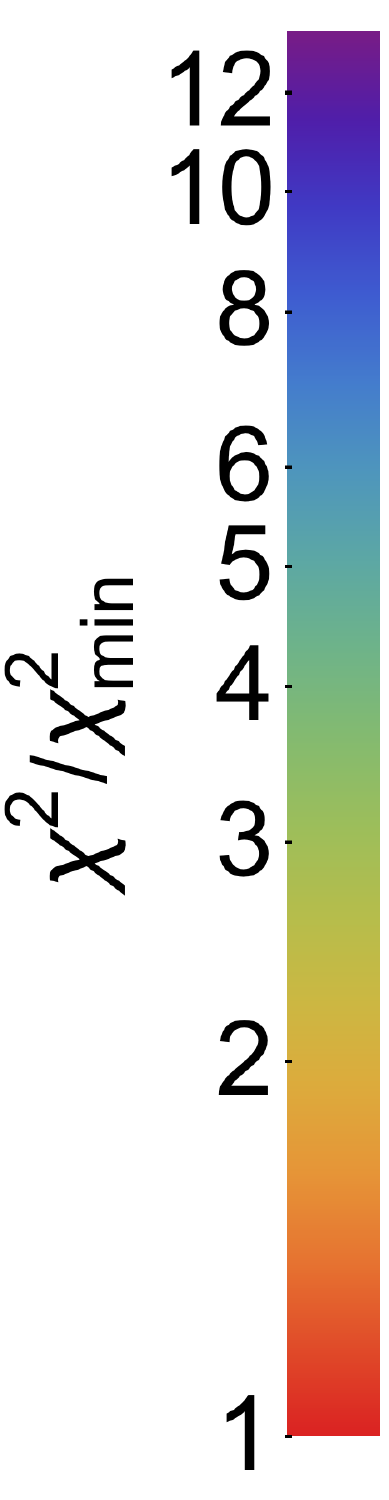}}
\end{minipage}
\begin{minipage}{4cm}
\includegraphics[width=4cm]{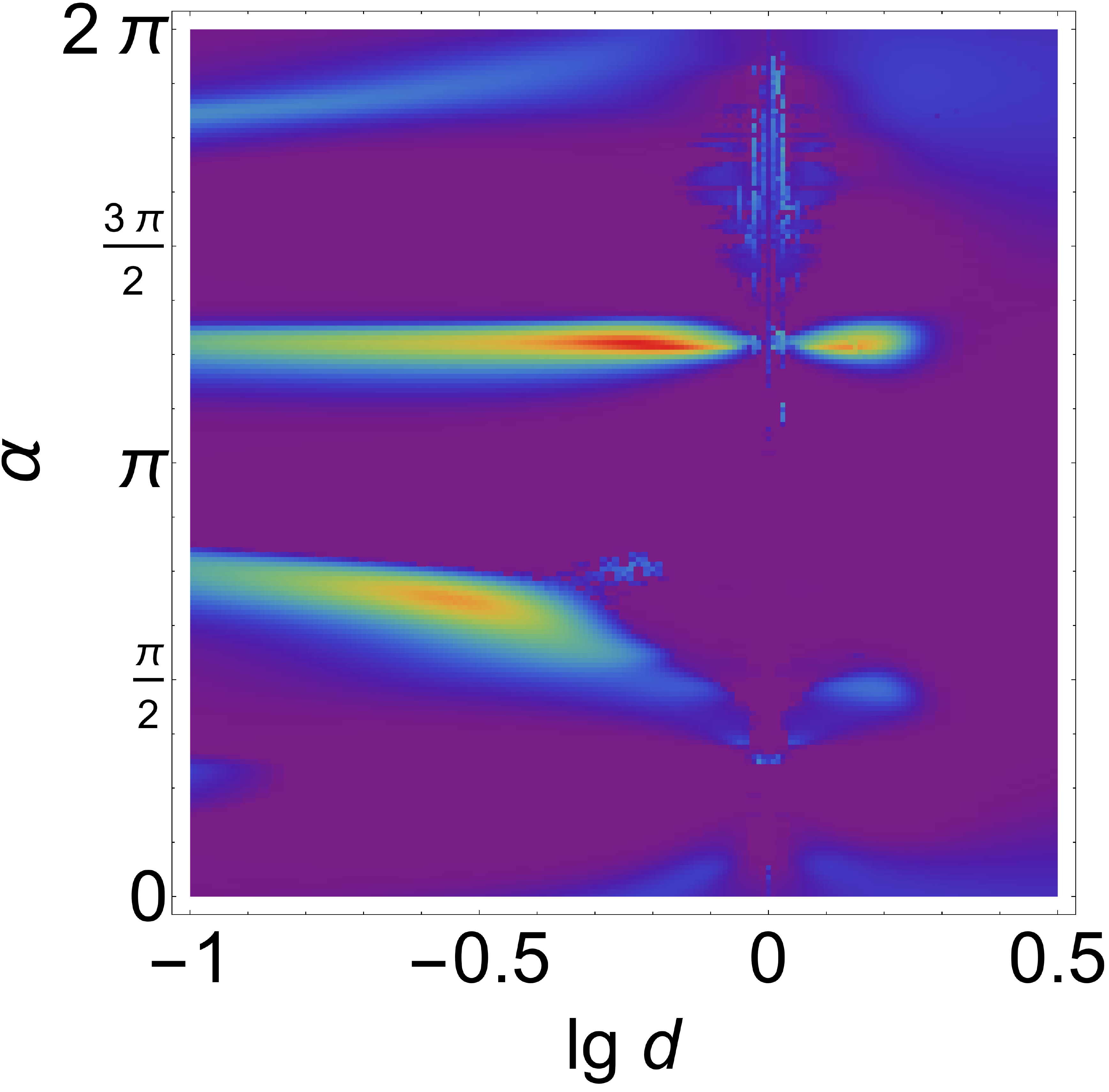}
\end{minipage}
\hfill
\begin{minipage}{4cm}
\hspace*{0.6cm}
\includegraphics[width=3.3cm]{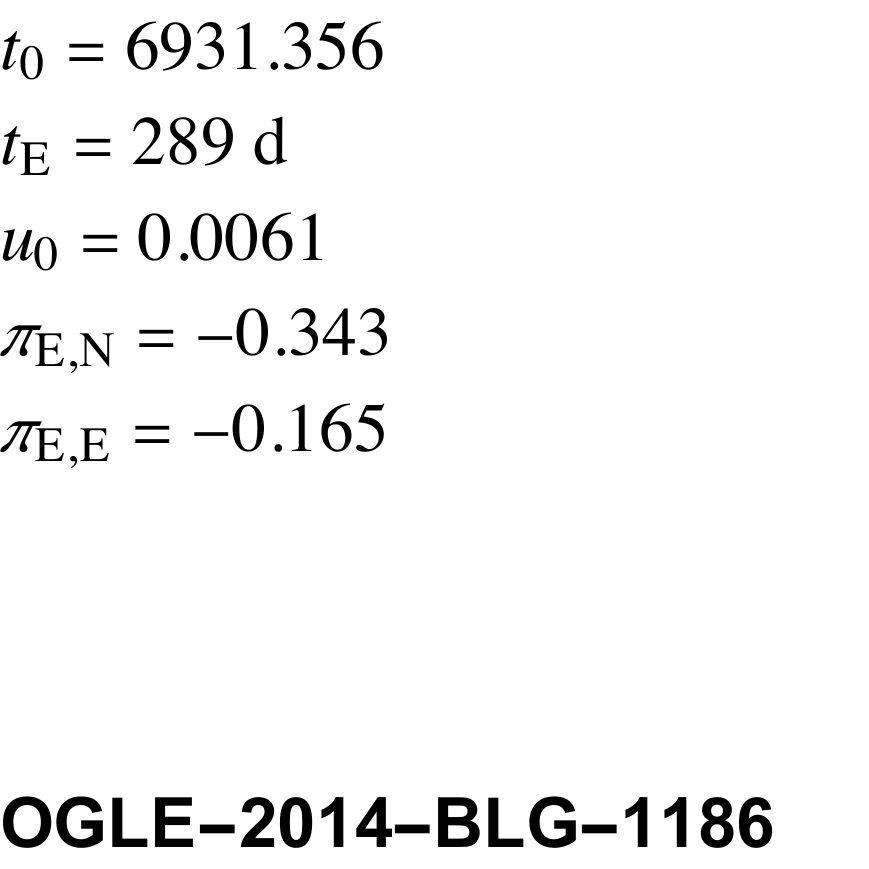}
\hspace*{-1.6cm}
\raisebox{5.3mm}{\includegraphics[height=3.34cm]{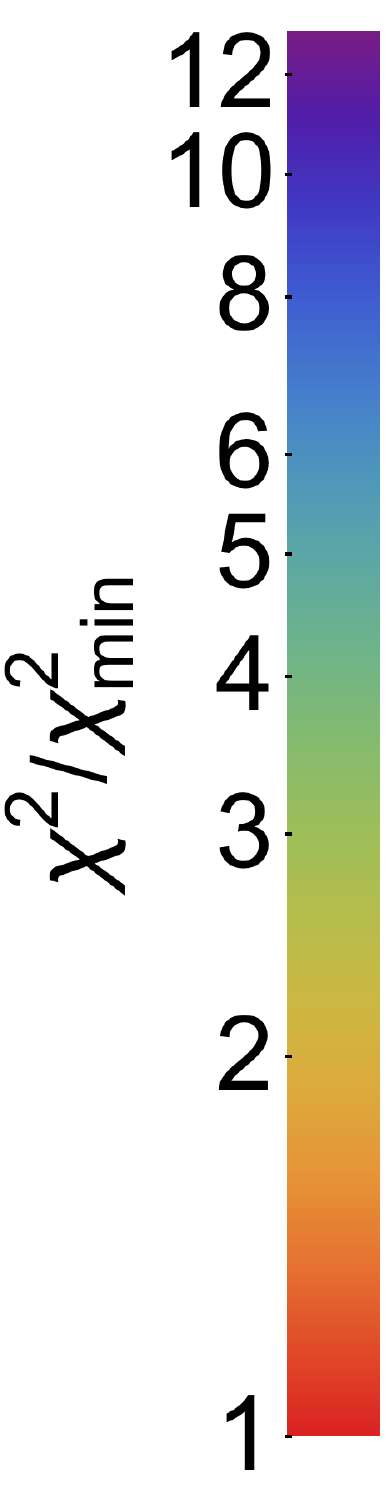}}
\end{minipage}
\begin{minipage}{4cm}
\includegraphics[width=4cm]{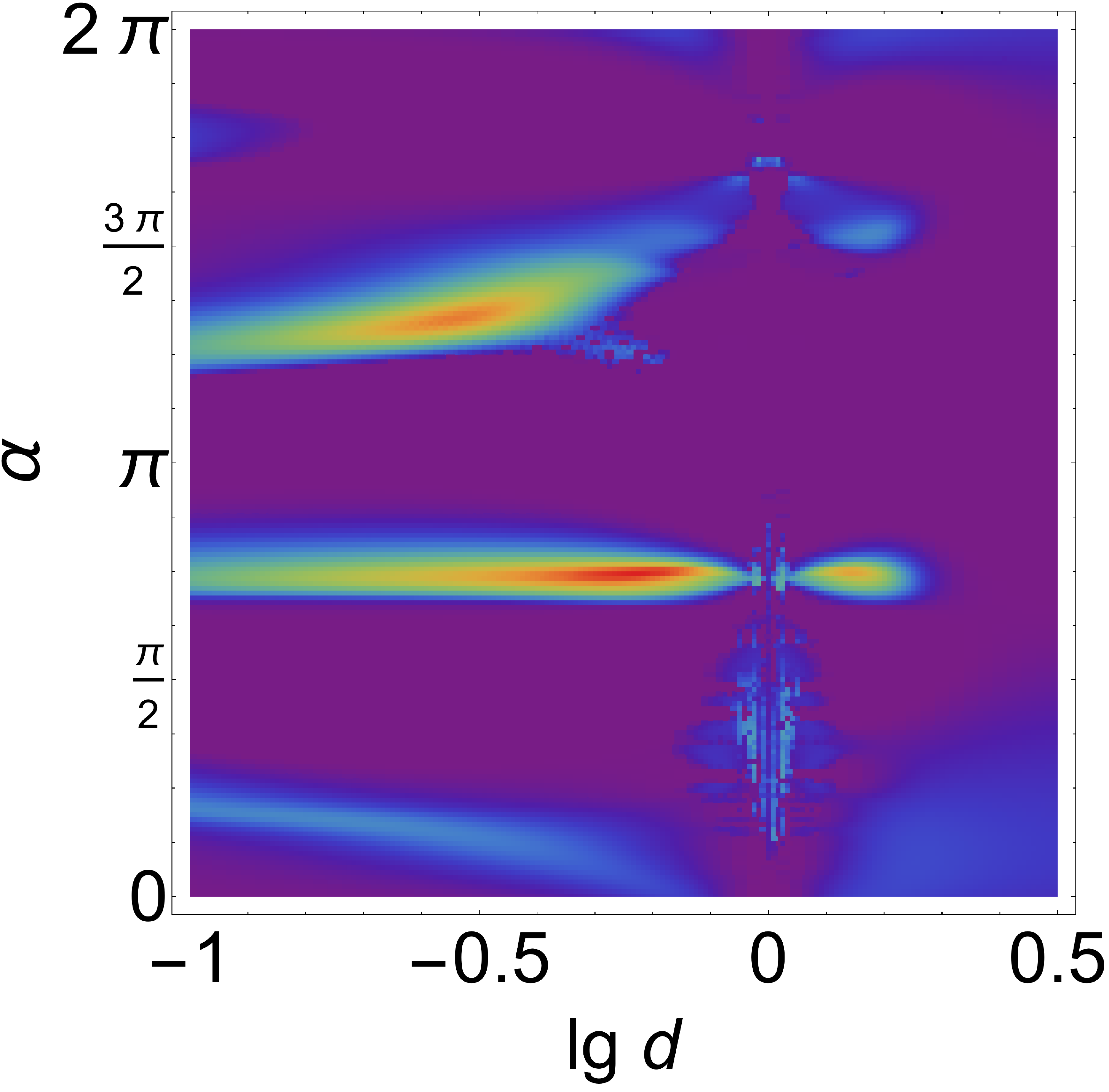}
\end{minipage}\\
\caption{Exploration of binary-lens parameter space. Colour-coded values of $\chi^2$ for the peak data as a function of the binary-lens parameters $(d,q,\alpha)$ as three diagrams of $\chi^2$ as a function of  $(\lg q,\alpha)$, $(\lg d,\lg q)$, and $(\lg d,\alpha)$, where each reported value of $\chi^2$ corresponds to the minimum over the remaining third parameter. These diagrams have been positioned so that all three parameters corresponding to local minima can readily be identified. The parameters $(t_0,t_\mathrm{E},u_0,\pi_{\mathrm{E},\mathrm{N}},\pi_{\mathrm{E},\mathrm{E}})$ have been kept fixed to values suggested by single-lens point-source parallax models for the off-peak data for $u_0 < 0$ or $u_0 > 0$, respectively (c.f.\ Tables~\ref{Tab:Models1} and~\ref{Tab:Models2}), and only the baseline fluxes $F^{[j]}_\mathrm{base}$ and the blend ratios $g^{[j]}$ have been adjusted for each $(d,q,\alpha)$ in order to minimise $\chi^2$. The colour scale has been normalized, so that the absolute minimum corresponds to the red end, while a single lens or any configuration with a larger $\chi^2$ corresponds to the purple end. While for small mass ratios $q$, one finds an ambiguity $d \leftrightarrow d^{-1}$ for the separation parameter, the valleys distinguished by the trajectory angle $\alpha$ correspond to the three possible types for the specific morphology observed in the light curve, illustrated in Fig.~\ref{Fig:types}.}
\label{Fig:binsearch}
\end{figure*}

With the parameters $(t_0,u_0,t_\mathrm{E},\pi_{\mathrm{E},\mathrm{N}},\pi_{\mathrm{E},\mathrm{E}})$ already being reasonably well determined from the off-peak data, we searched the complementary parameter sub-space $(d,q,\alpha)$, characterising the lens binarity, for viable models incorporating the peak data. In fact, for fixed $(t_0,u_0,t_\mathrm{E},\pi_{\mathrm{E},\mathrm{N}},\pi_{\mathrm{E},\mathrm{E}})$ and the adopted scaling of error bars (according to Table~\ref{Tab:Scales}), we evaluated $\chi^2$ for a dense grid of  $(d,q,\alpha)$ for the peak data, just adjusting the baseline fluxes $F^{[j]}_\mathrm{base}$ and the blend ratios $g^{[j]}$, so that $\chi^2$ is minimised. The resulting $\chi^2$ maps for the both cases $u_0 <0$ and $u_0>0$ are shown in Fig.~\ref{Fig:binsearch}.

\begin{figure}
\begin{center}
\includegraphics[width=4.5cm]{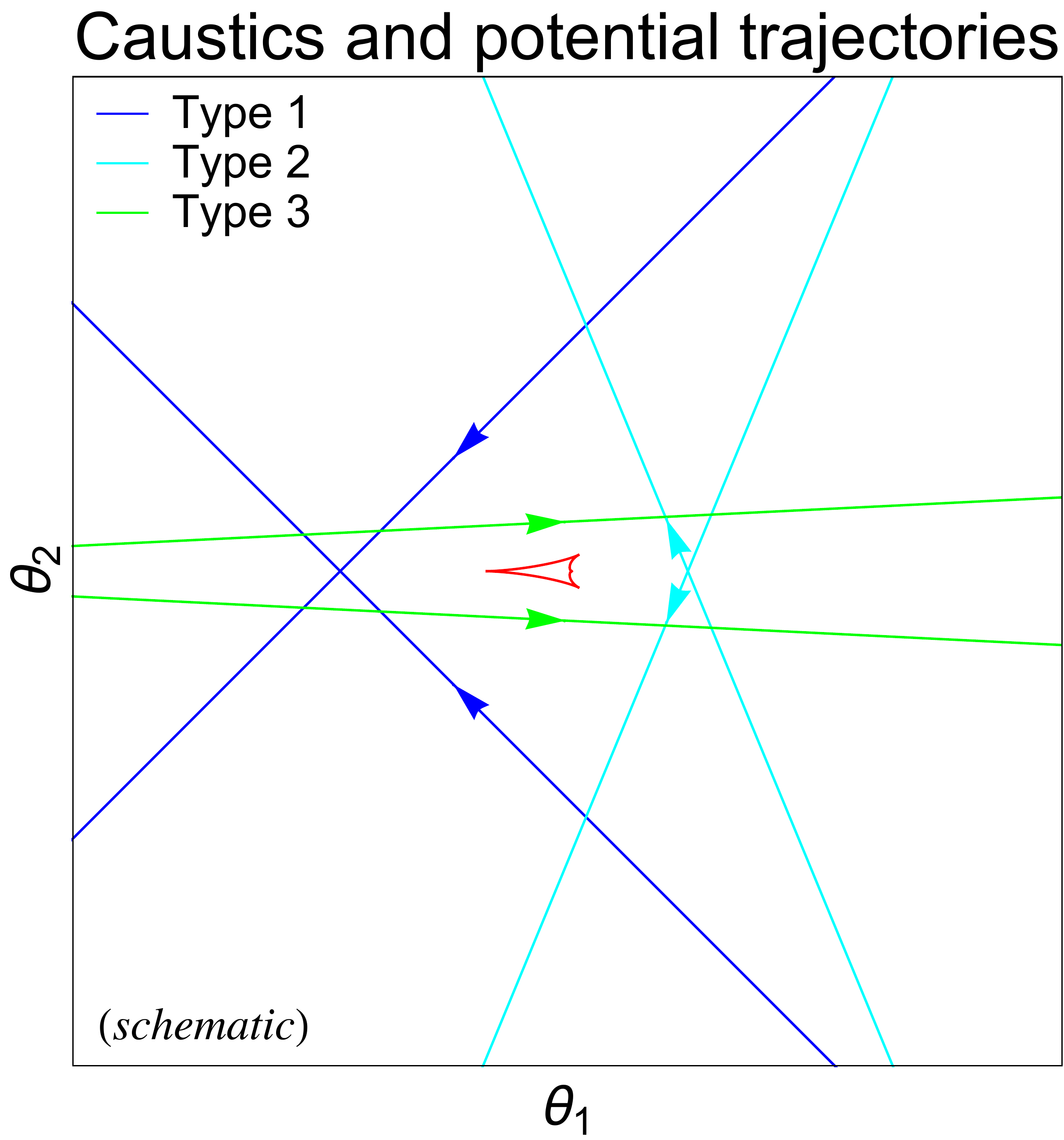}
\end{center}
 \caption{Schematic illustration of the three possible types of trajectories that can produce a light curve with the observed features. These are to arise from a very close approach of the source to a central caustic (first peak), with a subsequent approach to one of its cusps producing the second peak. The small impact parameter $u_0$ rules out intermediate topologies.  For each type of trajectory, there are two realisations for the impact parameter $u_0$ and trajectory angle $\alpha$, distinguished by $(u_0,\alpha) \leftrightarrow (-u_0,-\alpha)$. The three different types are clearly seen in the $\chi^2$ plots exploring the binary-lens parameter space, as shown in Fig.~\ref{Fig:binsearch}.}
 \label{Fig:types}
\end{figure}

Moreover, the binary-lens parameter space can be constrained straightforwardly from the morphology of the light curve. While we find an impact parameter $u_0 < 0.01$, the observed light curve does not exhibit any
strong features arising from the source passing over a caustic. This immediately rules out any configuration with an intermediate caustic, while the size of the central caustic for a close or wide binary is restricted by the small impact parameter. Moreover, the shape of the anomaly over the peak suggests that the source first reaches a closest approach to the central caustic, producing the first (main) peak, and then passes close to one of the cusps of the central caustic, producing the further second peak. As illustrated in Fig.~\ref{Fig:types}, this leaves us with only three options for the angle of the source trajectory with respect to the binary axis for each $u_0 <0$ and $u_0>0$, which are identifiable as $\chi^2$ valleys in Fig.~\ref{Fig:binsearch}. Namely, the second peak can arise from the source passing near the cusp on the binary axis at the `pointy end' towards the secondary (type~1), or the source passing near the off-axis cusp, with the trajectory either close to perpendicular to the binary axis (type~2) or close to parallel to the binary axis (type~3). The acquired data rule out configurations for which the source trajectory gets near the cusp on the binary axis that is opposite the secondary, because such would hit the caustic near at least one of the off-axis cusps. The $\chi^2$ maps (Fig.~\ref{Fig:binsearch}) also 
 explicitly reveal the ambiguity $d \leftrightarrow d^{-1}$ between close- and wide-binary models for small mass ratios $q$.

\subsubsection{The only viable binary-lens models and parameter ambiguities}

With our $\chi^2$ maps for $(d,q,\alpha)$ and our further assessment of possible configurations, viable models must reside within a local minimum that corresponds to one of the 12 options given by $u_0 < 0$ or $u_0 >0$, $d < 1$ or $d > 1$, and one of the three trajectory types shown in Fig.~\ref{Fig:types}. Local $\chi^2$ optimisation of the full parameter space $(t_0,t_\mathrm{E},u_0,\pi_{\mathrm{E},\mathrm{N}},\pi_{\mathrm{E},\mathrm{E}},d,q,\alpha,\vec{F}_\mathrm{base},\vec{g}) $ for all data shows that type~2 and type~3 trajectories cannot reasonably account for the data, given that best-fitting model light curves are clearly visually off the data, leaving us with the four models listed in Tables~\ref{Tab:Models1} and~\ref{Tab:Models2}, whose light curves are shown together with the peak data in Fig.~\ref{Fig:BPLC}, and no further possible options. Type~2 and type~3 trajectories fail on the requirement that in order to match the data, the impact parameter $u_0$, the trajectory angle $\alpha$, and the time-scale $t_\mathrm{E}$ must meet the size of the caustic and the time interval between the two observed peaks. We find that the values $(t_\mathrm{E},u_0,\pi_{\mathrm{E},\mathrm{N}},\pi_{\mathrm{E},\mathrm{E}})$ are essentially identical to what we estimated from the off-peak data, passing the check of robustness and consistency of our approach.

Visual inspection of the model light curves and the peak data (as shown in Fig.~\ref{Fig:BPLC}) reveals a few low-level discrepancies: 1) Most significantly, over the second peak, the slope of the model light curve is not in agreement with what two data sets (COJ~B and FTS) independently suggest, 2) between the two peaks, the model favours the LSC~B and LSC~C data, while substantially disfavouring the OGLE data, 3) the CPT~C  and OGLE data over the main peak are systematically above the model light curve, 4) the OGLE, LSC B, LSC B, and  FTS data just ahead of the main peak are all below the model light curve, 
5) the FTS data just after the first peak are all above the model light curve.



At this stage, we looked into the effect of the finite size of the source star on the light curve, which becomes significant for strong differential magnification with substantial second derivatives. 
It can be described by means of a  dimensionless parameter $\rho_\star$, where $\rho_\star\,\theta_\mathrm{E}$ is the angular source radius,
and to first order  the star can be approximated as being uniformly bright. 
For the evaluation of the magnification for given model parameters, we have adopted a
contour integration algorithm \citep{Do:Dipl,GG:contour,Do:contour2} improved with parabolic correction, optimal sampling and accurate error estimates, as described in detail by \citet{Bozza:contour}.

\begin{figure*}
\begin{minipage}{4.25cm}
\begin{center}
\includegraphics[width=3.8cm]{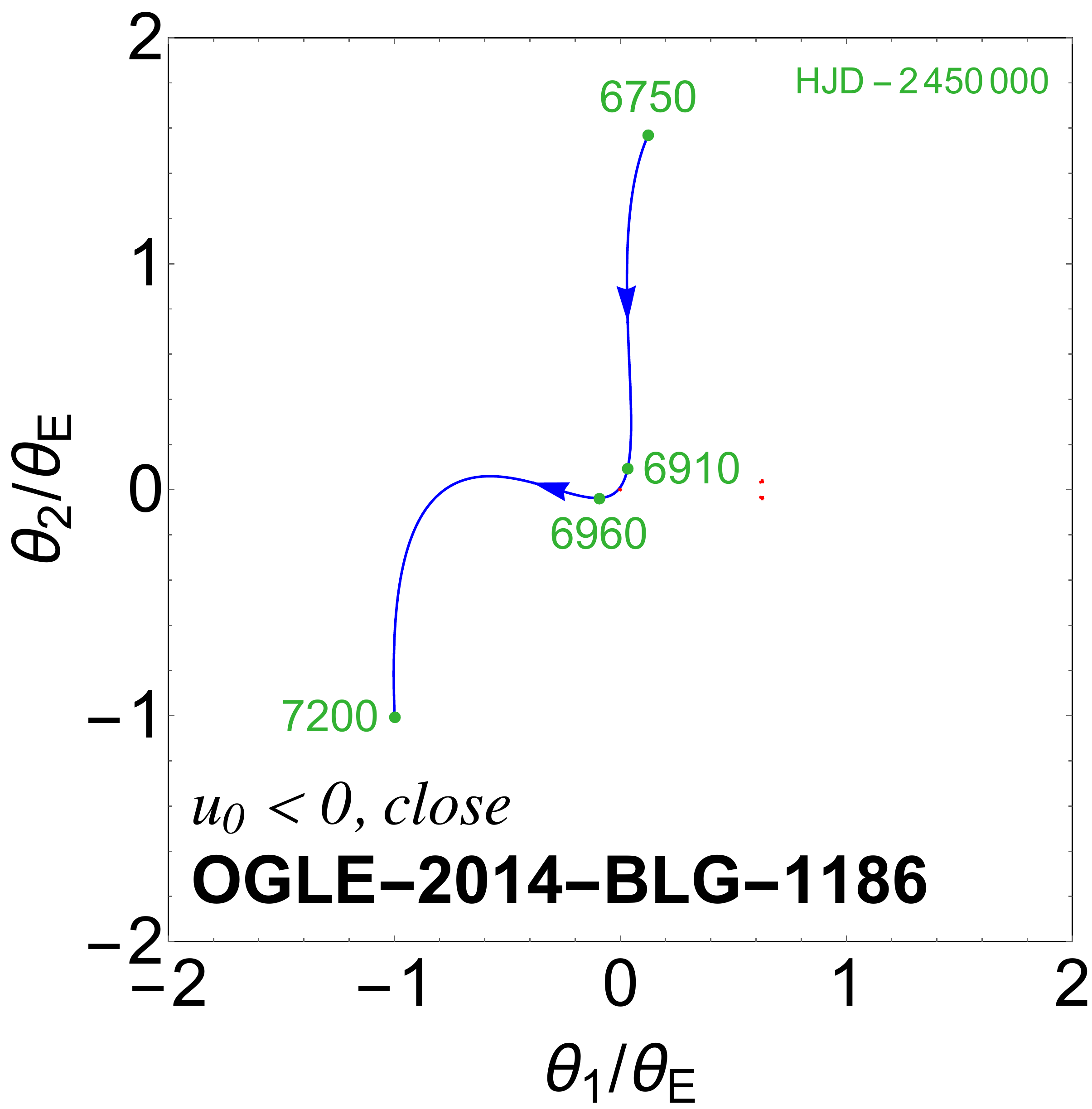}
\end{center}
\end{minipage}
\begin{minipage}{4.25cm}
\begin{center}
\includegraphics[width=3.8cm]{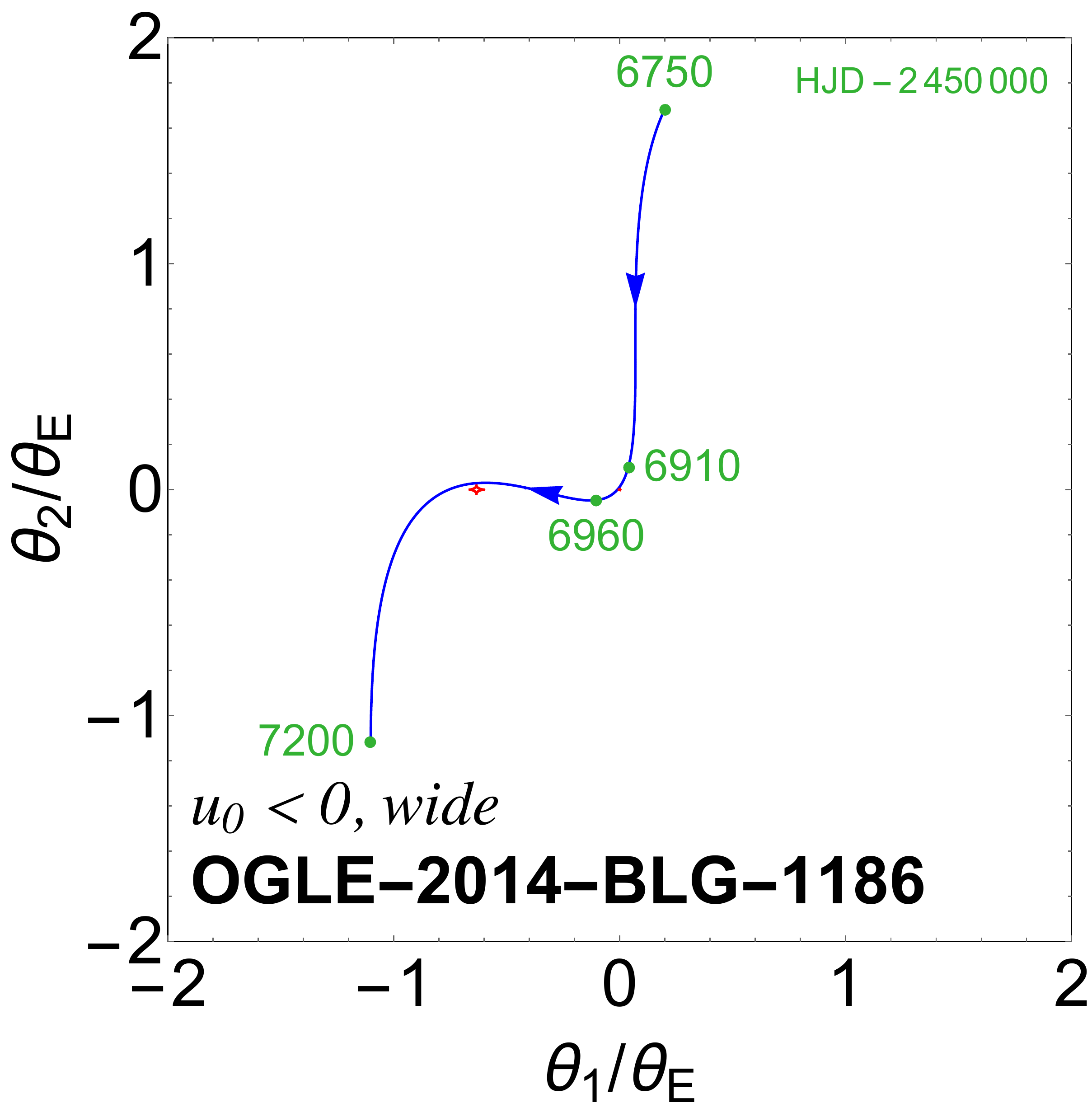}
\end{center}
\end{minipage}
\begin{minipage}{4.25cm}
\begin{center}
\includegraphics[width=3.8cm]{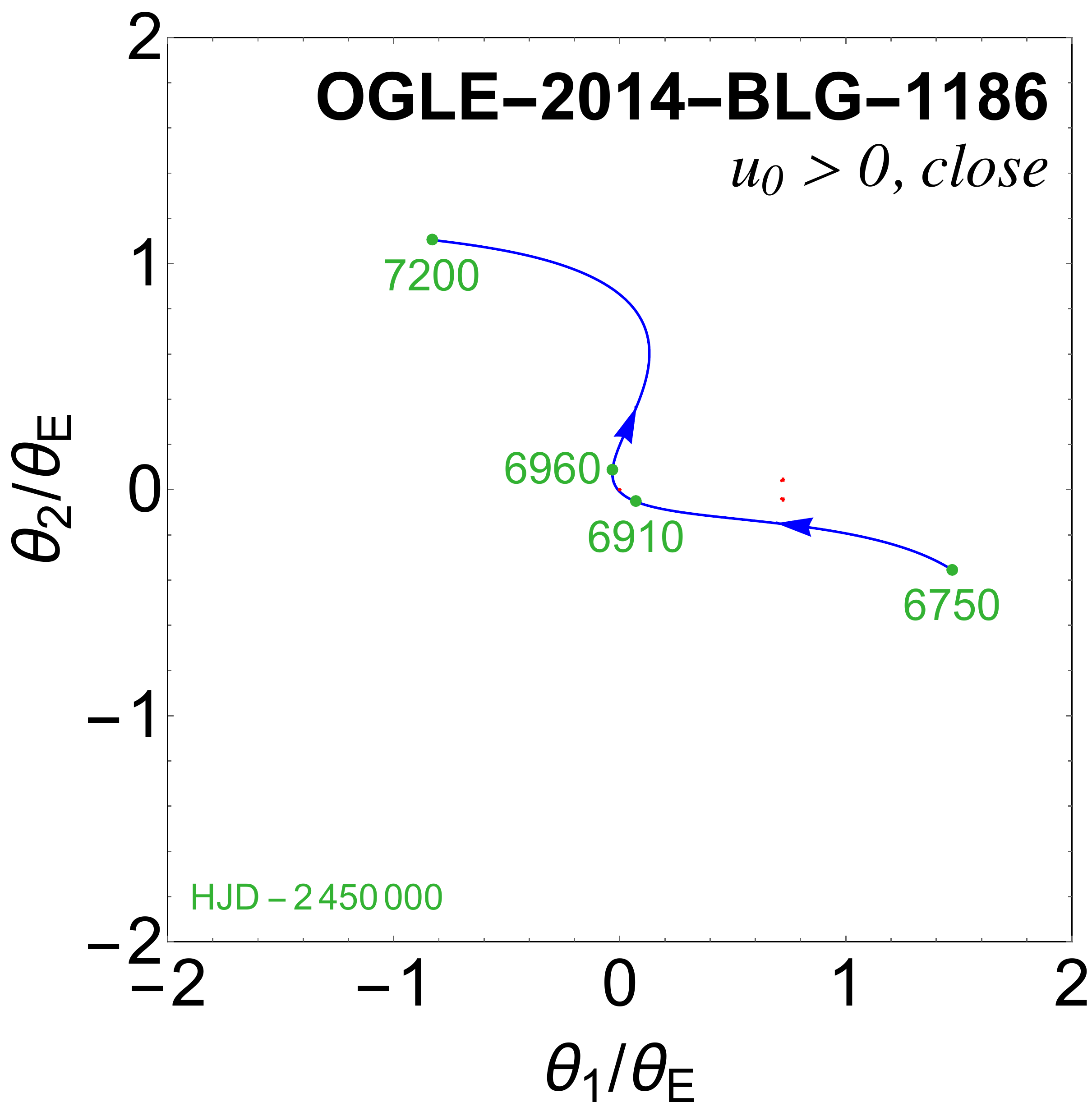}
\end{center}
\end{minipage}
\begin{minipage}{4.25cm}
\begin{center}
\includegraphics[width=3.8cm]{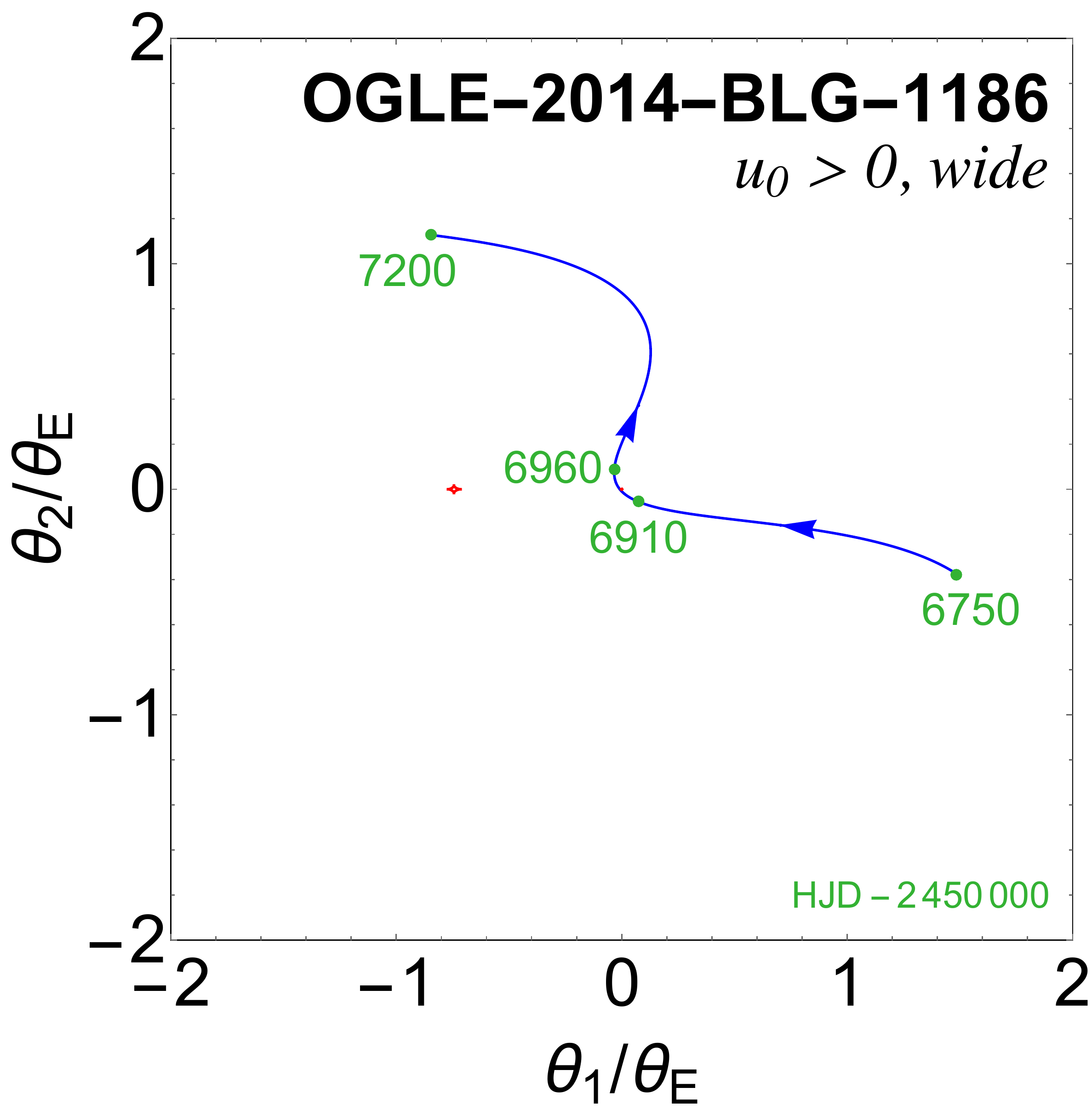}
\end{center}
\end{minipage}\\
\includegraphics[width=4.25cm]{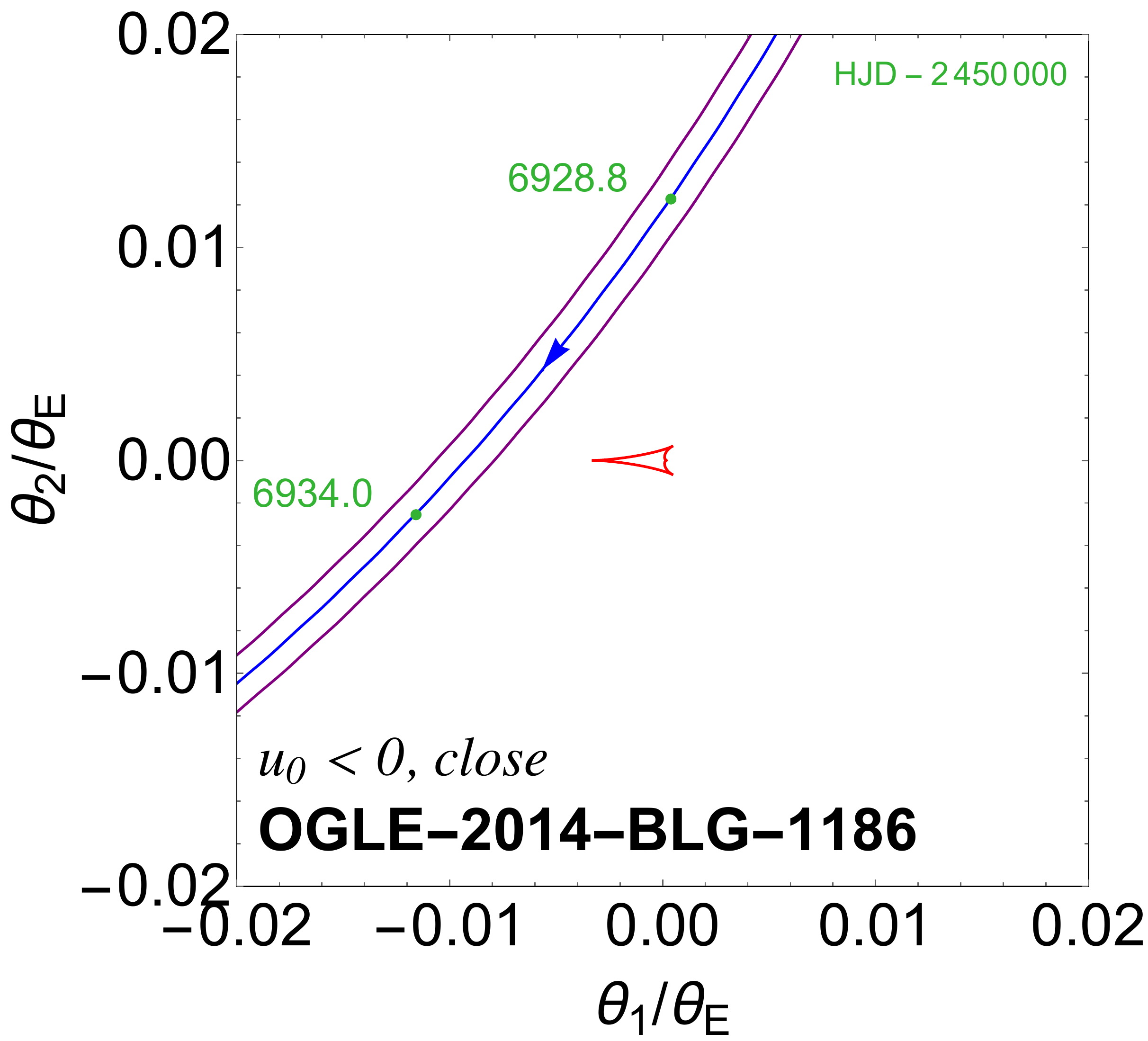}
\includegraphics[width=4.25cm]{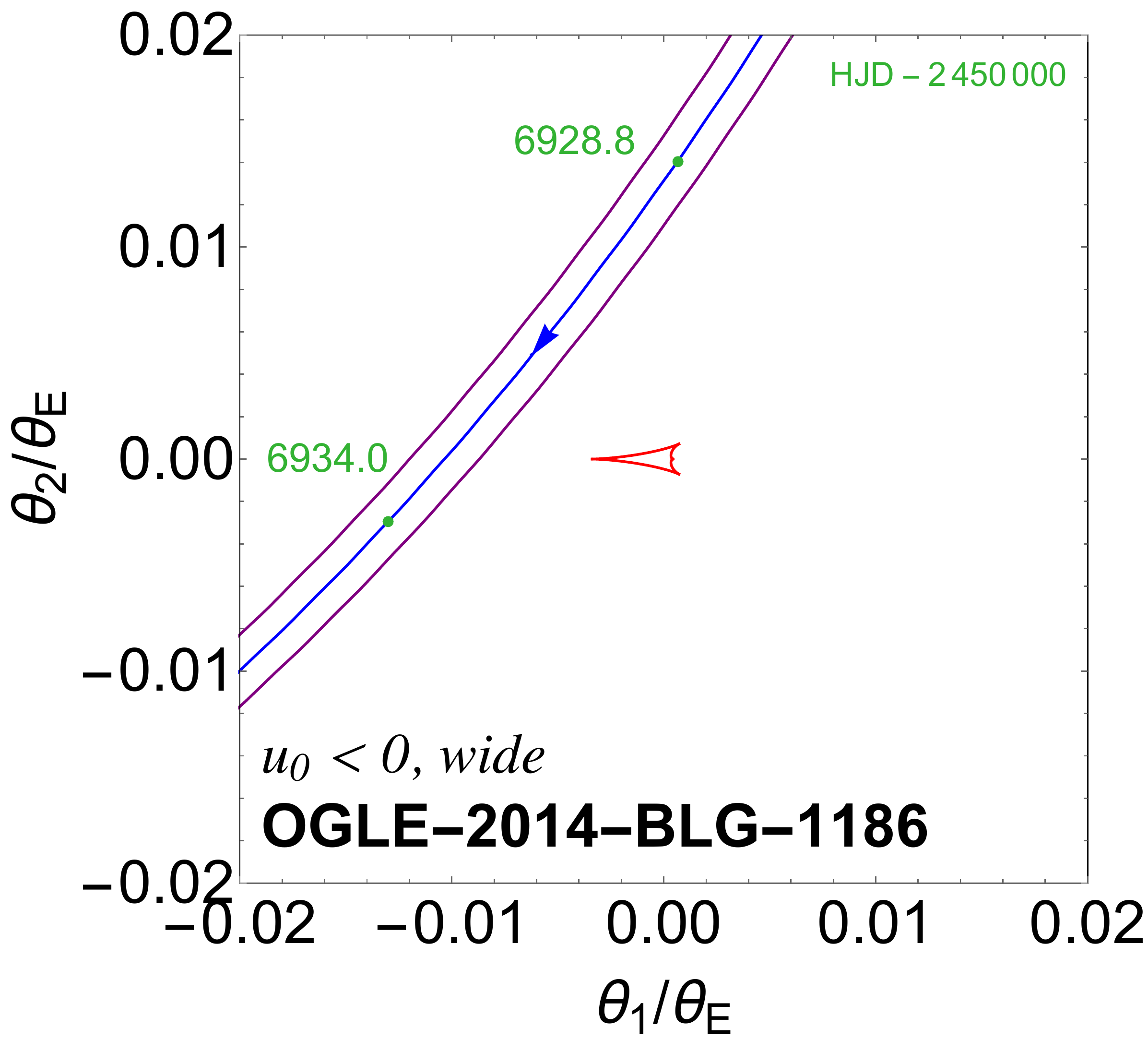}
\includegraphics[width=4.25cm]{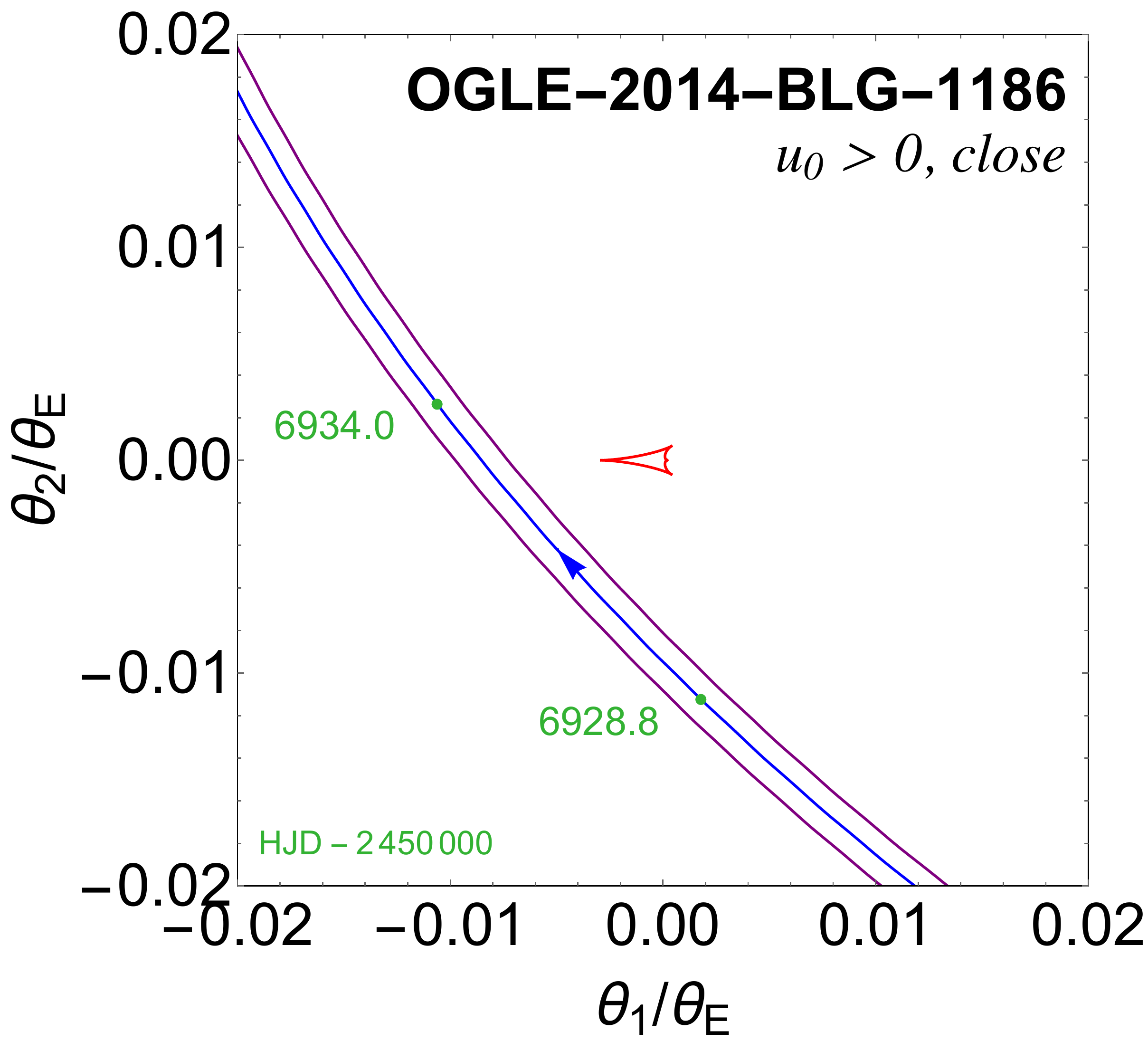}
\includegraphics[width=4.25cm]{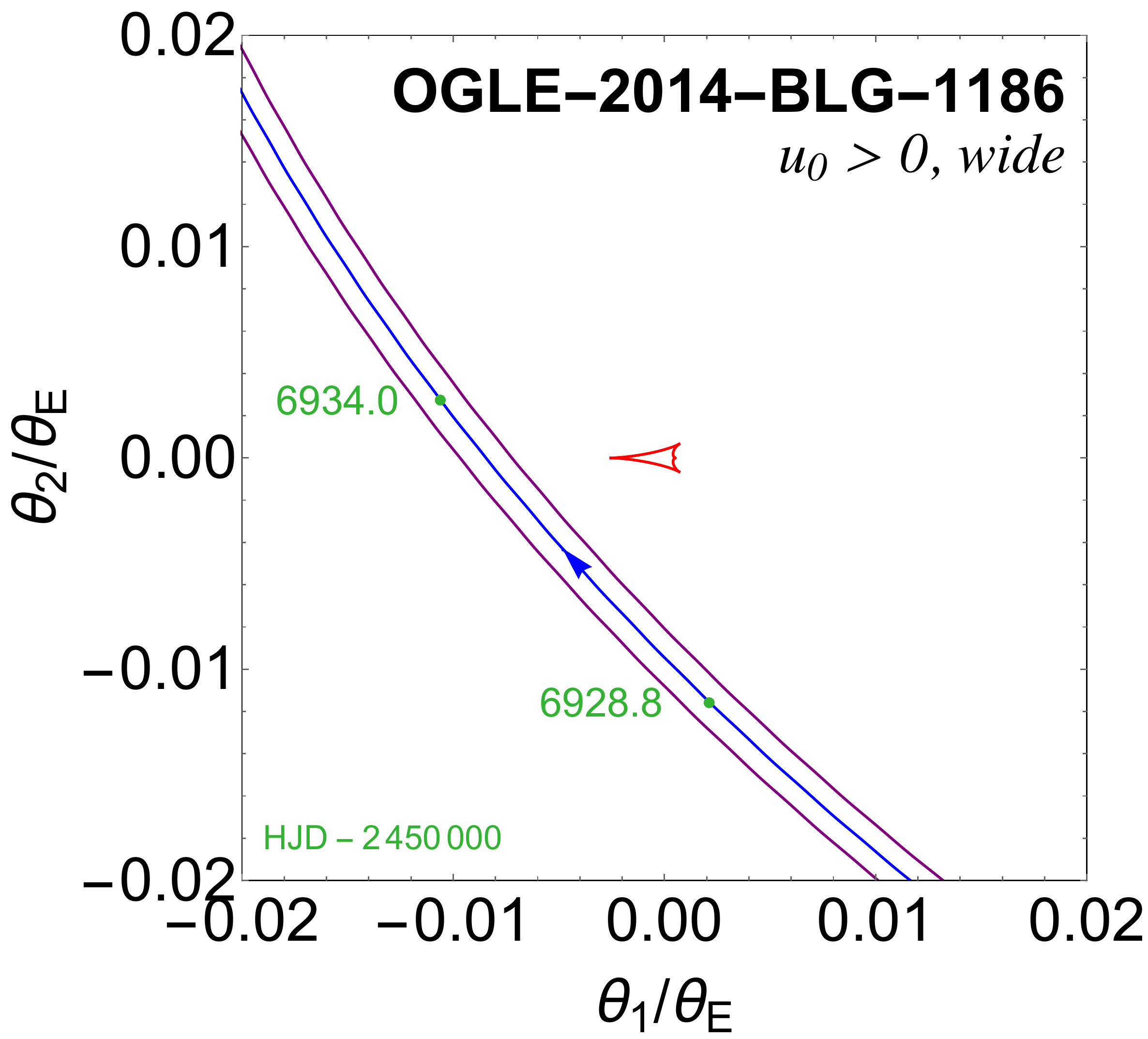}
\caption{Binary-lens caustics (in red) and source trajectory (in blue, indicating the finite source size by magenta lines) for the four binary-lens finite-source parallax models whose respective parameters are listed in Table~\ref{Tab:ModelsFinite}. Specific epochs are marked by green dots, and the direction of the source along the trajectory is indicated by arrows. The four-fold model ambiguity corresponds to solution with $u_0 < 0$ or $u_0 > 0$ on one hand, as well as close or wide binaries ($d < 1$ or $d> 1$) on the other hand. With the sign of the impact parameter $u_0$, the sign of the trajectory angle $\alpha$ gets inverted as well ($\alpha \leftrightarrow -\alpha$ or $\alpha \leftrightarrow 2\pi -\alpha$). The parallax effect is prominent in the wings of the light curve, which determine the parallax parameters $(\pi_{\mathrm{E},\mathrm{N}},\pi_{\mathrm{E},\mathrm{E}})$, while the local effective acceleration of the source near the peak is small, with opposite curvature of the source trajectory for the $u_0 < 0$ and $u_0 > 0$ cases. For the $u_0 <0$ wide-binary model,
the source trajectory gets close to the planetary caustic, resulting in a further feature (see Fig.~\ref{fig:furtherfeature}).}
\label{fig:configs}
\end{figure*}

\begin{figure*}
\includegraphics[width=\columnwidth]{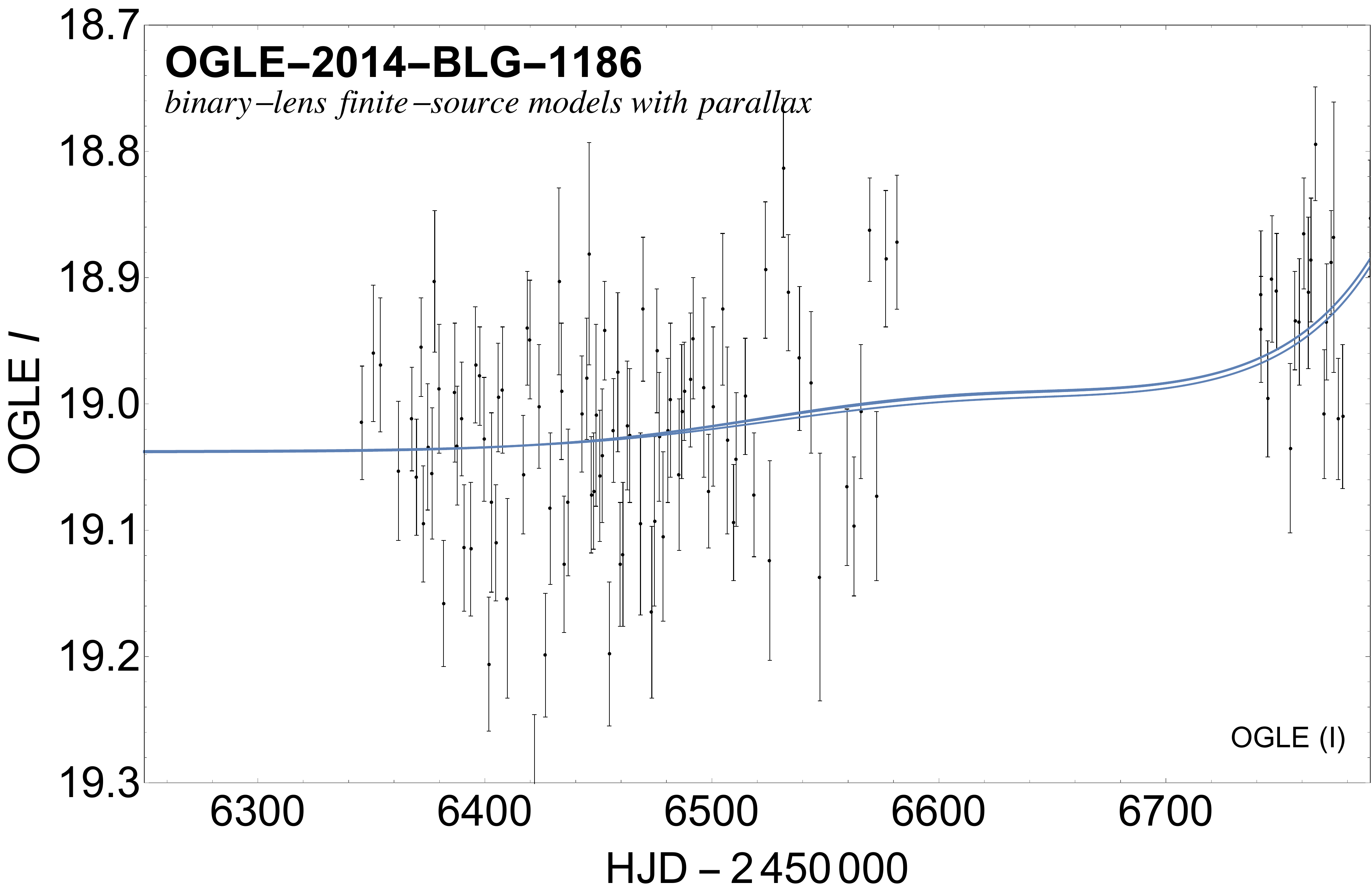}
\hfill
\includegraphics[width=\columnwidth]{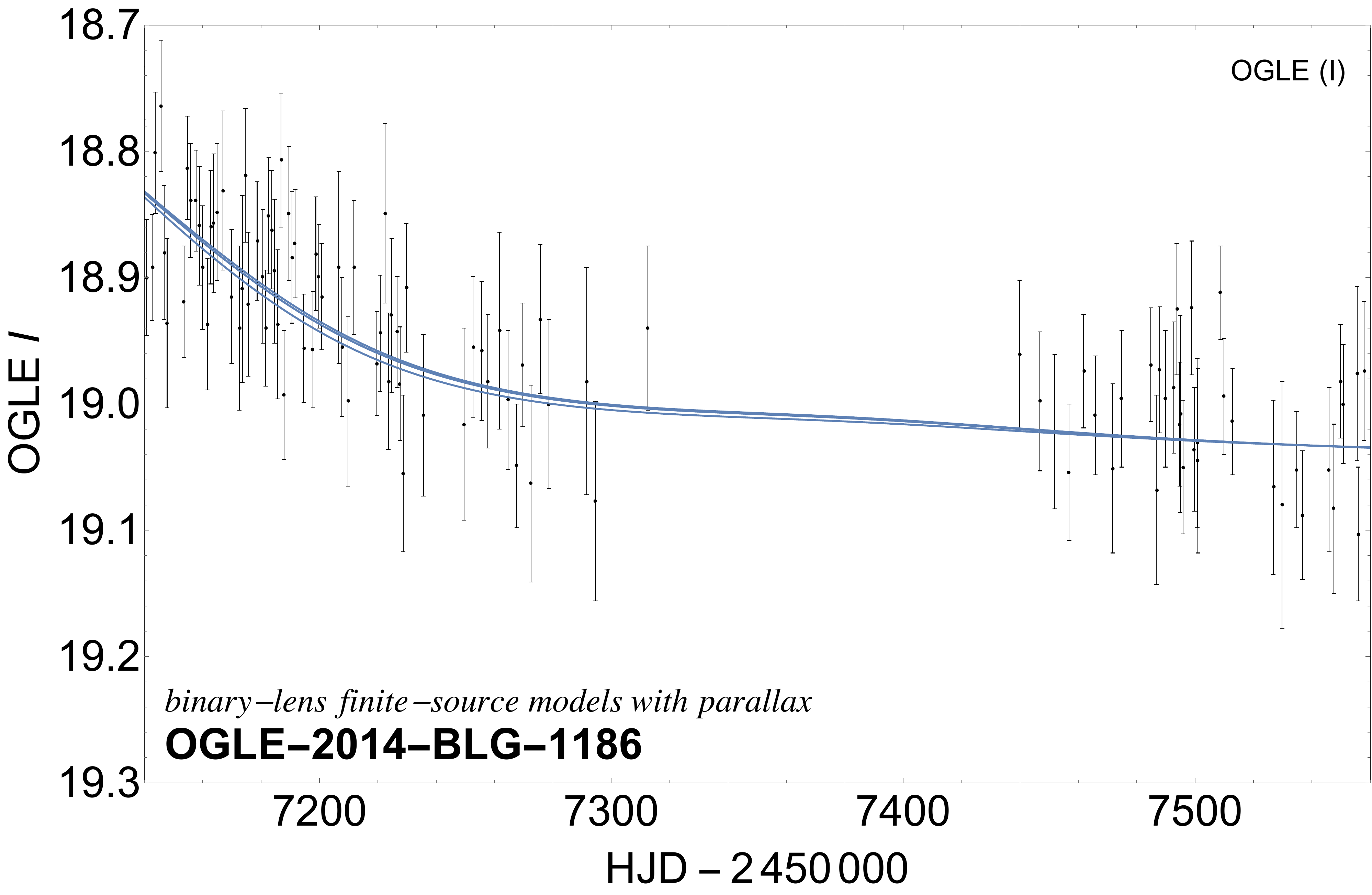}
 \caption{The effect of annual parallax on the light curve in early and late event phases close to the baseline magnitude, with light curves corresponding to each of the finite-source binary-lens parallax models listed in Table~\ref{Tab:ModelsFinite} shown. The left panel shows data from 2013 and 2014, whereas the right panel shows data from 2015 and 2016. The event OGLE-2014-BLG-1186 was evidently above its baseline magnitude over the course of 4~years.}
 \label{fig:far}
\end{figure*}

Considering the finite source star size with our binary-lens point-source parallax models, we find that the major differences arise over the second peak, which deforms into a shoulder at around $\rho_\star \sim 2\times 10^{-3}$, whereas a light curve for $5 \times 10^{-4}$ is rather close to the point-source case. We apply the pyLIMA software suite \citep{pyLIMA}, using differential evolution, to find the binary-lens finite-source parallax models, whose parameters are given in Table~\ref{Tab:ModelsFinite}. For these models, we also show the binary-lens caustics and the respective source trajectory in Fig.~\ref{fig:configs}. The four-fold ambiguity corresponds to close or wide binaries ($d < 1$ or $d > 1$), as well as $u_0 < 0$ or $u_0 > 0$, where $(u_0,\alpha,\vec{\pi}_\mathrm{E}) \leftrightarrow (-u_0,-\alpha,\vec{\pi}_\mathrm{E})$. We explicitly note
that we do not find any of the parallax ambiguities described by \citet{Skowron2011}, in particular not $(u_0,\alpha,\vec{\pi}_\mathrm{E}) \leftrightarrow (-u_0,-\alpha,-\vec{\pi}_\mathrm{E})$, which holds if the parallax affects the microlensing light curve mainly by a local effective acceleration near the peak. In contrast, we find this acceleration to be small and of opposite sign in our $u_0 < 0$ and $u_0 > 0$ models, while the parallax results in a substantial distortion of the wings of the light curve over 4 years. In fact, Fig.~\ref{fig:far} illustrates the effect of parallax in the 2013 and early 2014 data, as well as in the late 2015 and 2016 data.

\begin{figure*}
\begin{center}
\begin{minipage}[c]{4.25cm}
\includegraphics[width=4.25cm]{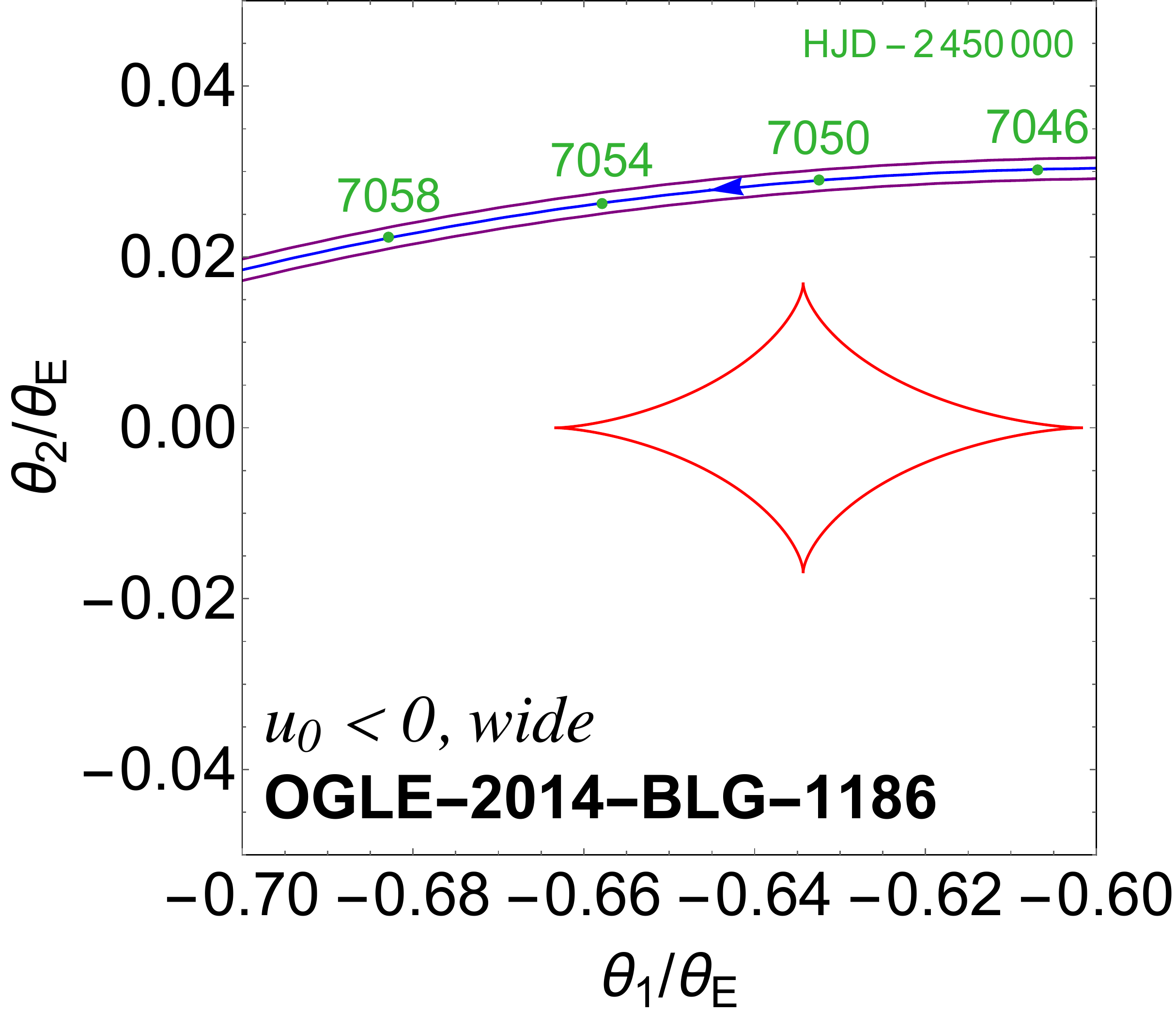}
\end{minipage}
\hspace*{3mm}
\begin{minipage}[c]{\columnwidth}
\resizebox{\columnwidth}{!}{\includegraphics{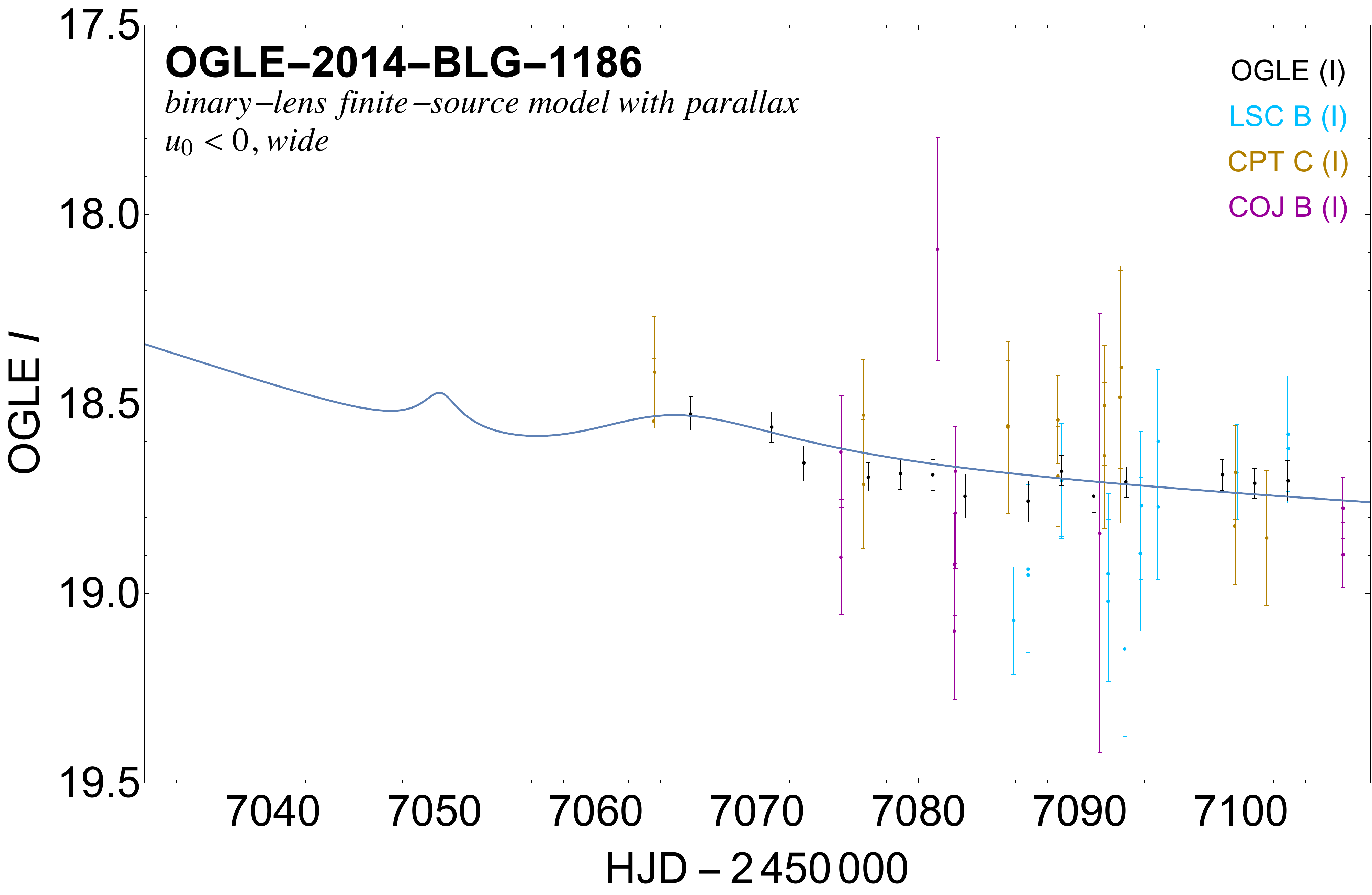}}
\end{minipage}
\end{center}
 \caption{Bump in the light curve for the $u_0 < 0$ finite-source wide-binary model with parallax (c.f.\ Table~\ref{Tab:ModelsFinite}), arising from the source approaching the vicinity of the planetary caustic. The left panel shows the planetary caustic (in red) together with the effective source trajectory (in blue), with the source size  indicated by the purple lines. Specific epochs are marked by green dots. The right panel shows the model light curve around the bump together with the acquired data.  While there is a lack of photometric data around the epoch at which this feature shows, the neglected but existent orbital motion of the planet can alter it and make it essentially disappear. It is therefore unsuitable to provide a distinction between the four presented models.}
 \label{fig:furtherfeature}
\end{figure*}

For the $u_0 <0$ wide-binary model, the source trajectory gets close to the planetary caustic, resulting in a further small feature (see Fig.~\ref{fig:furtherfeature}), most of it falling into a gap of data coverage. For this reason, this model appears to stand out slightly from the others with respect to the parameters. However, the details of the approach to the planetary caustic depend on the orbital motion which is present but cannot be reliably determined. Therefore, this potential feature does not provide us with an opportunity to distinguish between the four models.

\begin{figure*}
\resizebox{\columnwidth}{!}{\includegraphics{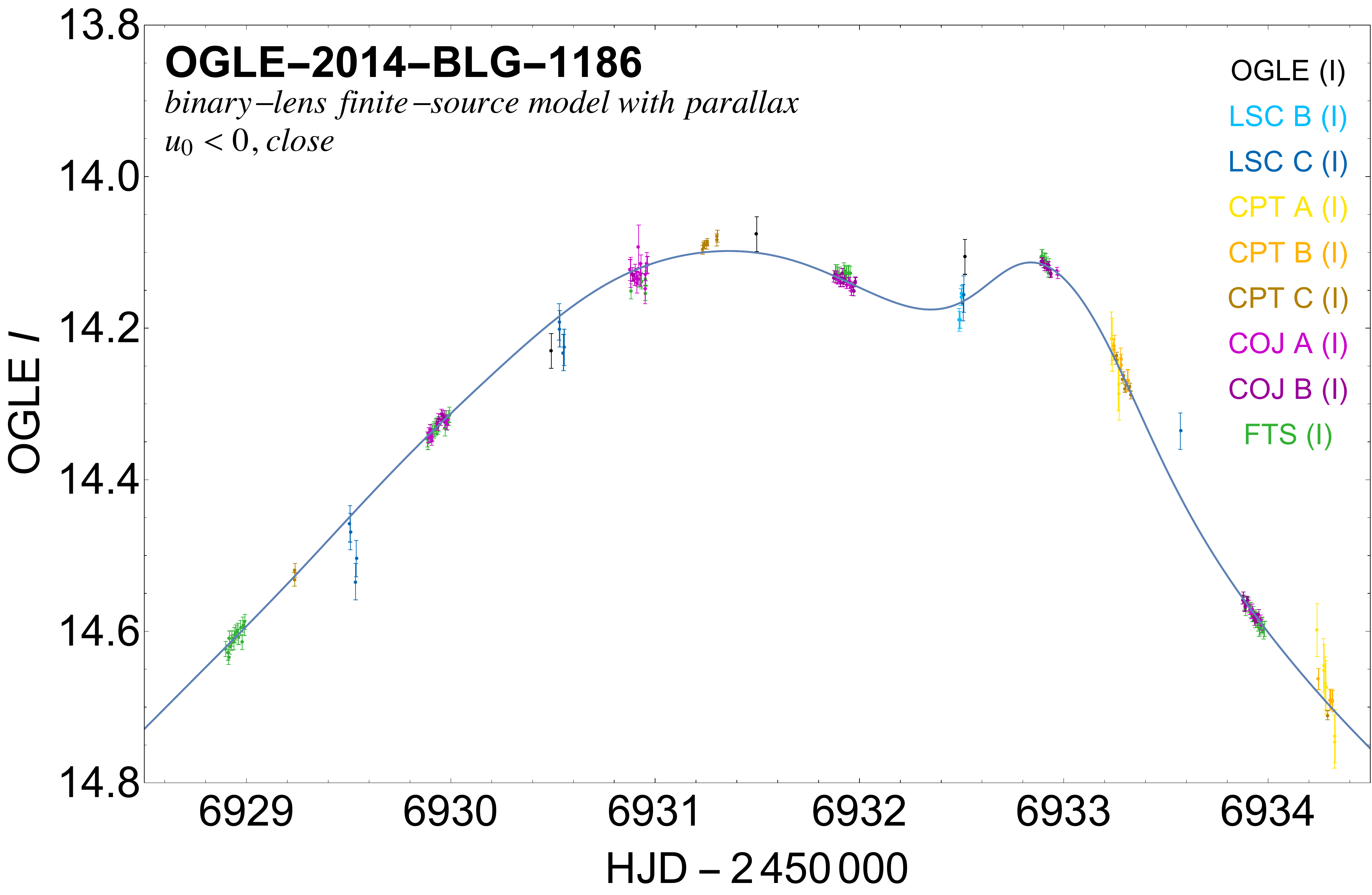}}
\hspace*{3mm}
\resizebox{\columnwidth}{!}{\includegraphics{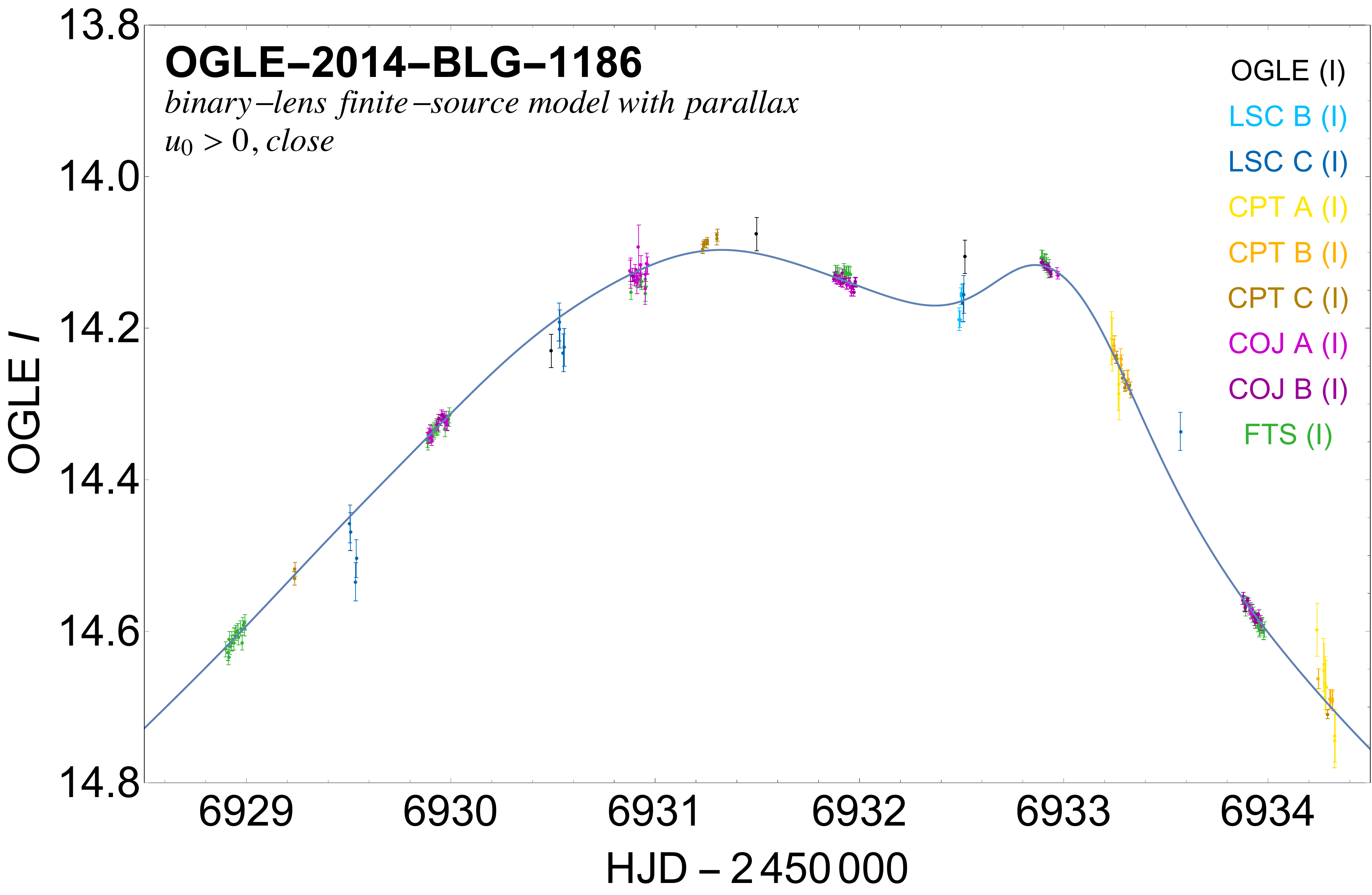}}
\\[2mm]
\resizebox{\columnwidth}{!}{\includegraphics{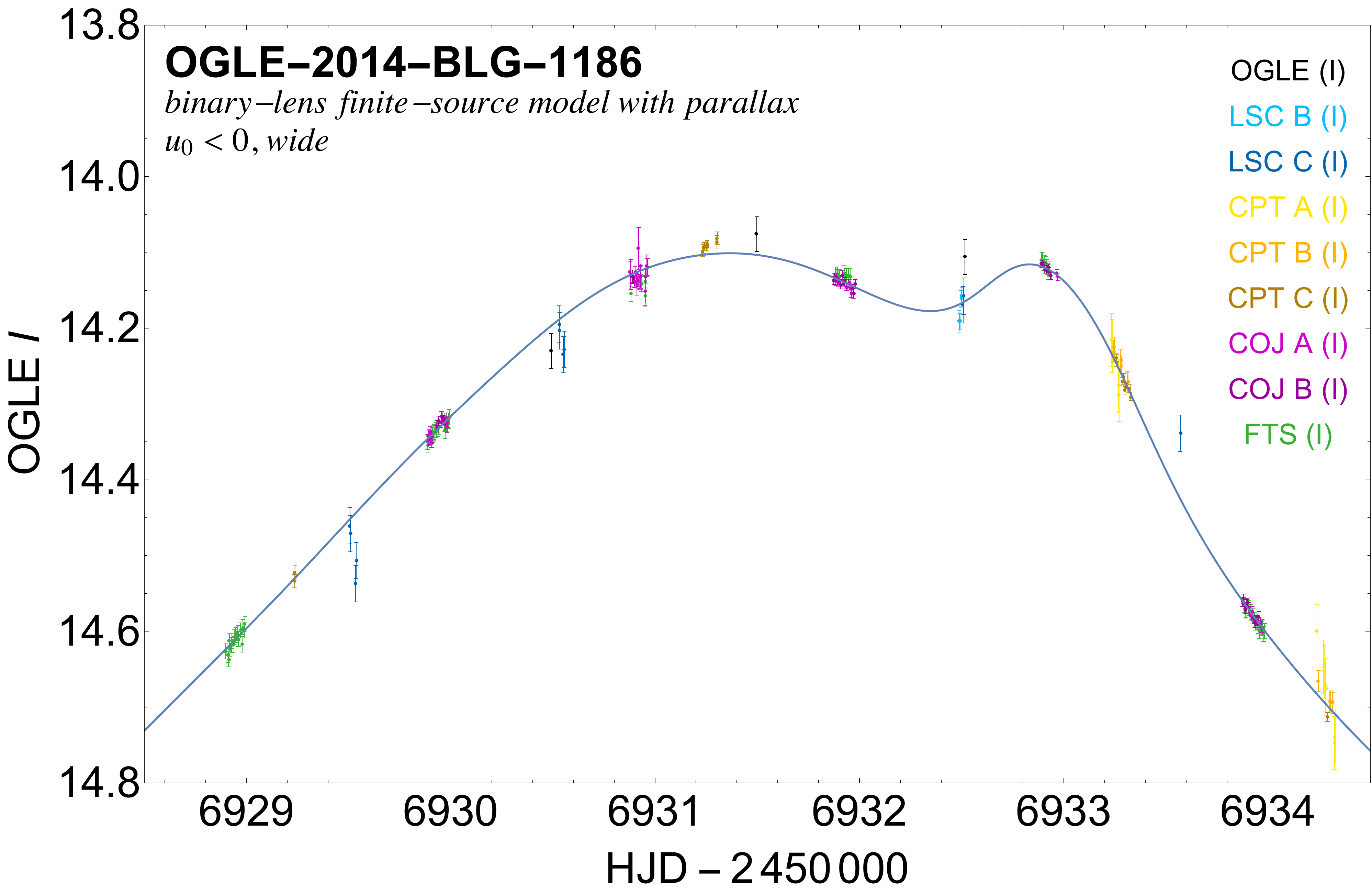}}
\hspace*{3mm}
\resizebox{\columnwidth}{!}{\includegraphics{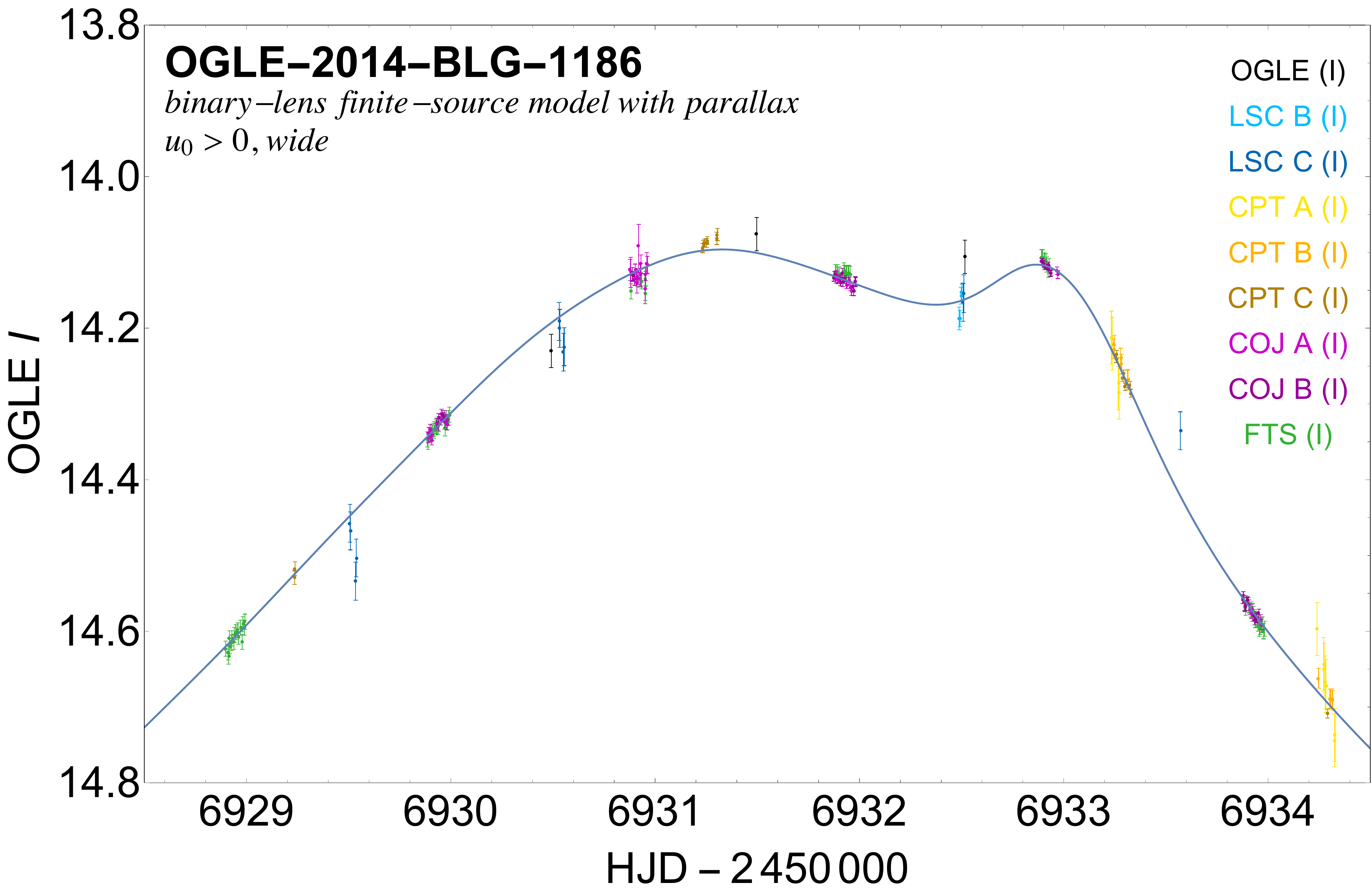}}
 \caption{Photometric data in peak region along with the four different binary-lens finite-source models with parallax whose parameters are listed in
 Table~\protect\ref{Tab:ModelsFinite}. The different underlying geometries produce pretty much the same light curve, which is moreover quite close to those
 found with the point-source models (Tables~\ref{Tab:Models1} and~\ref{Tab:Models2}, Fig.~\ref{Fig:BPLC}), the most visible difference being the slope through the data points near the second peak (close to $\mbox{HJD} - 2\,450\,000 = 6,933$).}
 \label{Fig:Etifinal}
\end{figure*}

The model light curves over the peak region are shown in Fig.~\ref{Fig:Etifinal}, which do not exhibit any visible differences amongst the four ambiguous models. Comparing the models with finite source size with those with a point-like source star, we find that considering the finite size of the star successfully removes the previously found problem with the wrong slope over the second peak. Moreover, the discrepancy of the OGLE point just before the 2nd peak is reduced. However, the finite size of the source star has little effect on the first peak. We have neglected any orbital motion or effects from any further massive bodies within the lens system. These would cause only quite small changes to the photometric light curve, at a level potentially comparable with systematic noise, preventing a reliable measurement of the underlying parameters. Given that we cannot do any better within the adopted model, we regard the model parameters as robust, with the $u_0 < 0$ vs $u_0 >0$ and $d < 1$ vs $d > 1$ ambiguities present.

\begin{table*}
\begin{center}
\begin{tabular}{ccccc}
\hline
Model &  \multicolumn{4}{c}{binary, parallax, finite source} \\
Data selection &  \multicolumn{4}{c}{all} \\
Data sets &\multicolumn{4}{c}{all}\\
Data scaling &  $u_0 < 0$ off-peak &  $u_0 < 0$ off-peak  & $u_0 > 0$ off-peak &  $u_0 > 0$ off-peak   \\
Minimisation & \multicolumn{4}{c}{ML}\\
Option &$u_0 < 0$, close & $u_0 <0$, wide & $u_0 > 0$, close & $u_0 > 0$, wide \\
\hline
$t_0$ &  $6931.455 \pm 0.004$ & $6931.495 \pm 0.004$  & $6931.421\pm 0.004$ & $6931.479 \pm 0.004$ \\
$t_E$ [d]& $271 \pm 19$ & $237 \pm 10$ & $277  \pm 19 $ & $270 \pm 18$\\
$u_0$ &  $-0.0071 \pm 0.0005$ & $-0.0079 \pm 0.0003$ & $0.0065 \pm 0.0005$ & $0.0065 \pm 0.0004$\\
$\pi_{\mathrm{E},\mathrm{N}}$ & $-0.364 \pm 0.009$ &  $-0.370 \pm 0.008$ & $-0.355 \pm 0.010$ & $-0.356  \pm 0.009$\\
$\pi_{\mathrm{E},\mathrm{E}}$ & $-0.191\pm 0.010$ & $-0.205\pm 0.009$ & $-0.176 \pm 0.009$ & $-0.179 \pm 0.009$\\
$d$ & $0.734 \pm 0.008$ & $1.366\pm 0.012$ & $0.702 \pm 0.009$ & $1.439 \pm 0.015$\\
$q$&   $(3.4 \pm 0.3) \times 10^{-4}$ & $(3.8 \pm 0.2) \times 10^{-4}$ & $(4.1 \pm 0.3) \times 10^{-4}$& $(4.2 \pm 0.3) \times 10^{-4}$\\
$\alpha$ & $4.045 \pm 0.004$& $4.046 \pm 0.004$ &$2.312 \pm 0.005$& $2.311 \pm 0.004$ \\
$\rho_\star$ & $(10.1 \pm 1.6) \times 10^{-4}$ &$(12.3\pm 1.5) \times 10^{-4}$ & ($9.7 \pm 1.8) \times 10^{-4}$ & $(9.6 \pm 1.5) \times 10^{-4}$\\
\hline
$\chi^2$ & 1716 & 1722 & 1702 & 1701\\
\hline
\end{tabular}
\caption{Parameters of the four successful binary-lens finite-source parallax models, distinguished by the side on which the source passes relative to the lens near the peak ($u_0 < 0$ or $u_0 > 0$),
and whether the binary lens is in a 'close' ($d < 1$) or 'wide' ($d > 1$) configuration. We adopted error bars arising from a scaling based on the off-peak data, and obtained a simple maximum-likelihood (ML) estimate on all data. The respective value of $\chi^2$ is reported for reference.
}
\label{Tab:ModelsFinite}
\end{center}
\end{table*}

\subsection{Binary-source models}

Double-peaked microlensing events can also arise if the source rather than the lens object is a binary \citep{GriHu92}.
We should therefore carefully consider a binary-lens interpretation of the observed data as an alternative to our binary-lens models.

The gravitational magnification of a binary source is straightforwardly given as the linear superposition of the magnification of its components, i.e.
\begin{eqnarray}
A_\mathrm{BS}({\vec u}^{(1)},{\vec u}^{(2)},\omega_\lambda, \vec{\pi}_\mathrm{E}) & = & (1-\omega_\lambda)\;A\left[u(t,t_0^{(1)},u_0^{(1)},t_\mathrm{E},\vec{\pi}_\mathrm{E})\right] \nonumber \\
& & \hspace{-5mm} + \omega_\lambda\;A\left[u(t,t_0^{(2)},u_0^{(2)},t_\mathrm{E},\vec{\pi}_\mathrm{E})\right]\,,
\end{eqnarray}
with $u(t;t_0,u_0,t_\mathrm{E},\vec{\pi}_\mathrm{E})$ given by Eq.~(\ref{eq:trajectoryparallax}) and
\begin{equation}
\omega_\lambda = \frac{L_{2,\lambda}}{L_{1,\lambda}+L_{2,\lambda}}
\end{equation}
being the luminosity offset ratio depending on the wavelength filter used, while $L_{1,\lambda}$ and $L_{2,\lambda}$ denote the luminosities of
the two source stars. 
For a uniformly bright source of angular radius $\rho_\star\,\theta_\mathrm{E}$, the magnification $A(u,\rho_\star)$ due to a point-mass lens can be computed efficiently in terms of complete elliptic integrals \citep{WM94}.

Given that all data acquired over the peak of event OGLE-2014-BLG-1186 are in $I$-band and the effect of binarity on the photometric light curve is negligible for other epochs,
we describe single-lens binary-finite-source models with annual parallax by the parameter vector $\vec p = (t_0^{(1)},t_0^{(2)},t_\mathrm{E},u_0^{(1)},u_0^{(2)},\pi_{\mathrm{E},\mathrm{N}},\pi_{\mathrm{E},\mathrm{E}},\omega_I,\rho_\star^{(1)},\rho_\star^{(2)})$, explicitly 
defining the reference epoch for parallax $t_0$, Eqs.~(\ref{eq:parallax}) and~(\ref{eq:pargam}), to refer to the $I$-band photocentre
\begin{equation}
t_0 \equiv (1-\omega_I)\, t_0^{(1)} + \omega_I\, t_0^{(2)}\,.
\end{equation}
This leads to the four sets of best-fitting model parameters listed in Table~\ref{Tab:ModelsBinSource}, which are distinguished by all combinations of the respective signs 
of $u_0^{(1)}$ and $u_0^{(2)}$. In the absence of significant parallax effects, binary-source models become blind to whether the two source stars are on the same side of the effective lens trajectory, i.e. $u_0^{(1)} u_0^{(2)} > 0$ (`cis' confuguration), or on opposite sides, i.e. $u_0^{(1)} u_0^{(2)} < 0$ (`trans' configuration)  \citep[][Appendix~C]{DoHi}.
We find the two source stars being separated by an angle $\lambda\,\theta_\mathrm{E}$,
where
\begin{equation}
\lambda = \sqrt{\left(\frac{t_0^{(2)}-t_0^{(1)}}{t_\mathrm{E}}\right)^2 + \left(u_0^{(2)}-u_0^{(1)}\right)^2}\,,
\label{eq:lambda1}
\end{equation}
which for our models evaluates to
\begin{eqnarray}
\lambda^\mathrm{cis} & = & 0.0087 \pm 0.0008\,,\nonumber\\
\lambda^\mathrm{trans} & = & 0.0107 \pm 0.0009\,.
\label{eq:lambda2}
\end{eqnarray}
Our binary-source models involve a brighter larger source star dominating the (earlier) main peak, while the (later) secondary peak is due to a fainter smaller source star that passes the lens star at a smaller minimal separation. 

\begin{figure*}
\resizebox{\columnwidth}{!}{\includegraphics{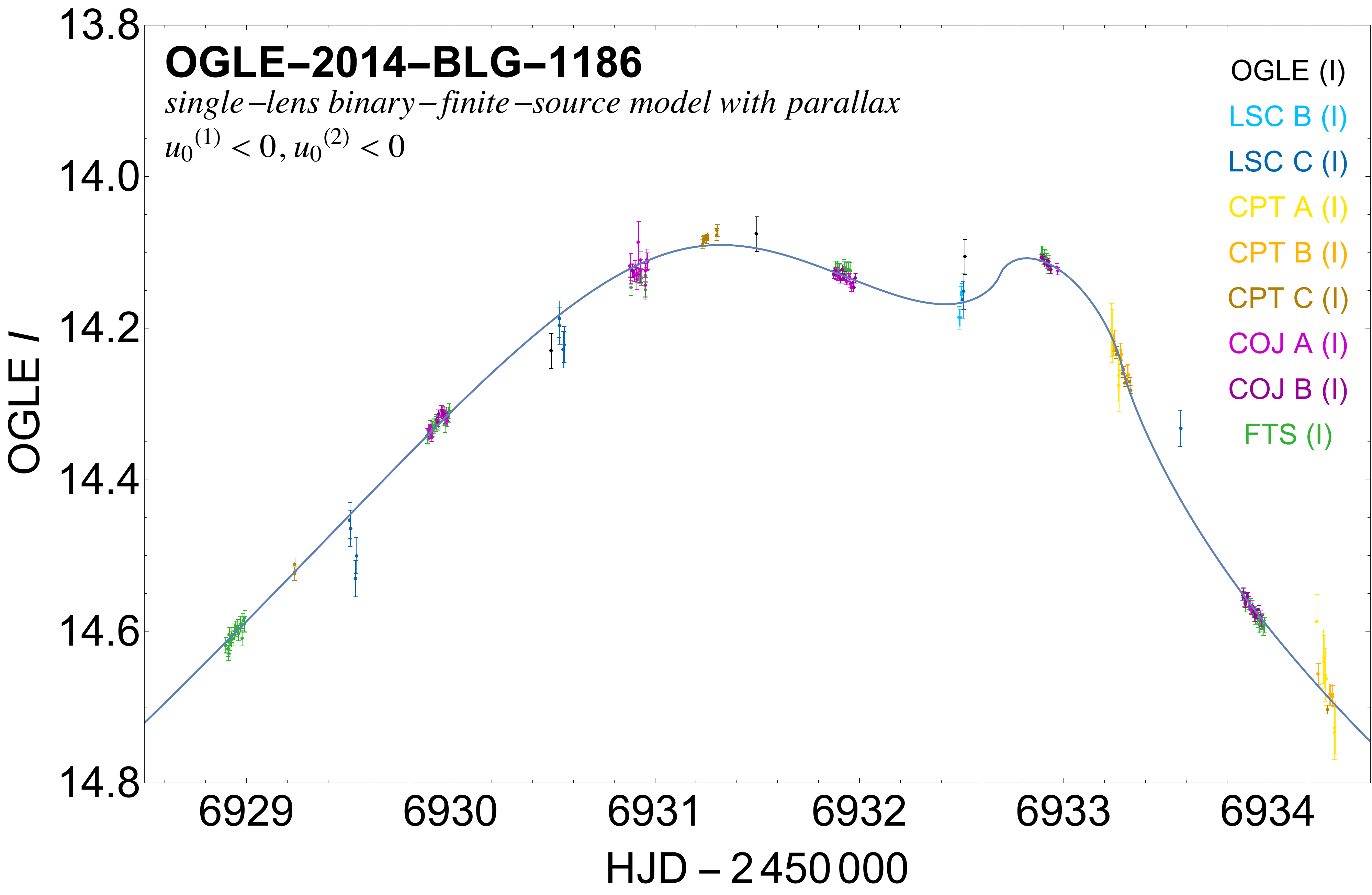}}
\hspace*{3mm}
\resizebox{\columnwidth}{!}{\includegraphics{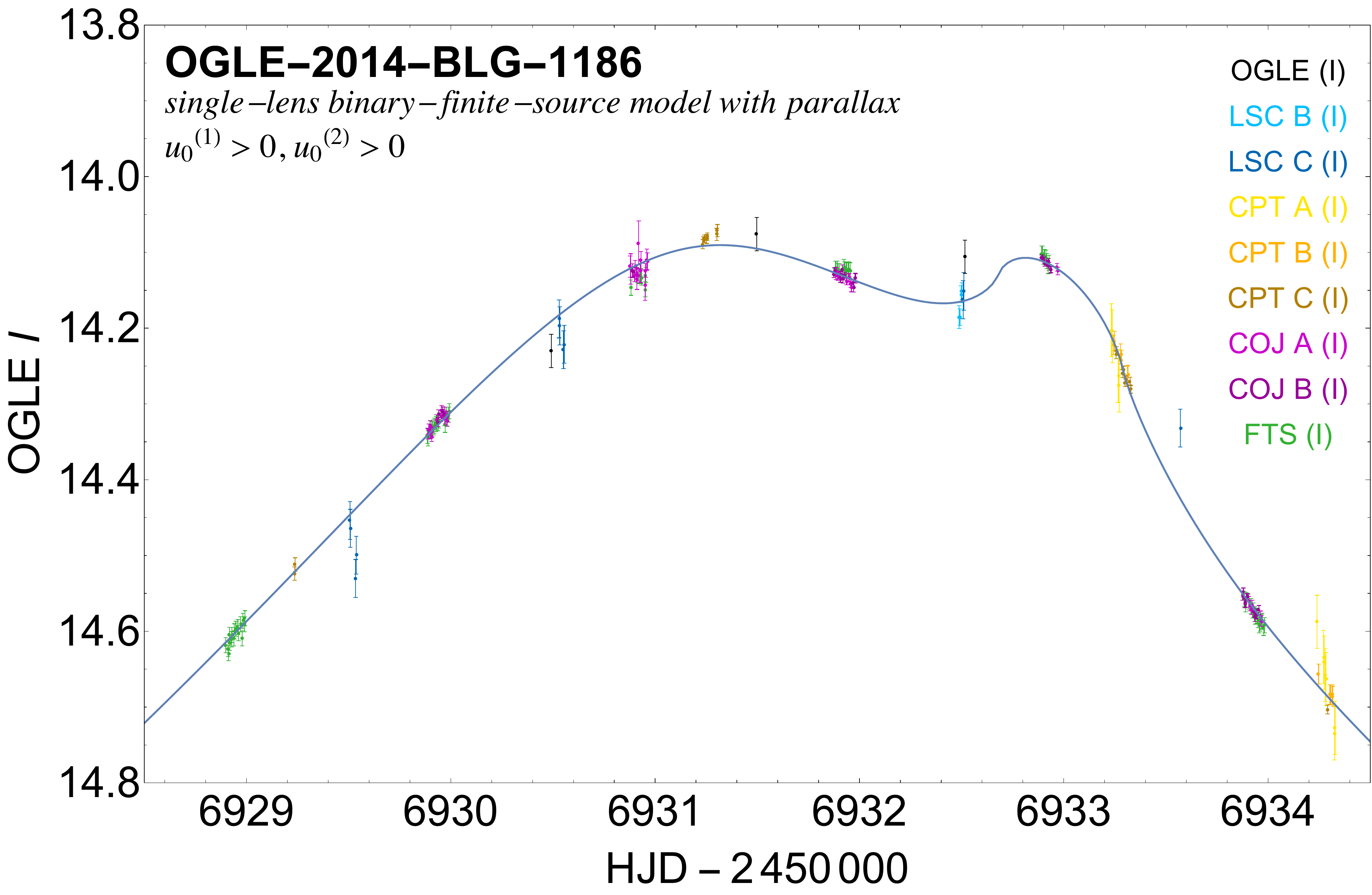}}
\\[2mm]
\resizebox{\columnwidth}{!}{\includegraphics{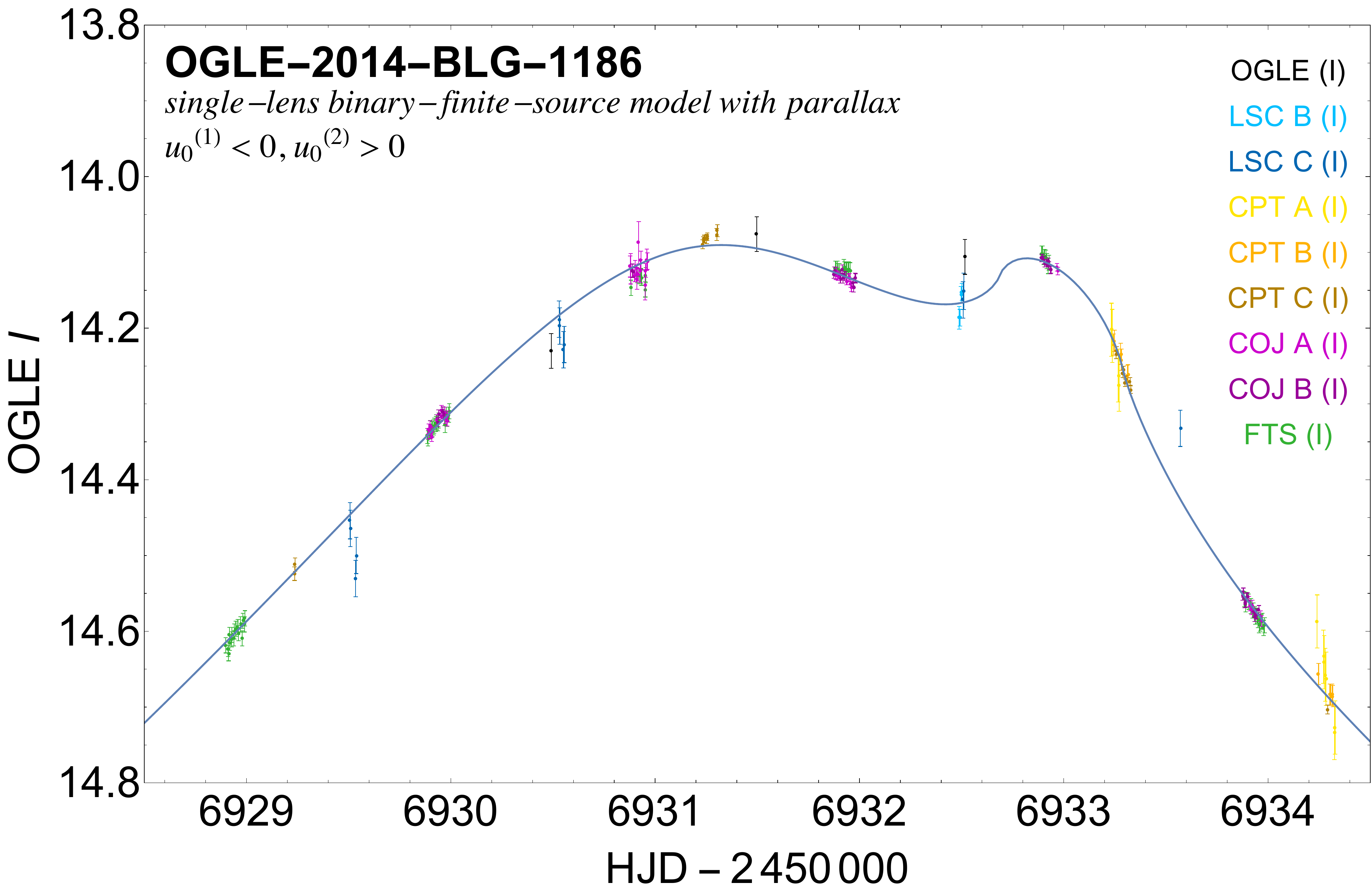}}
\hspace*{3mm}
\resizebox{\columnwidth}{!}{\includegraphics{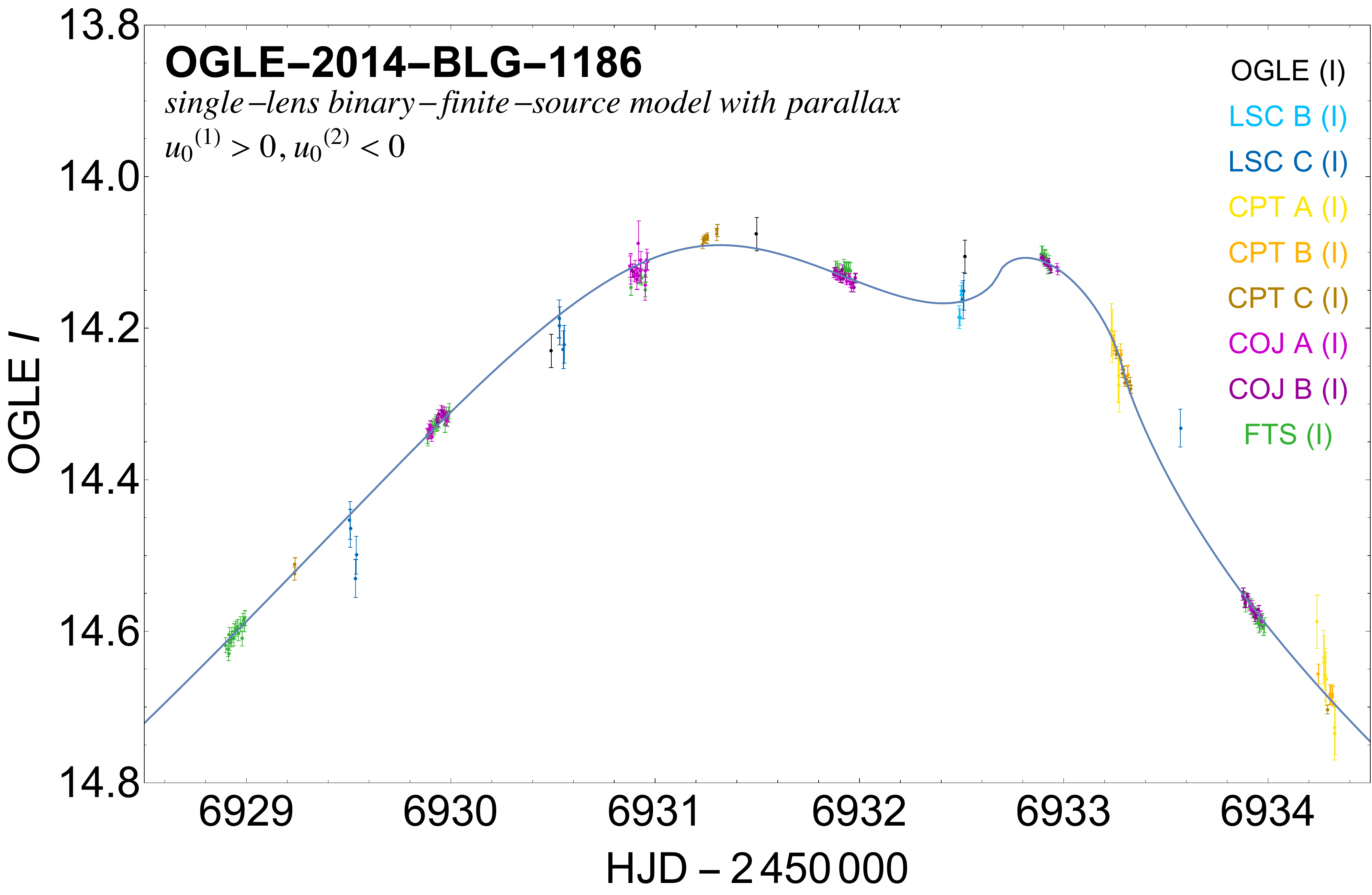}}
 \caption{Photometric data covering the peak region along with the four single-lens binary-finite-source models with parallax whose parameters are listed in
 Table~\protect\ref{Tab:ModelsBinSource}. These produce almost identical light curves, which moreover visibly differ from those produced by the binary-lens finite-source models (Table~\ref{Tab:ModelsFinite} and Fig.~\ref{Fig:Etifinal}) only on the shape of the secondary peak during epochs not covered by data.}
 \label{Fig:Binsource}
\end{figure*}

\begin{figure}
\hspace*{1.35mm}\resizebox{8.32cm}{!}{\includegraphics{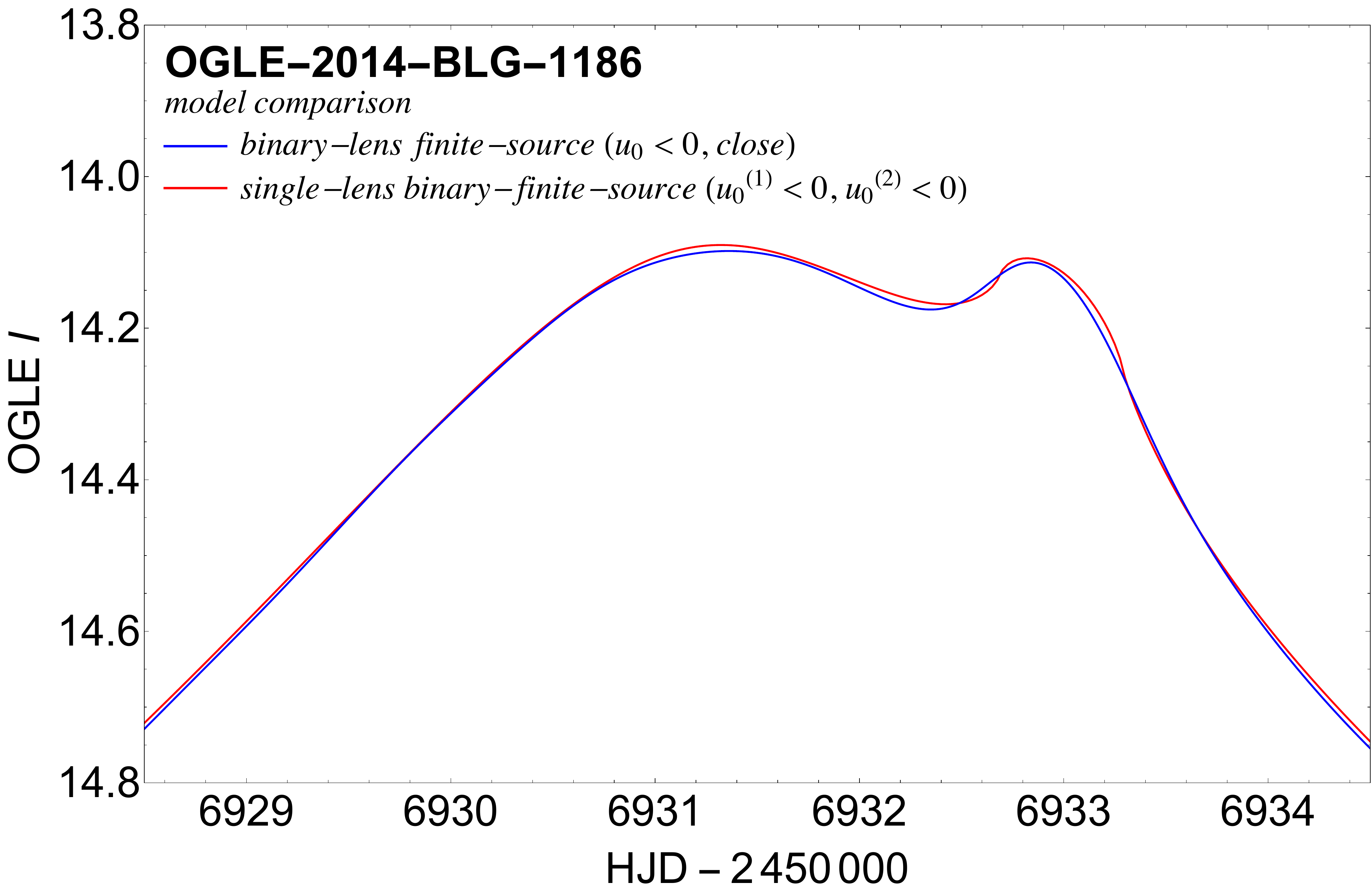}}
\\[2mm]
\resizebox{8.5cm}{!}{\includegraphics{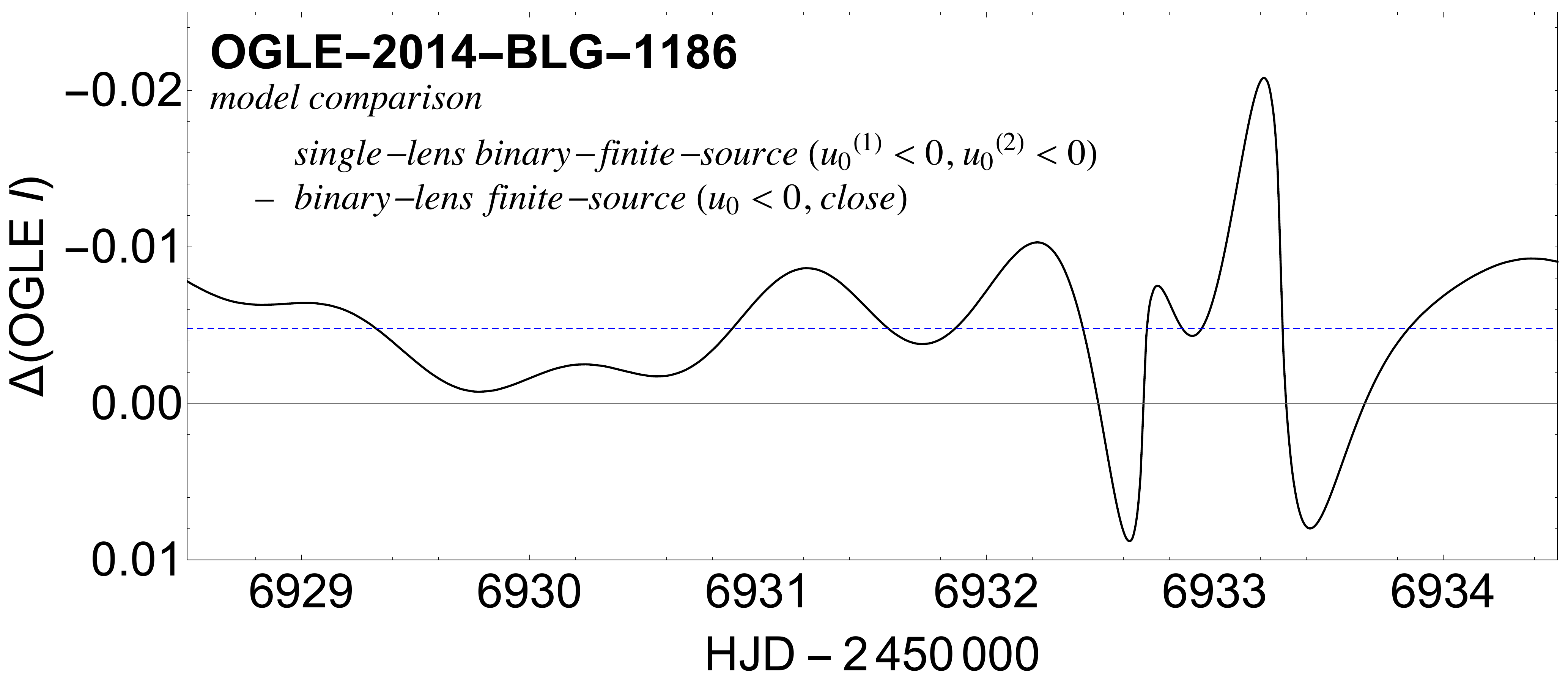}}
 \caption{Difference between light curves for the binary-lens finite-source models and the single-lens binary-finite-source models expressed in OGLE $I$-magnitudes. While the photometric uncertainties of peak OGLE data are large in comparison, models adjust to other data with slightly different baseline magnitudes and blend ratios, corresponding to an average shift of about 5~mmag over the peak (indicated by the blue dashed line). The largest difference in shape occurs around the secondary peak during epochs over which no data were acquired.}
 \label{Fig:ModelCompare}
\end{figure}

The respective light curves are shown in Fig.~\ref{Fig:Binsource}, which are apparently hardly distinguishable from those corresponding to the identified viable binary-lens models (Fig.~\ref{Fig:Etifinal}). 
In particular, the difference between the two presented models is not larger than the differences between model and data. If we were to trust our data at that level (excluding that any residuals are due to systematic uncertainties), we would need to reject both models. If we accept that there are systematic uncertainties at that level, we would need to accept both. We explicitly show the difference between the light curves 
of two of our binary-lens and binary-source models in Fig.~\ref{Fig:ModelCompare}. Taking into account a difference in the blend ratio relative to the OGLE data, the difference between the models is almost always
below 5 mmag, except for short epochs near the second peak that are not or poorly covered by data.

We continue our discussion of the viability of the binary-lens and binary-source models in Sect.~\ref{sec:lensorsource} after having investigated what the inferred model parameters mean for the physical nature of the lens and source systems.

\begin{table*}
\begin{center}
\begin{tabular}{ccccc}
\hline
Model &  \multicolumn{4}{c}{single lens, parallax, binary finite source} \\
Data selection &  \multicolumn{4}{c}{all} \\
Data sets &\multicolumn{4}{c}{all}\\
Data scaling &  $u_0 < 0$ off-peak &  $u_0 < 0$ off-peak  & $u_0 > 0$ off-peak &  $u_0 > 0$ off-peak   \\
Minimisation & \multicolumn{4}{c}{ML}\\
Option &$u_0^{(1)} < 0$, $u_0^{(2)} < 0$ & $u_0^{(1)} < 0$, $u_0^{(2)} > 0$  & $u_0^{(1)} > 0$, $u_0^{(2)} > 0$  & $u_0^{(1)} >0$, $u_0^{(2)} < 0$  \\
\hline
$t_0^{(1)}$ &  $6931.228 \pm 0.007$ & $6931.229 \pm 0.007$  & $6931.234\pm 0.007$ & $6931.234 \pm 0.007$ \\
$t_0^{(2)}$ &  $6932.989 \pm 0.007$ & $6932.945 \pm 0.007$  & $6932.944\pm 0.007$ & $6932.987 \pm 0.007$ \\
$t_E$ [d]& $306 \pm 19$ & $306 \pm 19$ & $311  \pm 19 $ & $311 \pm 19$\\
$u_0^{(1)}$ &  $-0.0082 \pm 0.0005$ & $-0.0082 \pm 0.0005$ & $0.0075 \pm 0.0005$ & $0.0075 \pm 0.0005$\\
$u_0^{(2)}$ &  $-0.00113 \pm 0.00008$ & $0.00142 \pm 0.00008$ & $0.00142 \pm 0.00008$ & $-0.00116 \pm 0.00008$\\
$\pi_{\mathrm{E},\mathrm{N}}$ & $-0.363 \pm 0.009$ &  $-0.363 \pm 0.009$ & $-0.347 \pm 0.009$ & $-0.347  \pm 0.009$\\
$\pi_{\mathrm{E},\mathrm{E}}$ & $-0.178\pm 0.008$ & $-0.178\pm 0.008$ & $-0.158 \pm 0.008$ & $-0.158 \pm 0.008$\\
$\omega_{I}$ & $0.040 \pm 0.002$ & $0.040\pm 0.002$ & $0.044 \pm 0.002$ & $0.045 \pm 0.002$\\
$\rho_\star^{(1)}$ & $(8.0 \pm 0.5) \times 10^{-3}$ &$(7.9\pm 0.5) \times 10^{-3}$ & ($7.3 \pm 0.5) \times 10^{-3}$ & $(7.3 \pm 0.5) \times 10^{-3}$\\
$\rho_\star^{(2)}$ & $(1.6 \pm 0.1) \times 10^{-3}$ &$(1.6\pm 0.1) \times 10^{-3}$ & ($1.6 \pm 0.1) \times 10^{-3}$ & $(1.6 \pm 0.1) \times 10^{-3}$\\
\hline
$\chi^2$ & 1715 & 1715 & 1716 & 1716\\
\hline
\end{tabular}
\caption{Parameters of four successful single-lens binary-finite-source parallax models, distinguished by the side on which each of the source stars passes relative to the lens near the peak. Given that all peak data have been acquired in $I$-band and the binarity does not significantly affect the light curve outside the peak region, we use a single luminosity offset ratio $\omega_I$ characteristic for $I$-band.
 We adopted error bars arising from a scaling based on the off-peak data, and obtained a simple maximum-likelihood (ML) estimate on all data. The respective value of $\chi^2$ is reported for reference. 
}
\label{Tab:ModelsBinSource}
\end{center}
\end{table*}

\section{Interpretation}

\begin{figure}
\includegraphics[width=\columnwidth]{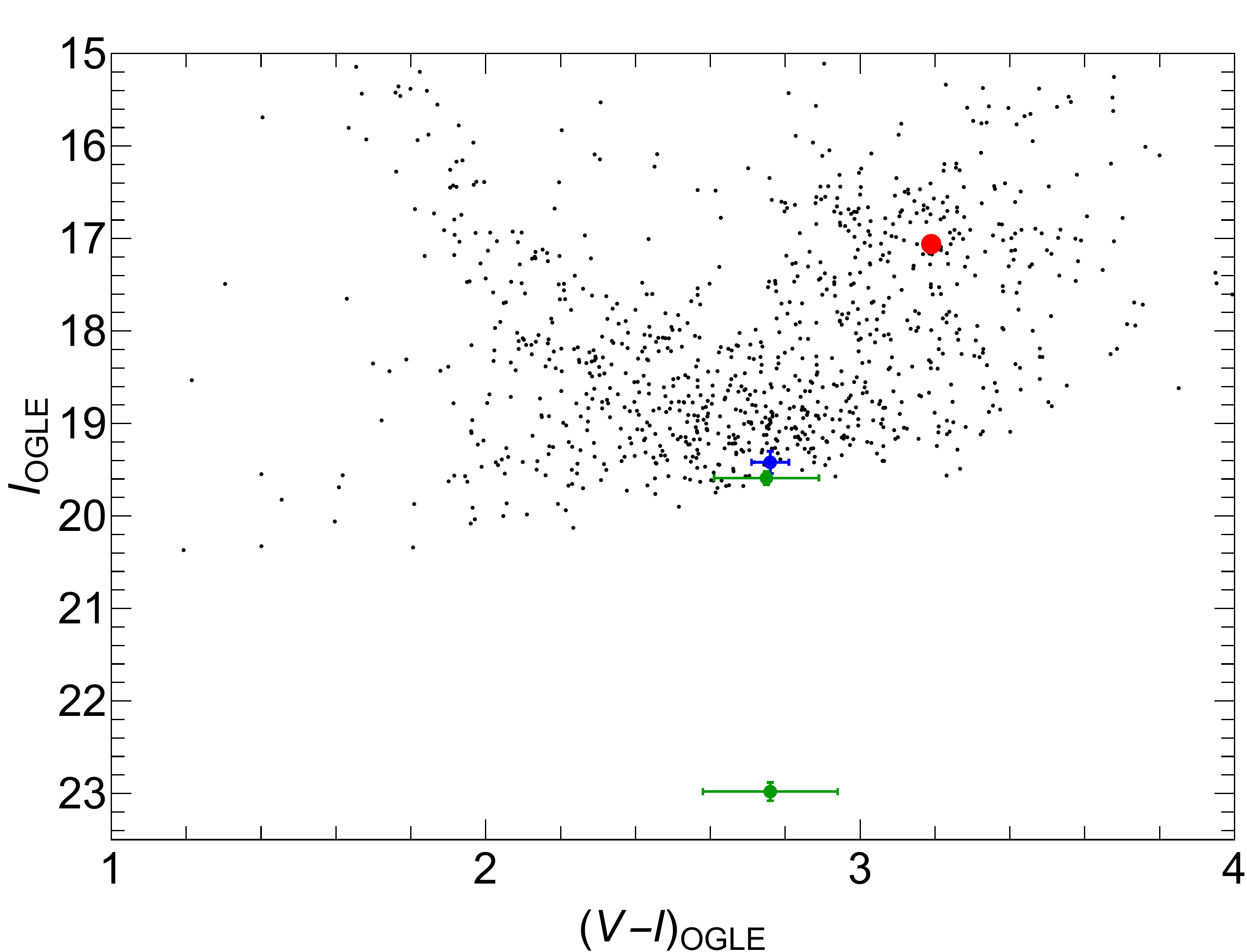}
 \caption{Instrumental colour-magnitude diagram for the OGLE field containing the microlensing event OGLE-2014-BLG-1186, marking the location of the centroid of the giant clump (red), as well as the
 source star(s) for our binary-lens models (blue) or our binary-source models (green).}
 \label{fig:CMD}
\end{figure}

\label{sec:interpretation}

\subsection{Lens binary}



Following the approach suggested by \citet{MB9741}, we use the de-reddened colour $(V-I)_0$ and the brightness $I_0$ of the source star
to estimate its angular radius $\theta_\star$. Exploiting the fact that OGLE monitors its fields not only in $I$, but also more sparsely in $V$, 
we construct an instrumental colour-magnitude diagram (CMD), shown in Fig.~\ref{fig:CMD}. We find the source star 
at $(V-I,I)=(2.76 \pm 0.05,19.42 \pm 0.12)$, where a major uncertainty arises from the blend ratio, where $I$- and $V$-band blend ratios are strongly
correlated with the event time-scale $t_\mathrm{E}$ and with each other. The centroid of the Galactic bulge red clump (RC) is at $(V-I,I)_\mathrm{GC}=(3.19\pm 0.01,17.06\pm 0.02)$.
According to \citet{Bensby2011} and \citet{Nataf2013}, the de-reddened colour and brightness of the red clump are $(V-I,I)_{\mathrm{RC},0} = (1.06,14.62)$ 
for the Galactic longitude of the target $l \sim 5^\circ$. Consequently, we find for our source star $(V-I,I)_0 = (0.63 \pm 0.05,16.98 \pm 0.12)$, indicative of an F-type dwarf or a G-type subgiant. 

For such stars, \citet{KervellaFouq} provide a direct empirical relation to estimate the angular source radius $\theta_\star$ from
$V$ and Cousins $I$ measurements (matching the OGLE filters), namely
\begin{equation}
\lg\left(\frac{2\,\theta_\star}{1~\mbox{mas}}\right) = 0.4992 + 0.4895 (V-I)_0 - 0.0657 (V-I)_0^2 - 0.2\, I_0\,,
\label{eq:KVeq}
\end{equation}
so that we obtain
\begin{eqnarray}
\theta_\star & = &(1.21 \pm 0.11)~\umu\mbox{as}\,,
\end{eqnarray}
including a typical uncertainty of 5.6 per cent for the empirical relation. With the assumption of the source star being close to the Galactic bulge,
\begin{eqnarray}
D_\mathrm{S} & = & (8.5 \pm 2.0)~\mbox{kpc}\,,
\label{eq:Ds}
\end{eqnarray}
we estimate the physical radius of the source star $R_\star = D_\mathrm{S}\,\theta_\star$ to be
\begin{eqnarray}
R_\star & = & (2.2 \pm 0.6)~R_\odot\,.
\end{eqnarray}

From our models, we find the parallax parameter $\pi_\mathrm{E} = \pi_\mathrm{LS}/\theta_\mathrm{E}$ and the source size parameter $\rho_\star = \theta_\star/\theta_\mathrm{E}$ as
\begin{eqnarray}
\pi_\mathrm{E} & = & 0.41 \pm 0.01\,,\\
\rho_\star & = & (10.4 \pm 1.6) \times 10^{-4}\,,
\end{eqnarray}
while Eq.~(\ref{eq:thetaE}) gives the total mass as
\begin{equation}
M = \frac{c^2}{4G}\,(1~\mbox{AU})\,\frac{\theta_\star}{\pi_\mathrm{E}\,\rho_\star}\,,
\label{eq:calcmass}
\end{equation}
which evaluates to
\begin{eqnarray}
M & = & (0.35 \pm 0.06)~M_{\odot}\,.
\end{eqnarray}
With the mass ratio
\begin{eqnarray}
q & = & (3.9 \pm 0.3)\times 10^{-4}
\end{eqnarray}
we then find the mass of the planet as
\begin{eqnarray}
M_2 & = & (45 \pm 9)~M_\oplus\,,
\end{eqnarray}
about 3 times the mass of Neptune or about half the mass of Saturn.
The uncertainty in the mass measurement is dominated by the uncertainty in the source size parameter $\rho_\star$ (about 15 per cent).

From $\theta_\star$ and $\rho\star = \theta_\star/\theta_\mathrm{E}$, we obtain the angular Einstein radius as
\begin{eqnarray}
\theta_\mathrm{E} & = & (1.2 \pm 0.2)~\mbox{mas}\,,
\end{eqnarray}
and with the event time-scale
\begin{eqnarray}
t_\mathrm{E}  & = & (264 \pm 17)~\mbox{d}\,,
\end{eqnarray}
where $t_\mathrm{E} = \theta_\mathrm{E}/\mu$, we find the effective proper motion as
\begin{eqnarray}
\mu & = & (4.4 \pm 0.8)~\umu\mbox{as}\;\mbox{d}^{-1}\nonumber\\
& = & (1.6 \pm 0.3)~\mbox{mas}\;\mbox{yr}^{-1} \,.
\end{eqnarray}

With $\pi_\mathrm{E} = \pi_\mathrm{LS}/\theta_\mathrm{E}$ and $\theta_\mathrm{E}$, we find
\begin{eqnarray}
\pi_\mathrm{LS} & = & (0.48 \pm 0.09)~\mbox{mas}\,,
\end{eqnarray}
so that
with $D_\mathrm{S}$ as given by Eq.~(\ref{eq:Ds}),
or equivalently
\begin{eqnarray}
\pi_\mathrm{S} & = & (0.12 \pm 0.03)~\mbox{mas}\,,
\label{eq:srcparallax}
\end{eqnarray}
one obtains
\begin{eqnarray}
\pi_\mathrm{L} & = & (0.60 \pm 0.09)~\mbox{mas}\,,
\end{eqnarray}
equivalent to 
\begin{eqnarray}
D_\mathrm{L} & = & (1.7 \pm 0.3)~\mbox{kpc}\,.
\end{eqnarray}

The effective proper motion then implies an
effective perpendicular lens velocity $v_\perp = D_\mathrm{L}\mu$ of
\begin{eqnarray}
v_\perp & = & (13 \pm 3)~\mbox{km}\;\mbox{s}^{-1}\,.
\end{eqnarray}
We moreover find the Einstein radius $r_\mathrm{E} = D_\mathrm{L}\,\theta_\mathrm{E}$, evaluating to
\begin{eqnarray}
r_\mathrm{E} & =  & (2.0 \pm 0.5)~\mbox{AU}\,,
\end{eqnarray}
and with the binary separation parameters for the close or wide binary case,
\begin{eqnarray}
 d^{(\mathrm{c})} & = & 0.718 \pm 0.009\,,\nonumber\\
 d^{(\mathrm{w})}  & = & 1.403 \pm 0.014\,,
\end{eqnarray}
the projected separation at epoch $t_0$ becomes
\begin{eqnarray}
r_{0,\perp}^{(\mathrm{c})} & = & (1.4 \pm 0.3)~\mbox{AU}\,,\nonumber\\
r_{0,\perp}^{(\mathrm{w})} & = & (2.7 \pm 0.6)~\mbox{AU}\,.
\end{eqnarray}
Consequently,
we can estimate the minimal orbital period
\begin{equation}
P_\mathrm{min} = 2 \upi\,\sqrt{\frac{r_{0,\perp}^3}{8GM}}
\end{equation} 
as
\begin{eqnarray}
P_\mathrm{min}^{(\mathrm{c})} & = & (1.0 \pm 0.4)~\mbox{yr}\,,\nonumber\\
P_\mathrm{min}^{(\mathrm{w})} & = & (2.7 \pm 1.0)~\mbox{yr}\,.
\end{eqnarray}

The inferred properties and the underlying collated model parameters are comprehensively listed in Table~\ref{Tab:Allpars}.

\citet{KP94} originally suggested that $\sim 60$ per cent of all Galactic bulge microlensing events would be caused by bulge stars and $\sim 40$ per cent by disk stars, with large uncertainties due to the simplicity of the adopted models and the uncertainty of their model parameters. As noted by \citet{Penny:devoid}, it turned out that planet detections reported from observed microlensing events show a strong preference for nearby stars, suggesting that the Galactic bulge stars might be devoid of planets as compared to the Galactic disk stars. The small lens distance $D_\mathrm{L} \sim 1.7~\mbox{kpc}$ further supports this, pointing to a lens star in the disk rather than the bulge. Moroever, this is even substantially less than the average distance of a disk lens star. However, the fact that the Galactic disk is structured into spiral arms, specifically favouring certain ranges of lens distances along the line of sight, should be taken into account. In fact, it is a key goal of observations of microlensing events with the Spitzer space telescope \citep{SCN:Spitzer,Zhu:Spitzer} to shed light on the distance distribution of microlensing events by combining these with ground-based photometry and thereby measuring the microlensing parallax parameter $\pi_\mathrm{E}$.
 The event time-scale of $t_\mathrm{E} \sim 300~\mbox{d}$ is much larger than
a median of $t_\mathrm{E} \sim 20~\mbox{d}$ \citep[e.g.][]{Do:esti} expected with best guesses of the stellar mass function \citep{Chabrier:mass}. A rather low effective transverse velocity should therefore be expected, and given the large width of the velocity distribution, substantial deviations from the average are within reason.

\begin{table*}
\begin{center}
\begin{tabular}{lcc}
\hline
& Lens binary & Source binary\\
\hline
\underline{Collated model parameters}  & \hspace*{4.5cm} & \hspace*{4.5cm} \\[0.5ex]
microlensing parallax parameter & $\pi_\mathrm{E}  =  0.41 \pm 0.01$ & $\pi_\mathrm{E}  =  0.39 \pm 0.01$ \\
source size parameter & $ \rho_\star  =  (10.4 \pm 1.6) \times 10^{-4}$ & $\rho_\star^{\mbox{\tiny (1)}}  =  (7.6 \pm 0.5) \times 10^{-3}$\\
& & $\rho_\star^{\mbox{\tiny (2)}}  =  (1.6 \pm 0.1) \times 10^{-3}$\\
mass ratio & $q  =  (3.9 \pm 0.3)\times 10^{-4}$ & --- \\
luminosity offset ratio & --- & $\omega_I = 0.042 \pm 0.003$ \\
event time-scale & $t_\mathrm{E}  =  (264 \pm 17)~\mbox{d}$ & $t_\mathrm{E}  =  (309 \pm 19)~\mbox{d}$\\
binary separation parameter  & $d^{(\mathrm{c})} = 0.718 \pm 0.009$ & $\lambda^\mathrm{cis} = 0.0087 \pm 0.0008$ \\
  & $d^{(\mathrm{w})} = 1.403 \pm 0.014$ & $\lambda^\mathrm{trans} = 0.0107 \pm 0.0009$ \\
source distance & \multicolumn{2}{c}{ $D_\mathrm{S} = (8.5 \pm 2.0)~\mbox{kpc}$}\\
baseline magnitude & \multicolumn{2}{c}{$I_\mathrm{base}^{\mathrm{OGLE}}= 19.04$}\\
\hline
\multicolumn{2}{l}{\underline{Lens star (system)}}\\[0.5ex]
mass of star & $M_1  =  (0.35 \pm 0.06)~M_{\odot}$ &  $M  =  (0.046 \pm 0.007)~M_{\odot}$\\
mass of planet & $M_2  = (45 \pm 9)~M_\oplus$ & --- \\
angular Einstein radius & $\theta_\mathrm{E}  =  (1.2 \pm 0.2)~\mbox{mas}$ & $\theta_\mathrm{E}  =  (0.15 \pm 0.02)~\mbox{mas}$\\
effective proper motion & $\mu  =  (4.4 \pm 0.8)~\umu\mbox{as}\;\mbox{d}^{-1}$  & $\mu  = (0.47 \pm 0.08)~\umu\mbox{as}\;\mbox{d}^{-1}$\\
 & $= (1.6 \pm 0.3)~\mbox{mas}\;\mbox{yr}^{-1}$ & $= (0.17 \pm 0.03)~\mbox{mas}\;\mbox{yr}^{-1}$\\
lens-source parallax & $\pi_\mathrm{LS}  =  (0.48 \pm 0.09)~\mbox{mas}$ & $\pi_\mathrm{LS}  =   (0.057 \pm 0.009)~\mbox{mas}$\\
lens distance & $D_\mathrm{L}  =  (1.7 \pm 0.3)~\mbox{kpc}$ & $D_\mathrm{L}  =  (5.7 \pm 0.9)~\mbox{kpc}$ \\
effective lens velocity & $v_\perp = (13 \pm 3)~\mbox{km}\;\mbox{s}^{-1}$ & $v_\perp = (4.7 \pm 1.1)~\mbox{km}\;\mbox{s}^{-1}$\\
Einstein radius & $r_\mathrm{E} = (2.0 \pm 0.5)~\mbox{AU}$ & $r_\mathrm{E} = (0.84 \pm 0.19)~\mbox{AU}$\\
current projected separation  & $r_{0,\perp}^{(\mathrm{c})} = (1.4 \pm 0.3)~\mbox{AU}$ & --- \\
& $r_{0,\perp}^{(\mathrm{w})} = (2.7 \pm 0.6)~\mbox{AU}$ & --- \\
minimal orbital period & $P_\mathrm{min}^{(\mathrm{c})} = (1.0 \pm 0.4)~\mbox{yr}$& ---\\
 & $P_\mathrm{min}^{(\mathrm{w})} = (2.7 \pm 1.0)~\mbox{yr}$ & --- \\
\hline
\underline{Source star (system) / microlensing target}\\[0.5ex]
right ascension (J2000) & \multicolumn{2}{c}{$\mbox{RA} = 17\fh41\fm59\fs63$}\\
declination (J2000) & \multicolumn{2}{c}{$\mbox{Dec}=-34\fdg17\farcm18\farcs1$}\\
deteddened red clump colour/mag & \multicolumn{2}{c}{$(V-I,I)_{\mathrm{RC},0} = (1.06,14.62)$ }\\
red clump colour/mag & \multicolumn{2}{c}{$(V-I,I)_\mathrm{RC}=(3.19 \pm 0.01,17.06 \pm 0.02)$}\\
source colour/mag & $(V-I,I) = (2.76 \pm 0.05,19.42 \pm 0.12)$ & $(V-I,I) = (2.75\pm 0.05, 19.54 \pm 0.06)$\\
dereddened source colour/mag &$(V-I,I)_0 = (0.63 \pm 0.05,16.98 \pm 0.12)$ &  $(V-I,I)_0 = (0.62 \pm 0.05,17.10 \pm 0.07)$ \\
dereddened source colour/mag (1) & --- & $(V-I,I)_0^{\mbox{\tiny (1)}}=(0.62\pm 0.14, 17.15\pm 0.07)$\\
dereddened source colour/mag (2) & --- & $(V-I,I)_0^{\mbox{\tiny (2)}}=(0.63 \pm 0.18, 20.54\pm 0.10)$\\
type of source & F~V, G~IV & F~V, G~IV / G~VI\\
angular radius of source & $\theta_\star = (1.21 \pm 0.11)~\umu\mbox{as}$ & $\theta_\star^{\mbox{\tiny (1)}} = (1.11 \pm 0.16)~\umu\mbox{as}$\\
 & & $\theta_\star^{\mbox{\tiny (2)}}  = (0.23 \pm 0.04)~\umu\mbox{as}$\\
physical radius of source & $R_\star = (2.2 \pm 0.6)~R_\odot$ & $R_\star^{\mbox{\tiny (1)}} = (2.0 \pm 0.6)~R_\odot$ \\
& &  $R_\star^{\mbox{\tiny (2)}} = (0.43 \pm 0.13)~R_\odot$ \\
angular separation of constituents & --- & $\beta^\mathrm{cis} = (1.3 \pm 0.2)~\mbox{$\umu$as}$\\
  & --- & $\beta^\mathrm{trans} = (1.6 \pm 0.3)~\mbox{$\umu$as}$\\
current projected separation & --- & $\rho_\perp^\mathrm{cis}  =(0.011 \pm 0.03)~\mbox{AU} $ \\
 & --- & $\rho_\perp^\mathrm{trans}  = (0.013 \pm 0.04)~\mbox{AU} $ \\
 minimal orbital period & --- &  $P_\mathrm{S,min}^\mathrm{cis}  =  (0.10 \pm 0.05)~\mbox{d}$\\
 & --- & $P_\mathrm{S,min}^\mathrm{trans}  =  (0.14 \pm 0.06)~\mbox{d}$\\
\hline
\end{tabular}
\caption{Overview of collated model parameters and arising physical properties of the lens and source systems.}
\label{Tab:Allpars}
\end{center}
\end{table*}

\subsection{Source binary}

With the binary-source models and the binary-lens models having similar blend ratios (within the uncertainties), we find $(V-I,I) = (2.75\pm 0.05, 19.54 \pm 0.06)$ for the combined light of the two source stars.
A calibration using the position of the red clump then gives $(V-I,I)_0 = (0.62 \pm 0.05,17.10 \pm 0.07)$. With a luminosity offset ratio $\omega_I = 0.042 \pm 0.003$,
one finds immediately $I_0^{(1)} = 17.15 \pm 0.07$ and $I_0^{(2)} = 20.54 \pm 0.10$. Moreover, the ratio between the angular radii is given by
$\rho_\star^{(1)}/\rho_\star^{(2)} = 4.8 \pm 0.4$.
Strikingly, the model parameters suggest the ratio between the $I$-band luminosities being roughly the square of the ratio between the 
angular radii. This implies that the two stars have similar colours. Explicitly, one finds with Eq.~(\ref{eq:KVeq}), neglecting the quadratic term,
\begin{equation}
\lg \frac{\theta_\star^{(1)}}{\theta_\star^{(2)}} = 0.4895\,\left[(V-I)_0^{(1)}-(V-I)_0^{(2)}\right] - 0.2 \left(I_0^{(1)}-I_0^{(2)}\right)\,,
\end{equation}
leading to
$
(V-I)_0^{(1)}-(V-I)_0^{(2)} = 0.00 \pm 0.09
$
with the estimated values, and consequently to 
\begin{eqnarray}
(V-I,I)_0^{(1)} & = & (0.62\pm 0.14, 17.15\pm 0.07)\,,\nonumber\\
(V-I,I)_0^{(2)} & = & (0.63 \pm 0.18, 20.54\pm 0.10)\,.
\end{eqnarray}
With Eq.~(\ref{eq:KVeq}), we then obtain the individual angular source radii as\footnote{In fact, we directly find $\theta_\star^{(2)} = \theta_\star^{(1)}( \rho_\star^{(2)}/\rho_\star^{(1)})$.}
\begin{eqnarray}
\theta_\star^{(1)} & = &(1.11 \pm 0.16)~\umu\mbox{as}\,,\nonumber\\
\theta_\star^{(2)} & = &(0.23 \pm 0.04)~\umu\mbox{as}\,.
\end{eqnarray}
With a source distance of 
$D_\mathrm{S}  =  (8.5 \pm 2.0)~\mbox{kpc}$,
these correspond to physical radii
\begin{eqnarray}
R_\star^{(1)} & = &(2.0 \pm 0.6)~R_\odot\,,\nonumber\\
R_\star^{(2)} & = &(0.43 \pm 0.13)~R_\odot\,.
\end{eqnarray}
The brighter source therefore appears compatible with an F~V or G~IV star, while the fainter source appears compatible with a G~VI star.

Given that our models provide the source size parameters $\rho_\star^{(1)} = \theta_\star^{(1)}/\theta_\mathrm{E}$ and $\rho_\star^{(2)} = \theta_\star^{(2)}/\theta_\mathrm{E}$,
where
\begin{eqnarray}
\rho_\star^{\mbox{\tiny (1)}}  & = &  (7.6 \pm 0.5) \times 10^{-3}\,,\nonumber\\
\rho_\star^{\mbox{\tiny (2)}}  & =  & (1.6 \pm 0.1) \times 10^{-3}\,,
\end{eqnarray}
the angular Einstein radius is estimated to be
\begin{eqnarray}
\theta_\mathrm{E}  & =  & (0.15 \pm 0.02)~\mbox{mas}\,.
\end{eqnarray}

We find that the angular separation between the source stars is close to the sum of their radii, i.e.
\begin{eqnarray}
\lambda^\mathrm{cis} - (\rho_\star^{\mbox{\tiny (1)}} + \rho_\star^{\mbox{\tiny (2)}}) & = -0.0005 \pm 0.0010\,,\nonumber \\
\lambda^\mathrm{trans} - (\rho_\star^{\mbox{\tiny (1)}} + \rho_\star^{\mbox{\tiny (2)}}) & = 0.0015 \pm 0.0010\,,
\end{eqnarray}
with $\lambda^{\mathrm{cis}/\mathrm{trans}}$  given by Eq.~(\ref{eq:lambda2}), which suggests that the source could be a (partially) eclipsing binary,
but the two stars could also miss each other.
We find an angular separation $\lambda\,\theta_\mathrm{E}$ of
\begin{eqnarray}
\beta^\mathrm{cis}  & = & (1.3 \pm 0.2)~\mbox{$\umu$as}\,,\nonumber\\
\beta^\mathrm{trans} & = & (1.6 \pm 0.3)~\mbox{$\umu$as} 
\end{eqnarray}
and a separation perpendicular to the line of sight
$\rho_\perp = \lambda\,D_\mathrm{S}\,\theta_\mathrm{E}$ of 
\begin{eqnarray}
\rho_\perp^\mathrm{cis} & = & (0.011 \pm 0.03)~\mbox{AU}\,,\nonumber\\
\rho_\perp^\mathrm{trans} & = & (0.013 \pm 0.04)~\mbox{AU}\,.
\end{eqnarray}

From the derived stellar types (G IV, F V / G VI), we broadly estimate the masses of the source stars as
\begin{eqnarray}
M^{(1)} = (1.4 \pm 0.2)~M_\odot\,,\nonumber\\
M^{(2)} = (0.7 \pm 0.1)~M_\odot\,,
\end{eqnarray}
leading to a total mass $M_\mathrm{S} = (2.1 \pm 0.2)~M_\odot$. We therefore obtain a miminal orbital period
\begin{equation}
P_\mathrm{S,min} = 2\pi\,\sqrt{\frac{\rho^3}{8GM_\mathrm{S}}}\,,
\end{equation}
evaluating to
\begin{eqnarray}
P_\mathrm{S,min}^\mathrm{cis} & = & (0.10 \pm 0.05)~\mbox{d}\,,\nonumber\\
P_\mathrm{S,min}^\mathrm{trans} & = & (0.14 \pm 0.06)~\mbox{d}\,.
\end{eqnarray}

For the parallax parameter $\pi_\mathrm{E} = \pi_\mathrm{LS}/\theta_\mathrm{E}$, the models give
\begin{eqnarray}
\pi_\mathrm{E} & = & 0.39 \pm 0.01\,,
\end{eqnarray}
so that with Eq.~(\ref{eq:calcmass}), one obtains the mass of the lens as
\begin{eqnarray}
M  & =  & (0.046 \pm 0.007)~M_{\odot}\,,
\end{eqnarray}
compatible with a brown dwarf. Moreover, with 
\begin{eqnarray}
\pi_\mathrm{LS}  & =  & (0.057 \pm 0.009)~\mbox{mas}
\end{eqnarray}
and $\pi_\mathrm{S}$ as given by Eq.~({\ref{eq:srcparallax}), we find
\begin{eqnarray}
\pi_\mathrm{L} & = & (0.17 \pm 0.03)~\mbox{mas}\,,
\end{eqnarray}
equivalent to
\begin{eqnarray}
D_\mathrm{L}  & =  & (5.7 \pm 0.9)~\mbox{kpc}\,.
\end{eqnarray}

The Einstein radius $r_\mathrm{E} = D_\mathrm{L}\,\theta_\mathrm{E}$ therefore
becomes
\begin{eqnarray}
r_\mathrm{E} & = & (0.84 \pm 0.19)~\mbox{AU}\,,
\end{eqnarray}
and with the event time-scale
\begin{eqnarray}
t_\mathrm{E}  & =  & (309 \pm 19)~\mbox{d}\,,
\end{eqnarray}
defined as $\theta_\mathrm{E}/\mu$, we obtain the proper motion
\begin{eqnarray}
 \mu  & =  & (0.47 \pm 0.08)~\umu\mbox{as}\;\mbox{d}^{-1}\nonumber\\
  & = & (0.17 \pm 0.03)~\mbox{mas}\;\mbox{yr}^{-1}\,,
\end{eqnarray}
so that the effective perpendicular lens velocity $v_\perp = D_\mathrm{L}\,\mu$ reads
\begin{eqnarray}
v_\perp & = & (4.7 \pm 1.1)~\mbox{km}\;\mbox{s}^{-1}\,.
\end{eqnarray}

\subsection{Lens binary or source binary?}
\label{sec:lensorsource}

Not only do the acquired photometric data fail to provide sufficient evidence for distinguishing between our binary-lens and binary-source models, but moreover neither of these alternatives lead to an obviously implausible physical nature of the lens or source system, taking into account that the event time-scale of $t_\mathrm{E}$ makes the event unusual.

Unfortunately, we missed out on the opportunity to obtain multi-band photometry over the peak, but even if we had done so, the discrimination power would have been limited, given that the binary-source models are compatible with the absence of significant colour effects. However, a positive detection of colour differences in the light curve over the peak could have ruled out the binary-lens interpretation. 

While orbital motion provides further freedom for both the binary-lens and the binary-source models  \citep{Do:rotating}, a substantial difference lies in the fact that plausible orbital periods are of the order of years for the binary-lens models, but of the order of days for the binary-source models. This means that the flexibility of binary-lens models over the peak is pretty much exhausted, whereas orbital motion can significantly affect the photometric light curve for binary-source models over the peak. However, if we are not certain that the remaining residuals are not due to low-amplitude systematics, we are running the risk that further model tuning would correspond to modelling noise. Moreover, the large number of additional model parameters for fixing a small discrepancy is likely to result in severe ambiguities in an intricate parameter space.
 
Despite the caveat that the ratios between the angular source radii and the luminosities of the source stars obtained for the binary-source models might not be somewhat misestimated due to orbital motion 
being mistaken for a contribution to source size, they are remarkably consistent, while one could have easily ended up with implausible properties of the constituents of the source binary, not matching any populated regions of the colour-magnitude diagram. This provides some support for the credibility of the binary-source interpretation. It is also interesting that the binary-source model parameters suggest that the source might be a (partially) eclipsing binary. We would definitely know that the source is a binary if (partial) eclipses were found in photometric data. Without (partial) eclipses, there will not be periodicities in the light curve, given that it is well explained by a point-source point-lens model outside the peak region.



The binary-source and binary-lens models also differ in the nature of the observed blended light, despite the fact that its amount does not differ significantly, given that in both cases the model needs to match the off-peak photometric data which is not affected by binarity.  For the binary-source models, the mass of the lens star is suggestive of a brown dwarf, so that the blended light would presumably arise from another star rather than the lens itself. For the binary-lens models, this looks different.
Assuming that the lens star of $M \sim 0.4~M_\star$ is an M dwarf (M2 V), it would have approximately $M_{V,\mathrm{L}} = 10.2$ and $(V-I)_\mathrm{L} = 2.16$, i.e. $M_{I,\mathrm{L}} = 8.04$.
For the source star, we found  $I = 16.98 \pm 0.12$. The source distance $D_\mathrm{S} = (8.5 \pm 2.0)~\mbox{kpc}$ corresponds to a distance modulus
$m-M = 14.6 \pm 0.5$, so that $M_I = 2.3 \pm 0.5$. Moreover, for a lens distance  $D_\mathrm{L} = (1.7 \pm 0.3)~\mbox{kpc}$, we find
$(m-M)_\mathrm{L} = 11.1 \pm 0.3$, resulting in a relative distance modulus $\Delta(m-M) = 3.5 \pm 0.6$. 
This gives 
\begin{equation}
\Delta I = M_I - M_{I,\mathrm{L}} + \Delta(m-M) = -2.2 \pm 0.8\,,
\end{equation}
suggesting that the lens star is fainter than the source star before considering extinction.
A blend ratio
\begin{equation}
g_I = 0.41 \pm 0.13
\end{equation}
provides the constraint
\begin{equation}
(\Delta I)_0 \leq 2.5~\mbox{lg}\;g_I  = -1.0 \pm 0.3
\end{equation}
which means that the extinction needs to be
\begin{equation}
(\Delta I)_0 - \Delta I  < 1.2 \pm 0.9\,,
\end{equation}
which appears to be compatible with an average $A_I = 1.96$ towards the direction of the observed target \citep{Nataf2013}, and some extinction caused by dust between 
the observer and the lens star. Hence, the lens star is not too bright and might be the main contributor to the blended light. Observing a star compatible with
the predicted brightness of the lens star for our binary-lens models that furthermore separates from the source star at a proper motion  $\mu = (1.6 \pm 0.3)~\mbox{mas}\;\mbox{yr}^{-1}$
would give strong support to the binary-lens interpretation and constitute evidence against the binary-source interpretation, which has a much fainter lens with a much smaller proper motion of
$\mu = (0.17 \pm 0.03)~\mbox{mas}\;\mbox{yr}^{-1}$ relative to the source star.

\section{Conclusions}
\label{sec:conclusions}

The power of inferred planet population statistics from gravitational microlensing campaigns greatly increases with the ability to distinguish low-amplitude signals from the noise floor of photometric data.


Separating model parameters and subsets of data has been demonstrated to be a generic and powerful approach for characterising localised effects in photometric light curves. In particular, this allows us to build effective models of the photometric noise on data that do not contain the putative signal under investigation, and thereby enables a meaningful probabilistic assessment of the significance of such a signal under the assumption that the data for epochs not covering the signal are reasonably well understood. Hence, signals of planets that are otherwise missed become detectable.


While we laid the groundwork for a detailed assessment of the feasibility of potential alternative model interpretations of the observed data, it turned out that for the concrete case of the microlensing event OGLE-2014-BLG-1186, we can straightforwardly rule out any binary-lens alternatives to the four configurations presented. Rather than claiming that our models are the right ones because no viable alternatives have been found, an analysis of the underlying mathematical properties of potential solutions that can provide matching morphologies enabled us to restrict all viable alternatives within the adopted model framework to a small finite number of prototypes, similar to what was suggested by \citet{Liebig}, which then either turned out to lead to a match to the data that cannot be improved, or an obvious mismatch. However, we can only check the adopted model framework for plausibility and consistency, whereas it is fundamentally impossible to rule out the existence of further plausible interpretations beyond the adopted model framework, given that our knowledge will always remain limited and incomplete.

In fact, we experienced that close binary source stars pose a challenge for claiming the detection of planets by microlensing in events where the source trajectory passes close to the central caustic near the lens star hosting the planet \citep{GS98}. This is different from the ambiguity between binary-lens and binary-source interpretations discussed by \citet{Gaudi:source}, which relates to planetary signatures arising from approaching planetary caustics. We note that while in this case a small luminosity offset ratio $\omega \la 10^{-2}$ is required, such a restriction does not hold for the type of ambiguity that we encountered. Close binary-source models come with a large number of degrees of freedom, involving two source size parameters as well as parameters that describe the orbital motion, which is likely to significantly affect the light curve over the peak due to orbital periods of the order of days. Binary-source interpretations must not be discarded prematurely on the basis of comparing binary-lens models with static binary-point-source models.

In order to resolve such ambiguities, uninterrupted high-cadence multi-band photometric observations over the peak would be useful. Simultaneous or quasi-simultaneous observations with different bandpass filters can not only measure chromaticity, but moreover increase the statistical significance of signals due to correlations \citep{DoHi,Street:Luhman}.
Source binarity could also be indicated by means of spectra taken at either peak. Moreover, the astrometric signature of binary-lens and binary-source events with similar photometric signature is substantially different \citep{Han:astrometry1,Han:astrometry2,Han:astrometry3}. \citet{Calchi:resolve} also recently discussed a case of binary-lens vs binary-source ambiguity for an event that shows an anomaly signature both from ground- and space-based photometric observations, providing complementary information due to the different lines of sight.

We finally note that gravitational microlensing events such as OGLE-2014-BLG-1186 for which both the source size parameter $\rho_\star$ and the parallax parameter $\pi_\mathrm{E}$ can be reliably measured provide a valuable sample for testing models that describe the mass distribution and kinematics of the Milky Way, given that with an estimate of the angular size $\theta_\star$ of the source star from a colour-magnitude diagram, one directly obtains the mass $M$ of the lens system (as well as the individual masses of its constituents), its distance $D_\mathrm{L}$ from the observer, as well as the effective proper motion $\mu$.

\section*{Acknowledgements}
This publication was made possible by grants NPRP-X-019-1-006 and NPRP-09-476-1-78 from the Qatar National Research Fund (a member of Qatar Foundation).
GD acknowledges Regione Campania for support from POR-FSE Campania 2014-2020.
M.P.G.H. acknowledges support from the Villum Foundation. Work by C.H. was supported by the grant (2017R1A4A1015178) of
National Research Foundation of Korea.
This work makes use of observations from the LCOGT network, which includes three ``SUPA\-scopes" owned by the University of St Andrews. The RoboNet programme is an LCOGT Key Project using time allocations from the University of St Andrews, LCOGT and the University of Heidelberg together with time on the Liverpool Telescope through the Science and Technology Facilities Council (STFC), UK. This research has made use of the LCOGT Archive, which is operated by the California Institute of Technology, under contract with the Las Cumbres Observatory.
OGLE Team thanks Profs.\ M.~Kubiak and G.~Pietrzy{\'n}ski, former
members of the OGLE team, for their contribution to the collection of
the OGLE photometric data over the past years.
The OGLE project has received funding from the National Science Centre,
Poland, grant MAESTRO 2014/14/A/ST9/00121 to AU. L.M. acknowledges support from the Italian Minister of Instruction,
University and Research (MIUR) through FFABR 2017 fund. L.M.
acknowledges support from the University of Rome Tor Vergata through
``Mission: Sustainability 2016'' fund. K.H. acknowledges support from STFC grant ST/R000824/1.


\bibliographystyle{mnras}
\bibliography{OB141186}

\bsp	
\label{lastpage}
\end{document}